\documentclass[12pt,preprint]{aastex}
\usepackage{graphicx}

\begin{document}
\title{GAMMA-RAY BURST JET BREAKS REVISITED}
\author{Xiang-Gao Wang\altaffilmark{1,2,3}, Bing Zhang\altaffilmark{2}, En-Wei Liang\altaffilmark{1,3},Rui-Jing Lu\altaffilmark{1,3},Jing Li\altaffilmark{1,3},Long Li\altaffilmark{1,3}}

\altaffiltext{1}{GXU-NAOC Center for Astrophysics and Space Sciences, Department of Physics, Guangxi University, Nanning 530004, China; wangxg@gxu.edu.cn; lew@gxu.edu.cn}
\altaffiltext{2}{Department of Physics and Astronomy, University of Nevada Las Vegas, NV 89154, USA; zhang@physics.unlv.edu}
\altaffiltext{3}{Guangxi Key Laboratory for the Relativistic Astrophysics, Nanning 530004, China}

\begin{abstract}
Gamma-ray Burst (GRB) collimation has been inferred with the observations of achromatic steepening in GRB light curves, known as jet breaks.
Identifying a jet break from a GRB afterglow lightcurve allows a measurement of the jet opening angle and true energetics of GRBs.
In this paper, we reinvestigate this problem using a large sample of GRBs that have an optical jet break which is consistent with being achromatic in the X-ray band. Our sample includes 99 GRBs from February 1997 to March 2015 that have optical and, for {\em Swift} GRBs, X-ray lightcurves that are consistent with the jet break interpretation.
Out of 99 GRBs we have studied, 55 GRBs are found to have temporal and spectral behaviors both before and after the break consistent with the theoretical predictions of the jet break models, respectively. These include 53 long/soft (Type II) and 2 short/hard (Type I) GRBs. Only 1 GRB is classified as the candidate of a jet break with energy injection. Another 41 and 3 GRBs are classified as the candidates with the lower and upper limits of the jet break time, respectively.
Most jet breaks occur at {\bf 90 ks}, with a typical opening angle $\theta_j = (2.5 \pm 1.0)^{\rm o}$. This gives a typical beaming correction factor $f_b^{-1} \sim 1000$ for Type II GRBs, suggesting an even higher total GRB event rate density in the universe. Both isotropic and jet-corrected energies have a wide span in their distributions: $\log (E_{\rm \gamma,iso} / {\rm erg}) = 53.11$ with $\sigma = 0.84$; $\log (E_{\rm K,iso} / {\rm erg}) = 54.82$ with $\sigma = 0.56$; $\log (E_{\rm \gamma} / {\rm erg}) = 49.54$ with $\sigma = 1.29$; and $\log (E_{\rm K} / {\rm erg}) = 51.33$ with $\sigma = 0.58$. We also investigate several empirical correlations (Amati, Frail, Ghirlanda and Liang-Zhang) previously discussed in the literature. We find that in general most of these relations are less tight than before. The existence of early jet breaks and hence small opening angle jets, which were detected in the {\em Swfit era}, is most likely the source of scatter. If one limits the sample to jet breaks later than $10^4$ s, the Liang-Zhang relation remains tight and the Ghirlanda relation still exists. These relations are derived from Type II GRBs, and Type I GRBs usually deviate from them.

\end{abstract}

\section{Introduction\label{sec:intro}}

Gamma ray bursts (GRBs) are the most luminous phenomena observed in the universe, with an isotropic $\gamma$-ray energy up to
$E_{\rm \gamma,iso} \sim 10^{55}$ erg \citep[][]{kumarzhang15}.
They signify the birth of a stellar-mass black hole or a rapidly rotating magnetized neutron star during the core collapse of
massive stars (Type II GRBs) or mergers of compact objects (Type I GRBs) \citep[e.g.,][]{woosley93,paczynski98,woosleyb06,gehrels05,berger14,zhang06b,zhang07,zhangz09}. Phenomenologically, GRBs are classified based on the burst durations  \citep[][]{kouveliotou93}, with long GRBs (LGRBs, $T_{90} > 2$ s) mostly correspond to Type II and short GRBs (SGRBs, $T_{90} < 2$ s) mostly correspond to Type I. An important result from the pre-Swift era observations is that Type II GRBs are highly collimated with a typical opening angle of $\sim 5^{\rm o}$ \citep[e.g.][]{frail01,bloom03,berger03}. Some empirical correlations, several involving jet opening angles, have been discussed in the literaure
\citep[e.g.][]{frail01,amati02,ghirlanda04a,liangz05,wangfy11}.

Theoretically, GRB afterglow is essentially independent of the progenitor and central engine, and invokes the interaction between the fireball that produced the GRB and an circumburst medium (CBM) with a density profile generally described as $n(r)\propto r^{-k}$. A generic synchrotron external shock model has been well established to interpret the broad-band afterglow data
 \citep[e.g.,][]{meszarosrees97,sari98,chevalier00,gao13}.
Our recent study \citep{wang15} suggests that the simplest external forward shock models can account for
the multi-wavelength afterglow data of at least half of the GRBs. When more advanced modeling (e.g., long-lasting reverse shock, structured jets) is invoked, up to $90 \%$ of the afterglows may be interpreted within the framework of the external shock models.

An achromatic, steepening temporal break observed in some afterglow lightcurves suggests that the GRB outflows are collimated. In the fireball external shock model, the burst ejecta moves with a relativistic speed and is assumed to form a conical jet with half opening angle $\theta_j$. As the burst ejecta are decelerated by the ambient, the relativistic beaming angle $1/\Gamma$ continues to increase with time. When$1/\Gamma > \theta_j$ is satisfied, a steepening break in the afterglow lightcurve (known as the jet break) is predicted. This is mostly due to an edge effect, which is purely geometric: the $1/\Gamma$ cone is no longer filled with emission beyond the jet break time (when $1/\Gamma > \theta_j$). This gives a reduction of flux by $\theta_{j}^{2}/(1/\Gamma)^{2}=\Gamma^{2}\theta_{j}^{2}$. It has been suggested that a maximized sideways expansion effect may further steepen the light curve \citep{rhoads99,sari99}. This theory suggests that sound waves in the jet would cross the jet in the transverse direction when $1/\Gamma > \theta_j$. The cross section of the jet would increase with time, leading to an exponential deceleration of the jet. Later numerical simulations suggested that the sideways expansion effect is not significant, but the post-jet-break decay index could be similar to that predicted in the sideways expansion models \citep[e.g.,][]{zhangw09, granotp12}.

Extensive studies on the jet break phenomenon have been carried out. In the pre-\emph{Swift} era, several cases of jet break have been observed in the optical band at several days after the GRB trigger \citep[e.g.,][]{rhoads99,sari99,huang00,halpern00,frail01,bloom01,jaunsen01,wei02,wu04,gaow05,panaitescu05,starling05,yonetoku05,zeh06,gorosabel06,liang08}. However, the achromatic behavior of the break, a prediction of the jet model, could not be confirmed with the optical data only. A rich database of broadband afterglow lightcurves are accumulating after the \emph{Swift} satellite was launched. Many investigations to search for and to study the statistical properties of jet breaks have been carried out based on the XRT data \citep[e.g.,][]{burrows06,grupe06,wangxy06,burrowsr07,daix07,jin07,nava07,panaitescu07,willingale07,liang08,kocevskib08,depasquale09,evans09,kamble09,racusin09,urata09,gaoy10,tanvir10,zheng10,guelbenzu11,fong12,fong14}, the optical data \citep[e.g.,][]{daix07,daix08,panaitescu07,kruhler09,tanvir10,afonso11,filgas11,guelbenzu11,fong14},
and the radio data \citep[e.g.,][]{sheth03,vanderhorst05,fong14}. The X-ray lightcurves of some GRBs did not show a clear jet break at very late times \citep{grupe06,grupe07}. Some argued that a jet break may be hidden in the low signal-to-noise ratio (S/N) lightcurves \citep[e.g.,][]{shao07,sato07,curran08}. More late time optical observations are needed to reveal late jet breaks and to constrain GRB collimation and energetics \citep{zhang11}.
Indeed,  X-ray observations with {\em Chandra} X-Ray Telescope have led to detections of some late jet breaks, which allowed a study of the off-axis effect of GRB jets \citep{zhangbb15}.
 Among the GRBs with optical afterglow detections, only 1/3 were also detected in the radio band. However, there is a lack of GRBs with high quality lightcurves in the radio band to conduct jet break searches.

Based on a rich database of broad-band afterglow up to 2015, this paper aims at a systematic analysis of the jet break features in GRBs.
The sample selection and data analysis are described in \S2. We use the {\em closure relations} of the external forward shock model to select the jet break candidates, and the results are presented in \S3. A statistical analysis of energetics and luminosity correlations of the jet break sample is presented in \S4.
Our results are summarized in \S5 with some discussion.

We characterize the dependence of the afterglow flux on time and frequency as $F(t,\nu)\propto t^{-\alpha} \nu^{-\beta}$, where $\alpha$ is the temporal decay index, and $\beta$ is the spectral index. We use the $\rm \Lambda CDM$ model with cosmological parameters of $\rm \Lambda_{M}=0.27$, $\rm \Omega_{\Lambda}=0.27$, and $H_{0}=71$ km s$^{-1}$ Mpc$^{-1}$ to calculate the energetics of the GRBs.

\section{Data Sample\label{sec:data}}
To systematically investigate jet breaks, we collect all the optical afterglow data from the first GRB optical afterglow detected from February 1997 to March 2015. This includes 17 pre-\emph{Swift} GRBs from \cite{liang08} that have been studied extensively. A sample of $\sim$ 260 optical lightcurves are compiled from the published papers \citep[e.g.][]{kann10,kann11} or GCN Circulars after the \emph{Swift} launch. The UVOT data are not included in our sample. For the \emph{Swift} data, we obtain a sample of 85 well-observed GRBs with light curves in both X-ray and optical bands and the constrained spectral indices. Out of these 85 bursts, 82 GRBs have been graded as the achromatic sample consistent with the external shock model (i.e. Gold  and Silver samples defined in \citealt{wang15}). We thus select 82 GRBs from the achromatic sample for the purpose of this work. As a result, altogether 99 GRBs are included in our final sample.

 {Most observations were carried out in the R-band. For those observations carried out in other bands, we correct them to the R band with the optical spectral indices ($\beta_{\rm O}$, with the convention $F_\nu \propto \nu^{-\beta_{\rm O}}$) collected from the literature assuming that there is no spectral evolution.
The correction due to Galactic extinction is taken into account using the reddening map presented by \cite{schlegel98}.
Because of large uncertainties, we do not make corrections to the extinction in the GRB host galaxies.

The optical light curves are usually composed of one or more power-law segments along with some humps, flares or rebrightening features \citep[e.g.][]{lil12,liang13,wang13,wang15}. To decompose the rich temporal features of GRB light curves, we fit the light curves with a model of multiple components. Similar to \cite{wang15}, we decompose the lightcurves into several basic components, i.e. a single power-law (SPL) function

\begin{equation}
 F_1 = F_{01} t^{-\alpha},
\end{equation}
or a smooth broken power-law (BPL) function
\begin{equation}
F_2 =  F_{02} \left[\left(\frac{t}{t_{\rm
b}}\right)^{\alpha_1\omega} +\left(\frac{t}{t_{\rm
b}}\right)^{\alpha_2\omega}\right]^{-1/\omega}, \label{F2}
\end{equation}
where $\alpha$, $\alpha_1$, $\alpha_2$ are the temporal slopes,
$t_{\rm b}$ is the break time, and $\omega$ measures the sharpness
of the break, or a smooth triple-power-law (TPL) function that catches the canonical shape of X-ray lightcurves \citep{zhang06,nousek06}, i.e.
\begin{equation}\label{STPL}
F_3=(F_{2}^{-\omega_2}+F_4^{-\omega_2})^{-1/\omega_2}
\end{equation}
where $\omega_2$ is the sharpness factor of the second break at
$t_{b,2}$, and
\begin{equation}\label{PL}
F_4=F_2(t_{\rm b,2})\left(\frac{t}{t_{b,2}}\right)^{-\alpha_3}.
\end{equation}

We perform best fits to the data using the subroutine
MPFIT \cite[][]{markwardt09}. The sharpness parameter $\omega$ and $\omega_2$ are usually adopted as 3 or 1 in our
fitting. A minimum number of components (SPL, BPL, or TPL)
are introduced initially based on eye inspection.

If the reduced $\chi^2$ is much larger than 1, we then continue to add more components
into the fitting, until the reduced $\chi^2$ becomes close to 1 (usually less than 1.5). We'd like to stress that one may not solely based on $\chi^2$ to evaluate whether a jet break is robust.
This is because some GRBs (e.g., GRB 050730, 060729, 090926A) show erratic fluctuations in the lightcurves
with small error bars. The reduced $\chi^2$ values of these bursts are much larger than 1.  Inspecting their light curves, the large $\chi^2$ are caused by the complicated features in the light curves (such as small flares and fluctuations). However, the PL and BPL fits in any case catch the general features of these lightcurves.  Even though adding more comments can reach better reduced $\chi^2$, we do not add them since we are not interested in the fine-details of the lightcurves. The $\chi^2$ values of these fits remain much greater than 1.
Furthermore, to avoid the additional features (e.g., steep decay phase, flares, re-brightening features) to affect the fits, we just perform the best fits in the time interval around the jet break. In some cases, the reduced $\chi^2$ values for the lightcurve fittings are much smaller than 1. It means that some model parameters are poorly constrained. For these cases, we fix some parameters and redo the fits until the reduced $\chi^2$ becomes close to 1. According to the MPFIT documentation, the error estimates produced by MPFITFUN/EXPR would not be correct if the data points with large errors are not be properly weighted. In this case, we set the ``Error'' term to unity and proceed with the fit\footnote{http://www.physics.wisc.edu/$\sim$craigm/idl/mpfittut.html.}. We call the PERROR routine in the MPFITFUN to obtain the parameter errors, and use the 2$\sigma$ parameter errors in our analysis.

\section{Selection criteria and Jet Break Candidates}
\subsection{Jet break light curves}

After the launch of the \emph{Swift} satellite in 2004 \citep {gehrels04}, a rich data base of the light curves have been collected, which allowed a systematic analysis of
the emission components of the broadband light curves, especially for the X-ray lightcurves \citep[e.g.][]{barthelmy05,fan05,tagliaferri05,zhang06,nousek06,liang07, liang08,racusin09,margutti10}
and the optical lightcurves \citep[e.g.][]{liang06,nardini06,kann06,kann10,kann11,panaitescu08,panaitescu11,lil12}.
``Synthetic'' lightcurves of X-ray and optical emission have been summarized by \cite{zhang06} and \cite{lil12}.
In both bands, one prominent feature in the late afterglow phase is the existence of a jet break feature. In principle, one can have two types of jet breaks \citep[e.g.][]{wang15}:

 \begin{itemize}
   \item Standard jet break:
This corresponds to the transition from the normal decay phase (standard afterglow component) to the post-jet-break phase in the canonical
lightcurve. Lightcurves of such a category are caused by an edge-effect or with a contribution of sideways expansion. The post-break decay index is required to be steeper than 1.5 for this model. The post-break index can be as steep as the electron energy index $p$ as predicted by the sideways expansion models \citep{sari99}.
   \item Jet break with energy injection:
This corresponds to the case of an extended energy injection phase which extends to a duration longer than the jet break time. As a result, the jet break is expressed in terms of a shallow decay phase followed by a steeper decay phase with the break consistent with being due to a jet edge effect. Both before and after the break, the afterglow can be delineated by an afterglow model with a continuous energy injection defined by a long-lasting central engine activity history $L(t)=L_{0}(\frac{t}{t_{0}})^{-q}$ (Zhang \& M\'esz\'aros 2001), where $q$ is the energy injection parameter. In principle, energy injection can be interpreted as either a long-lasting central engine
\citep{dai98,zhangmeszaros01} or a Lorentz-factor-stratified ejecta \citep{rees98,sari00,uhm12}. These two scenarios are equivalent with each other, and can be both delineated with the parameter $q$ \citep{zhang06}.

 \end{itemize}

\subsection{Selection criteria}
We make use of the standard synchrotron external shock models of GRB afterglow to select the jet
 break sample. The criteria are the relationship between the temporal index $\alpha$ and spectral
 index $\beta$ (with the convention $F_{\nu} \propto t^{-\alpha} \nu^{-\beta}$) as predicted by
 various external shock models, known as the ``closure relations
 '' \citep[e.g.,][]{zhangmeszaros04,zhang06,gao13}. The indices $\alpha$ and $\beta$ can be
  directly measured from the observational data. The predictions of the $\alpha-\beta$ relation
   depend on the sub-models (ISM vs. wind, adiabatic vs. radiative, whether or not there is energy
    injection, etc.), dynamical regimes (reverse shock crossing phase, self-similar deceleration
    phase, post-jet-break phase, Newtonian phase, etc.), and spectral regimes (different orders
    among the observed frequency ($\nu$) and several characteristic frequencies (minimum injection
    frequency $\nu_m$, cooling frequency $\nu_c$, and self-absorption frequency $\nu_a$). More
    details can be found in the comprehensive review by \cite{gao13}. Generally, the optical band
    is in either the spectral regime  $\nu>\nu_{c}$ (Regime I, $\beta=p/2$) or $\nu_{m}<\nu<\nu_{c}$ (Regime II, $\beta=(p-1)/2$) in the simplest analytical model \citep{sari98}.
    Due to the smoothness of the spectral breaks, the transition between the two regimes (regime I - II, $(p-1)/2 < \beta < p/2$) may take several orders of magnitude in observer time. This period may be defined as a ``grey zone'' \citep{zhang06,uhm14a}, during which the $\alpha-\beta$ relation does not need to strictly satisfy the Regime I and Regime II regimes. The parameter space between the two closure relation lines is allowed by the theory. Data points falling into this grey zone should be regarded as consistent with the model.

We employ the $\alpha-\beta$ closure relations (Table \ref{Tab:alpha-beta}) for the ISM ($k=0$) or wind ($k=2$) medium models and with or without energy injection. For a steepening break, we require that the same model applies to both pre- and post-break phase, with the post-break decay defined either by the edge effect of sideways expansion effect. We have assumed $\nu_a< {\rm min}(\nu_m, \nu_c)$ and $\nu_{\rm O} > \nu_{m}$ ($\nu_{O}$ is the frequency of optical band), which is usually satisfied for optical afterglow emission for typical GRB parameters. A GRB to be included in our jet break sample needs to satisfy the following criteria:

\begin{itemize}
  \item The optical lightcurves should satisfy closure relations of the same circumburst medium type (ISM or wind) in both pre- and post-break temporal segments, and the inferred electron spectral index $p$ from both pre- and post-break segments should be consistent with each other within error.
  \item For a jet break without energy injection, the light curves should satisfy the closure relations for the constant-energy ISM or wind models before the break, and the corresponding jet model for either edge or sideways expansion effect after the break.
  \item For a jet break with energy injection, the light curves should satisfy the closure relations for the energy-injection ISM or wind models in both pre- and post-break phases, with the energy injection $q$ parameter consistent with each other within error.
  \item The X-ray afterglow lightcurves of these bursts are found to be consistent with the same jet break model, as have been studied in detail in \cite{wang15}.
   We only plot X-ray lightcurves in Figure \ref{jetgrade}-\ref{jetupper} for a self-consistency check, without repeating the closure relation analysis for X-rays.
\end{itemize}

 For those GRBs that cannot be identified as jet break candidates with the $\alpha-\beta$ closure relation, we classify them as either lower limit or upper limit candidates. Some GRBs satisfy the $\alpha-\beta$ closure relations of pre-jet break phase, and no break is observed at the last observational data point. These bursts are included in the lower limit jet break sample. Some other GRBs have their temporal slopes steeper than the normal decay and satisfy the $\alpha-\beta$ closure relations of post-jet break phase. However, no jet break is identified at the first point of this lightcurve segment (there might be complicated components before that). We classify these as the upper limit jet break sample.

\subsection{Jet break candidates}
We find that 55 out of 99 GRBs can satisfy the jet break criteria. We then characterize these GRBs as the jet break candidates (as shown in Table \ref{table:sample} and Figure \ref{jetgrade}).
Figure \ref{Jetclosure} shows the $\alpha-\beta$ values measured from the lightcurves as compared against the closure relations. Another 41 and 3 GRBs are classified as the candidates with the lower and upper limits of the jet break time, respectively (as shown in Table \ref{table:lowsample} - \ref{table:upersample} and Figure \ref{jetlower} - \ref{jetupper}).

Among the 55 jet break GRB candidates, 53 are Type II (long) GRBs and 2 are Type I (short) GRB (GRB 051221A, and 130603B). Three GRBs (GRB 080319B, 080413B, 090426)
have two jet breaks and are consistent with the two-component jet model \citep[e.g.][]{racusin08,guelbenzu11,filgas11}. Only one GRB (GRB 030723) shows jet break with energy injection.

There are some GRBs that are consistent with more than one closure relation given their error bars. 53/55 and 40/55 GRBs can be consistent with ISM and wind model, respectively. The ISM model applies to more bursts that the wind model, which is consistent with the previous results \citep[e.g.,][]{panaitescu02,yost03,zhang06,schulze11,wang15}. {\bf 38/55} GRBs are located in the grey zone between regime I and regime II. The median electron spectral index $p$ of the jet break GRBs is $p=2.39\pm0.48$ (Figure \ref{pvalue}), which is very consistent with the previous studies \citep[e.g.,][]{achterberg01,ellison02,shen06,liang07,liang08,curran10,wang15}. The jet break time distribution can be roughly fit by a Gaussian function, with a typical value $ t_{b}=90.06\pm84.36$ ks (Figure \ref{breaktime}). The early jet break for the two-component jets has a distribution of $t_{b}=0.2\sim2$ ks (Figure \ref{breaktime}).

\section{Jet Angle Distribution and GRB Energetics}
With the jet break time, one can calculate the half opening angle of the GRB jet, i.e.
\begin{equation}\label{jetISM}
%\begin{split}
\theta_{j}=0.07 ~{\rm rad}~\left(\frac{t_{b}}{1\ \rm
day}\right)^{3/8} \left(\frac{1+z}{2}\right)^{-3/8}
\left(\frac{E_{\rm K,iso}}{10^{53}\ \rm ergs}\right)^{-1/8}
%(\frac{\hat\eta_{\rm \gamma}}{0.2})^{1/8}
\left(\frac{n}{0.1\ \rm cm^{-3}}\right)^{1/8}
%\end{split}
\end{equation}
for a constant density ISM medium \citep{rhoads99,sari99,frail01}, and
\begin{equation}\label{jetISM}
%\begin{split}
\theta_{j}=0.1 ~{\rm rad}~\left(\frac{t_{b}}{1\ \rm
day}\right)^{1/4} \left(\frac{1+z}{2}\right)^{-1/4}
\left(\frac{E_{\rm K,iso}}{10^{53}\ \rm ergs}\right)^{-1/4}
%(\frac{\hat\eta_{\rm \gamma}}{0.2})^{1/8}
\left(\frac{A_{\ast}}{1\ \rm cm^{-3}}\right)^{1/4}
%\end{split}
\end{equation}
for a wind medium \citep{chevalier00,bloom03}. Notice that $\theta_j$ depends on the isotropic kinetic energy of the blastwave, $E_{\rm K,iso}$, rather than the isotropic $\gamma$-ray energy, $E_{\rm \gamma,iso}$. In some works, $\theta_j$ is expressed in terms of $E_{\rm \gamma,iso}$ through an efficiency parameter, which is assumed for a typical value. In order to more precisely estimate $\theta_j$, in this work we infer $E_{\rm K,iso}$ directly from the data.

%\subsection{Energetics and jet opening angle}
The isotropic $\gamma$-ray energy $E_{\rm \gamma,iso}$ of a GRB is calculated as
\begin{equation}\label{Eiso}
E_{\rm \gamma,iso}=\frac{4 \pi D^{2}_{\rm L}S_{\gamma}k}{1+z},
\end{equation}
where $S_{\rm \gamma}$ is the gamma-ray fluence in the instrument band, $D_{\rm L}$ is the luminosity distance of the source at redshift $z$, and the parameter $k$ is a factor to correct the observed $\gamma$-ray energy in a given band pass to a broad band (e.g., $1-10^4$ keV in the rest frame) with the observed GRB spectra \citep{bloom01}. We assume a Band function shape of the GRB spectrum \citep{band93} and use the fitted spectral parameters to do the extrapolation.
The isotropic kinetic energy $E_{\rm K,iso}$ is calculated based on the standard afterglow models \citep[][]{zhang07,gao13,wang15}. More specifically we use Equations (13-41) of \cite{wang15} to calculate $E_{\rm K,iso}$ based on the medium type, spectral regime, and the value of $p$ ($>2$ or $<2$) inferred from the data. Since the optical band is typically in the regime $\nu_m < \nu < \nu_c$, we adopt Eqs. (20), (25), (34) and (39) of \cite{wang15} to perform the calculations. If the GRB is consistent with more than one closure relation, we choose the ISM model for the calculation of jet angles and energetics. The model parameters are taken as typical values: $\epsilon_e = 0.1$, $\epsilon_B=10^{-5}$, $n=1$ or $A_\ast=1$, and $Y=1$ \citep{wang15}.

With inferred $\theta_j$, one can derive the geometrically corrected $\gamma$-ray energy
\begin{equation}\label{Eiso}
E_{\gamma}=f_b E_{\rm \gamma,iso} = (1-\cos\theta_{j})E_{\rm \gamma,iso},
\end{equation}
and kinetic energy
\begin{equation}\label{Ek}
E_{\rm K}=f_b E_{\rm K,iso}=(1-\cos\theta_{j})E_{\rm K,iso},
\end{equation}
through the beaming correction factor defined as
\begin{equation}
 f_b = (1-\cos\theta_j).
\end{equation}

For all the bursts in this sample,
 we collect the measured $z$ and $E_{p}$ from the literature (as listed in Tables \ref{table:sample}-\ref{table:upersample}). We derive jet angles and energetics of GRBs using the available $z$ and $E_{p}$. Within the jet break, lower limit, and upper limit samples, 50, 38 and 3 GRBs have  measured $z$ from the literature, respectively. Since  \emph{Swift}-BAT has a narrow energy bandpass, the best fit spectrum is generally a power-law. The $E_{p}$ values derived from the \emph{Swift}-BAT data alone are limited, since $E_{p}$ may be measured only when it falls approximately between 15 and 150 keV. In our sample, we only adopted the $E_{p}$ derived from \emph{Swift}-BAT for GRB 050416A \citep{Sakamoto05,Nava08,Nava12}, 050502A \citep{Nava08,Nava12} and 060206 \citep{Cenko06,Nava08,Nava12}. If there are multiple instruments/missions (e.g., \emph{CGRO}-BATSE, \emph{Fermi}-GBM, \emph{HETE-2}-FREGATE,\emph{INTEGRAL}-SPI/IBIS, \emph{Konus}-Wind, \emph{RHESSI}, \emph{Suzaku}-WAM) that detected a same burst, the instruments/missions with the best fit Band function spectrum are adopted as the source of $E_{p}$ and $S_{\rm \gamma}$ (also listed in Tables \ref{table:sample}-\ref{table:upersample}).

Our data suggest $\theta_{j}=(2.5\pm1.0)^{\rm o}$ (as shown in Figure \ref{jetangle}), and the typical  jet beaming factor $f_{b}^{-1} \sim 1000$.
Notice that the typical jet half opening angle is smaller than the value ($\sim 5^{\rm o}$) inferred before \citep[e.g.][]{frail01,bloom03}. The main reason is that recent studies have suggested a relatively small value of $\epsilon_B$,  of the order of $10^{-5}-10^{-7}$ \citep[e.g.][]{kumar09,santana14,wang15,gao15,zhangbb15}, which suggests a relatively larger $E_{\rm K,iso}$ than estimated before assuming a fixed GRB efficiency\footnote{ \cite{zhangbb15} focused on GRBs with long-lasting X-ray light curves that require very late-time \emph{Chandra} observations, and therefore selected against bursts with small jet opening angles.}. The smaller half jet opening angle also suggests a smaller beaming correction factor  $f_{b}^{-1}$, suggesting a somewhat higher event rate density of GRB progenitor systems.

Figure \ref{efficiency} shows the radiative efficiency calculated at $t_{\rm b}$. Most GRBs show a small radiative efficiency with less than 10\%. With a smaller value of $\epsilon_B \sim 10^{-5}$, the derived $E_{\rm K,iso}$ values are systematically larger, so that the derived efficiency values are somewhat smaller than the values derived in previous work \citep[e.g.][]{zhang07,racusin11}. Nonetheless, the efficiency derived at the end of prompt emission (beginning of the shallow decay case) remain tens of percent for most GRBs, which demands a contrived setup for the internal shock models \citep[e.g.,][]{beloborodov00,kobayashisari01,zhangyan11,deng15,gaozhang15}.

Table \ref{table:result} and Figure \ref{Eiso} displays the inferred mean value of various energies. For Type II GRBs, one has $\log (E_{\rm \gamma,iso} \rm /erg)=53.11\pm0.84$,  $\log (E_{\rm K,iso} \rm /erg)=54.82\pm0.56$, $\log (E_{\rm \gamma} \rm /erg)=49.54\pm1.29$, and $\log (E_{\rm K} \rm /erg)=51.33\pm0.58$. For Type I GRBs, the energies are typically lower, i.e. $\log (E_{\rm \gamma,iso} \rm /erg)\sim51$, $\log (E_{\rm K,iso} \rm /erg)\sim53$, $\log (E_{\rm \gamma} \rm /erg)\sim49$ and $\log (E_{\rm K} \rm /erg)\sim (50-51)$.

The results of the lower limit and upper limit samples are presented in Table \ref{table:result1} - \ref{table:result2} and Figure \ref{Eiso} - \ref{efficiency}. For the lower limit sample, the epoch of last data point is used to calculate the lower limit of the jet opening angle. For the upper limit sample, the first data point that marks the transition to the post-jet break phase (from a complicated component, e.g. flare) is used to set the upper limit of the jet opening angle (as shown in Figure \ref{jetupper}).

\section{Luminosity Correlations}

In order to investigate several GRB luminosity correlations claimed in previous papers, we compile the necessary parameters of the GRBs in our sample in Table \ref{table:sample}. Their derived parameters are presented in Table \ref{table:result}.

We test four correlations, i.e. the $E_{\rm p,z}-E_{\rm \gamma,iso}$ (\emph{Amati}) \citep{amati02,amati06}, $E_{\rm p,z}-E_{\rm \gamma}$ (\emph{Ghirlanda}) \citep{ghirlanda04a}, $E_{\rm p,z}-E_{\rm \gamma,iso}-t_{\rm b,z}$ (\emph{Liang-Zhang}) \citep{liangz05}, and $E_{\rm \gamma,iso}-f_{\rm b}$ (\emph{Frail})\citep{frail01} relations. We write the correlations of $E_{\rm p,z}-E_{\rm \gamma,iso}$ (\emph{Amati}) or $E_{\rm p,z}-E_{\rm \gamma}$ (\emph{Ghirlanda}) in the form of
\begin{eqnarray}
\frac{E_{\rm p,z}}{\rm 100 keV}= C \left(\frac{E_{\rm \gamma,iso}(E_{\rm \gamma})}{10^{52} \rm erg} \right)^a,
\label{fp5}
\end{eqnarray}
and the $E_{\rm p,z}-E_{\rm \gamma,iso}-t_{\rm b,z}$ correlation (\emph{Liang-Zhang}) in the form of
\begin{eqnarray}
\frac{E_{\rm p,z}}{\rm 100 keV}= C \left(\frac{E_{\rm \gamma,iso}}{10^{52} \rm erg} \right)^a \left(\frac{t_{\rm b,z}}{\rm day}\right)^b,
\label{fp5}
\end{eqnarray}
where $E_{\rm p,z}$ and $t_{\rm b,z}$ are the peak energy and jet break time in the rest-frame with $E_{\rm p,z}=(1+z)E_{p}$ and $t_{\rm b,z}=t_{b}/(1+z)$, respectively; $C$, $a$ and $b$ are the correlation indices. When conducting both single and multiple variable regression analyses to look for correlations, one may find discrepancy of the dependencies among variables by specifying different dependent variables for a given data set, especially when the data have large error bars or large scatter. To avoid specifying independent and dependent variables in the best linear fits, in principle the algorithm of the bisector of two ordinary least-squares may be adopted. We therefore use the Spearman correlation analysis to search for correlations among these parameters, and adopt the stepwise regression analysis method to perform a multiple regression analysis for multiple parameters \citep[][]{liang15}.

Type II GRBs in our sample can be fit with a tight \emph{Amati} relation, with $\frac{E_{\rm p,z}}{\rm 100 keV}\simeq(0.63\pm0.31)(\frac{E_{\rm \gamma,iso}}{10^{52} \rm erg})^{(0.69\pm0.07)}$ (Talbe \ref{table:relations} and Figure \ref{Amatirelation}). The previous studies suggested that $C\sim(0.8-1)$ and $m\sim(0.4-0.6)$ \citep{amati02,amati06,sakamoto06,frontera12}. Our slope is slightly steeper. One GRB (GRB 120729A) deviates from the relation. Type I GRBs (GRB 051221A and 1306033B) generally deviate from this relation with a relative low energy and high $E_{\rm p,z}$.

The $E_{\rm p,z}-E_{\rm \gamma}$ (\emph{Ghirlanda}) (Table \ref{table:relations} and Figure \ref{Ghirlandarelation}) has large scatter, with $\frac{E_{\rm p,z}}{\rm 100 keV}\simeq(7.9\pm4.8)(\frac{E_{\rm \gamma}}{10^{51} \rm erg})^{(0.44\pm0.17)}$. Even though the results are generally consistent with \cite{ghirlanda04a}, $\frac{E_{\rm p,z}}{\rm 100 keV}\simeq8(\frac{E_{\rm \gamma}}{10^{51} \rm erg})^{0.7}$, the large dispersion in the normalization parameter $C$ suggests that it is not as tight as claimed before. We separate our sample to GRBs with jet break time earlier and later than $10^4$ s, and found that they are well separated in the $E_{\rm p,z}-E_{\rm \gamma}$ plane. Limiting the sample to the late jet break ones, one gets a tighter \emph{Ghirlanda}-relation.

Similar to \emph{Ghirlanda} relation,  the early time jet break GRBs also introduce scatter to the $E_{\rm p,z}-E_{\rm \gamma,iso}-t_{\rm b,z}$ (\emph{Liang-Zhang}) correlation. Limiting to the late jet break sample, we get a tight $E_{\rm p,z}-E_{\rm \gamma,iso}-t_{\rm b,z}$ relation,  $\frac{E_{\rm p,z}}{\rm 100 keV}=(1.2\pm0.3) (\frac{E_{\rm \gamma,iso}}{10^{52} \rm erg})^{(0.56\pm0.07)}(\frac{t_{\rm b,z}}{\rm day})^{(0.67\pm0.08)}$. The correlation coefficient is 0.85 and the dispersion is $\delta=0.15$ with a chance probability $p<10^{-4}$ (Talbe \ref{table:relations} and Figure \ref{Liang-zhangrelation}). This is a tight correlation as claimed by \cite{liangz05}. Including early jet breaks, the correlation is less tight with $\frac{E_{\rm p,z}}{\rm 100 keV}=(1.3\pm0.4) (\frac{E_{\rm \gamma,iso}}{10^{52} \rm erg})^{(0.49\pm0.07)}(\frac{t_{\rm b,z}}{\rm day})^{(-0.08\pm0.05)}$, and regression leads to a correlation with different indices, i.e. correlation coefficient $\sim$ 0.67, dispersion $\sim$ $\delta=0.22$ and chance probability $p<10^{-4}$.

The $E_{\rm \gamma,iso}-f_{\rm b}$ relation (\emph{Frail}) remains loose (Figure \ref{Frailrelation}(a)). We also extend it to $E_{\rm K,iso}-f_{\rm b}$ \citep{berger03} (Figure \ref{Frailrelation}(b)). Limiting to Type II GRBs, we get a scatter relation with $E_{\rm K,iso}\propto f_{\rm b}^{-0.8}$.

\section{Conclusions and Discussion}

The collimation of GRB jets is an important subject, and there have been many investigations in the past. After more than 10 years of successful operation of the {\em Swift} satellite, the sample of GRB afterglow expands significantly. It is therefore justified to revisit the jet break problem with a much larger sample, especially with the bursts with multi-wavelength data to confirm the predicted achromatic feature of jet breaks. In this paper, we have systematically studied the optical jet breaks of all the GRBs detected from February 1997 to March 2015, with most of them having X-ray data showing the consistency with the achromatic prediction. Making use of the standard external shock model, we identified 55 out of 99 GRBs that display a clear jet break in the optical light curves, which include 53 type II and 2 type I GRBs. Among them 3 GRBs show two jet breaks, 1 GRBs show jet break with energy injection.

Some interesting conclusions are obtained from our analysis:
\begin{itemize}
 \item Most GRBs in the jet break sample are generally consistent with the ISM model. Only one jet break with energy injection is identified, suggesting that the time when energy injection ceases is typically earlier than the jet break time\footnote{ Similar to X-ray lightcurves that usually show an early shallow decay phase \citep{zhang06,nousek06}, a shallow decay phase is also seen in the optical band in some GRBs \citep{lil12,wang15}, which requires energy injection into the blastwave.}.
 \item The jet break time has a distribution $\rm t_{b} =90.06\pm84.36$ ks, which gives a jet half-opening-angle distribution $\theta_j =(2.5\pm1.0)^{\rm o}$ and the beaming correction factor $\log f_{b}^{-1}=3.00\pm0.48$. The typical angle is smaller than the previous claimed $5^{\rm o}$, which is caused by a more general treatment of the afterglow kinetic energy, which is derived to be larger due to the small $\epsilon_B$ inferred by many recent studies.
 \item The typical jet correction factor $f_b^{-1} \sim 1000$ is larger than the previously inferred values \citep{frail01,guetta05,liang07,racusin09}, suggesting a factor of two higher event rate density of GRB events.
 \item With the inferred jet opening angles, one can derive the distributions of various energies, which read $\log (E_{\rm \gamma,iso} \rm /erg)=53.11\pm0.84$, $\log (E_{\rm K,iso} \rm /erg)=54.82\pm0.56$, $\log (E_{\rm \gamma} \rm /erg)=49.54\pm1.29$ and $\log (E_{\rm K} \rm /erg)=51.33\pm0.58$. They generally all have large scatter, even for jet corrected values. This suggests that GRBs do not have a standard energy reservoir as speculated before.
 \item The derived electron spectral index has a distribution  $p=2.39\pm0.48$, which is consistent with earlier results.
 \item A fraction of GRBs have lower limits of jet break time. However, due to the sensitivity limits, these lower limits are generally consistent with the observed jet break time distribution. The typical jet half-opening-angle of the true distribution may be consistent with or somewhat larger than the value inferred from this paper.
\end{itemize}

We also revisited several previously claimed luminosity correlations. Following results are obtained:
 \begin{itemize}
   \item  $E_{\rm p,z}-E_{\rm \gamma,iso}$ (\emph{Amati} relation):  The relation remains tight even though with a slightly steeper power law index, i.e. $\frac{E_{\rm p,z}}{\rm 100 keV}\simeq(0.63\pm0.31)(\frac{E_{\rm \gamma,iso}}{10^{52} \rm erg})^{(0.69\pm0.07)}$.
   \item $E_{\rm p,z}-E_{\rm \gamma}$ (\emph{Ghirlanda} relation): This relation has much larger scatter than claimed before, i.e.
%   Dependence of the derived $E_{\rm \gamma}$, the $E_{\rm p,z}-E_{\rm \gamma}$  relation is quite scatter, with
$\frac{E_{\rm p,z}}{\rm 100 keV}\simeq(7.9\pm4.8)(\frac{E_{\rm \gamma}}{10^{51} \rm erg})^{(0.44\pm0.17)}$. The existence of early jet breaks discovered in the {\em Swift} era is likely the origin of the scatter.
   \item $E_{\rm p,z}-E_{\rm \gamma,iso}-t_{\rm b,z}$ (\emph{Liang-Zhang} relation): Similar to the \emph{Ghirlanda} relation, early jet breaks introduce significant scatter to the correlation. Limited to the late jet break sample ($t_b > 10^4$ s), the correlation remains tight, i.e. $\frac{E_{\rm p,z}}{\rm 100 keV}=(1.2\pm0.3)(\frac{E_{\rm \gamma,iso}}{10^{52} \rm erg})^{(0.56\pm0.07)}$ $ (\frac{t_{\rm b,z}}{\rm day})^{(0.67\pm0.08)}$. Including early jet breaks, the correlation is less tight with $\frac{E_{\rm p,z}}{\rm 100 keV}=(1.3\pm0.4) (\frac{E_{\rm \gamma,iso}}{10^{52} \rm erg})^{(0.49\pm0.07)}(\frac{t_{\rm b,z}}{\rm day})^{(-0.08\pm0.05)}$.
   \item $E_{\rm \gamma}-f_{\rm b}$ (\emph{Frail} relation): Both $E_{\rm \gamma,iso}-f_{\rm b}$ and $E_{\rm K,iso}-f_{\rm b}$ relations have very large scatter, suggesting that GRBs do not have a standard energy reservoir.
 \item Type I GRBs usually deviate from these correlations, which are derived from Type II GRBs.
 \end{itemize}

We note that there are jet breaks happening in the early time in the optical band, e.g.,  GRB 051111 $\sim$ 2.7 ks, GRB 070419A $\sim$ 1.5 ks, GRB 080413A $\sim$ 1 ks, GRB 120729A $\sim$ 5.6 ks. The $t_{b}$ of the first jet beak in the two-component jets are 0.2 $\sim$ 2 ks. Usually, the data between $\sim$ 1 $\sim$ 3 ks are missed in the XRT, which may bury the early jet break phenomena.
The results are differ from the observational strategies for searching GRBs jet break, which are focused on the late time afterglows \citep[e.g.][]{zhangbb15}. In the past few years, \emph{swift} has changed their GRB follow-up strategies, generally following GRBs for a lot less duration than earlier in the mission. However, there is not much change since optical observers only need the accurate position of GRBs, which usually provided by XRT.

The GRBs studied in this paper are chosen to have both measure X-ray and optical afterglows and their spectral indices. In order to study their energetics, redshift information is needed. As a result, the studied sample is a small fraction of all the detected GRBs, especially for short GRBs and low-luminosity GRBs.
In order to address many open questions in GRB physics \citep[e.g.][]{zhang11}, such as GRB prompt emission and afterglow physics, central engine, cosmological setting, more advanced multi-wavelength instruments with higher sensitivity, larger field of view, and wider energy bandpass  are needed. Many in-progress observational facilities such as, Space-based Variable Objects Monitor (SVOM) \citep{weijy16}, Advanced Telescope for High ENergy Astrophysics (ATHENA) \citep{nandra13}, The Transient High Energy Sources and Early Universe Surveyor (THESEUS) \citep{amati17}, enhanced X-ray Timing and Polarimetry mission (eXTP) \citep{zhangsn16},  Einstein Telescope \citep{hild08}, Transient Astrophysics Observer on the International Space Station (ISS-TAO) \footnote{ https://heasarc.gsfc.nasa.gov/docs/heasarc/missions/concepts.html}, and Transient Astrophysics Probe (TAP) \footnote{ https://asd.gsfc.nasa.gov/tap/index.html}, will usher in an exciting era of GRB study.}

\acknowledgments
We thank an anonymous referee for helpful suggestions. This work is supported by the National Basic Research Program (973 Programme) of China (Grant No. 2014CB845800),  the National Natural Science Foundation of China (Grants 11673006, 11533003, 11303005), the Guangxi Science Foundation (2016GXNSFFA380006, AD17129006).

\clearpage

%\bibliography{ms}

%%%%%%%%%%%%%%%%%%%%%%%%%%%%%%%%%%%%%%%First sample%%%%%%%%%%%%%%%%%%%%%%%%%%%%%%%%%%%%%%%%%%%%%%%%%%%%%%%%%%%%%%%

\clearpage \thispagestyle{empty} \setlength{\voffset}{-18mm}
\begin{deluxetable}{ccccccc}
\tabletypesize{\scriptsize}
\tablewidth{0pt}

\tablecaption{The temporal decay index $\alpha$ and spectral index
$\beta$ in different afterglow models.}
%\tablewidth{0pt}
%\rotate
\tablehead{ \colhead{CMB}& \colhead{Spectral regime}&
\colhead{$\beta(p)$}&
 %\colhead{$\alpha(p)/\alpha(p,q)$}&
\colhead{$\alpha(\beta)/\alpha(\beta,q)$}&
%\colhead{$\alpha(p)/\alpha(p,q)$}&
\colhead{$\alpha(\beta)/\alpha(\beta,q)$}}
%\begin{tabular}{lllllll}
%\begin{tabular}{ccccccccccccccccccccccccc}
 \startdata
& &  \multicolumn{2}{c}{$p>2$}& \multicolumn{2}{c}{$1<p<2$}& \\
\hline
     & \multicolumn{4}{c}{  Pre-jet break phase without energy injection}                    \\
\hline ISM II & $\nu_m<\nu<\nu_c$   &  ${{p-1 \over 2}}$  &
%${3(p-1)\over 4}$
 ${3\beta \over 2}$ &
%${3(p+2)\over 16}$
 ${6\beta+9 \over 16}$ \\
ISM I  &$\nu>\nu_c$   &  ${{p\over 2}}$   &
%${3p-2 \over 4}$
 ${3\beta-1 \over 2}$ &
%${3p+10 \over 16}$
 ${3\beta+5 \over 8}$\\

Wind II & $\nu_m<\nu<\nu_c$   &  ${p-1\over 2}$  &
%${3p-1\over 4}$ &
${3\beta+1 \over 2}$ &
%${p+8\over 8}$    &
${2\beta+9 \over 8}$  \\
 Wind I   & $\nu>\nu_c$   &  ${p\over 2}$   &
 %${3p-2\over 4}$    &
 ${3\beta-1
     \over 2}$
    % &   ${p+6\over 8}$
     &  ${2\beta+6 \over 8}$  \\
\hline
     & \multicolumn{4}{c}{  Pre-jet break phase with energy injection}                   \\
\hline
ISM II& $\nu_m<\nu<\nu_c$   &  ${{p-1 \over 2}}$ &
%${(2p-6)+(p+3)q \over 4}$
 $(q-1)+\frac{(2+q)\beta}{2}$&
%$-{12-18q-p(q+2) \over 16}$
 $\frac{19q-10}{16}+\frac{(2+q)\beta}{8}$\\
ISM I &$\nu>\nu_c$   &  ${{p\over 2}}$   &
%${(2p-4)+(p+2)q\over 4}$
 $\frac{q-2}{2}+\frac{(2+q)\beta}{2}$ &
%${14q+p(q+2)-4\over 16}$
 $\frac{7q-2}{8}+\frac{(2+q)\beta}{8}$\\

Wind II& $\nu_m<\nu<\nu_c$   &  ${p-1\over 2}$ &
%${(2p-2)+(p+1)q \over 4}$ &
$\frac{q}{2}+\frac{(2+q)\beta}{2}$ &
%${4+(p+4)q \over 8}$ &
 $\frac{5q+4}{8}+\frac{\beta q}{4}$\\
Wind I   & $\nu>\nu_c$   & ${p\over 2}$  &
%${(2p-4)+(p+2)q\over 4}$  &
$\frac{q-2}{2}+\frac{(2+q)\beta}{2}$&
%${(6+p)q\over 8}$  &
$\frac{(\beta+3)q}{4}$\\

\hline
     & \multicolumn{4}{c}{   Post-jet break phase without energy injection}                   \\
\hline ISM II& $\nu_m<\nu<\nu_c$   &  ${{p-1 \over 2}}$  &
%${3p \over
%4}$
     ${6\beta+3 \over 4}$ &
 %  ${3(p+6) \over 16}$
     ${3(2\beta+7) \over 16}$\\
ISM I& $\nu>\nu_c$   &  ${{p\over 2}}$   &
%${3p+1 \over 4}$  &
${6\beta+1 \over 4}$ &
%${3p+22 \over 16}$   &
${3\beta+11 \over 8}$ \\
Wind II&$\nu_m<\nu<\nu_c$   &  ${p-1\over 2}$   &
%${3p+1\over 4}$    &
 ${3\beta+2 \over 2}$ &
 %${p+12 \over 8}$ &
 ${2\beta+13 \over 8}$\\
Wind I&$\nu>\nu_c$   &  ${p\over 2}$   &
 %${3p\over 4}$    &
 ${3\beta \over 2}$   &
% ${p+10 \over 8}$  &
 ${\beta+5 \over 4}$\\
\hline
     & \multicolumn{4}{c}{   Post-jet break phase with energy injection}                   \\
\hline

ISM II& $\nu_m<\nu<\nu_c$   &  ${{p-1 \over 2}}$  &
%${p(q+2)-4(1-q)
%\over 4}$
     $\frac{5q-2}{4}+\frac{(2+q)\beta}{2}$ &
  %  ${22q-4+p(q+2) \over 16}$
      $\frac{11q-2}{8}+\frac{(2+q)\beta}{8}$\\

ISM I& $\nu>\nu_c$   &  ${{p\over 2}}$   &
%${3q-2+p(q+2) \over 4}$  &
${3q-2+2\beta(q+2) \over 4}$ &
%${18q+4+p(q+2) \over 16}$   &
${9q+2+\beta(q+2) \over 8}$ \\

Wind II & $\nu_m<\nu<\nu_c$   &  ${p-1\over 2}$   &
%${3q-2+p(q+2)\over 4}$    &
$q+\frac{(2+q)\beta}{2}$ &
% ${pq+8q+4 \over 8}$ &
 $\frac{1}{2}+\frac{(2\beta+9)q}{8}$\\
Wind I & $\nu>\nu_c$   &  ${p\over 2}$   &
%${p(q+2)-4(1-q)\over 4}$    &
${\beta(q+2)-2(1-q)\over 2}$   &
%${(p+10)q \over 8}$  &
$
{(\beta+5)q \over 4}$\\

\hline

%\end{tabular}
\enddata

 \label{Tab:alpha-beta}
\end{deluxetable}

\begin{deluxetable}{ccccccccccccccccccccccccc}

%\tabletypesize{\scriptsize}
\tabletypesize{\tiny}
\tablecaption{Temporal and spectral parameters of GRBs with jet break features. }
\tablewidth{0pt}
\tabcolsep=5.5pt
\rotate \tabletypesize{\tiny}

\tablehead{ \colhead{GRB}& \colhead{$\beta_{O}$}&
\colhead{$\alpha_{1}$}& \colhead{$\alpha_{2}$}&
%\colhead{$\omega$}&
\colhead{$t_{b}$\tablenotemark{a}}&
\colhead{$z$}&
\colhead{Ref. for $\beta_{O}$ and $z$\tablenotemark{b}}&
\colhead{$E_{p}$\tablenotemark{c}}&
\colhead{INST\tablenotemark{d}}&
\colhead{INST and Ref. for $E_{p}$}\tablenotemark{e}}

\startdata		
980703	&	1.01	$\pm$	0.02	&	1.11 			&	2.83 			&	214.9	$\pm$	10.2	&	0.966	&	(1);(1)	&	254	$\pm$	50.8	&	SAX/CGO	&	CGO,(63),(64),(65)	 \\
990123	&	0.75	$\pm$	0.07	&	0.99 	$\pm$	0.12 	&	1.93 	$\pm$	0.04 	&	123.0 	$\pm$	13.7 	&	1.6	&	(1);(1)	&	781 	$\pm$	62 	&	SAX/CGO/KON	&	 SAX/CGO/KON,(63),(65),(63)	\\
990510	&	0.55			&	0.86 	$\pm$	0.03 	&	1.95 	$\pm$	0.14 	&	101.9	$\pm$	12.5	&	1.619	&	(1);(1)	&	162 	$\pm$	16 	&	SAX/CGO	&	 SAX,(63),(65),(63)	\\
990712	&	0.99	$\pm$	0.02	&	0.97 			&	2.32 			&	1000			&	0.434	&	(1);(1)	&	93 	$\pm$	15 	&	SAX	&	SAX,(63),(65),(63)	\\
991216	&	0.57	$\pm$	0.08	&	1.00 	$\pm$	0.05 	&	1.80 	$\pm$	0.05 	&	103.7 	$\pm$	41.1 	&	1.02	&	(2);(3)	&	317 	$\pm$	63 	&	SAX/CGO/KON	&	 CGO/KON,(63)£¬(64),(65)	\\
000301C	&	0.7			&	1.04 			&	2.97 			&	562.9	$\pm$	18.7	&	2.03	&	(1);(1)	&	326 	$\pm$	137 	&	CGO/KON	&	CGO/KON,(63)	\\
000926	&	1	$\pm$	0.18	&	1.48 			&	2.49 			&	175.2	$\pm$	4.6	&	2.07	&	(4);(1)	&	101 	$\pm$	7 	&	SAX/KON	&	KON,(63)	\\
011211	&	0.8	$\pm$	0.15	&	0.95 	$\pm$	0.05 	&	2.11 	$\pm$	0.05 	&	134.7 	$\pm$	1.9 	&	2.14	&	(5);(6)	&	59 	$\pm$	8 	&	SAX	&	SAX,(63)	\\
020405	&	1.43	$\pm$	0.08	&	1.40 	$\pm$	0.05 	&	1.95 	$\pm$	0.05 	&	147.7 	$\pm$	53.9 	&	0.69	&	(1);(1)	&	193 	$\pm$	54 	&	SAX/KON	&	 SAX/KON,(65)£¬(66)	 \\
021004	&	0.39			&	0.82 	$\pm$	0.02 	&	1.39 	$\pm$	0.05 	&	300.3			&	2.335	&	(1);(1)	&	80 	$\pm$	35 	&	HET	&	HET,(63)	\\
030226	&	0.7	$\pm$	0.03	&	0.77 	$\pm$	0.05 	&	1.99 	$\pm$	0.05 	&	89.9 	$\pm$	11.0 	&	1.986	&	(1);(1)	&	97 	$\pm$	20 	&	HET	&	HET,(63)	\\
030323	&	0.89	$\pm$	0.04	&	1.29 			&	2.11 			&	400			&	3.37	&	(1);(1)	&	62 	$\pm$	26 	&	HET	&	HET,(63)	\\
030329	&	0.5			&	0.84 	$\pm$	0.08 	&	1.89 	$\pm$	0.10 	&	41.0 	$\pm$	4.0 	&	0.1685	&	(1);(1)	&	68 	$\pm$	3 	&	HET/KON	&	 HET/KON,(65)£¬(63),(67)	\\
030429	&	0.75			&	0.72 	$\pm$	0.03 	&	2.72 			&	158.7			&	2.65	&	(1);(1)	&	35	$\pm$	9	&	HET	&	HET,(65),(63),(67)	\\
030723	&	0.66	$\pm$	0.21	&	0.05 	$\pm$	0.06 	&	2.01 			&	103.2	$\pm$	5	&		&	(7);(-)	&				&	HET	&		\\
050319	&	0.74	$\pm$	0.42	&	0.66 	$\pm$	0.14 	&	1.86 	$\pm$	0.00 	&	319.8 	$\pm$	13.7 	&	3.2425	&	(8);(1)	&				&	SWI	&		\\
050408	&	0.28 	$\pm$	0.33 	&	0.55 	$\pm$	0.09 	&	1.49 	$\pm$	0.05 	&	73.1 	$\pm$	2.3 	&	1.2357	&	(9);(8)	&				&	HET	&		\\
050502A	&	0.76 	$\pm$	0.16 	&	0.90 	$\pm$	0.05 	&	1.76 	$\pm$	0.07 	&	8.4 	$\pm$	0.5 	&	3.793	&	(10);(1)	&				&	INT/SWI	&		\\
050525A	&	0.52	$\pm$	0.08	&	0.95 	$\pm$	0.13 	&	1.95 	$\pm$	0.19 	&	27.4 	$\pm$	2.2 	&	0.606	&	(1);(1)	&	84 	$\pm$	2 	&	KON/SWI	&	 SWI,(68)	\\
050730	&	0.52	$\pm$	0.05	&	0.66 	$\pm$	0.12 	&	1.61 	$\pm$	0.13 	&	15.0 	$\pm$	1.5 	&	3.96855	&	(10);(1)	&				&	SWI	&		\\
050801	&	1.00 	$\pm$	0.16 	&	1.08 	$\pm$	0.06 	&	2.01 	$\pm$	0.09 	&	14.2 	$\pm$	1.0 	&	1.56	&	(11);(12)	&				&	SWI	&		\\
050820A	&	0.72	$\pm$	0.03	&	0.82 	$\pm$	0.12 	&	1.67 	$\pm$	0.09 	&	194.1 	$\pm$	9.0 	&	2.6147	&	(10);(1)	&	367 	$\pm$	77 	&	SWI/KON	&	 KON,(63),(69)	 \\
050922C	&	0.51	$\pm$	0.05	&	0.92 	$\pm$	0.05 	&	1.62 	$\pm$	0.35 	&	45.6 	$\pm$	0.7 	&	2.2	&	(10);(13)	&	130 	$\pm$	3 	&	HET/KON/SWI	&	 HET,(63)	\\
051109A	&	0.7	$\pm$	0.05	&	0.76 	$\pm$	0.12 	&	1.65 	$\pm$	0.14 	&	26.0 	$\pm$	6.5 	&	2.346	&	(14);(12)	&	161 	$\pm$	58 	&	KON/SWI	&	 KON,(63)	\\
051111	&	0.76	$\pm$	0.07	&	0.79 	$\pm$	0.12 	&	1.77 	$\pm$	0.09 	&	2.7 	$\pm$	0.3 	&	1.55	&	(15);(12)	&	255 	$\pm$	156 	&	SWI/SUZ	 &		\\
051221A	&	0.64	$\pm$	0.05	&	0.94 	$\pm$	0.09 	&	1.68 	$\pm$	0.11 	&	100.4 	$\pm$	2.0 	&	0.55	&	(16);(17)	&	402 	$\pm$	93 	&	KON/SWI/SUZ	 &	KON,(63)	 \\
060111B	&	0.70 	$\pm$	0.10 	&	0.80 	$\pm$	0.07 	&	1.55 	$\pm$	0.08 	&	7.2 	$\pm$	0.5 	&		&	(18) ;(-)	&				&		&		\\
060206	&	0.73	$\pm$	0.05	&	1.15 	$\pm$	0.12 	&	1.94 	$\pm$	0.10 	&	157.7 	$\pm$	7.8 	&	4.0479	&	(10);(18)	&	83 	$\pm$	35 	&	SWI	&	 SWI,(63),(68),(70)	 \\
060418	&	0.78	$\pm$	0.09	&	1.27 	$\pm$	0.12 	&	2.56 	$\pm$	0.47 	&	819.7	$\pm$	32.0 	&	1.5 	&	(5);(19)	&	230.0 			&	KON/SWI	&	 KON,(63),(71)	\\
060526	&	0.51	$\pm$	0.32	&	0.94 	$\pm$	0.07 	&	1.99 	$\pm$	0.23 	&	121.4	$\pm$	12.4 	&	3.2213	&	(5);(1)	&				&	SWI	&		\\
060605	&	1.06	$\pm$	0	&	0.88 	$\pm$	0.15 	&	3.22 	$\pm$	0.53 	&	25.7	$\pm$	6.7 	&	3.8	&	(20);(21)	&				&	SWI	&		\\
060729	&	0.78	$\pm$	0.03	&	1.28 	$\pm$	0.05 	&	2.45 	$\pm$	0.31 	&	82.1	$\pm$	12.3 	&	0.54	&	(22);(23)	&				&	SWI	&		\\
061126	&	0.82	$\pm$	0.09	&	0.96 	$\pm$	0.12 	&	2.31 	$\pm$	0.16 	&	204.2 	$\pm$	11.4 	&	1.5	&	(24);(25)	&	620 	$\pm$	20 	&	SWI/RHE	&	 SWI/RHE,(63),(72)	 \\
070411	&	0.75 			&	0.90 	$\pm$	0.08 	&	1.60 	$\pm$	0.11 	&	152.0 	$\pm$	74.3 	&	2.95 	&	(26);(27)	&				&	SWI	&		\\
070419A	&	0.80 			&	0.62 	$\pm$	0.10 	&	1.55 	$\pm$	0.08 	&	1.5 	$\pm$	0.3 	&	0.97 	&	(28);(29)	&				&	SWI	&		\\
070518	&	0.80 			&	0.90 	$\pm$	0.07 	&	1.65 	$\pm$	0.11 	&	40.0 	$\pm$	3.2 	&		&	(28);(-)	&				&	SWI	&		\\
071003	&	1.25	$\pm$	0.09	&	0.90 	$\pm$	0.07 	&	1.81 	$\pm$	0.13 	&	0.2 	$\pm$	0.0 	&	1.6044	&	(30);(31)	&	410 	$\pm$	190 	&	KON/SWI	 &	KON,(73)	 \\
071003	&	1.25	$\pm$	0.09	&	-1.20 	$\pm$	0.21 	&	2.05 	$\pm$	0.18 	&	55.0 	$\pm$	6.0 	&	1.6044	&	(30);(31)	&	410 	$\pm$	190 	&	KON/SWI	 &		\\
080310	&	0.42	$\pm$	0.12	&	1.19 	$\pm$	0.13 	&	2.44 	$\pm$	0.18 	&	29.2 	$\pm$	3.2 	&	2.43 	&	(8);(32)	&				&	SWI	&		\\
080319B	&	0.51	$\pm$	0.26	&	1.03 	$\pm$	0.11 	&	1.69 	$\pm$	0.12 	&	1.7 	$\pm$	0.2 	&	0.937	&	(33);(34)	&	651 	$\pm$	14 	&	KON/SWI	&	 KON,(63),(74)	 \\
080319B	&	0.51	$\pm$	0.26	&	0.83 	$\pm$	0.16 	&	1.65 	$\pm$	0.09 	&	59.1 	$\pm$	7.6 	&	0.937	&	(33);(34)	&	651 	$\pm$	14 	&	KON/SWI	&		 \\
080413A	&	0.52	$\pm$	0.37	&	0.62 	$\pm$	0.14 	&	1.52 	$\pm$	0.16 	&	1.0 	$\pm$	0.1 	&	2.433	&	(18);(35)	&	170 	$\pm$	80 	&	SWI/SUZ	&	 SWI/SUZ,(75)	 \\
080413B	&	0.52	$\pm$	0.37	&	0.71 	$\pm$	0.06 	&	1.45 	$\pm$	0.13 	&	0.6 	$\pm$	0.2 	&	1.1	&		&	67 	$\pm$	13 	&	SWI/SUZ	&	SUZ,(76)	\\
080413B	&	0.25	$\pm$	0.07	&	0.05 	$\pm$	0.08 	&	1.50 	$\pm$	0.11 	&	75.1 	$\pm$	12.3 	&	1.1	&	(36);(37)	&	67 	$\pm$	13 	&	SWI/SUZ	&		\\
080603A	&	0.98	$\pm$	0.04	&	0.94 	$\pm$	0.00 	&	2.29 	$\pm$	0.00 	&	160.8 	$\pm$	12.9 	&	1.5635	&	(38);(39)	&	60 	$\pm$	10 	&	INT/SWI	&		 \\
080710	&	0.8	$\pm$	0.09	&	0.79 	$\pm$	0.08 	&	1.78 	$\pm$	0.01 	&	20.3 	$\pm$	1.4 	&	0.85 	&	(40);(41)	&				&	SWI	&		\\
081008	&	0.40 	$\pm$	0.23 	&	0.64 	$\pm$	0.06 	&	1.52 	$\pm$	0.09 	&	9.5 	$\pm$	0.8 	&	1.9685	&	(42);(43)	&				&	SWI	&		\\
081203A	&	0.596			&	1.10 	$\pm$	0.14 	&	1.87 	$\pm$	0.15 	&	10.1 	$\pm$	1.7 	&	2.1	&	(44);(45)	&	578 	$\pm$	290 	&	KON/SWI	&	 KON,(77),(78)	\\
090426	&	0.76 	$\pm$	0.14 	&	0.75 	$\pm$	0.12 	&	1.65 	$\pm$	0.08 	&	0.4 	$\pm$	0.1 	&	2.609	&	(46);(47)	&				&	SWI	&		\\
090426	&	0.76 	$\pm$	0.14 	&	0.24 	$\pm$	0.09 	&	1.59 	$\pm$	0.08 	&	25.0 	$\pm$	6.3 	&	2.609	&		&				&	SWI	&		\\
090618	&	0.5	$\pm$	0.05	&	0.67 	$\pm$	0.07 	&	1.60 	$\pm$	0.08 	&	31.1 	$\pm$	6.7 	&	0.54	&	(4);(48)	&	156 	$\pm$	12 	&	FER/KON/SWI/SUZ	 &	FER,(79)	 \\
090926A	&	0.72	$\pm$	0.17	&	1.20 	$\pm$	0.13 	&	2.50 	$\pm$	0.21 	&	1024.1 	$\pm$	42.2 	&	2.1062	&	(8);(49)	&	321 	$\pm$	12 	&	 FER/KON/SWI/SUZ	&	FER,(79)	 \\
091029	&	0.57			&	0.54 	$\pm$	0.11 	&	1.47 	$\pm$	0.09 	&	30.0 	$\pm$	6.8 	&	2.752	&	(4);(50)	&	36 	$\pm$	2 	&	SWI	&		\\
091127	&	0.43	$\pm$	0.1	&	0.69 	$\pm$	0.00 	&	1.47 	$\pm$	0.12 	&	55.0 	$\pm$	1.3 	&	0.49	&	(51);(52)	&	61 	$\pm$	18 	&	SWI/FER	&	 FER,(79)	\\
100219A	&	0.60 	$\pm$	0.12 	&	0.84 	$\pm$	0.08 	&	1.75 	$\pm$	0.12 	&	1.8 	$\pm$	0.4 	&	4.80 	&	(53);(54)	&				&	SWI	&		\\
100219A	&	0.66 	$\pm$	0.13 	&	-2.40 	$\pm$	0.17 	&	1.70 	$\pm$	0.14 	&	17.5 	$\pm$	5.4 	&	4.80 	&	(53);(54)	&				&	SWI	&		\\
110205A	&	1.12	$\pm$	0.24	&	1.45 	$\pm$	0.12 	&	2.31 	$\pm$	0.22 	&	103.2 	$\pm$	8.0 	&	2.22	&	(55);(56)	&	222 	$\pm$	74 	&	KON/SWI/SUZ	 &	KON,(80)	 \\
120729A	&	1	$\pm$	0.1	&	0.91 	$\pm$	0.12 	&	2.14 	$\pm$	0.19 	&	5.6 	$\pm$	0.5 	&	0.8	&	(57);(58)	&	311 	$\pm$	20 	&	FER/SWI	&	FER,(81)	 \\
130427A	&	0.69	$\pm$	0.01	&	1.01 	$\pm$	0.09 	&	1.88 	$\pm$	0.11 	&	127.9 	$\pm$	1.4 	&	0.34	&	(59);(60)	&	933 	$\pm$	112 	&	 KON/FER/SWI/RHE	&	FER,(82)	\\
130603B	&	-0.84	$\pm$	0.1	&	1.11 	$\pm$	0.11 	&	2.10 	$\pm$	0.21 	&	31.0 	$\pm$	7.7 	&	0.3564	&	(61);(62)	&	660 	$\pm$	100 	&	SWI/KON	&	 KON,(83)	\\

\enddata
\tablenotetext{a}{Break time, in unit of ks;}
\tablenotetext{b}{References for $\beta_{O}$ and $z$;}
\tablenotetext{c}{Spectral peak energy on observe-frame,, in unit of keV;}
\tablenotetext{d}{The instrument name of the experiment(s), or of the satellite(s), that provided the estimates of spectral parameters and energy, CGO=\emph{CGRO-BATSE}, FER=\emph{Fermi-GBM}, HET=\emph{HETE-2-FREGATE}, INT=\emph{INTEGRAL-SPI/IBIS}, KON=\emph{Konus-Wind}, RHE=\emph{RHESSI}, SAX=\emph{BeppoSAX}, SWI=\emph{Swift-BAT}, SUZ=\emph{Suzaku-WAM};}

\tablenotetext{e}{References for $E_p$:(1)\cite{Liang06b};
(2)\cite{Panaitescu05b};(3)\cite{Svensson10};(4)\cite{lil12};(5)\cite{Kann10a};
(6)\cite{Vreeswijk06b};(7)\cite{kann06};(8)\cite{kann10}
(9)\cite{urata07};(10)\cite{Mannucci11};(11)\cite{Pasquale07};(12)\cite{Robertson12};
(13)\cite{Price06};(14)\cite{Yost07};
(15)\cite{Guidorzi07b};
(16)\cite{Soderberg06};(17)\cite{Butler07}(18)\cite{Fynbo09};
(19)\cite{Prochaska07};(20)\cite{Ferrero09};
(21)\cite{Savaglio07};(22)\cite{Zafar11};(23)\cite{Thoene06};(24)\cite{Perley08};(25)\cite{Schady06};
(26)\cite{wang15};(27)\cite{Jakobsson07a};(28)\cite{Xin10};(29)\cite{Cenko07};
(30)\cite{Kruhler09b};(31)\cite{Perley08a};
(32)\cite{Prochaska08};(33)\cite{D'Elia09};(34)\cite{Vreeswijk08a};
(35)\cite{Thoene08a};(36)\cite{Robertson12};(37)\cite{Vreeswijk08b};
(38)\cite{guidorzi11};(39)\cite{Perley08b};(40)\cite{kann11};
(41)\cite{Perley08c};(42)\cite{Yuan10};(43)\cite{D'Avanzo08};(44)\cite{evans09};(45)\cite{Landsman08};
(46)\cite{nicuesa11};(147)\cite{Levesque09};(48)\cite{Cenko09a};(49)\cite{Malesani09};
(50)\cite{Chornock09a};(51)\cite{Vergani11};(52)\cite{Cucchiara09};
(53)\cite{Thone13};(54)\cite{Groot10};(55)\cite{Cucchiara11};(56)\cite{Cenko11};
(57)\cite{cano14b};(58)\cite{tanvir12a};(59)\cite{perley14};(60)\cite{Levan13};
(61)\cite{fong14};(62)\cite{Xu13};(63)\cite{amati08};
(64)\cite{Jimenez01};(65)\cite{liangz05};(66)\cite{Price03};(67)\cite{Sakamoto05c};(68)\cite{Nava12};(69)\cite{Cenko06};(70)\cite{Sakamoto08};(71)\cite{Golenetskii06a};
(72)\cite{Perley08};(73)\cite{Butler10};(74)\cite{Golenetskii08a};(75)\cite{Ohno08a};(76)\cite{Krimm09};(77)\cite{Golenetskii08b};(78)\cite{Kann10a};(79)\cite{guetta11};(80)\cite{Golenetskii11};(81)\cite{Cano14a};
(82)\cite{Kienlin13};(83)\cite{Golenetskii13};
}

\label{table:sample}
\end{deluxetable}

\begin{deluxetable}{ccccccccccccccccccccccccc}

%\tabletypesize{\scriptsize}
\tabletypesize{\tiny}
\tablecaption{Temporal and Spectral Parameters of GRBs with a Lower Limit of Jet Break Time. }
\tablewidth{0pt}
\tabcolsep=5.5pt
%\rotate \tabletypesize{\tiny}
\tablehead{ \colhead{GRB}& \colhead{$\beta_{O}$}&
\colhead{$\alpha$}&
\colhead{$z$}&
\colhead{Ref. for $\beta_{O}$ and $z$\tablenotemark{a}}&
\colhead{$E_{p}$}&
\colhead{INST}&
\colhead{INST and Ref. for $E_{p}$}\tablenotemark{a}}
\startdata		
020124	&	0.91	$\pm$		&	1.85	$\pm$	0.11	&	3.2	&	(1),(2)	&	87	$\pm$	15	&	HET/KON	&	HET/KON,(44),(45)	\\
020813	&	0.85	$\pm$		&	1.26	$\pm$	0.24	&	1.25	&	(1),(3)	&	142	$\pm$	13	&	HET/KON	&	HET/KON,(44),(45)	\\
050401	&	0.5	$\pm$	0.2	&	0.89	$\pm$	0.08	&	2.899	&	(4),(5)	&	132	$\pm$	16	&	SWI/KON	&	KON,(46),(47)	\\
050416A	&	0.92	$\pm$	0.3	&	1.31	$\pm$	0.09	&	0.6852	&	(4),(6)	&	15	$\pm$	5	&	SWI	&	SWI,(45),(47)	\\
050603	&	0.71	$\pm$	0.1	&	1.7	$\pm$	0.13	&	2.821	&	(4),(7)	&	349	$\pm$	28	&	SWI/KON/INT	&	KON,(47),(48)	\\
050721	&	1.16	$\pm$	0.35	&	1.09	$\pm$	0.11	&		&	(8),(-)	&				&	SWI	&		\\
051028	&	0.6	$\pm$	0	&	0.99	$\pm$	0.1	&	3.7	&	(4),(9)	&	298	$\pm$	73	&	HET/KON	&	KON,(49)	\\
060210	&	0.37	$\pm$	0.08	&	1.31	$\pm$	0.11	&	3.9133	&	(4),(9)	&				&	SWI	&		\\
060512	&	0.24	$\pm$	0.2	&	1.11	$\pm$	0.09	&	1.8836	&	(4),(10)	&	124	$\pm$	38	&	SWI/SUZ	&		\\
060714	&	1.02	$\pm$	0.05	&	1.62	$\pm$	0.08	&	1.2622	&	(4),(11)	&	399	$\pm$	19	&	RHE/SUZ/KON/SWI	&		\\
060904B	&	1.11	$\pm$	0.1	&	1.01	$\pm$	0.21	&	0.7029	&	(4),(12)	&				&	SWI	&		\\
060906	&	0.56	$\pm$	0.02	&	1.02	$\pm$	0.15	&	3.6856	&	(4),(13)	&				&	SWI	&		\\
060908	&	0.98	$\pm$	0.42	&	1.12	$\pm$	0.13	&	1.949	&	(4),(14)	&	307	$\pm$	92	&	SWI/KON	&	KON,(50),(8)	\\
060912A	&	0.6	$\pm$	0.15	&	0.94	$\pm$	0.01	&	0.937	&	(4),(15)	&				&	SWI/KON	&		\\
060927	&	0.61	$\pm$	0.05	&	1.3	$\pm$	0.15	&	5.467	&	(4),(16)	&				&	SWI	&		\\
061007	&	0.68	$\pm$	0.02	&	1.31	$\pm$	0.05	&	2.602	&	(4),(11)	&	485	$\pm$	67	&	SWI/KON	&	KON,(51)	\\
070110	&	0.55	$\pm$	0.04	&	1.2	$\pm$	0.15	&	2.3521	&	(4),(17)	&				&	SWI	&		\\
070306	&	0.43	$\pm$	0	&	0.87	$\pm$	0.01	&	2.2	&	(4),(18)	&				&	SWI	&		\\
070311	&	1	$\pm$	0.2	&	0.73	$\pm$	0.21	&		&	(4),(-)	&				&	INT	&		\\
070318	&	0.78	$\pm$	0.1	&	1.02	$\pm$	0.09	&	0.84	&	(4),(17)	&				&	SWI	&		\\
070611	&	0.73	$\pm$	0	&	0.58	$\pm$	0.12	&	2.0394	&	(4),(19)	&				&	SWI	&		\\
071025	&	0.96	$\pm$	0.14	&	1.28	$\pm$	0.12	&	4.8	&	(8),(20)	&				&	SWI	&		\\
071031	&	0.74	$\pm$	0.22	&	1.16	$\pm$	0.09	&	2.692	&	(4),(4)	&	451	$\pm$	73	&	SWI/KON	&		\\
071112C	&	0.63	$\pm$	0.29	&	0.94	$\pm$	0.15	&	0.8227	&	(4),(21)	&				&	SWI	&		\\
080319A	&	0.77	$\pm$	0.02	&	0.65	$\pm$	0.11	&	2.2	&	(4),(22)	&	654	$\pm$	14	&	SWI/INT	&		\\
080319C	&	0.85	$\pm$	0.05	&	0.94	$\pm$	0.05	&	1.95	&	(4),(23)	&	4400	$\pm$	400	&	SWI/KON/SUZ/FER	&	KON,(52)	\\
080721	&	0.36	$\pm$	0	&	1.27	$\pm$	0.05	&	2.591	&	(8),(24)	&				&	SWI	&		\\
080804	&	0.7	$\pm$	0.4	&	1.07	$\pm$	0.08	&	2.2045	&	(8),(25)	&	56	$\pm$	6	&	SWI/FER	&		\\
080913	&	0.79	$\pm$	0.03	&	0.98	$\pm$	0.11	&	6.7	&	(4),(26)	&	121	$\pm$	39	&	SWI/KON/FER/INT	&	SWI/KON,(53)	\\
080928	&	1.32	$\pm$	0.02	&	2.1	$\pm$	0.22	&	1.692	&	(4),(27)	&				&	SWI	&		\\
090102	&	0.74	$\pm$	0.22	&	1.49	$\pm$	0.12	&	1.547	&	(4),(28)	&	451	$\pm$	73	&	SWI/KON	&	KON,(54)	\\
090323	&	0.74	$\pm$	0.15	&	1.55	$\pm$	0.13	&	3.57	&	(4),(29)	&	416	$\pm$	76	&	KON/FER	&	KON,(55)	\\
090328	&	0.52	$\pm$	0.02	&	0.96	$\pm$	0.04	&	0.736	&	(8),(30)	&	1060	$\pm$	60	&	SWI/INT/FER/KON	&	FER,(56),(8)	\\
090510	&	0.68	$\pm$	0.05	&	0.81	$\pm$	0.05	&	0.4428	&	(31),(32)	&				&	SWI	&		\\
090812	&	0.44	$\pm$	0.04	&	1.11	$\pm$	0.15	&	2.452	&	(4),(33)	&				&	SWI	&		\\
100418A	&	0.7	$\pm$	0.1	&	1.22	$\pm$	0.11	&	0.624	&	(34),(35)	&				&	SWI	&		\\
101024A	&	0.64	$\pm$	0.05	&	0.79	$\pm$	0.05	&	2.69	&	(36),(4)	&				&	SWI	&		\\
110918A	&	0.63	$\pm$	0.29	&	0.95	$\pm$	0.11	&	0.982	&	(37),(38)	&				&	SWI	&		\\
120326A	&	0.75	$\pm$	0.08	&	1.65	$\pm$	0.14	&	1.798	&	(39),(40)	&	46	$\pm$	4	&	SWI/FER/SUZ	&	FER,(57)	\\
120711A	&	0.98	$\pm$	0.09	&	1.37	$\pm$	0.13	&	1.405	&	(41),(42)	&				&	SWI	&		\\
120815A	&	0.78	$\pm$	0.01	&	0.63	$\pm$	0.12	&	2.358	&	(4),(43)	&				&	SWI	&		\\
\enddata
\tablenotetext{a}{References for $E_p$: (1)\cite{Liang06b};(2)\cite{Firmani06};(3)\cite{Price02};(4)\cite{lil12};(5)\cite{Golenetskii05d};(6)\cite{Mannucci11};(7)\cite{Robertson12};(8)\cite{lil12};(9)\cite{Nava08};(10)\cite{Bloom06};
(11)\cite{Jakobsson06a};(12)\cite{Fugazza06};(13)\cite{Vreeswijk06b};(14)\cite{Rol06};(15)\cite{Jakobsson06b};(16)\cite{Fynbo06};(17)\cite{Jaunsen07a};(18)\cite{Jaunsen07b};(19)\cite{thoene07};(20)\cite{kann10};
(21)\cite{Jakobsson07};(22)\cite{Barthelmy08b};(23)\cite{Wiersema08};(24)\cite{Jakobsson08};(25)\cite{Thoene08};(26)\cite{Stamatikos08a};(27)\cite{Vreeswijk08};(28)\cite{Postigo09a};(29)\cite{Chornock09};(30)\cite{Cenko09};
(31)\cite{NicuesaGuelbenzu12};(32)\cite{Ukwatta09};(33)\cite{Postigo09b};(34)\cite{jia12};(35)\cite{Cucchiara10};(36)\cite{Gendre11};(37)\cite{Frederiks13};(38)\cite{Levan11};(39)\cite{urata14};(40)\cite{Tello12};
(41)\cite{Martin-Carrillo14};(42)\cite{Tanvir12b};(43)\cite{Malesani12};(44)\cite{liangz05};(45)\cite{Sakamoto05a};(46)\cite{Golenetskii05a};(47)\cite{amati06};(48)\cite{Golenetskii05b};(49)\cite{Golenetskii05g};(50)\cite{Krimm09};
(51)\cite{Golenetskii06c};(52)\cite{Golenetskii08c};(53)\cite{PalShin08};(54)\cite{Golenetskii09g};(55)\cite{Golenetskii09h};(56)\cite{guetta11a};(57)\cite{Collazzi12a}.}
\label{table:lowsample}
\end{deluxetable}

\begin{deluxetable}{ccccccccccccccccccccccccc}

%\tabletypesize{\scriptsize}
\tabletypesize{\tiny}
\tablecaption{Temporal and Spectral Parameters of GRBs with a Upper Limit of Jet Break Time. }
\tablewidth{0pt}
\tabcolsep=5.5pt
%\rotate \tabletypesize{\tiny}

\tablehead{ \colhead{GRB}& \colhead{$\beta_{O}$}&
\colhead{$\alpha$}&
\colhead{$z$}&
\colhead{Ref. for $\beta_{O}$ and $z$\tablenotemark{a}}&
\colhead{$E_{p}$}&
\colhead{INST}&
\colhead{INST and Ref. for $E_{p}$}\tablenotemark{a}}

\startdata		
070125	&	0.59	$\pm$	0.10	&	1.82805	$\pm$	0.07	&	1.5471	&	(1);(2)	&	367	$\pm$	65	&	KON/SUZ/INT	&	KON,(8),(9)	\\
071010A	&	0.61	$\pm$	0.12	&	2.1	$\pm$	0.23	&	0.985	&	(3);(4)	&				&	SWI	&		\\
100901A	&	0.52	$\pm$	0.10	&	1.51	$\pm$	0.07	&	1.408	&	 (5);(6)	&				&	SWI	&		\\

\enddata
\tablenotetext{a}{(1)\cite{lil12};(2)\cite{Fox07};(3)\cite{Greiner11};(4)\cite{Robertson12};(5)\cite{gorbovskoy12};(6)\cite{Chornock10},(8)\citep{golenetskii07}, (9)\citep{amati08}}

\label{table:upersample}
\end{deluxetable}

\begin{deluxetable}{ccccccccccccccccccccccccc}
%\tabletypesize{\scriptsize}
\tabletypesize{\tiny}
\tablecaption{Jet Break GRBs Candidates and Their Derived Parameters. }
\tablewidth{0pt}
\tabcolsep=2.5pt
%\rotate \tabletypesize{\tiny}

\tablehead{ \colhead{GRB}&
\colhead{$p$}&
\colhead{Model\tablenotemark{a}}&
\colhead{$E_{\rm \gamma,iso}$\tablenotemark{b}}&
\colhead{$E_{\rm K,iso}$\tablenotemark{b}}&
\colhead{$E_{\rm \gamma}$\tablenotemark{c}}&
\colhead{$E_{\rm K}$\tablenotemark{c}}&
\colhead{$\theta_{\rm j}^{\circ}$}&
\colhead{$\eta_{\rm \gamma}$}\tablenotemark{d}.}

\startdata		
980703	&	3.02 	$\pm$	0.04 	&	ISMI-ISMII,windI-windII	&	7.08 	$\pm$	0.68 	&	261.95 	$\pm$	48.94 	&	0.10 	$\pm$	0.03 	&	3.64 	$\pm$	0.96 	&	3.02 	 $\pm$	0.25 	 &	2.6 	\\
990123	&	2.5	$\pm$	0.14	&	ISMII,windI	&	436.52 	$\pm$	64.67 	&	487.54 	$\pm$	281.03 	&	0.77 	$\pm$	0.15 	&	0.86 	$\pm$	0.16 	&	3.41 	$\pm$	0.29 	 &	47.2 	\\
990510	&	2.50 	$\pm$	0.14 	&	ISMI-ISMII,windI-II	&	19.50 	$\pm$	2.38 	&	539.97 	$\pm$	346.41 	&	0.18 	$\pm$	0.06 	&	5.05 	$\pm$	1.74 	&	2.48 	 $\pm$	0.28 	&	 3.5 	\\
990712	&	2.98 	$\pm$	0.04 	&	ISMI-ISMII,windI-II	&	0.76 	$\pm$	0.04 	&	190.48 	$\pm$	4.33 	&	0.01 	$\pm$	0.01 	&	3.63 	$\pm$	1.55 	&	3.54 	 $\pm$	0.35 	&	 0.4 	\\
991216	&	2.48	$\pm$	0.1	&	ISMII,windI	&	70.79 	$\pm$	6.83 	&	789.43 	$\pm$	191.93 	&	0.12 	$\pm$	0.03 	&	1.31 	$\pm$	0.29 	&	3.31 	$\pm$	0.29 	 &	8.2 	\\
000301C	&	2.80 	$\pm$	0.04 	&	ISMI-ISMII,windI-II	&	199.00 	$\pm$	35.00 	&	1817.47 	$\pm$	129.78 	&	1.23 	$\pm$	0.41 	&	11.25 	$\pm$	3.75 	&	2.02 	 $\pm$	0.19 	 &	9.9 	\\
000926	&	3.00 	$\pm$	0.40 	&	ISMII,windI-windII	&	31.40 	$\pm$	6.80 	&	1550.66 	$\pm$	2712.40 	&	0.20 	$\pm$	0.14 	&	9.89 	$\pm$	7.01 	&	 2.05 	$\pm$	0.44 	 &	2.0 	\\
011211	&	2.48	$\pm$	0.1	&	ISMII	&	10.23 	$\pm$	0.99 	&	720.29 	$\pm$	223.44 	&	0.02 	$\pm$	0.01 	&	1.07 	$\pm$	0.22 	&	3.13 	$\pm$	0.59 	&	 1.4 	\\
020405	&	3.46	$\pm$	0.24	&	ISMI,windI	&	14.79 	$\pm$	0.34 	&	945.48 	$\pm$	482.02 	&	0.05 	$\pm$	0.01 	&	3.19 	$\pm$	0.48 	&	4.71 	$\pm$	 0.37 	&	1.5 	 \\
021004	&	1.78 	$\pm$	0.24 	&	ISMII,windI	&	3.80 	$\pm$	0.50 	&	10405.35 	$\pm$	8945.33 	&	0.05 	$\pm$	0.04 	&	138.43 	$\pm$	124.53 	&	2.96 	 $\pm$	2.12 	&	 0.0 	\\
030226	&	2.4	$\pm$	0.06	&	ISMI-ISMII,windI-windII	&	7.94 	$\pm$	0.97 	&	2392.17 			&	0.01 	$\pm$	0.01 	&	2.08 	$\pm$	0.31 	&	2.39 	$\pm$	 0.45 	&	0.3 	 \\
030323	&	2.78 	$\pm$	0.08 	&	ISMII,windI-windII	&	3.20 	$\pm$	1.00 	&	1093.91 	$\pm$	165.61 	&	0.02 	$\pm$	0.01 	&	5.84 	$\pm$	1.79 	&	1.87 	 $\pm$	0.16 	 &	0.3 	\\
030329	&	2.32	$\pm$	0	&	ISMI-ISMII,windI-windII	&	1.55 	$\pm$	0.15 	&	411.66 	$\pm$	180.15 	&	0.002 	$\pm$	0.001 	&	0.61 	$\pm$	0.16 	&	3.11 	 $\pm$	0.26 	&	 0.4 	\\
030429	&	2.50 	$\pm$	0.08 	&	ISMI-ISMII,windI-II	&	1.74 	$\pm$	0.30 	&	527.31 	$\pm$	152.79 	&	0.01 	$\pm$	0.00 	&	3.87 	$\pm$	1.11 	&	2.20 	 $\pm$	0.20 	&	 0.3 	\\
030723	&	2.32 	$\pm$	0.42 	&	ISMII,windII	&				&				&				&				&				&		\\
050319	&	2.48	$\pm$	0.84	&	ISMII	&				&				&				&				&				&		\\
050408	&	1.56	$\pm$	0.66	&	ISMII	&				&				&				&				&				&		\\
050502A	&	2.52	$\pm$	0.1	&	ISMI-ISMII,windI-windII	&				&				&				&				&				&		\\
050525A	&	2.04	$\pm$	0.16	&	ISMII,windI-windII	&	9.16 	$\pm$	0.80 	&	249.75 	$\pm$	565.53 	&	0.01 	$\pm$	0.02 	&	0.24 	$\pm$	0.14 	&	2.52 	 $\pm$	0.40 	&	 3.5 	\\
050730	&	2.04	$\pm$	0.1	&	ISMII	&				&				&				&				&				&		\\
050801	&	3	$\pm$	0.32	&	ISMI-ISMII,windI-windII	&				&				&				&				&				&		\\
050820A	&	2.44	$\pm$	0.06	&	ISMI-ISMII	&	112.05 	$\pm$	54.97 	&	1293.70 	$\pm$	305.85 	&	0.17 	$\pm$	0.02 	&	1.97 	$\pm$	0.23 	&	3.16 	$\pm$	 0.18 	&	8.0 	 \\
050922C	&	2.02	$\pm$	0.1	&	ISMII,windI-windII	&	9.93 	$\pm$	1.06 	&	2825.85 	$\pm$	4671.10 	&	0.004 	$\pm$	0.001 	&	1.18 	$\pm$	0.35 	&	1.66 	 $\pm$	0.19 	 &	0.4 	\\
051109A	&	2.4	$\pm$	0.1	&	ISMI-ISMII,windI-windII	&	8.52 	$\pm$	1.64 	&	667.87 	$\pm$	1074.51 	&	0.004 	$\pm$	0.001 	&	0.28 	$\pm$	0.10 	&	1.66 	 $\pm$	0.18 	&	 1.3 	\\
051111	&	1.52	$\pm$	0.14	&	ISMI-ISMII	&	10.99 	$\pm$	3.34 	&	81.60 	$\pm$	6455.62 	&	0.002 	$\pm$	0.004 	&	0.01 	$\pm$	0.03 	&	1.03 	$\pm$	 0.40 	&	11.9 	 \\
051221A	&	2.28	$\pm$	0.2	&	ISMII,windI-windII	&	0.31 	$\pm$	0.18 	&	22.24 	$\pm$	15.10 	&	0.002 	$\pm$	0.001 	&	0.11 	$\pm$	0.02 	&	5.64 	$\pm$	 0.47 	&	1.4 	 \\
060111B	&	1.4	$\pm$	0.2	&	ISMI	&				&				&				&				&				&		\\
060206	&	2.46	$\pm$	0.1	&	ISMII,windI-windII	&	4.78 	$\pm$	2.23 	&	1939.97 	$\pm$	2156.28 	&	0.004 	$\pm$	0.001 	&	1.78 	$\pm$	0.23 	&	2.45 	 $\pm$	0.18 	 &	0.2 	\\
060418	&	2.56	$\pm$	0.18	&	ISMII,windI-windII	&	9			&	186.78 	$\pm$	23.86 	&	0.11 	$\pm$	0.06 	&	2.37 	$\pm$	1.25 	&	2.89 	$\pm$	0.41 	 &	4.6 	\\
060526	&	2.02	$\pm$	0.64	&	ISMII,windI-windII	&				&				&				&				&				&		\\
060605	&	3.12	$\pm$	0	&	ISMI-ISMII,windI-windII	&				&				&				&				&				&		\\
060729	&	2.56	$\pm$	0.06	&	ISMII,windI-windII	&				&				&				&				&				&		\\
061126	&	2.64	$\pm$	0.18	&	ISMI-ISMII,windI-windII	&	15.18 	$\pm$	6.72 	&	173.56 	$\pm$	74.75 	&	0.06 	$\pm$	0.01 	&	0.67 	$\pm$	0.09 	&	5.02 	 $\pm$	0.34 	 &	8.0 	\\
070411	&	1.5	$\pm$	0	&	ISMI,windI	&				&				&				&				&				&		\\
070419A	&	1.82	$\pm$	0.34	&	ISMII	&				&				&				&				&				&		\\
070518	&	2.6	$\pm$	0	&	ISMI-ISMII,windI-windII	&				&				&				&				&				&		\\
071003	&	2.88	$\pm$	0.06	&	ISMI-ISMII,windI-windII	&	35.25 	$\pm$	4.46 	&	92.83 	$\pm$	21089.25 	&	0.001 	$\pm$	0.002 	&	0.002 	$\pm$	0.007 	&	 0.36 	$\pm$	0.12 	 &	27.5 	\\
071003	&	2.88	$\pm$	0.06	&	ISMI-ISMII,windI-windII	&	35.25 	$\pm$	4.46 	&	1771.25 	$\pm$	3737.62 	&	0.25 	$\pm$	0.05 	&	1.24 	$\pm$	0.24 	&	 2.14 	$\pm$	 0.20 	&	2.0 	\\
080310	&	2.74	$\pm$	0.06	&	ISMII,windI-windII	&				&				&				&				&				&		\\
080319B	&	2.02	$\pm$	0.52	&	ISMII,windI-windII	&	131.95 	$\pm$	9.08 	&	1918.72 	$\pm$	1682.47 	&	0.01 	$\pm$	0.09 	&	0.18 	$\pm$	1.24 	&	 0.80 	$\pm$	0.69 	 &	6.4 	\\
080319B	&	2.02	$\pm$	0.52	&	ISMII,windI-windII	&	131.95 	$\pm$	9.08 	&	1918.72 	$\pm$	492.02 	&	1.26 	$\pm$	0.20 	&	2.29 	$\pm$	0.36 	&	7.93 	 $\pm$	0.93 	 &	6.4 	\\
080413A	&	2.04	$\pm$	0.74	&	ISMII	&	7.83 	$\pm$	3.55 	&	4081.98 	$\pm$	1941.55 	&	0.0002 	$\pm$	0.0003 	&	0.10 	$\pm$	1.75 	&	0.39 	$\pm$	 0.69 	&	0.2 	 \\
080413B	&	1.56	$\pm$	0.14	&	ISMII	&	1.65 	$\pm$	0.46 	&	1050.09	$\pm$	980.53 	&	0.0001	$\pm$	0.0001	&	0.4	$\pm$	0.4	&	0.39 	$\pm$	0.23 	&	0.2 	 \\
080413B	&	1.56	$\pm$	0.14	&	ISMII	&	1.65 	$\pm$	0.46 	&	11952.73 	$\pm$	13544.90 	&	0.001 	$\pm$	0.001 	&	7.68 	$\pm$	3.19 	&	2.06 	$\pm$	 0.31 	&	0.01 	 \\
080603A	&	2.96	$\pm$	0.08	&	ISMI-ISMII,windI-windII	&	15.26 	$\pm$	3.07 	&	471.05 	$\pm$	158.86 	&	0.03 	$\pm$	0.01 	&	1.00 	$\pm$	0.12 	&	3.74 	 $\pm$	0.22 	 &	3.1 	\\
080710	&	2.6	$\pm$	0.18	&	ISMI-ISMII	&				&				&				&				&				&		\\
081008	&	1.8	$\pm$	0.46	&	ISMII	&				&				&				&				&				&		\\
081203A	&	2.2	$\pm$		&	windI-windII	&	36.13 	$\pm$	18.42 	&	3515.60 	$\pm$	4232.23 	&	0.01 	$\pm$	0.01 	&	0.51 	$\pm$	0.36 	&	0.98 	$\pm$	 0.15 	&	1.0 	 \\
090426	&	1.52	$\pm$	0.28	&	ISMI	&				&				&				&				&				&		\\
090426	&	1.52	$\pm$	0.28	&	ISMI	&				&				&				&				&				&		\\
090618	&	2	$\pm$	0.1	&	ISMII,windI-windII	&	25.30 	$\pm$	1.28 	&	171.38 	$\pm$	136.81 	&	0.03 	$\pm$	0.01 	&	0.21 	$\pm$	0.09 	&	2.82 	$\pm$	 0.37 	&	12.9 	 \\
090926A	&	2.44	$\pm$	0.34	&	windI-windII	&	185.16 	$\pm$	9.33 	&	1030.94 	$\pm$	358.63 	&	1.16 	$\pm$	0.10 	&	6.47 	$\pm$	0.54 	&	6.42 	 $\pm$	0.42 	&	 15.2 	\\
091029	&	1.36	$\pm$	0	&	ISMII	&	1.62 	$\pm$	0.03 	&	47406.78 	$\pm$	38926.07 	&	0.002 	$\pm$	0.002 	&	11.07 	$\pm$	9.98 	&	1.24 	$\pm$	 0.92 	&	0.0 	 \\
091127	&	1.98	$\pm$	0.24	&	ISMII	&	9.81 	$\pm$	4.60 	&	20287.63 	$\pm$	6160.40 	&	0.001 	$\pm$	0.003 	&	7.43 	$\pm$	3.24 	&	1.55 	$\pm$	 0.19 	&	0.0 	 \\
100219A	&	2.2	$\pm$	0.24	&	ISMII,windI-windII	&				&				&				&				&				&		\\
100219A	&	2.2	$\pm$	0.24	&	ISMII,windI-windII	&				&				&				&				&				&		\\
110205A	&	2.96	$\pm$	0.16	&	ISMII,windI-windII	&	61.72 	$\pm$	7.80 	&	592.20 	$\pm$	471.13 	&	0.08 	$\pm$	0.02 	&	0.75 	$\pm$	0.15 	&	2.88 	 $\pm$	0.24 	&	 9.4 	\\
120729A	&	2.01	$\pm$	0.2	&	ISMI-ISMII,windI-windII	&	1.24 	$\pm$	0.27 	&	716.77 	$\pm$	4647.10 	&	0.0003 	$\pm$	0.0004 	&	0.15 	$\pm$	0.22 	&	1.17 	 $\pm$	0.30 	 &	0.2 	\\
130427A	&	2.38	$\pm$	0.02	&	ISMII,windI-windII	&	81.33 	$\pm$	0.45 	&	910.72 	$\pm$	420.37 	&	0.33 	$\pm$	0.05 	&	0.72 	$\pm$	0.11 	&	4.98 	 $\pm$	0.32 	&	 8.2 	\\
130603B	&	2.68	$\pm$	0.2	&	ISMII,windI-windII	&	0.20 	$\pm$	0.02 	&	6.23 	$\pm$	24.10 	&	0.001 	$\pm$	0.001 	&	0.02 	$\pm$	0.01 	&	4.47 	$\pm$	 0.68 	&	3.2 	 \\

\enddata
\tablenotetext{a}{ISMI: the ISM model in the spectral regime I ($\nu>\nu_{c}$);
ISMII: the ISM model in the spectral regime II ($\nu_{m}<\nu<\nu_{c}$);
windI: the wind model in the spectral regime I;
windII: the wind model in the spectral regime II;}
%\tablenotetext{b}{cosmological rest-frame spectral peak energy, $E_{\rm p,z}=E_{\rm p,obs}\times(1+z)$, in unit of keV;}
\tablenotetext{b}{In units of $10^{52} erg$. $E_{\rm \gamma,iso}$ and $E_{\rm K,iso}$ is the isotropic $\gamma$-ray energy and kinetic energy, respectively;}
\tablenotetext{c}{In units of $10^{51} erg$. $E_{\rm \gamma}$ and $E_{\rm K}$ is jet-corrected $\gamma$-ray energy and kinetic energy, respectively;}
\tablenotetext{d}{In units of \%}

\label{table:result}
\end{deluxetable}

\begin{deluxetable}{ccccccccccccccccccccccccc}
%\tabletypesize{\scriptsize}
\tabletypesize{\tiny}
\tablecaption{GRBs with a Lower Limit Jet Break time and Their Derived Parameters. }
\tablewidth{0pt}
\tabcolsep=2.5pt
%\rotate \tabletypesize{\tiny}

\tablehead{ \colhead{GRB}&
\colhead{$p$}&
\colhead{Model\tablenotemark{a}}&
\colhead{$E_{\rm \gamma,iso}$}&
\colhead{$E_{\rm K,iso}$}&
\colhead{$E_{\rm \gamma}$}&
\colhead{$E_{\rm K}$}&
\colhead{$\theta_{\rm j}^{\circ}$}&
\colhead{$\eta_{\rm \gamma}$}.}

\startdata
020124	&	2.82 	$\pm$	0.28 	&	windII	&	10.23 	$\pm$	0.99 	&	2.05 	$\pm$	1.68 	&	0.07 	$\pm$	0.03 	&	3.83 	$\pm$	3.65 	&	2.05 	$\pm$	1.23 	 &	83.3 	\\
020813	&	2.70 	$\pm$	0.14 	&	ISMII,windI-windII	&	134.90 	$\pm$	19.99 	&	302.43 	$\pm$	55.23 	&	1.63 	$\pm$	0.70 	&	3.66 	$\pm$	1.57 	&	2.82 	 $\pm$	0.35 	&	 30.8 	\\
050401	&	2	$\pm$	0.4	&	ISMII,windI-windII	&	41.76 	$\pm$	2.15 	&	310.95 	$\pm$	147.79 	&	0.32 	$\pm$	0.03 	&	2.38 	$\pm$	0.22 	&	7.09 	$\pm$	 0.53 	&	11.8 	 \\
050416A	&	2.84	$\pm$	0.6	&	ISMII,windI-windII	&				&				&				&				&				&		\\
050603	&	2.42	$\pm$	0.2	&	windII	&	61.12 	$\pm$	1.18 	&	115.32 	$\pm$	79.53 	&	1.79 	$\pm$	1.64 	&	3.38 	$\pm$	2.85 	&	4.39 	$\pm$	3.59 	&	 34.6 	\\
050721	&	2.32	$\pm$	0.7	&	ISMI	&				&				&				&				&				&		\\
051028	&	2.2	$\pm$	0	&	ISMII,windI-windII	&	19.93 	$\pm$	3.31 	&	176.82 	$\pm$	55.27 	&	0.02 	$\pm$	0.01 	&	0.15 	$\pm$	0.03 	&	2.34 	$\pm$	 0.19 	&	10.1 	 \\
060210	&	1.74	$\pm$	0.16	&	windII	&				&				&				&				&				&		\\
060512	&	1.48	$\pm$	0.4	&	windII	&	6.92 	$\pm$	4.11 	&	5020.68 	$\pm$	4876.35 	&	0.02 	$\pm$	0.00 	&	11.51 	$\pm$	0.52 	&	3.88 	$\pm$	 0.14 	&	0.1 	 \\
060714	&	3.04	$\pm$	0.1	&	ISMII,windI-windII	&	99.80 	$\pm$	7.10 	&	151.53 	$\pm$	59.78 	&	0.30 	$\pm$	0.04 	&	0.46 	$\pm$	0.06 	&	4.47 	$\pm$	 0.28 	&	39.7 	 \\
060904B	&	2.22	$\pm$	0.2	&	ISMI,windI	&				&				&				&				&				&		\\
060906	&	2.12	$\pm$	0.04	&	ISMII,windI-windII	&				&				&				&				&				&		\\
060908	&	2.96	$\pm$	0.84	&	ISMI-ISMII,windI-windII	&	13.84 	$\pm$	3.85 	&	26.20 	$\pm$	20.44 	&	0.00 	$\pm$	0.01 	&	0.01 	$\pm$	0.02 	&	1.41 	 $\pm$	0.86 	 &	34.6 	\\
060912A	&	2.2	$\pm$	0.3	&	ISMII,windI-windII	&				&				&				&				&				&		\\
060927	&	2.22	$\pm$	0.1	&	windII	&				&				&				&				&				&		\\
061007	&	2.36	$\pm$	0.04	&	windI-windII	&	124.07 	$\pm$	11.96 	&	3024.47 	$\pm$	13.32 	&	0.47 	$\pm$	0.01 	&	11.55 	$\pm$	0.37 	&	5.01 	 $\pm$	0.15 	&	 3.9 	\\
070110	&	2.1	$\pm$	0.08	&	windI-windII	&				&				&				&				&				&		\\
070306	&	1.86	$\pm$	0	&	ISMII,windI-windII	&				&				&				&				&				&		\\
070311$^{a}$	&				&		&				&				&				&				&				&		\\
070318	&	2.56	$\pm$	0.2	&	ISMII	&				&				&				&				&				&		\\
070611$^{a}$	&				&		&				&				&				&				&				&		\\
071025	&	2.92	$\pm$	0.28	&	ISMII,windI-windII	&				&				&				&				&				&		\\
071031	&	2.48	$\pm$	0.44	&	ISMII,windI-windII	&	21.18 	$\pm$	2.05 	&	126.32 	$\pm$	46.73 	&	0.28 	$\pm$	0.02 	&	1.68 	$\pm$	0.14 	&	3.37 	 $\pm$	0.64 	&	 14.4 	\\
071112C	&	2.26	$\pm$	0.58	&	ISMII,windI-windII	&				&				&				&				&				&		\\
080319A$^{a}$	&		$\pm$		&		&				&				&				&				&				&		\\
080319C	&	1.7	$\pm$	0.1	&	ISMI,windI	&	13.83 	$\pm$	3.85 	&	26.20 	$\pm$	19.52 	&	0.00 	$\pm$	0.01 	&	0.01 	$\pm$	0.02 	&	1.41 	$\pm$	0.86 	&	 34.5 	\\
080721	&	1.72	$\pm$	0	&	windII	&				&				&				&				&				&		\\
080804	&	2.4	$\pm$	0.8	&	ISMII,windI-windII	&	9.62 	$\pm$	2.30 	&	1.73 	$\pm$	176.07 	&	0.01 	$\pm$	0.01 	&	0.22 	$\pm$	0.13 	&	2.94 	$\pm$	 0.78 	&	84.8 	 \\
080913	&	1.58	$\pm$	0.06	&	ISMI,windI	&	8.44 	$\pm$	1.55 	&	178.07 	$\pm$	142.89 	&	0.06 	$\pm$	0.03 	&	1.20 	$\pm$	0.54 	&	2.11 	$\pm$	 0.23 	&	4.5 	 \\
080928	&	3.64	$\pm$	0.04	&	ISMII,windI-windII	&				&				&				&				&				&		\\
090102	&	2.48	$\pm$	0.44	&	windII	&	22.74 	$\pm$	2.12 	&	1463.07 	$\pm$	864.83 	&	0.98 	$\pm$	0.05 	&	62.82 	$\pm$	3.01 	&	5.31 	$\pm$	 0.20 	&	1.5 	 \\
090323	&	2.48	$\pm$	0.3	&	windII	&	372.28 	$\pm$	16.86 	&	568.43 	$\pm$	278.20 	&	2.26 	$\pm$	1.90 	&	3.46 	$\pm$	2.91 	&	2.00 	$\pm$	0.45 	&	 39.6 	\\
090328	&	2.04	$\pm$	0.04	&	ISMII,windI-windII	&	177.51 	$\pm$	12.10 	&	12.00 	$\pm$	49.39 	&	0.38 	$\pm$	0.06 	&	0.48 	$\pm$	0.07 	&	3.77 	 $\pm$	0.24 	&	 93.7 	\\
090510	&	2.36	$\pm$	0.1	&	ISMI-ISMII	&				&				&				&				&				&		\\
090812	&	1.88	$\pm$	0.08	&	windI-windII	&				&				&				&				&				&		\\
100418A	&	2.4	$\pm$	0.2	&	ISMII,windI-windII	&				&				&				&				&				&		\\
101024A	&	2.28	$\pm$	0.1	&	ISMI-ISMII	&				&				&				&				&				&		\\
110918A	&	2.26	$\pm$	0.58	&	ISMII,windI-windII	&				&				&				&				&				&		\\
120326A	&	2.5	$\pm$	0.16	&	windII	&	3.18 	$\pm$	0.04 	&	149.75 	$\pm$	101.34 	&	0.03 	$\pm$	0.02 	&	1.31 	$\pm$	0.97 	&	2.40 	$\pm$	2.15 	&	 2.1 	\\
120711A	&	2.96	$\pm$	0.18	&	ISMII,windI-windII	&				&				&				&				&				&		\\
120815A$^{a}$	&				&		&				&				&				&				&				&		\\

\enddata
\tablenotetext{a}{The detected lightcurves still in the shallow decay phase}

%\tablenotetext{e}{The name of the experiment(s), or of the satellite(s), that provided the estimates of spectral parameters and energy, BA =BATSE, HET=HETE-2, KW=Konus-Wind, SAX=\emph{BeppoSAX}, SUZ=Suzaku;}
%\tablenotetext{f}{Here the spectral peak energy is on observe-frame, since no observed red-shift $z$ are available;}
%\tablenotetext{g}{The energy of the second jet component for the two-component jet bursts.}

\label{table:result1}
\end{deluxetable}

\begin{deluxetable}{ccccccccccccccccccccccccc}
%\tabletypesize{\scriptsize}
\tabletypesize{\tiny}
\tablecaption{GRBs with an Upper Limit Jet Break Time and Their Derived Parameters. }
\tablewidth{0pt}
\tabcolsep=2.5pt
%\rotate \tabletypesize{\tiny}

\tablehead{ \colhead{GRB}&
\colhead{$p$}&
\colhead{Model\tablenotemark{a}}&
\colhead{$E_{\rm \gamma,iso}$}&
\colhead{$E_{\rm K,iso}$}&
\colhead{$E_{\rm \gamma}$}&
\colhead{$E_{\rm K}$}&
\colhead{$\theta_{\rm j}^{\circ}$}&
\colhead{$\eta_{\rm \gamma}$}.}

\startdata
070125	&	2.18 	$\pm$	0.20 	&	ISMII,windI-II	&	93.92	$\pm$	10.13	&	209.85 	$\pm$	123.74 	&	3.20 	$\pm$	0.50 	&	7.04 	$\pm$	1.02 	&	4.70 	$\pm$	 &	0.29 	30.9 	 \\
071010A	&	2.22 	$\pm$	0.24 	&	ISMII,windII	&				&				&				&				&			&			\\
100901A	&	2.04 	$\pm$	0.20 	&	ISMII,windI-II	&				&				&				&				&			&			\\

\enddata
%\tablenotetext{a}{ISMI: the ISM model in the spectral regime I ($\nu>\nu_{c}$);
%ISMII: the ISM model in the spectral regime II ($\nu_{m}<\nu<\nu_{c}$);
%windI: the wind model in the spectral regime I;
%windII: the wind model in the spectral regime II;}
%%\tablenotetext{b}{cosmological rest-frame spectral peak energy, $E_{\rm p,z}=E_{\rm p,obs}\times(1+z)$, in unit of keV;}
%\tablenotetext{b}{In units of $10^{52} erg$. $E_{\rm \gamma,iso}$ and $E_{\rm K,iso}$ is the isotropic $\gamma$-ray energy and kinetic energy, respectively;}
\tablenotetext{c}{In units of $10^{51} erg$. $E_{\rm \gamma}$ and $E_{\rm K}$ is jet-corrected $\gamma$-ray energy and kinetic energy, respectively;}
%\tablenotetext{e}{The name of the experiment(s), or of the satellite(s), that provided the estimates of spectral parameters and energy, BA =BATSE, HET=HETE-2, KW=Konus-Wind, SAX=\emph{BeppoSAX}, SUZ=Suzaku;}
%\tablenotetext{f}{Here the spectral peak energy is on observe-frame, since no observed red-shift $z$ are available;}
%\tablenotetext{g}{The energy of the second jet component for the two-component jet bursts.}

\label{table:result2}
\end{deluxetable}

\begin{deluxetable}{ccccccccccccccccccccccccc}
%\tabletypesize{\scriptsize}
\tabletypesize{\tiny}
\tablecaption{Results of Our Linear Regression Analysis for the Luminosity Correlations. }
\tablewidth{0pt}
\tabcolsep=2.5pt
%\rotate \tabletypesize{\tiny}

\tablehead{ \colhead{Relations}&
\colhead{Expressions}&
\colhead{$r$\tablenotemark{c}}&
\colhead{$p$\tablenotemark{d}}&
\colhead{$\delta$\tablenotemark{e}}}

\startdata
$E_{\rm p,z}(E_{\rm \gamma,iso})$ (\emph{Amati})	&	$\frac{E_{\rm p,z}}{\rm 100 keV}\simeq(0.63\pm0.31)(\frac{E_{\rm \gamma,iso}}{10^{52} \rm erg})^{(0.69\pm0.07)}$	&	0.81	&	 $<10^{-4}$	&	 0.33	\\
$E_{\rm p,z}(E_{\rm \gamma})$ (\emph{Ghirlanda})	&	 $\frac{E_{\rm p,z}}{\rm 100 keV}\simeq(7.9\pm4.8)(\frac{E_{\rm \gamma}}{10^{51} \rm erg})^{(0.44\pm0.17)}$	&	0.43	&	$0.005$	 &	-	\\
$E_{\rm p,z}(E_{\rm \gamma,iso},t_{\rm b,z})$ (\emph{Liang-Zhang})$^{a}$	&	 $\frac{E_{\rm p,z}}{\rm 100 keV}=(1.2\pm0.3) (\frac{E_{\rm \gamma,iso}}{10^{52} \rm erg})^{(0.56\pm0.07)}(\frac{t_{\rm b,z}}{\rm day})^{(0.67\pm0.08)}$	&	0.85	&	$<10^{-4}$	&	0.15	\\
$E_{\rm p,z}(E_{\rm \gamma,iso},t_{\rm b,z})$ (\emph{Liang-Zhang})$^{b}$	&	$\frac{E_{\rm p,z}}{\rm 100 keV}=(1.3\pm0.4) (\frac{E_{\rm \gamma,iso}}{10^{52} \rm erg})^{(0.49\pm0.07)}(\frac{t_{\rm b,z}}{\rm day})^{(-0.08\pm0.05)}$	&	0.67	&	$<10^{-4}$	&	0.22	\\
\enddata
\tablenotetext{a}{relations for the late time jet break sample;}
\tablenotetext{b}{relations for the entire sample,including the early and late time jet break sample;}
\tablenotetext{c}{$r$ is the spearman correlation coefficient;}
\tablenotetext{d}{$p$ is the change probability;}
\tablenotetext{e}{$\delta$ is the dispersion.}

\label{table:relations}
\end{deluxetable}

\clearpage
\setlength{\voffset}{-18mm}
\begin{figure*}

\includegraphics[angle=0,scale=0.2,width=0.19\textwidth,height=0.18\textheight]{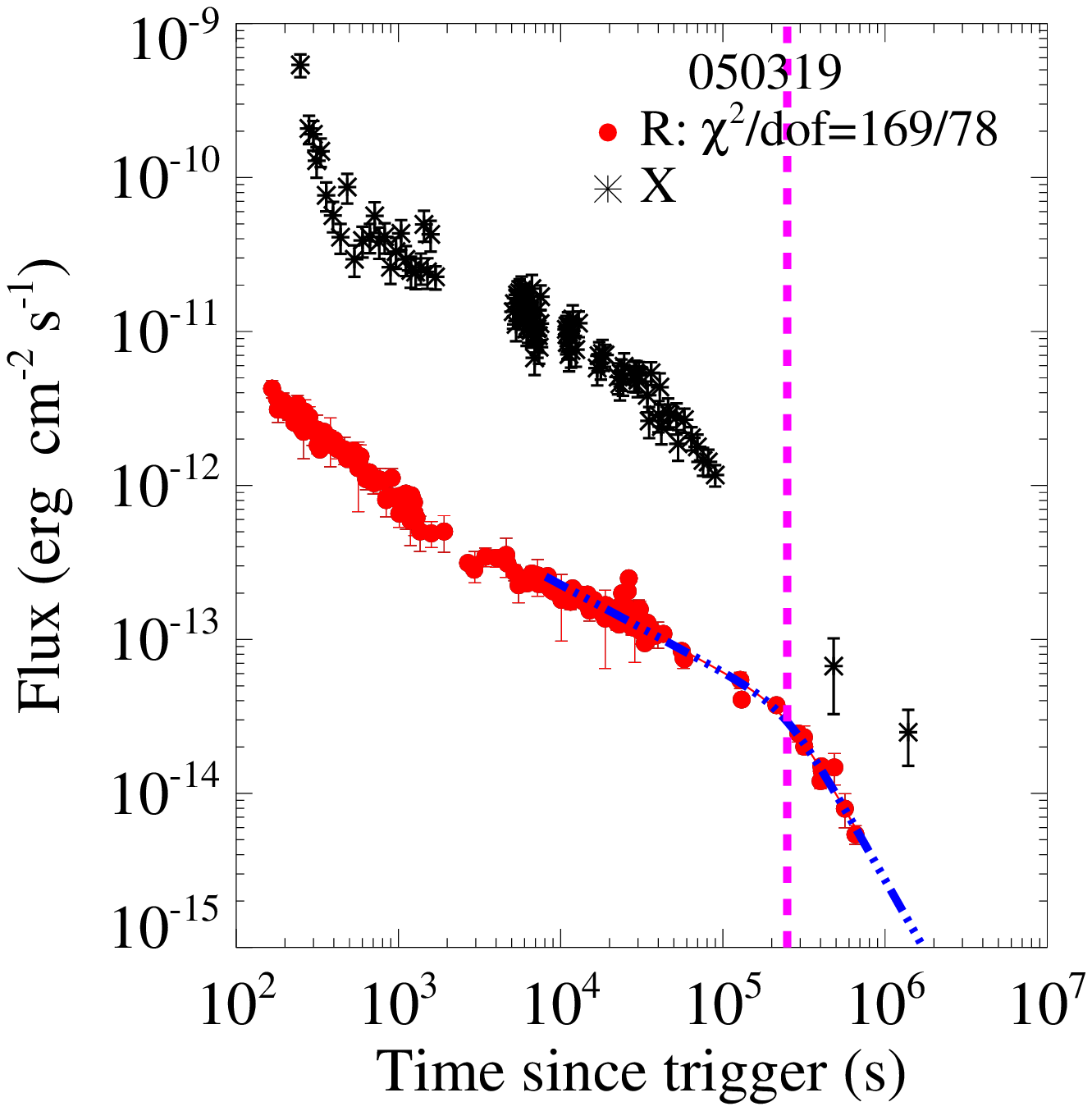}
\includegraphics[angle=0,scale=0.2,width=0.19\textwidth,height=0.18\textheight]{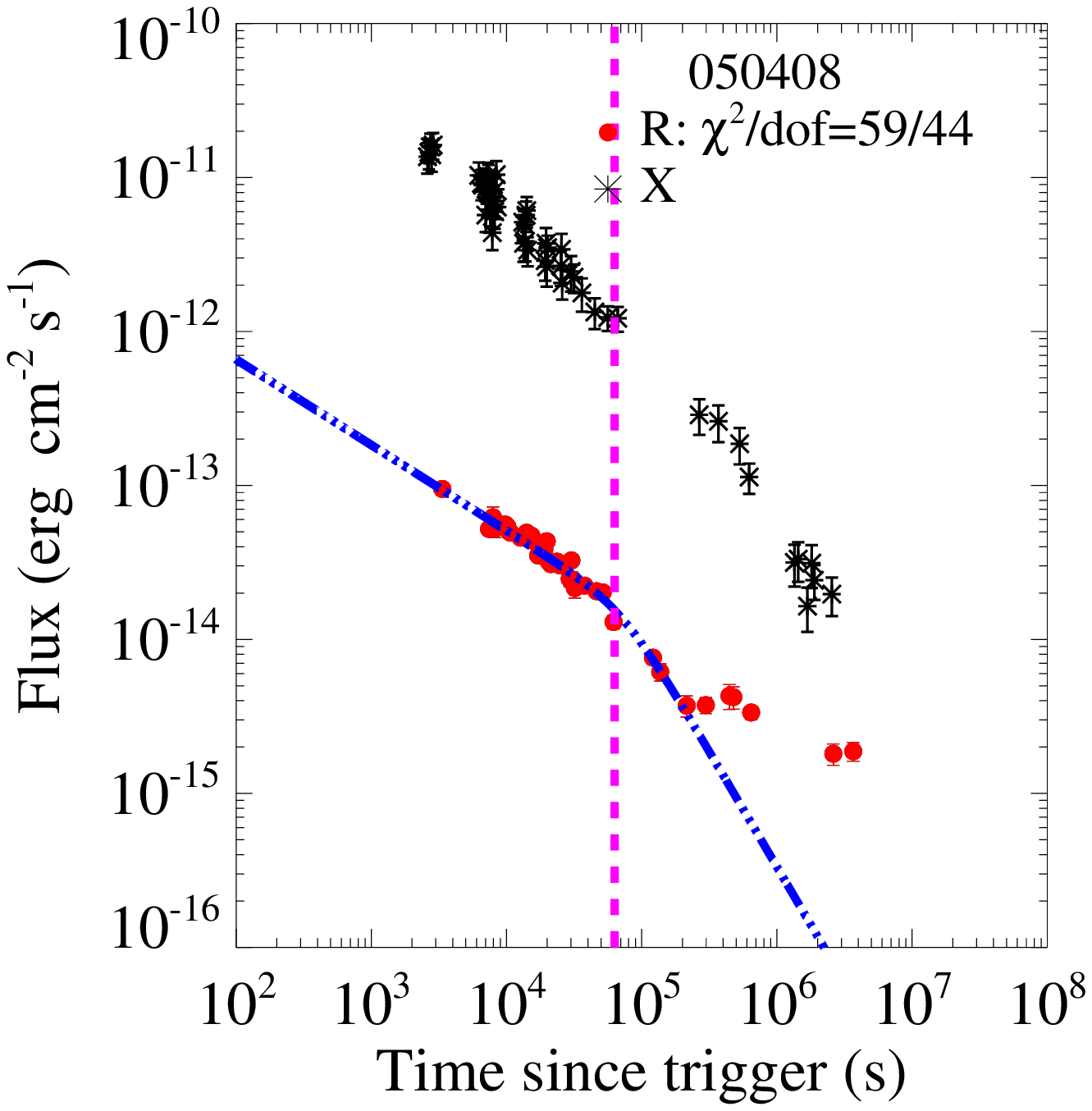}
\includegraphics[angle=0,scale=0.2,width=0.19\textwidth,height=0.18\textheight]{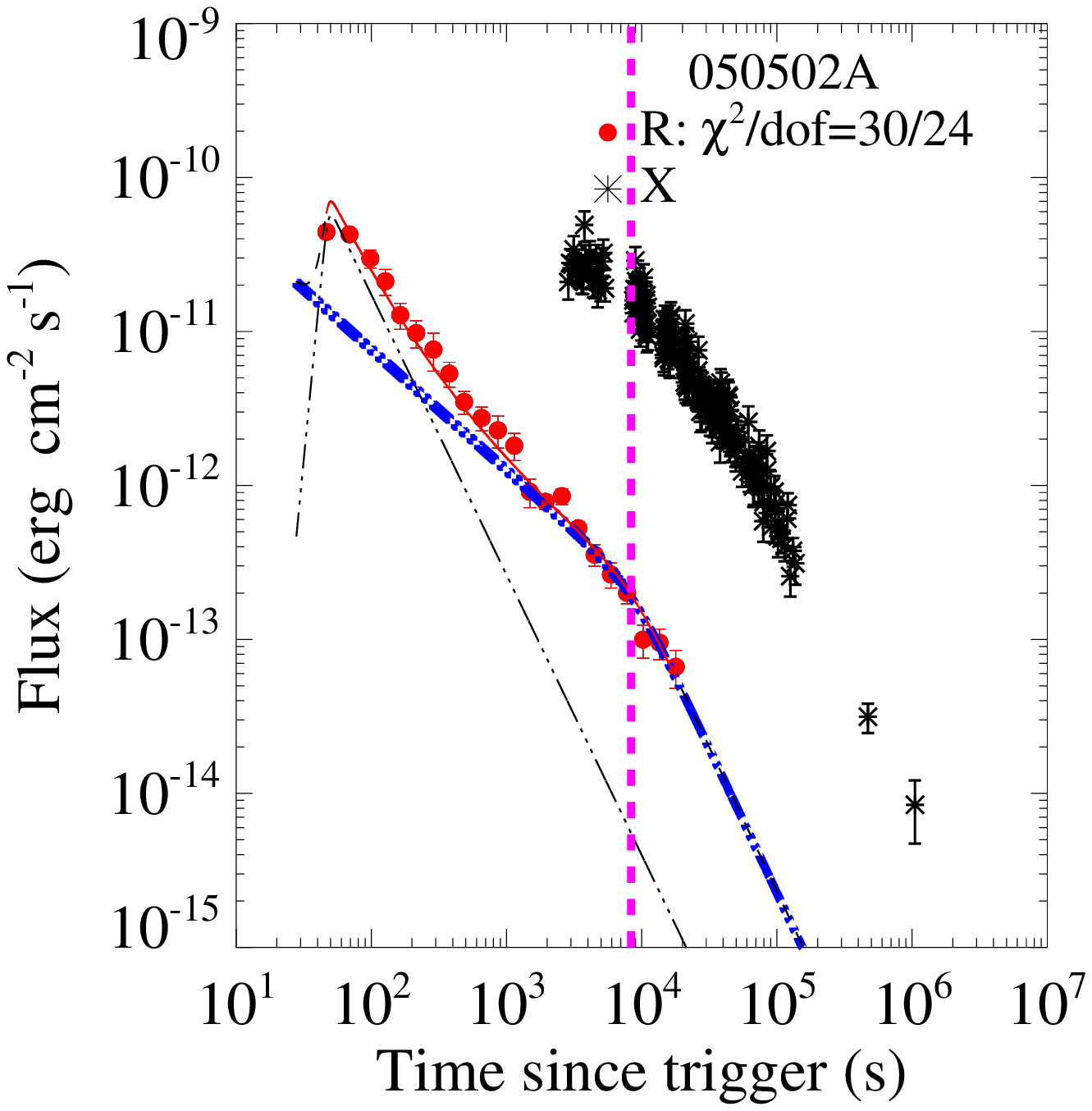}
\includegraphics[angle=0,scale=0.2,width=0.19\textwidth,height=0.18\textheight]{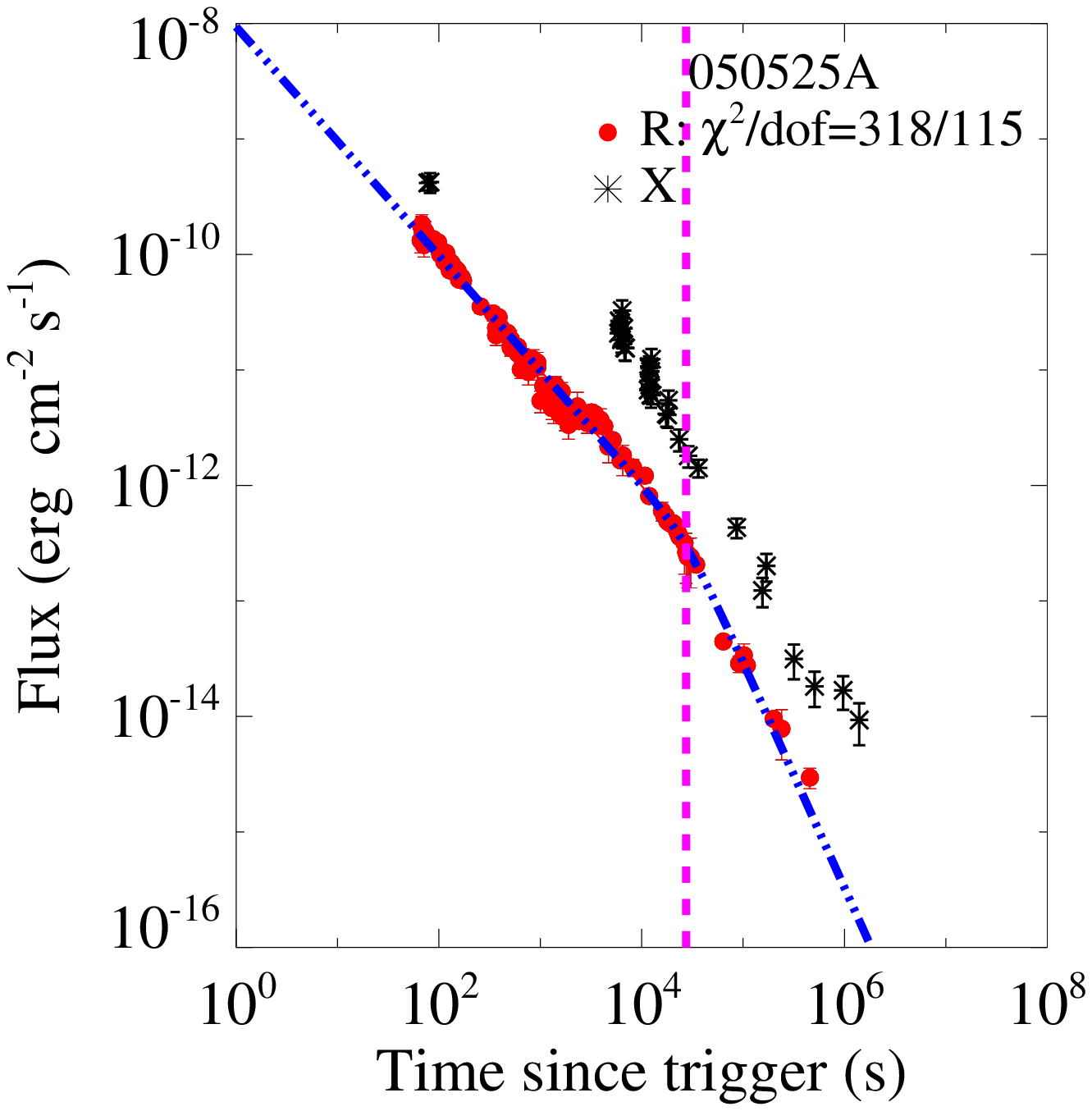}
\includegraphics[angle=0,scale=0.2,width=0.19\textwidth,height=0.18\textheight]{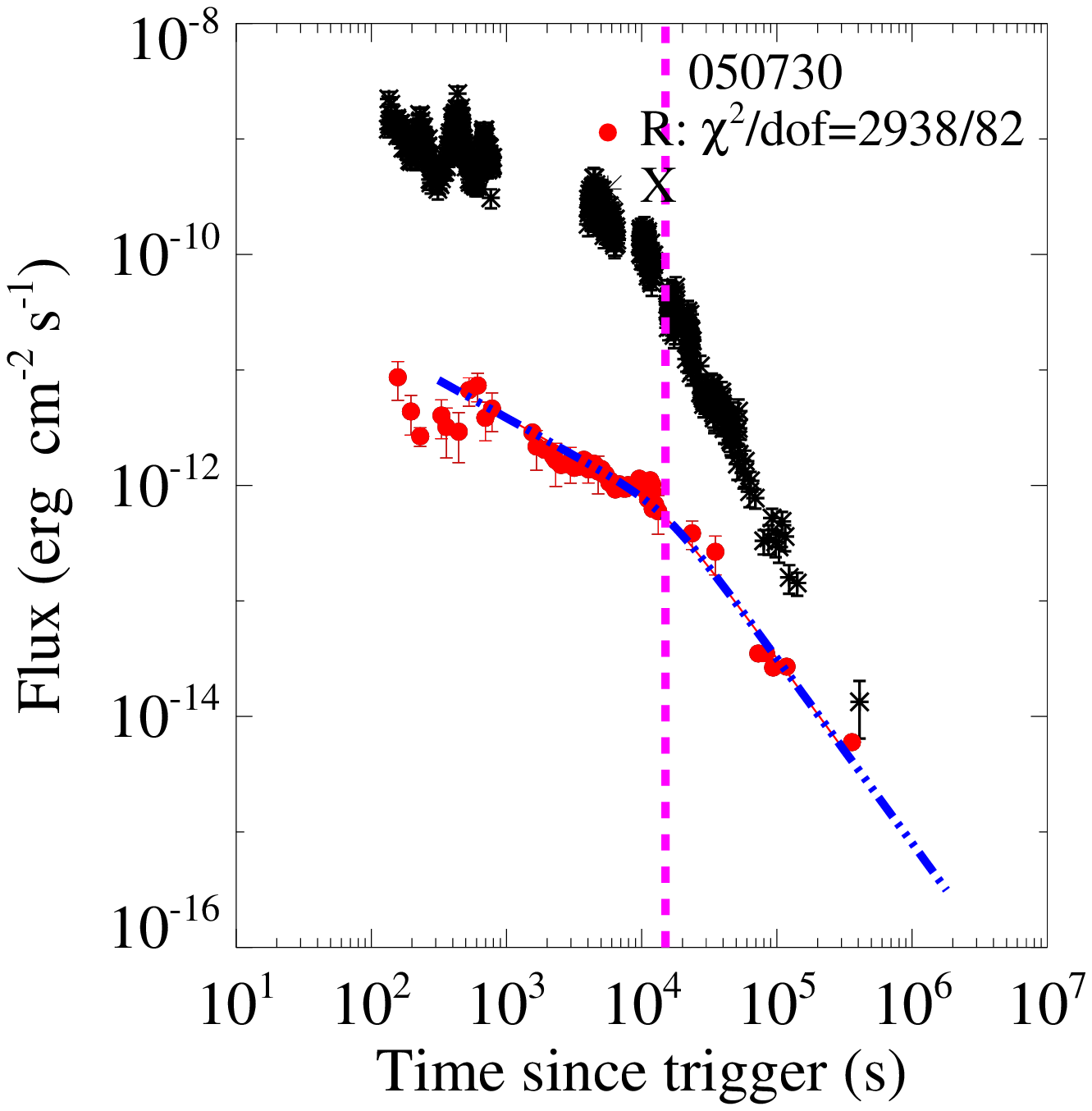}
\includegraphics[angle=0,scale=0.2,width=0.19\textwidth,height=0.18\textheight]{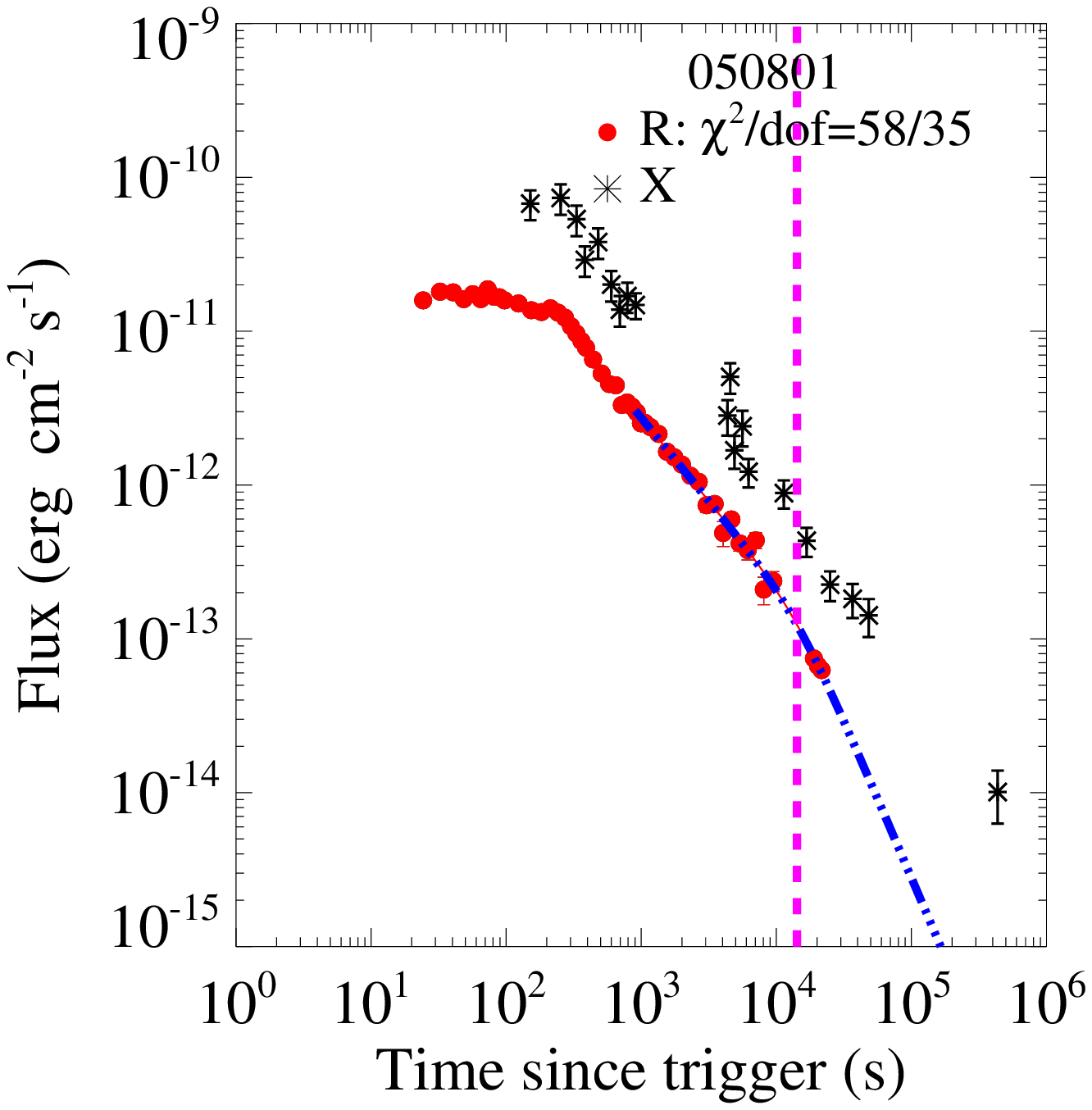}
\includegraphics[angle=0,scale=0.2,width=0.19\textwidth,height=0.18\textheight]{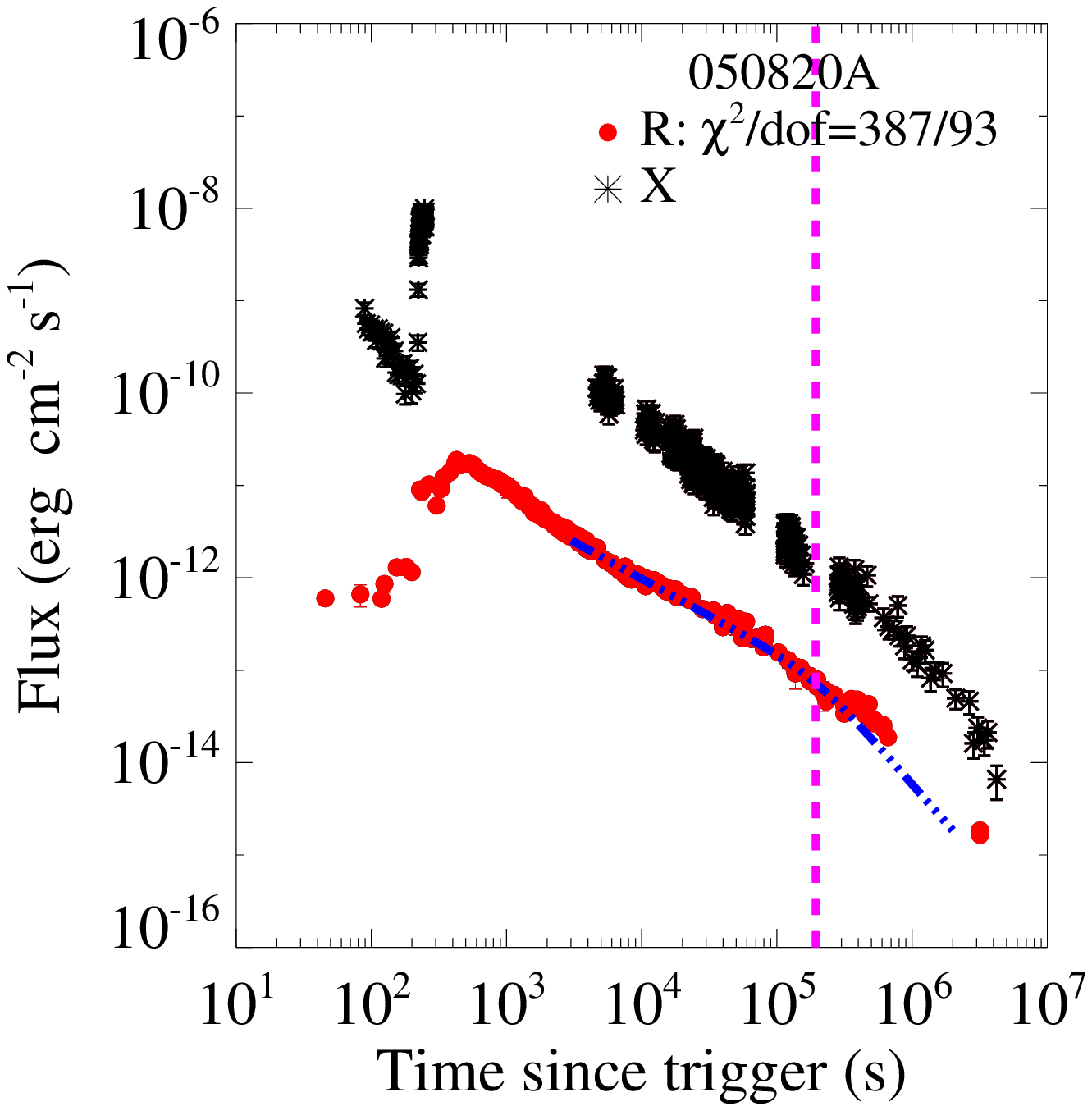}
\includegraphics[angle=0,scale=0.2,width=0.19\textwidth,height=0.18\textheight]{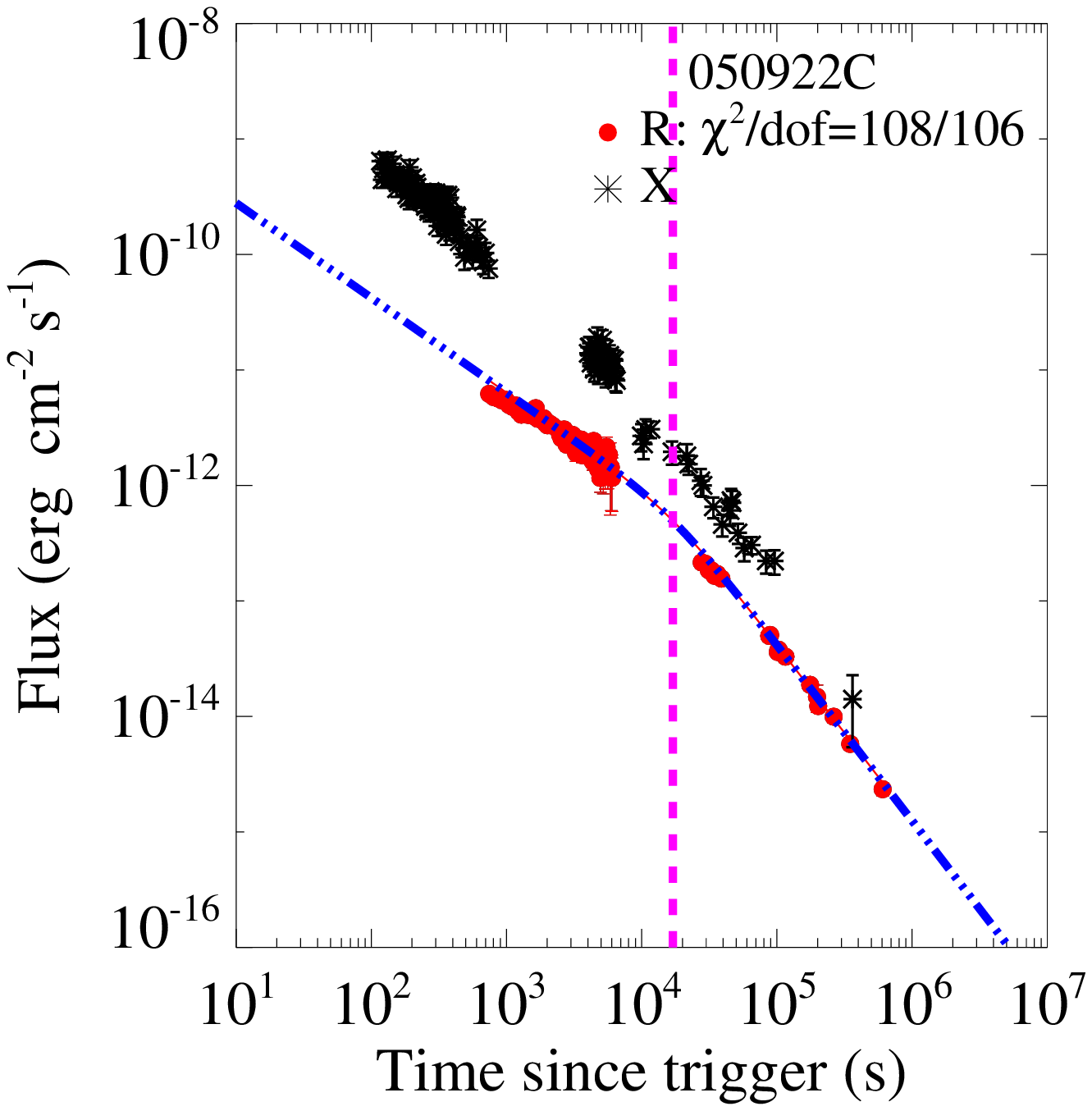}
\includegraphics[angle=0,scale=0.2,width=0.19\textwidth,height=0.18\textheight]{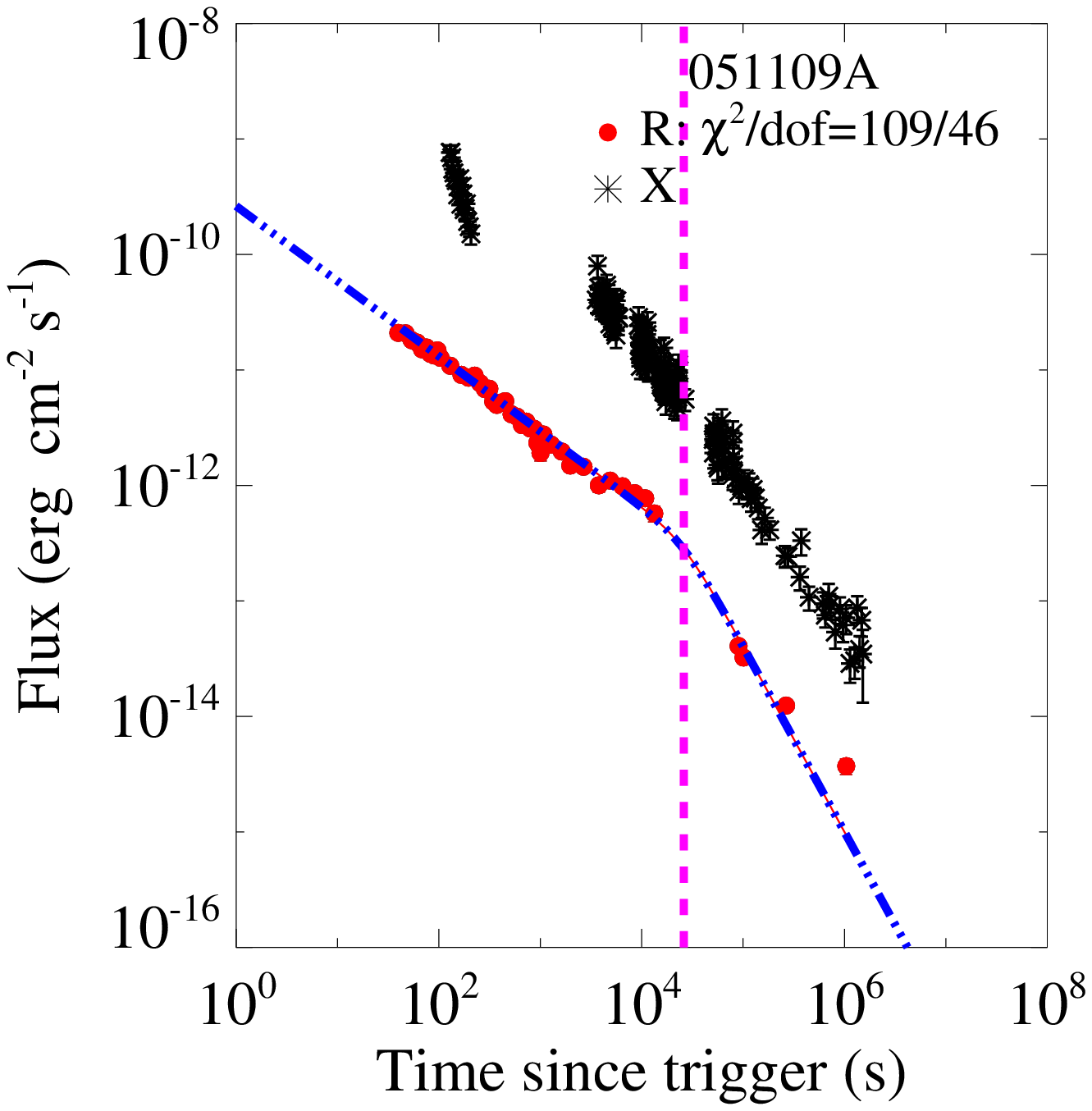}
\includegraphics[angle=0,scale=0.2,width=0.19\textwidth,height=0.18\textheight]{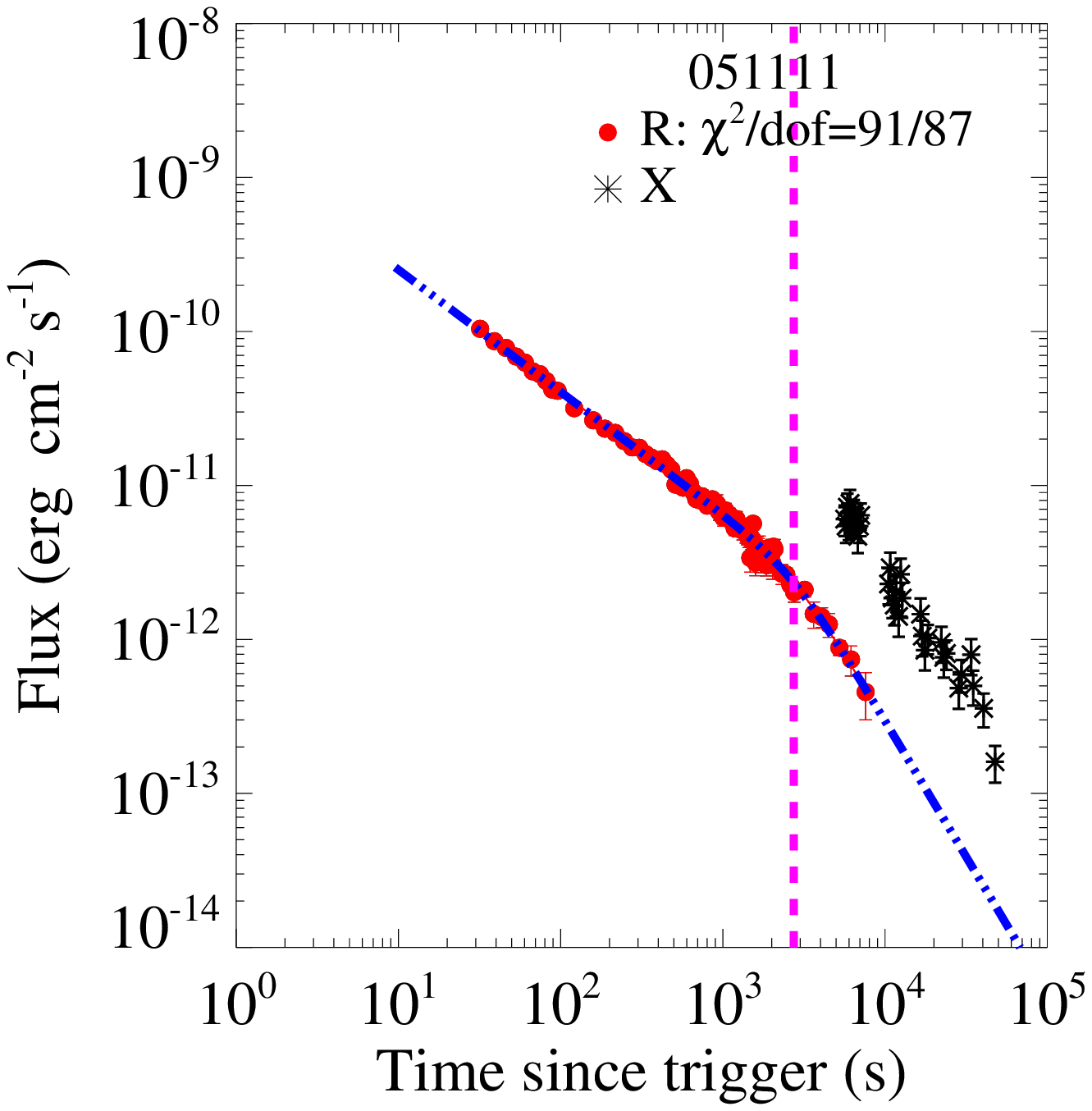}
\includegraphics[angle=0,scale=0.2,width=0.19\textwidth,height=0.18\textheight]{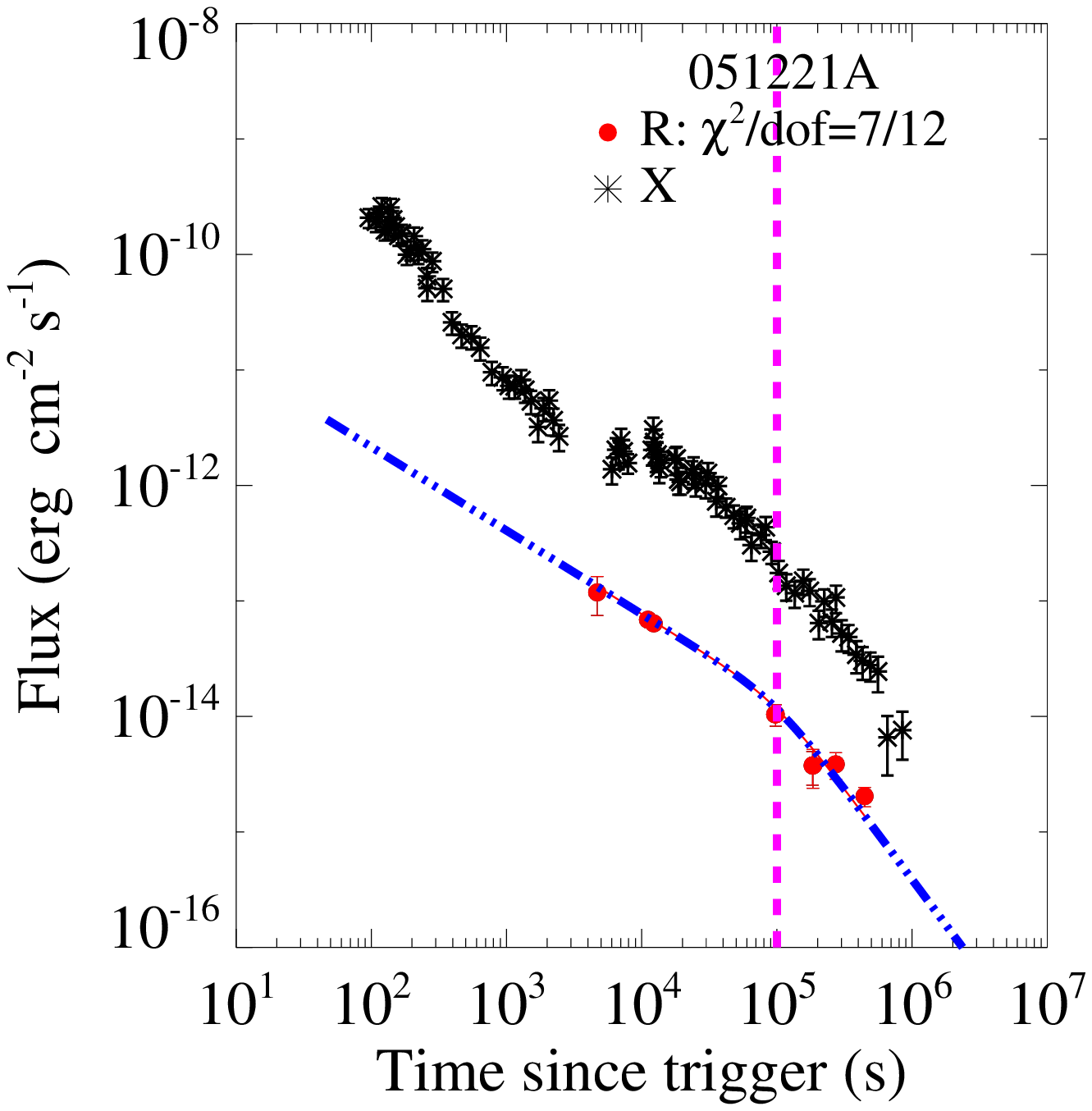}
\includegraphics[angle=0,scale=0.2,width=0.19\textwidth,height=0.18\textheight]{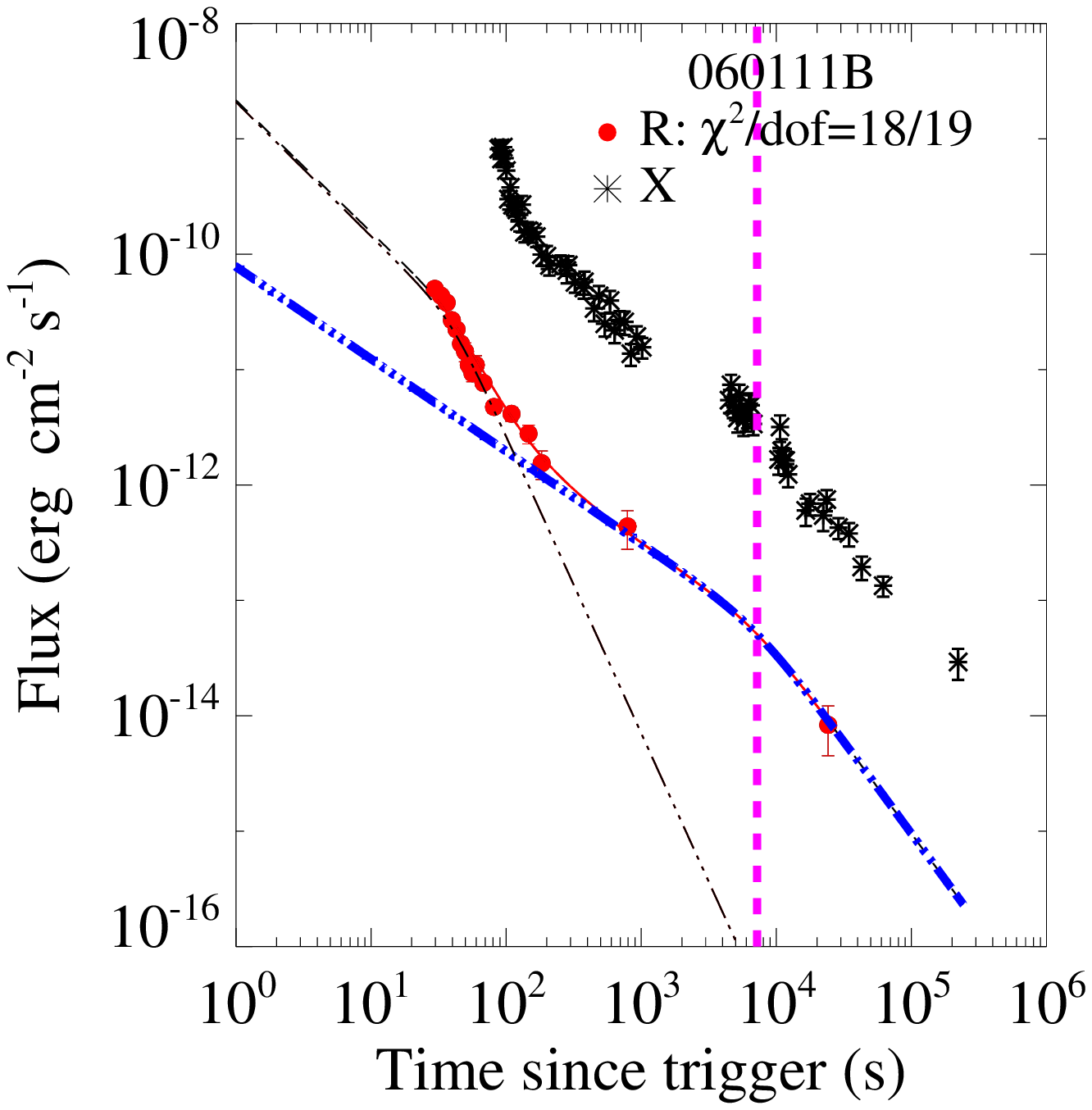}
\includegraphics[angle=0,scale=0.2,width=0.19\textwidth,height=0.18\textheight]{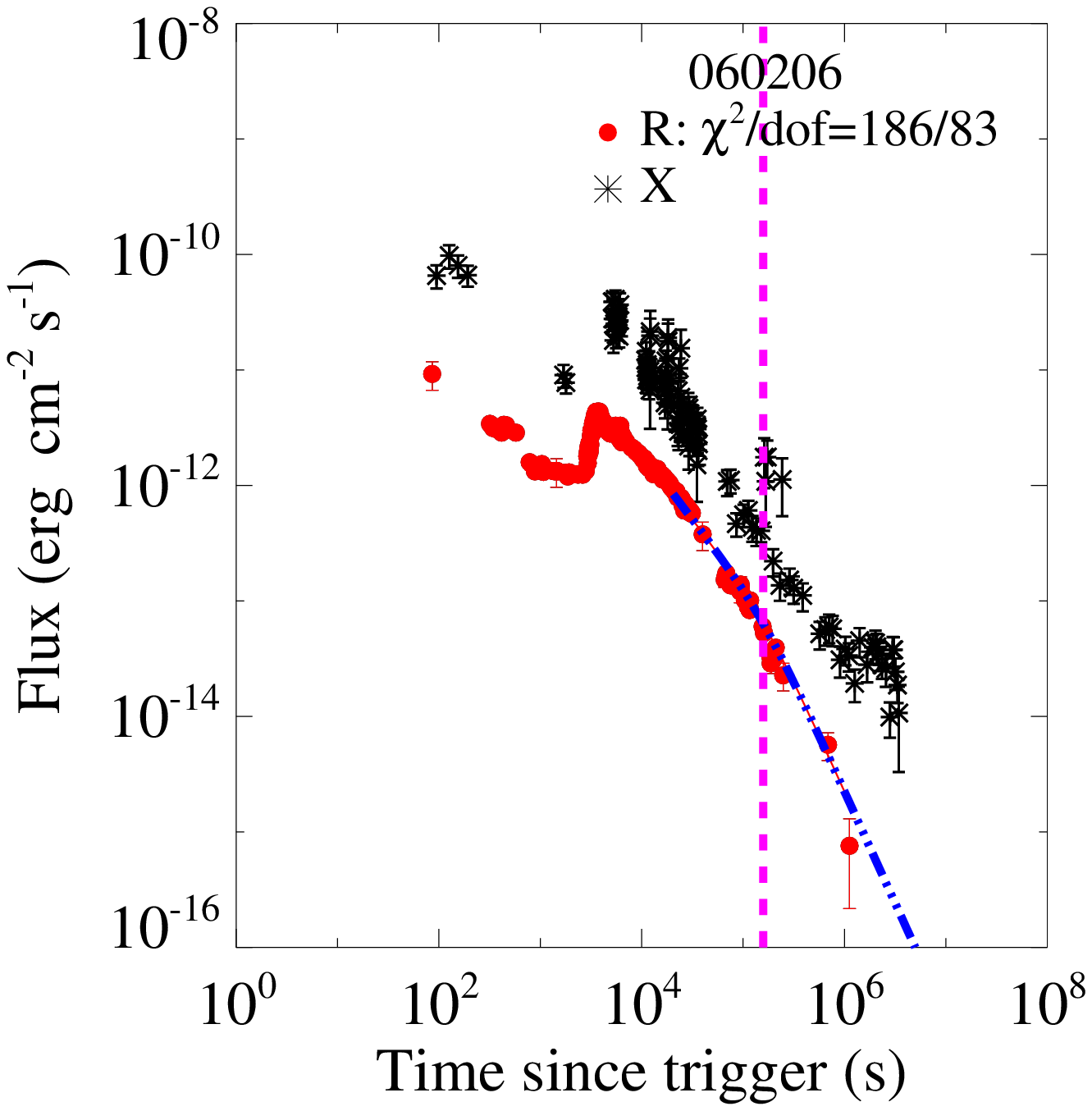}
\includegraphics[angle=0,scale=0.2,width=0.19\textwidth,height=0.18\textheight]{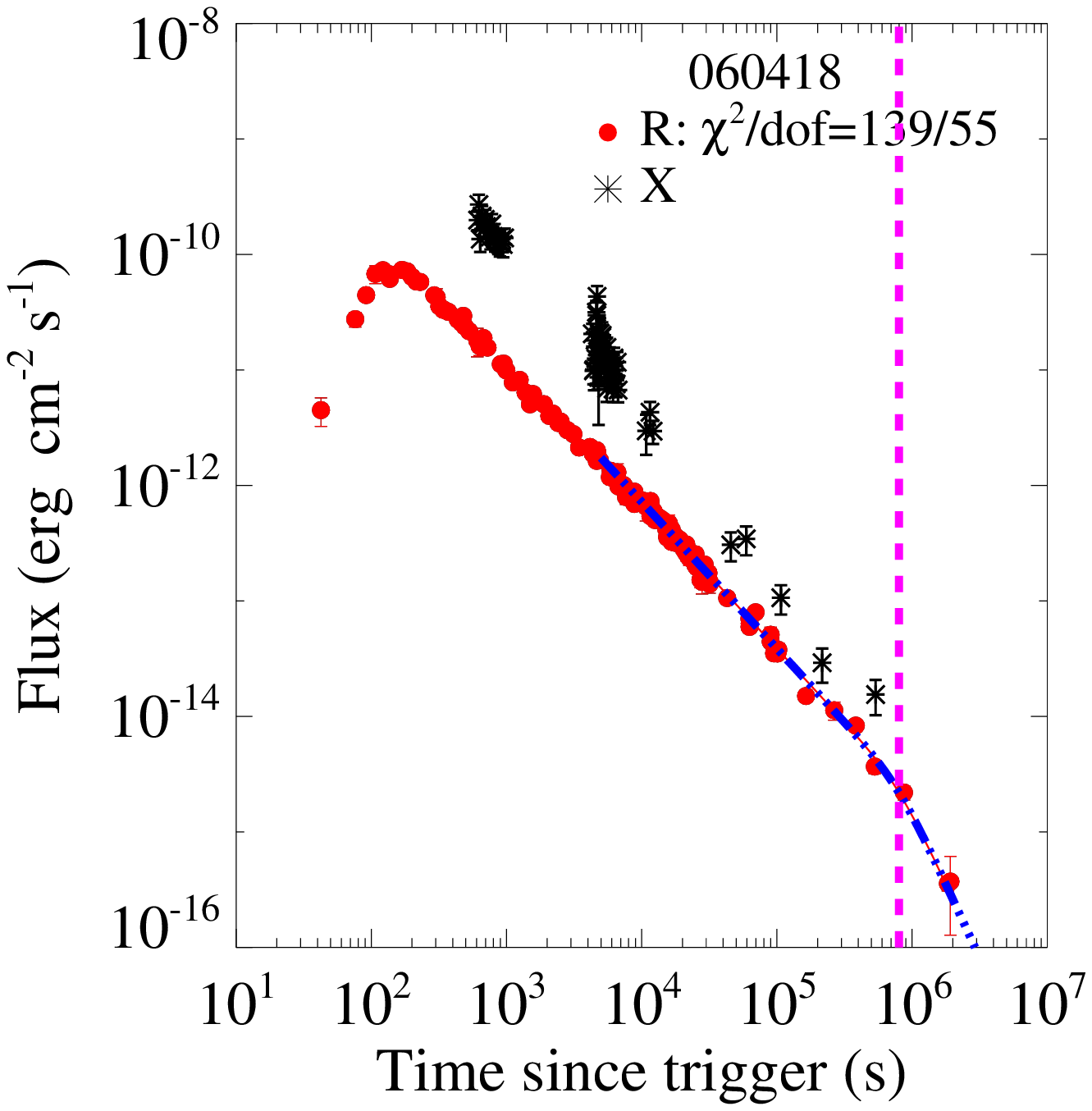}
\includegraphics[angle=0,scale=0.2,width=0.19\textwidth,height=0.18\textheight]{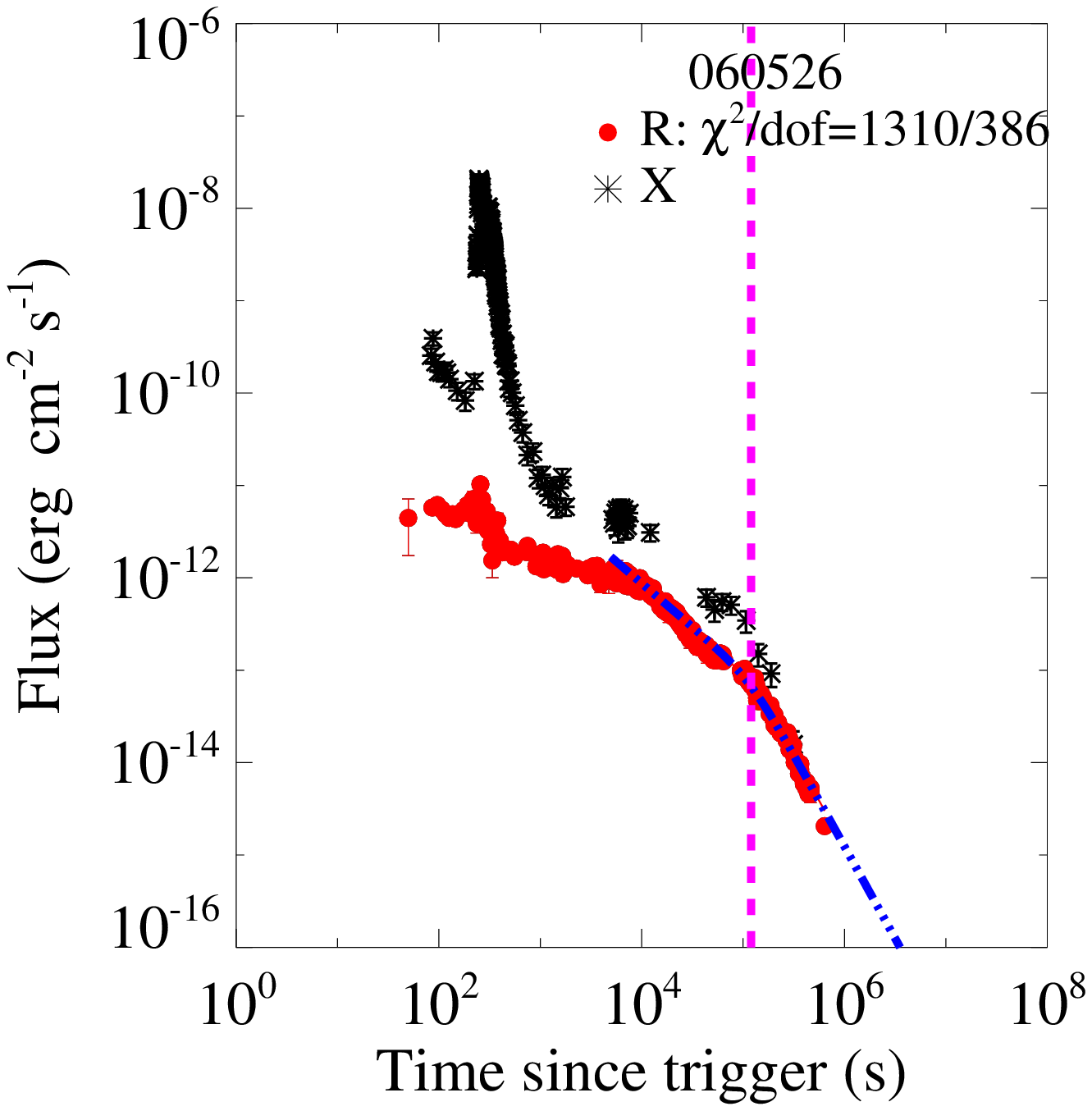}
\includegraphics[angle=0,scale=0.2,width=0.19\textwidth,height=0.18\textheight]{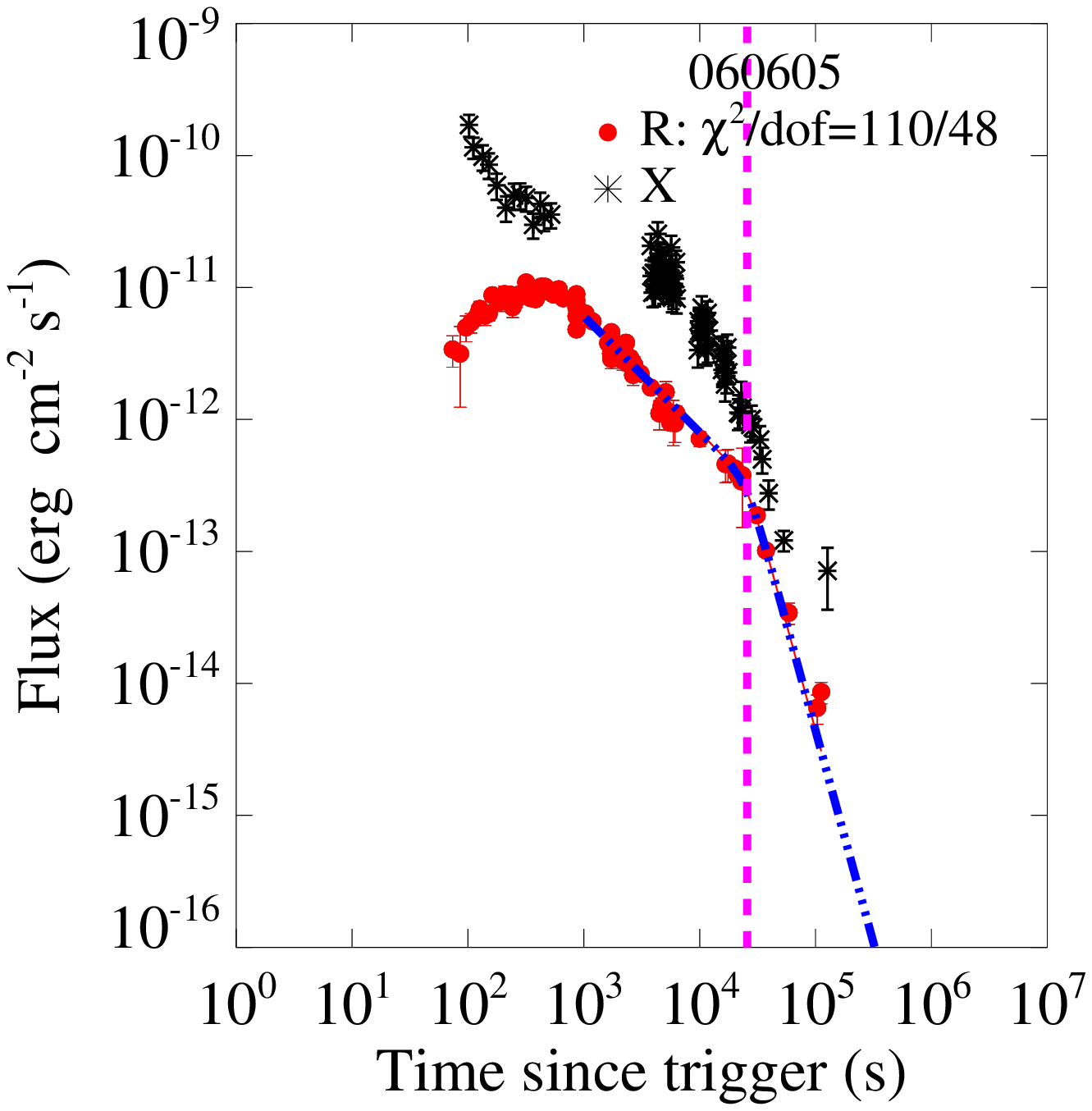}
\includegraphics[angle=0,scale=0.2,width=0.19\textwidth,height=0.18\textheight]{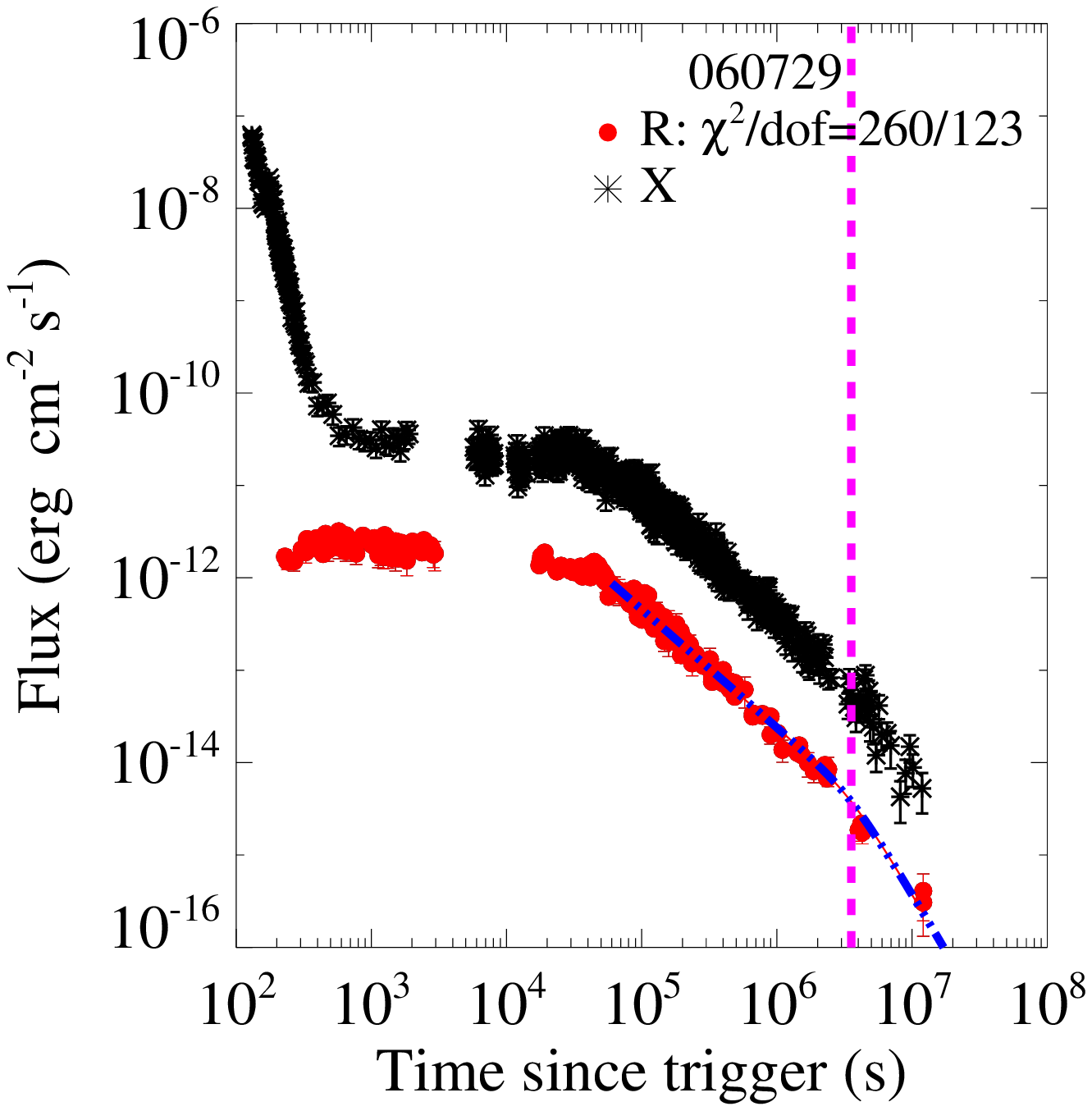}
\includegraphics[angle=0,scale=0.2,width=0.19\textwidth,height=0.18\textheight]{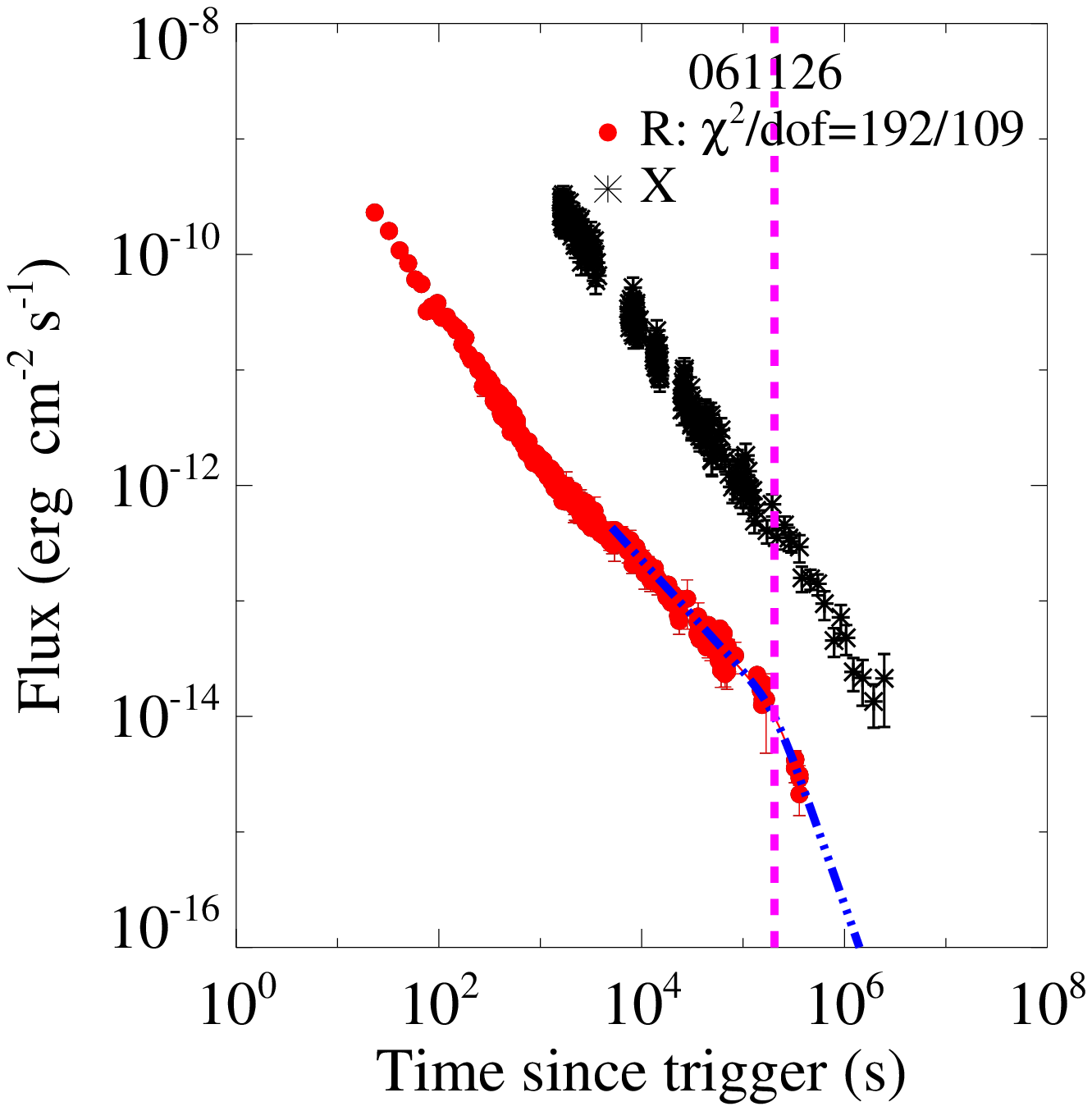}
\includegraphics[angle=0,scale=0.2,width=0.19\textwidth,height=0.18\textheight]{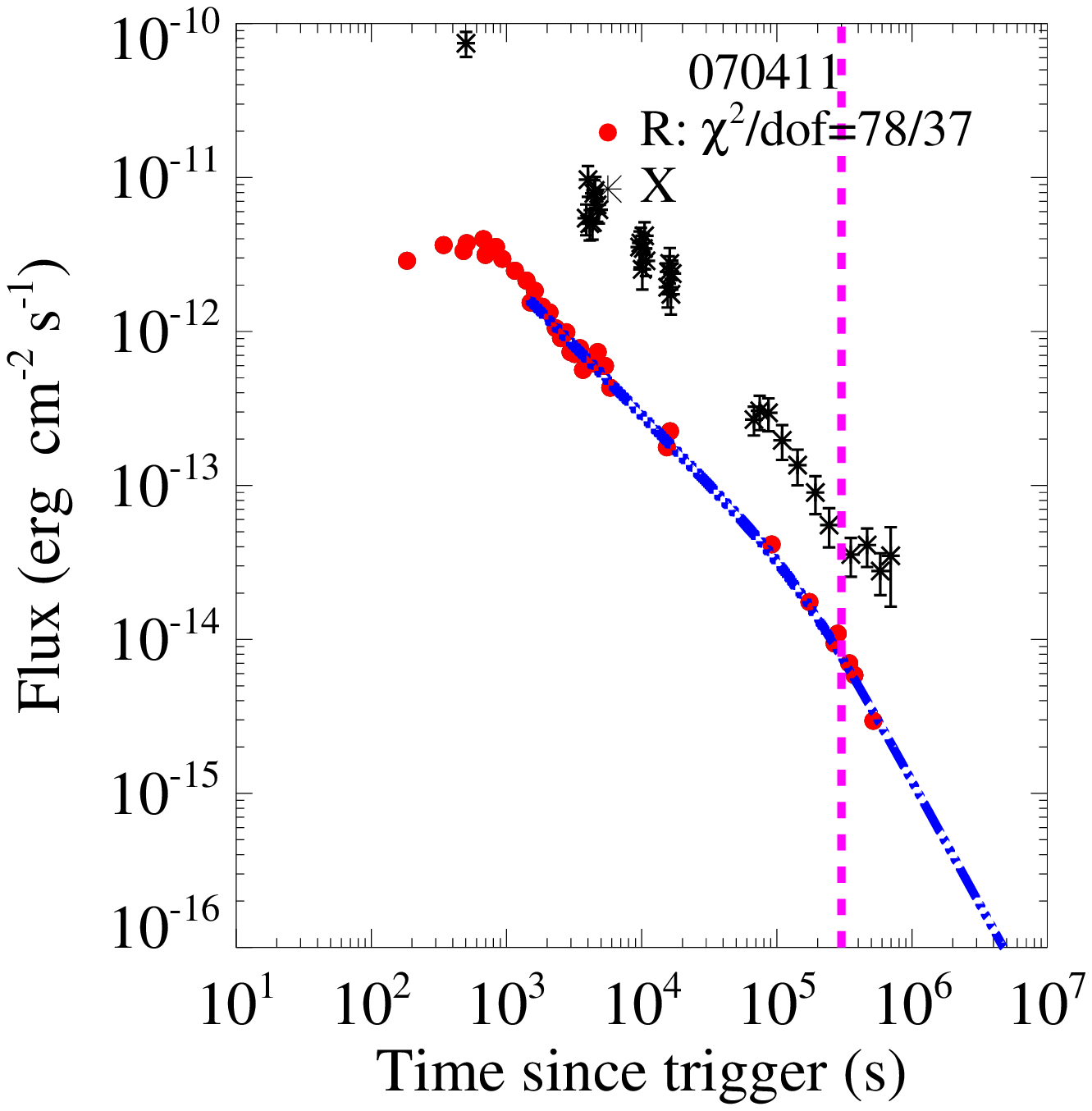}
\includegraphics[angle=0,scale=0.2,width=0.19\textwidth,height=0.18\textheight]{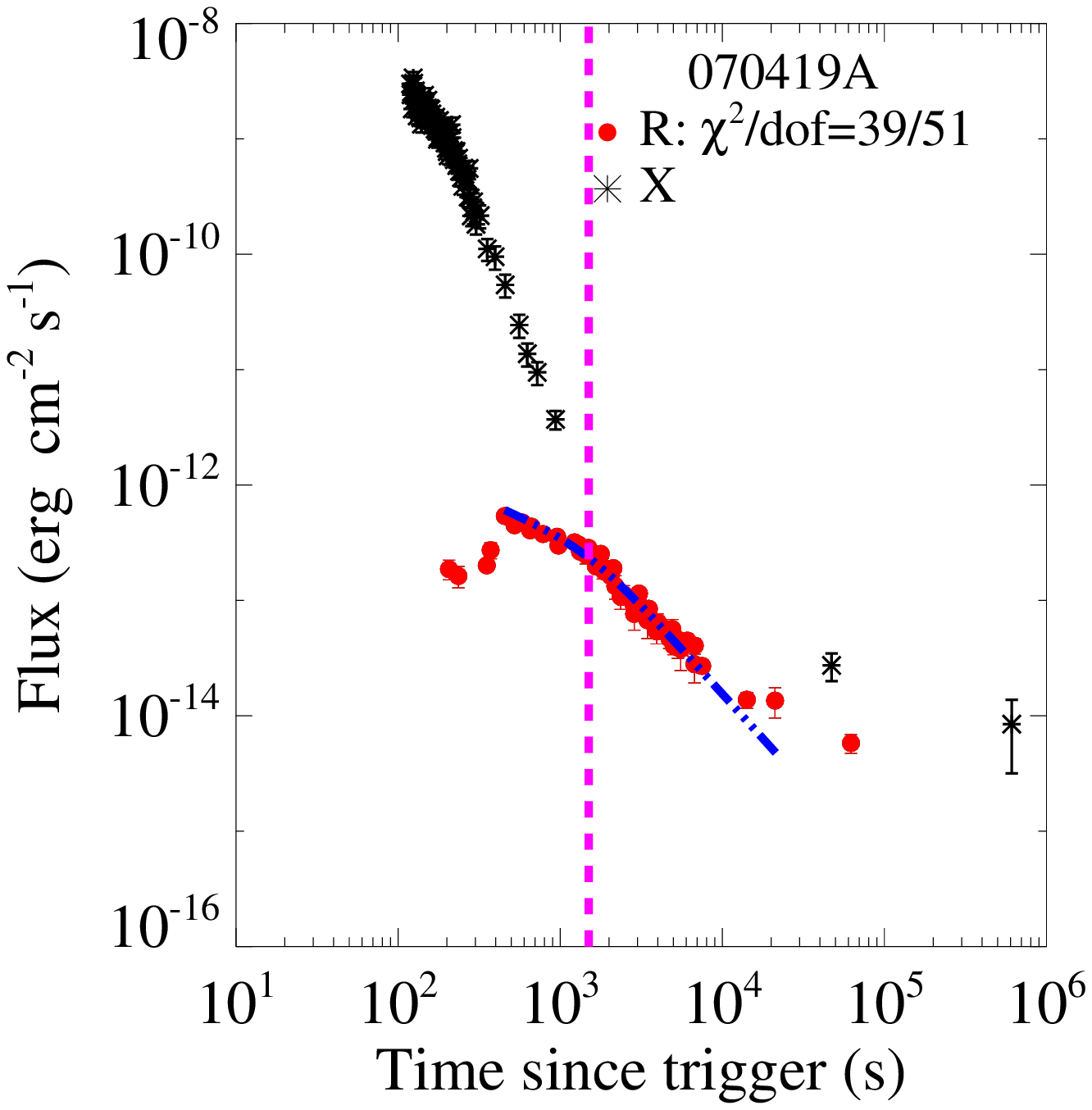}
\includegraphics[angle=0,scale=0.2,width=0.19\textwidth,height=0.18\textheight]{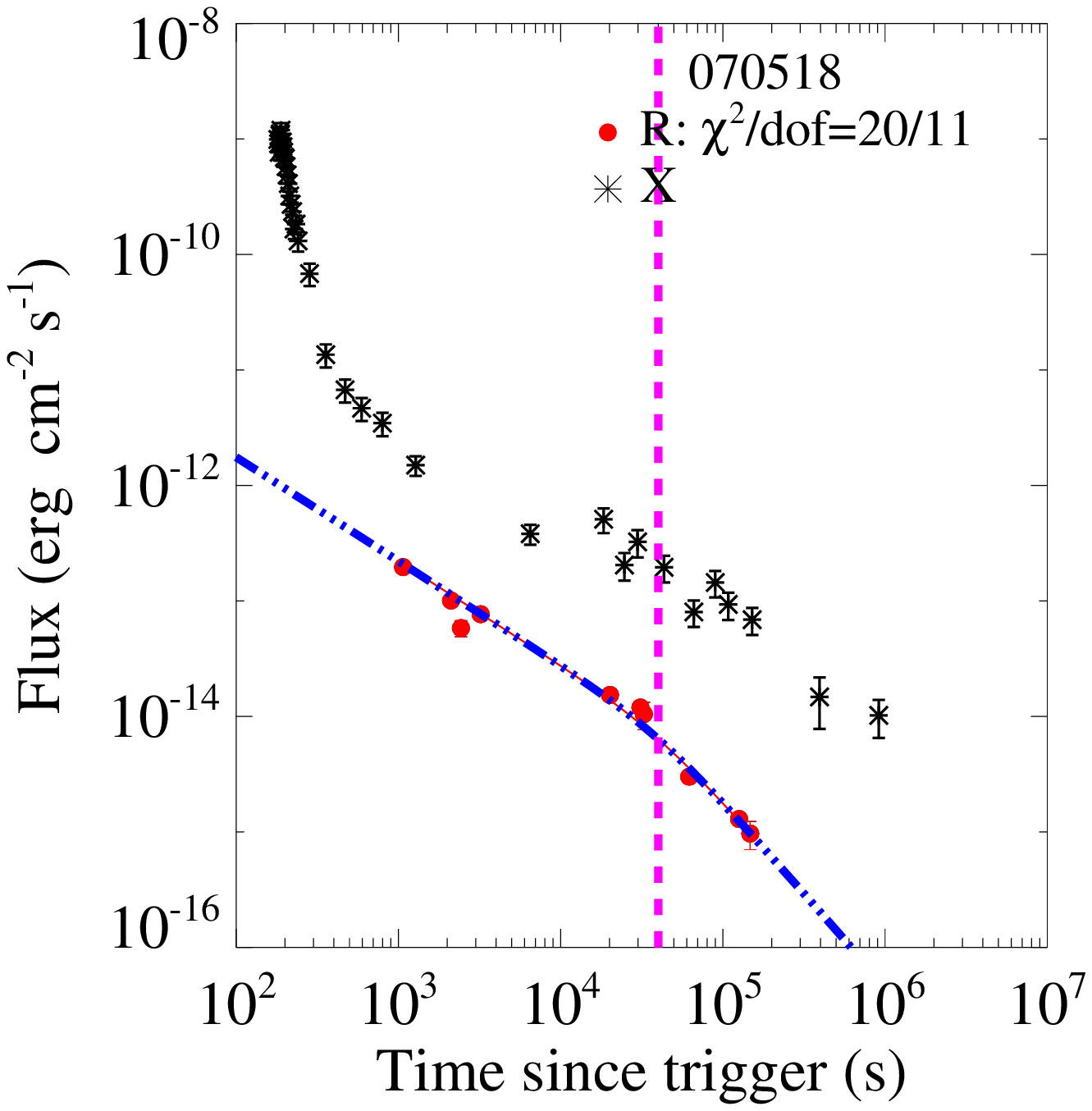}
\includegraphics[angle=0,scale=0.2,width=0.19\textwidth,height=0.18\textheight]{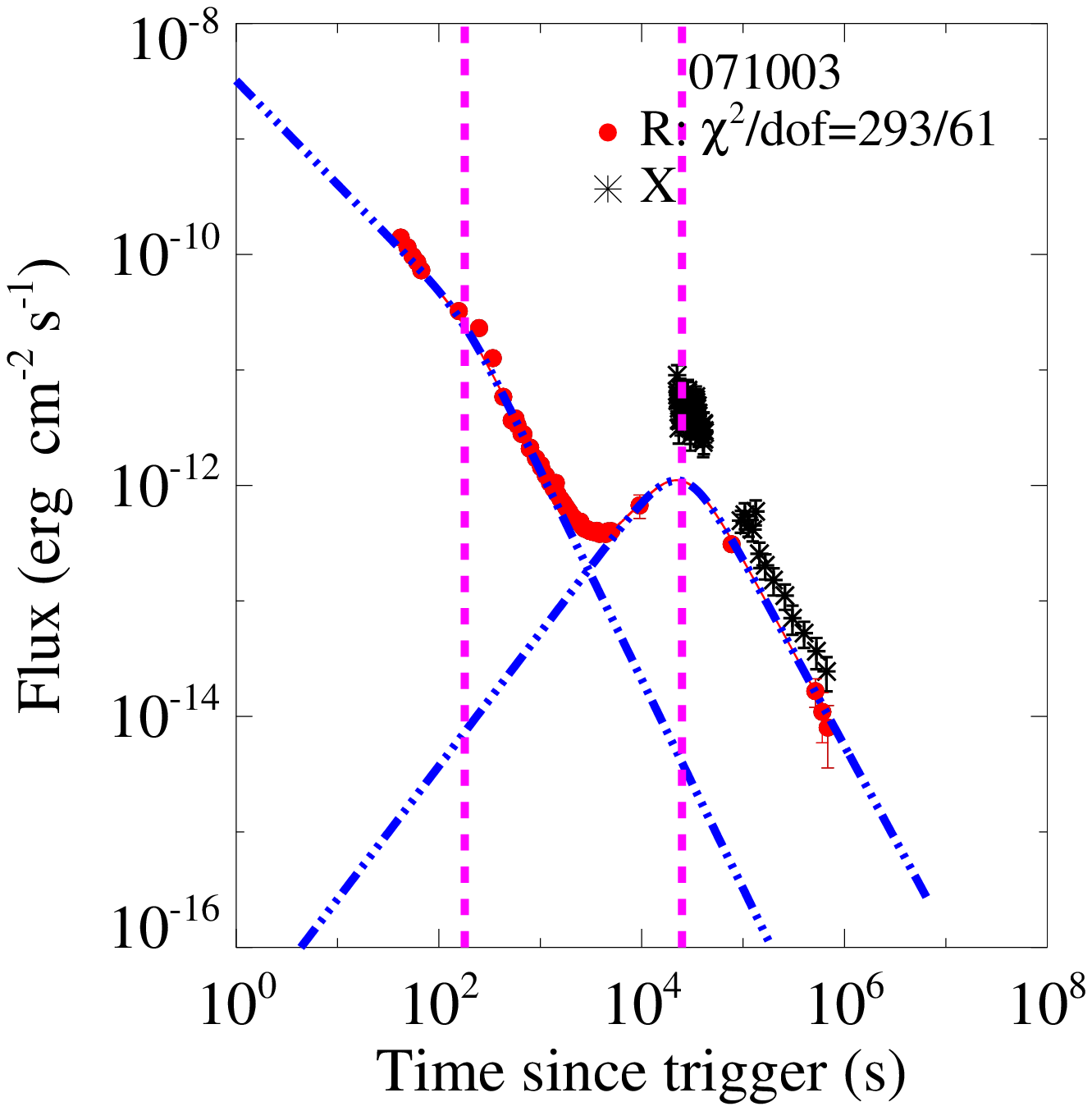}
\includegraphics[angle=0,scale=0.2,width=0.19\textwidth,height=0.18\textheight]{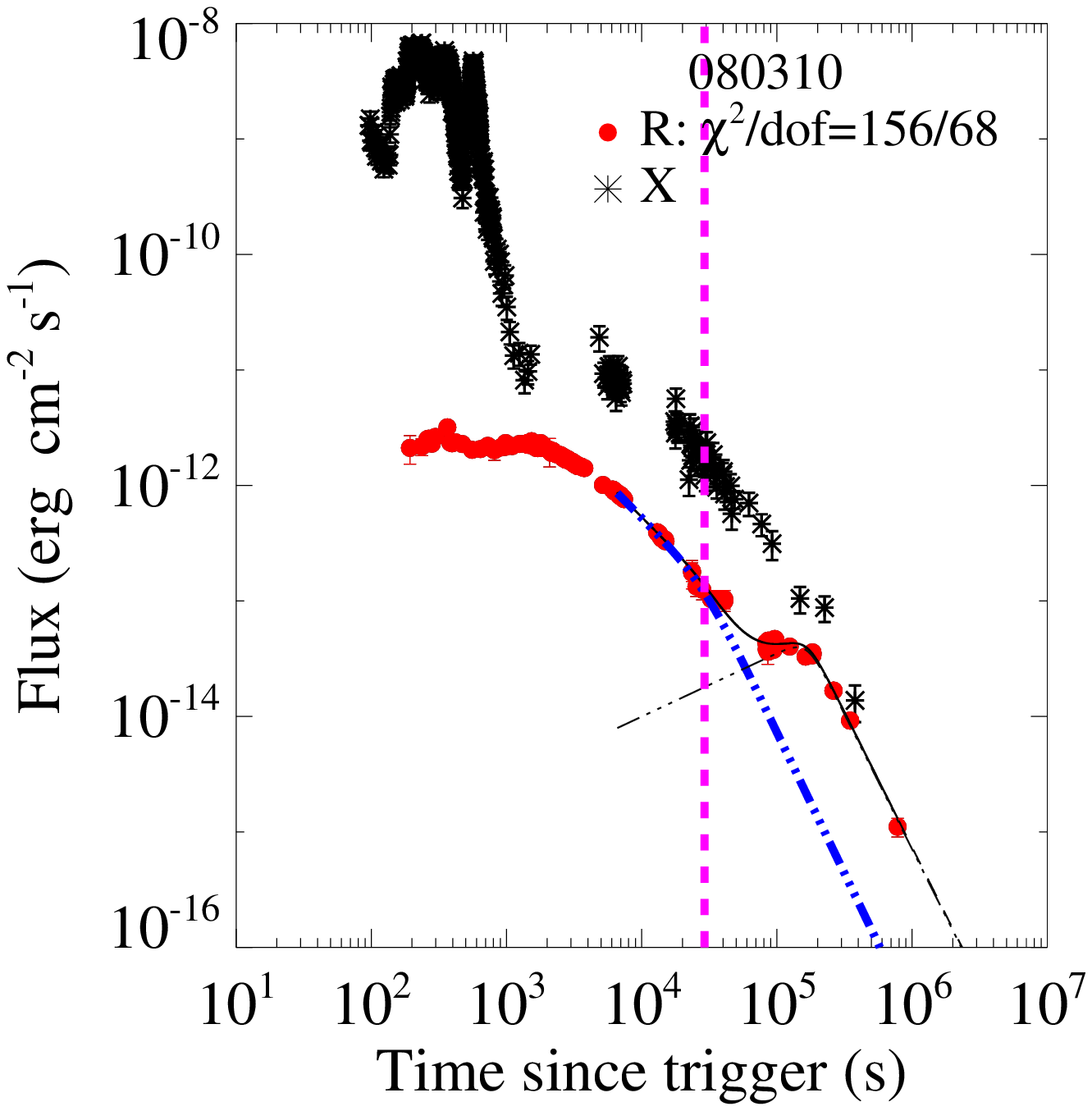}
\includegraphics[angle=0,scale=0.2,width=0.19\textwidth,height=0.18\textheight]{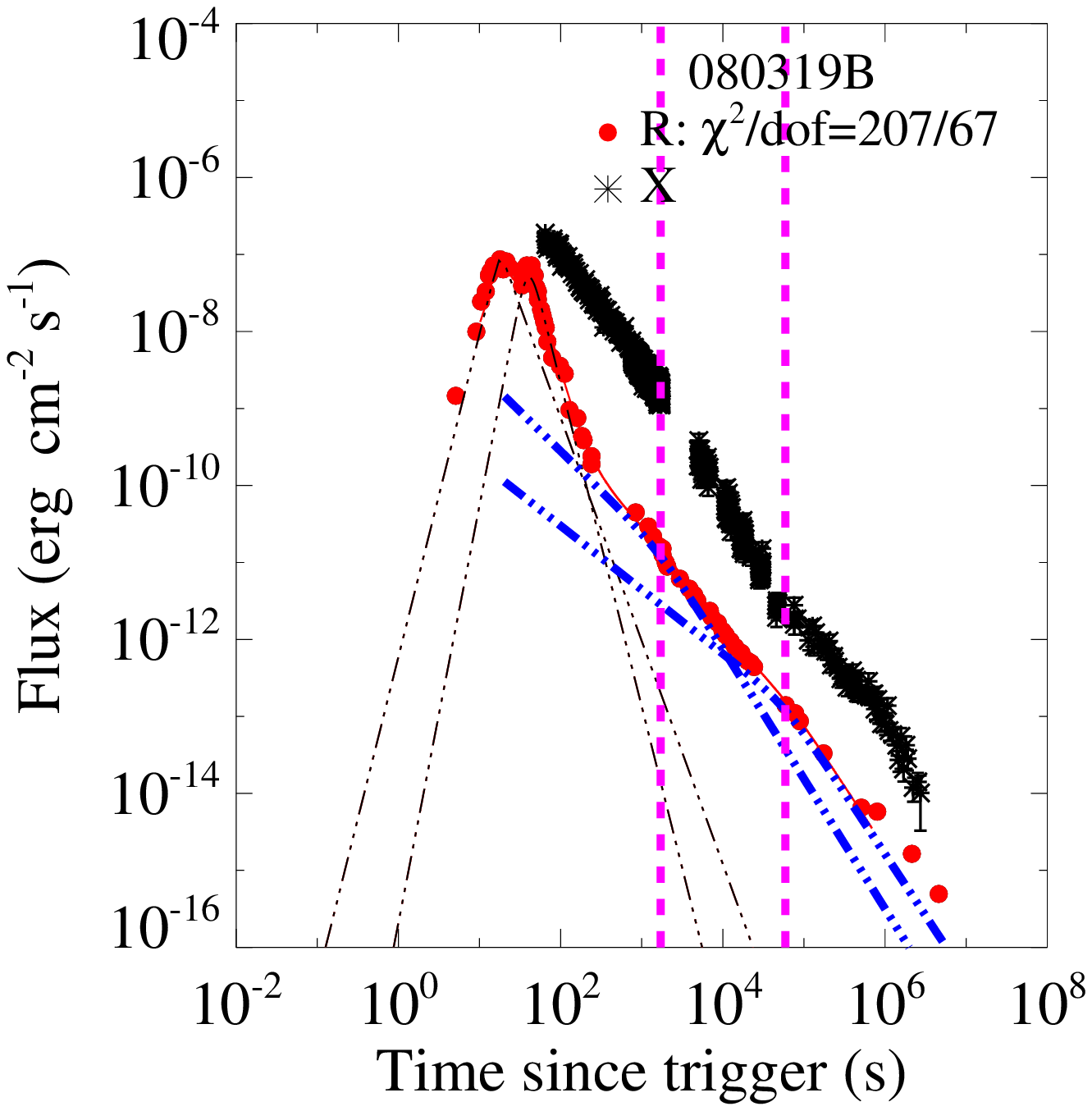}
\includegraphics[angle=0,scale=0.2,width=0.19\textwidth,height=0.18\textheight]{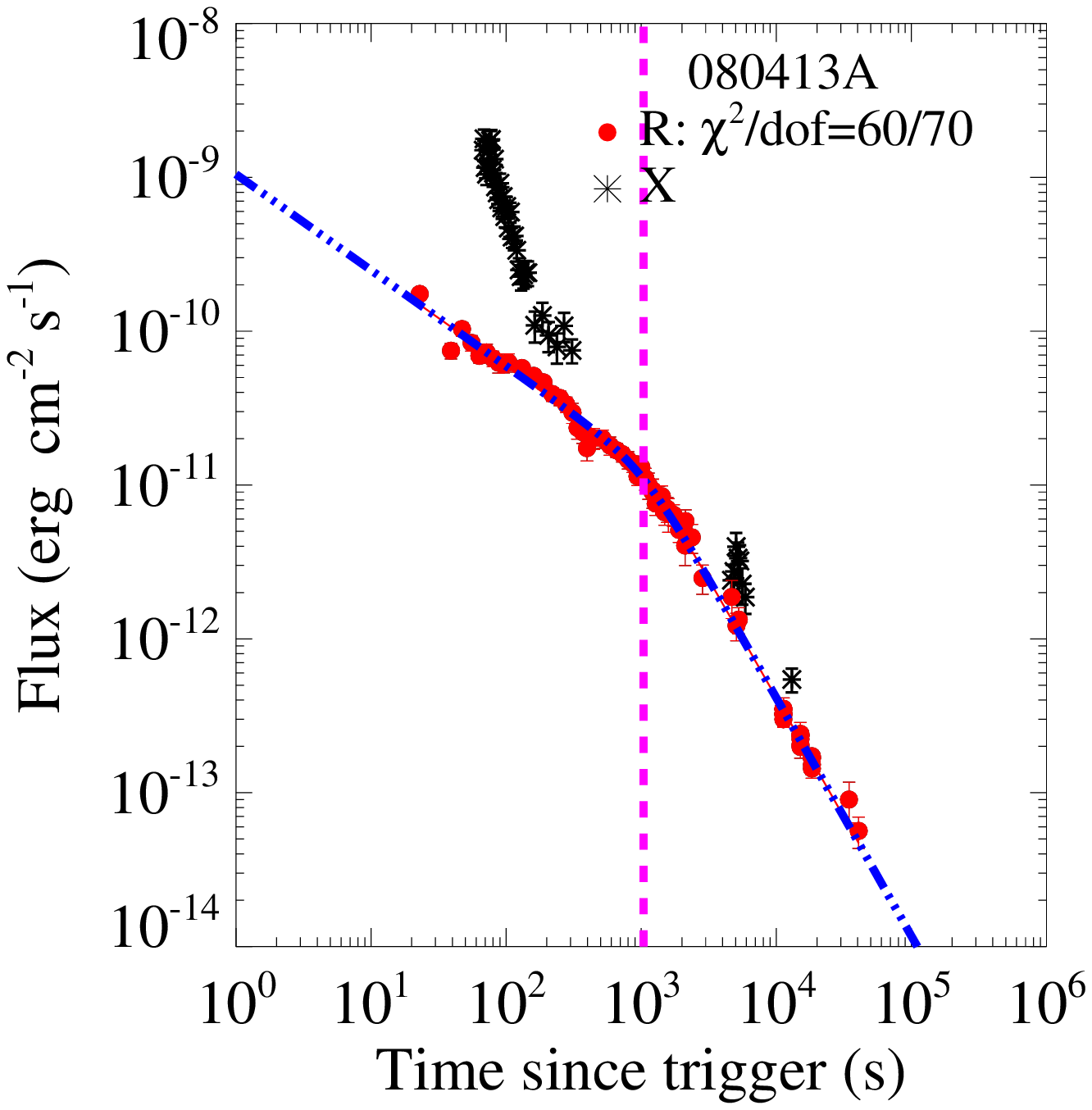}

\caption{The observed light curves of the GRBs with one and more identified jet break(s). The optical lightcurves (red) are used to derive the jet break(s). The lightcurves are fit with the blue dotted-dashed lines, and the jet break times are shown by the purple vertical dashed
lines. The X-ray lightcurves (black) are plotted for reference.} \label{jetgrade}
\end{figure*}

\clearpage
\setlength{\voffset}{-18mm}
\begin{figure*}
\includegraphics[angle=0,scale=0.2,width=0.19\textwidth,height=0.18\textheight]{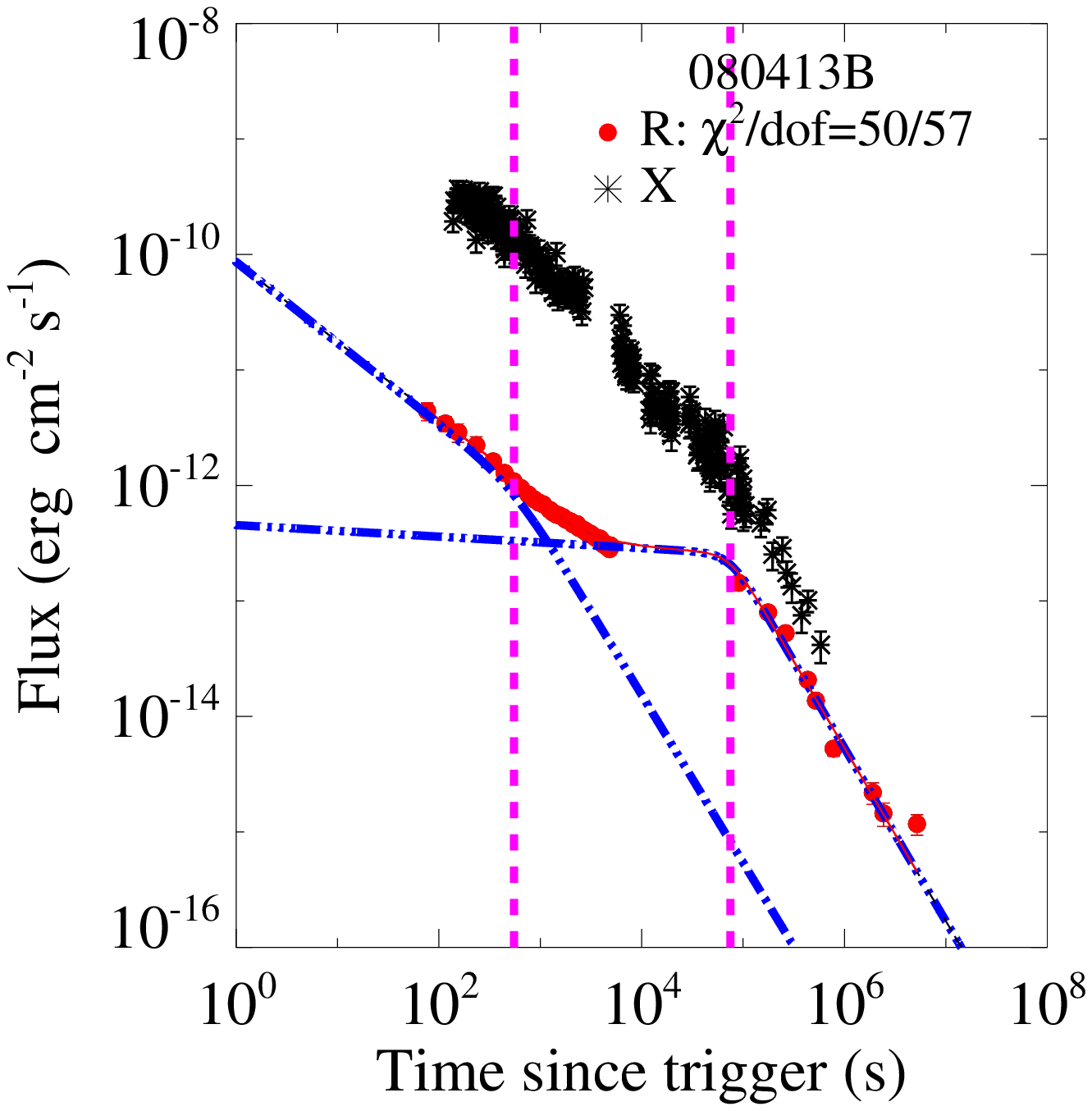}
\includegraphics[angle=0,scale=0.2,width=0.19\textwidth,height=0.18\textheight]{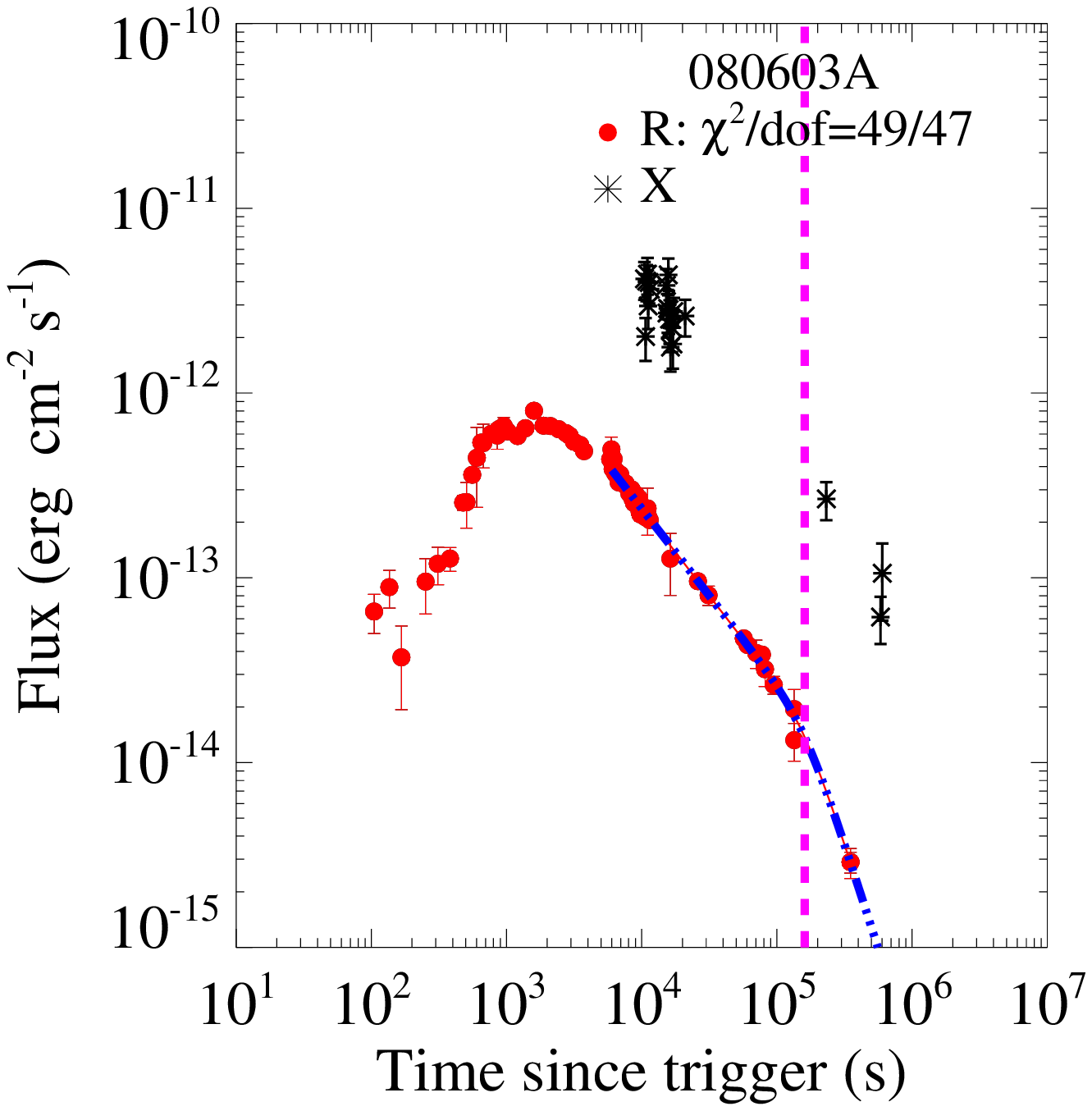}
\includegraphics[angle=0,scale=0.2,width=0.19\textwidth,height=0.18\textheight]{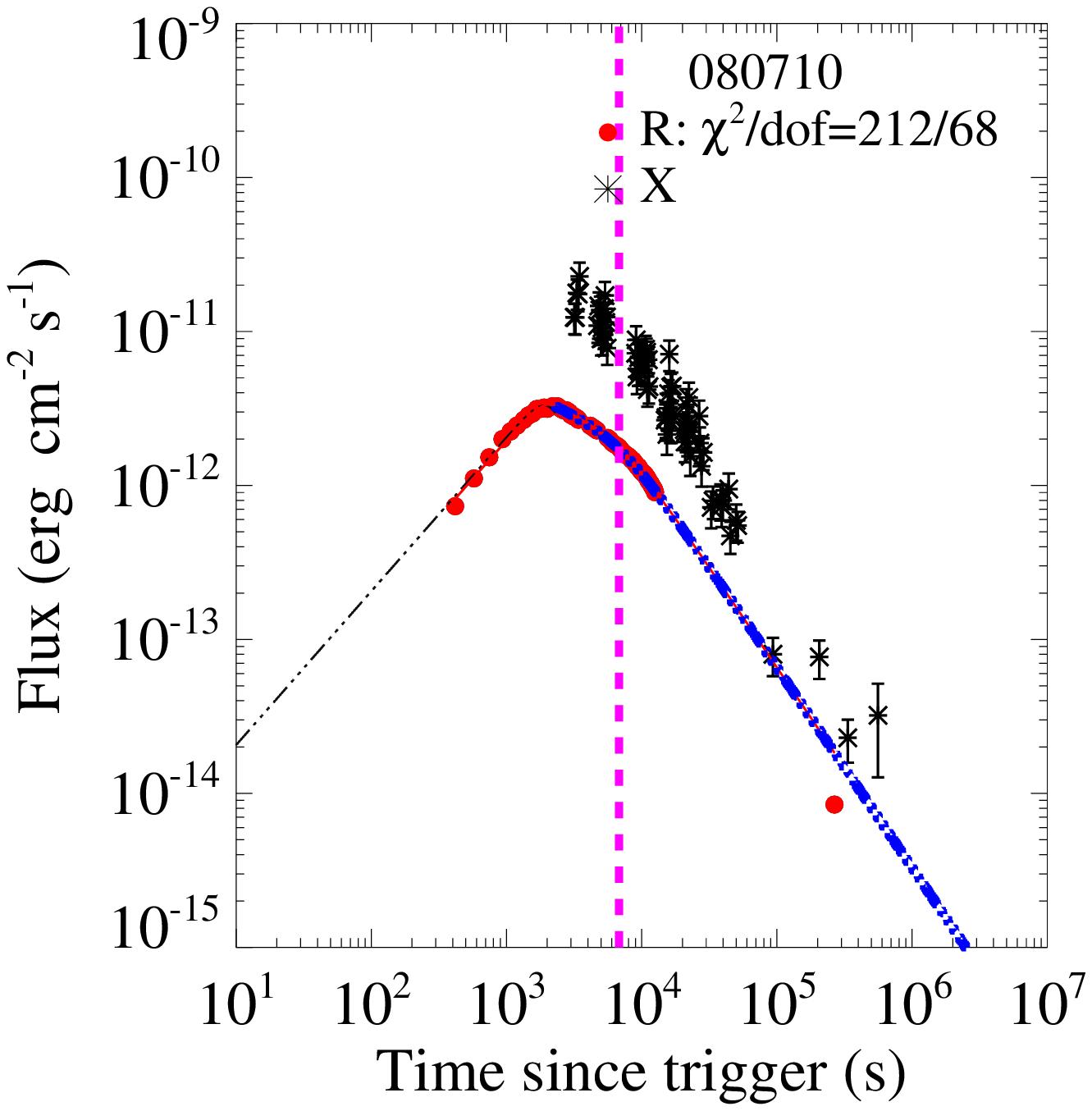}
\includegraphics[angle=0,scale=0.2,width=0.19\textwidth,height=0.18\textheight]{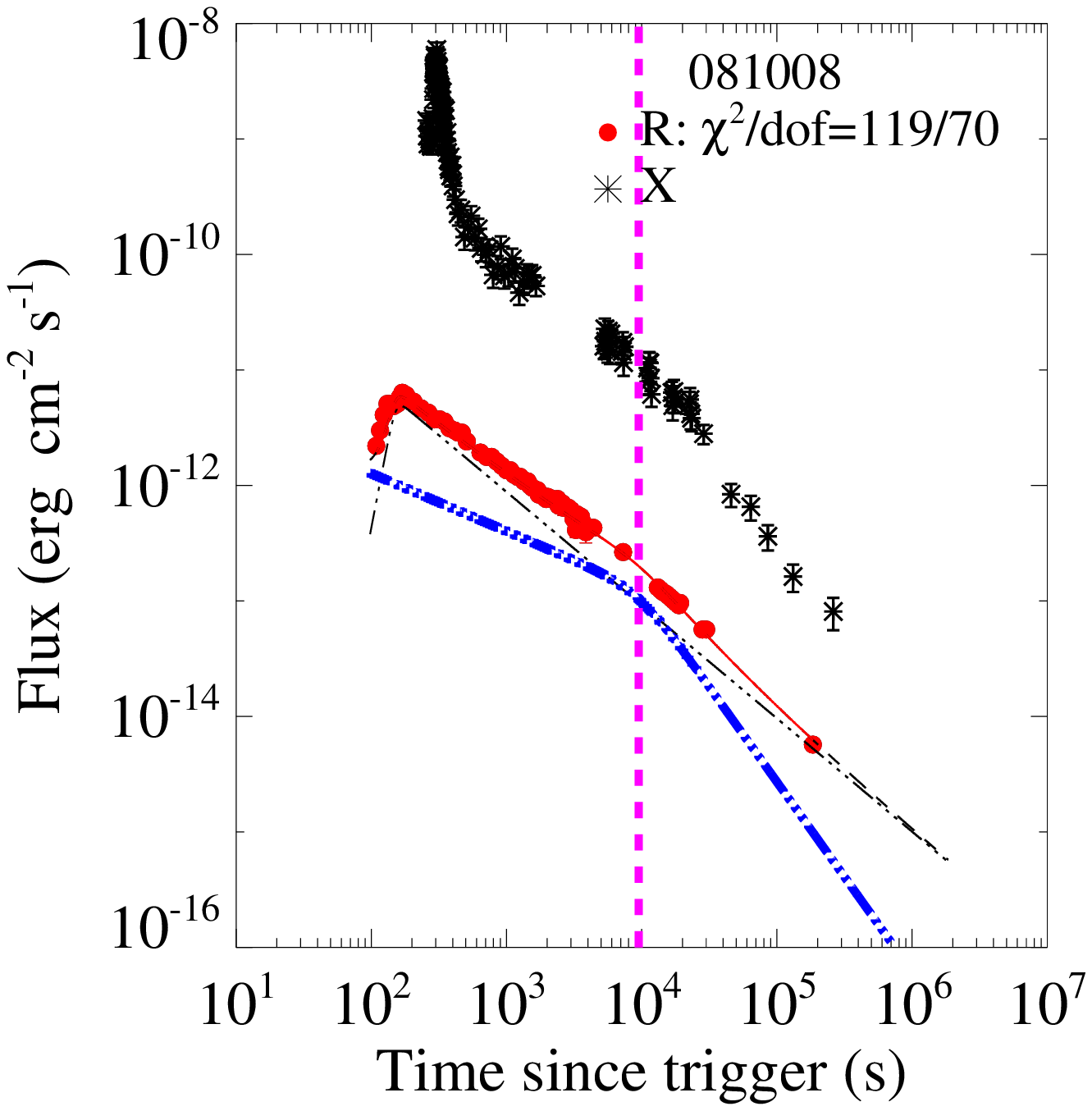}
\includegraphics[angle=0,scale=0.2,width=0.19\textwidth,height=0.18\textheight]{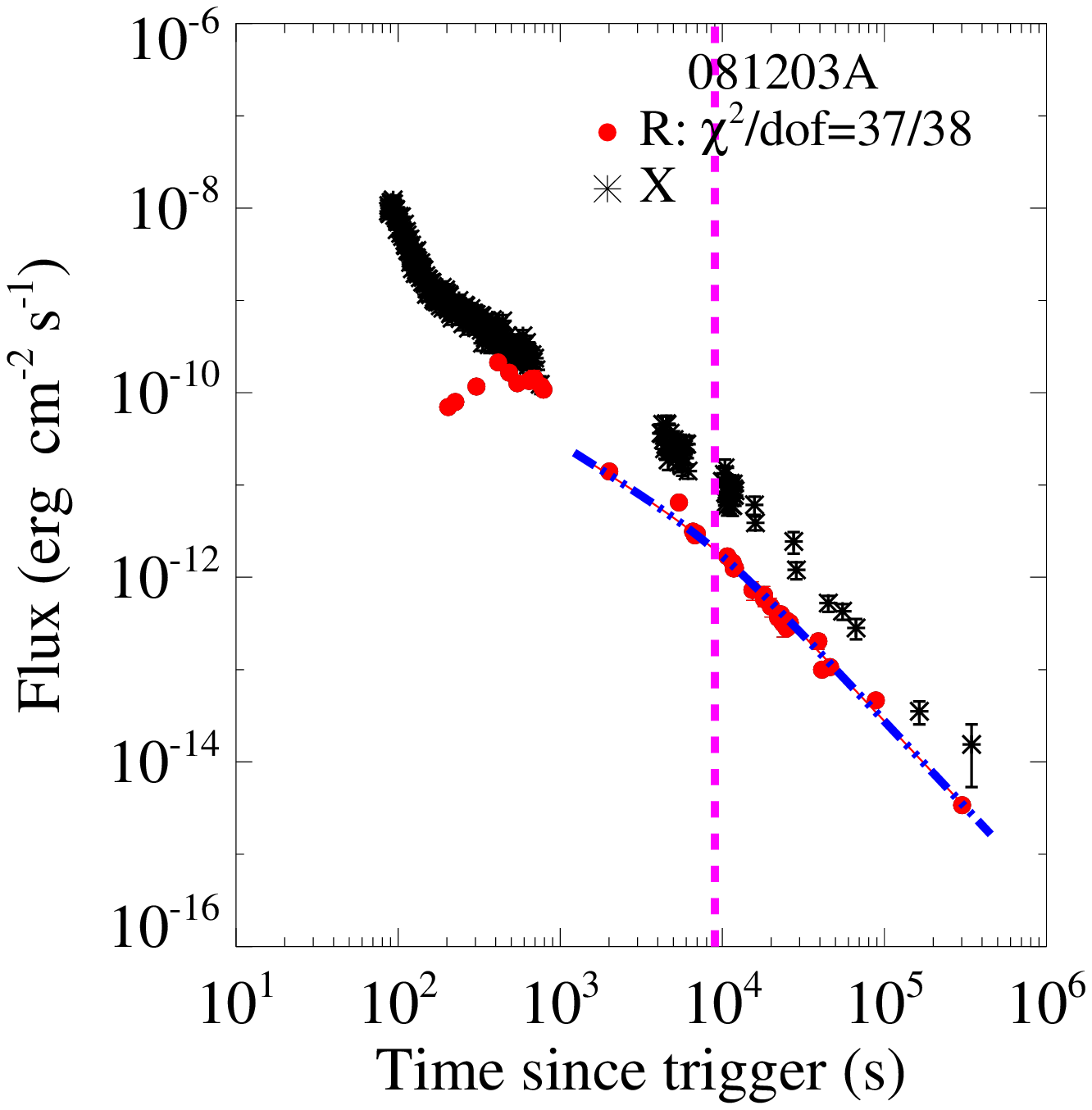}
\includegraphics[angle=0,scale=0.2,width=0.19\textwidth,height=0.18\textheight]{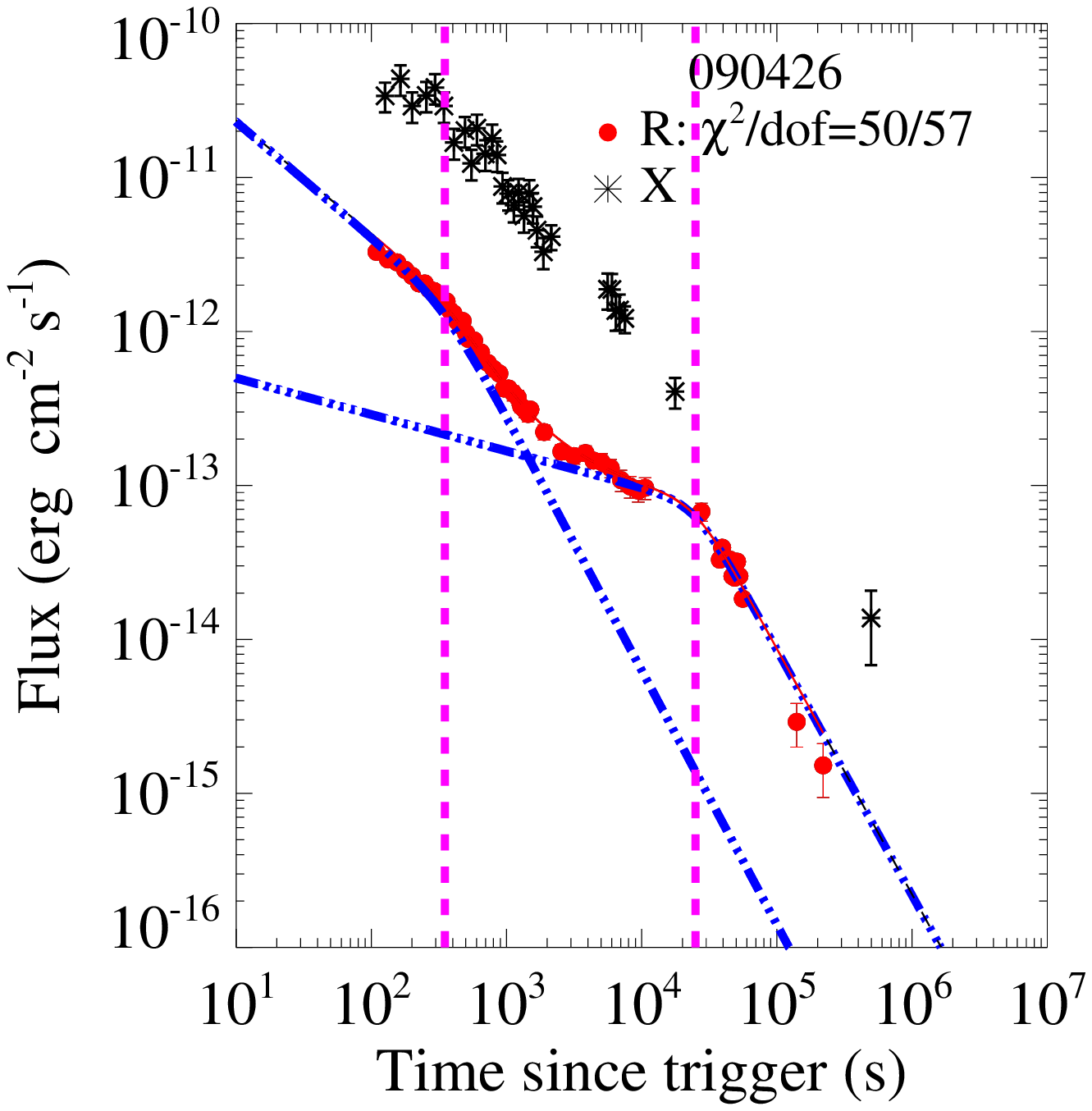}
\includegraphics[angle=0,scale=0.2,width=0.19\textwidth,height=0.18\textheight]{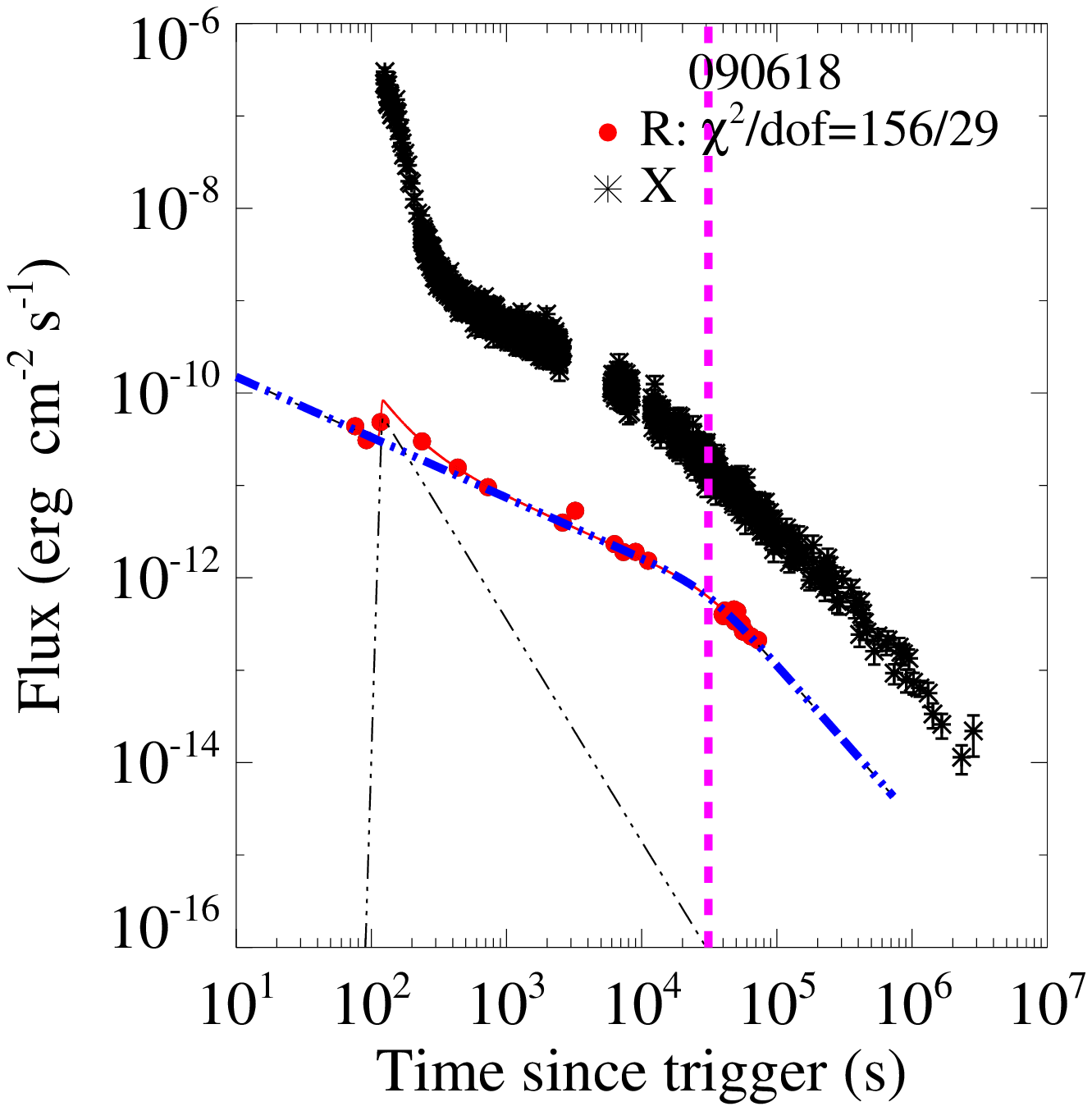}
\includegraphics[angle=0,scale=0.2,width=0.19\textwidth,height=0.18\textheight]{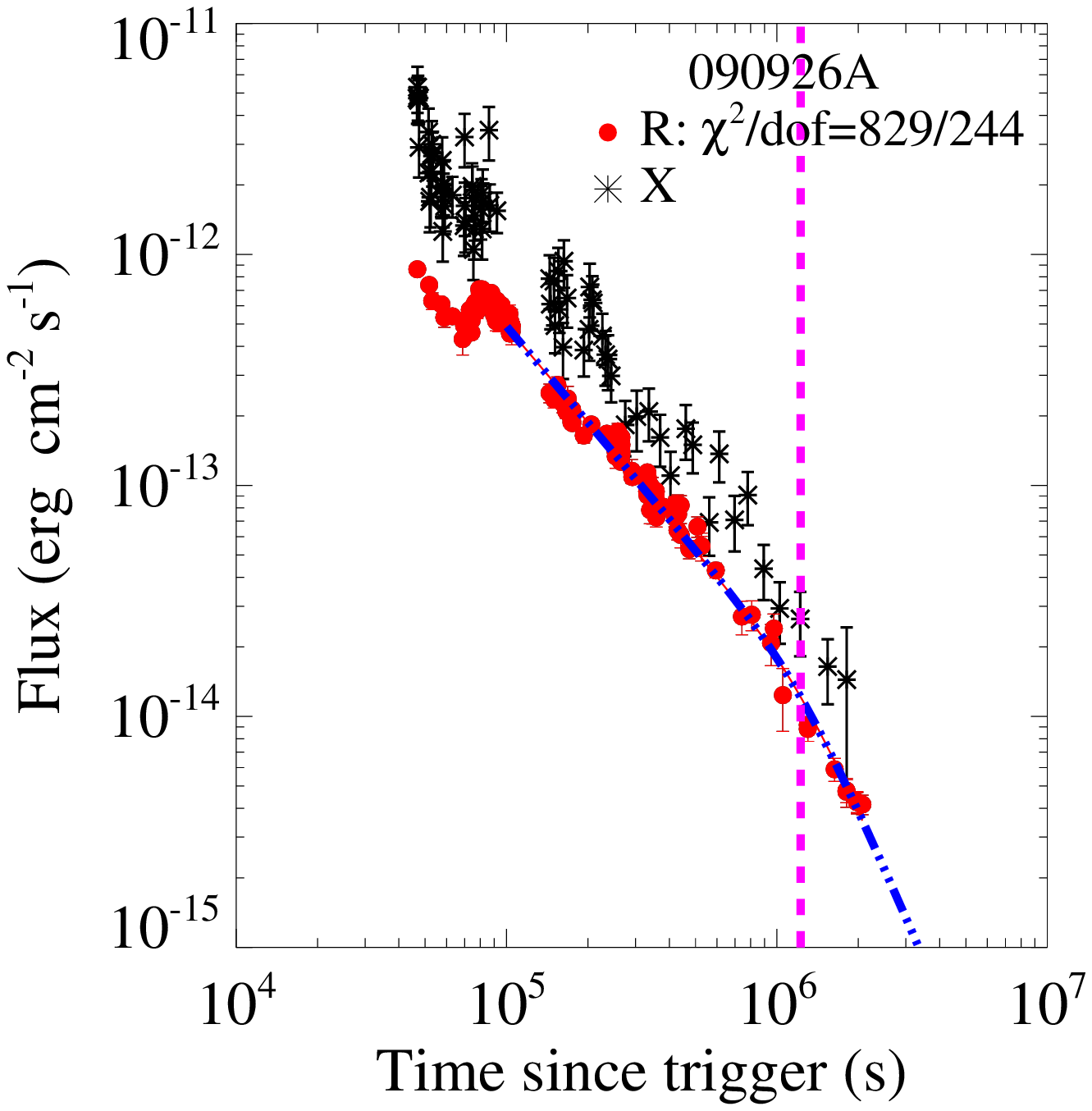}
\includegraphics[angle=0,scale=0.2,width=0.19\textwidth,height=0.18\textheight]{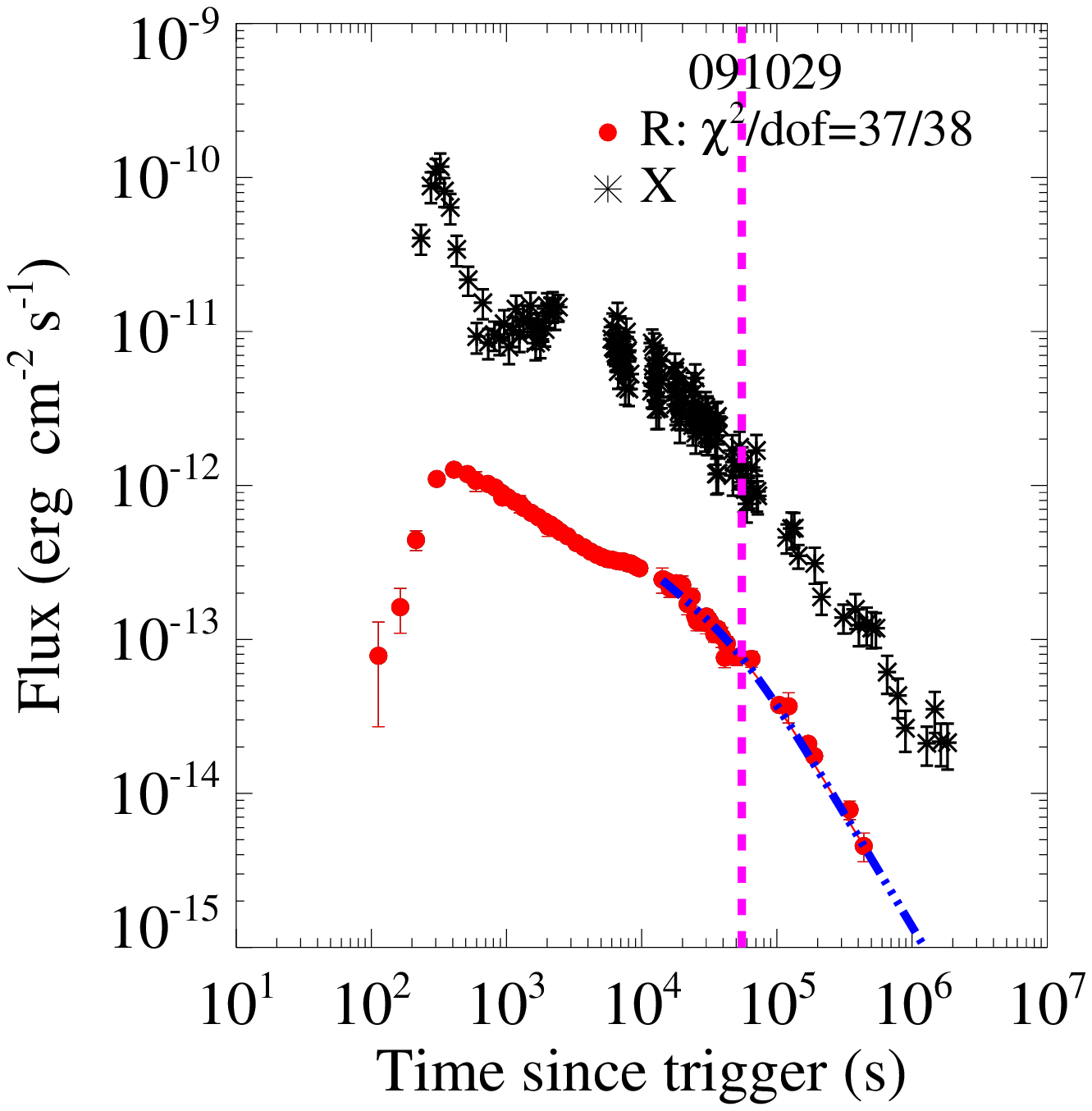}
\includegraphics[angle=0,scale=0.2,width=0.19\textwidth,height=0.18\textheight]{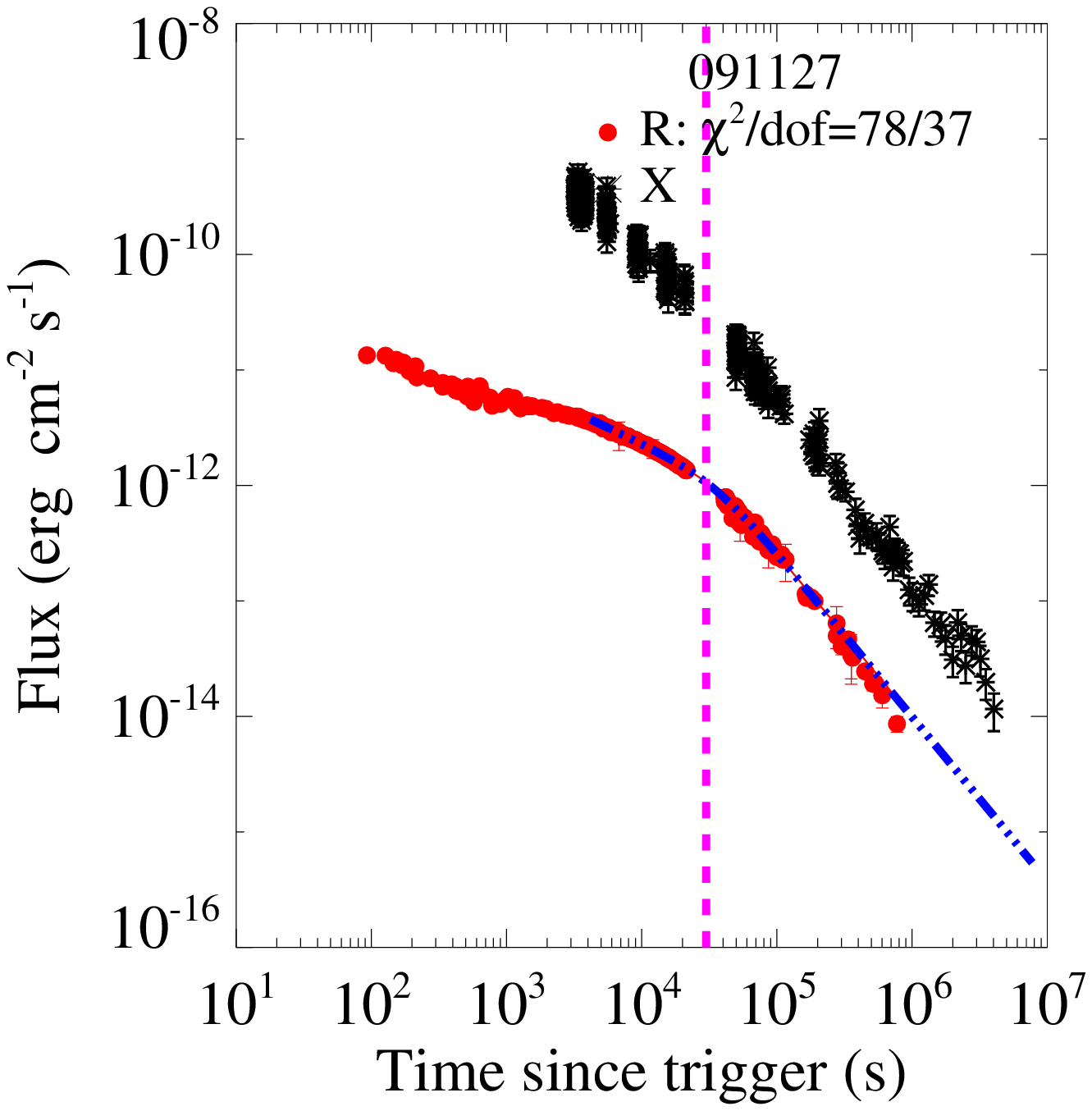}
\includegraphics[angle=0,scale=0.2,width=0.19\textwidth,height=0.18\textheight]{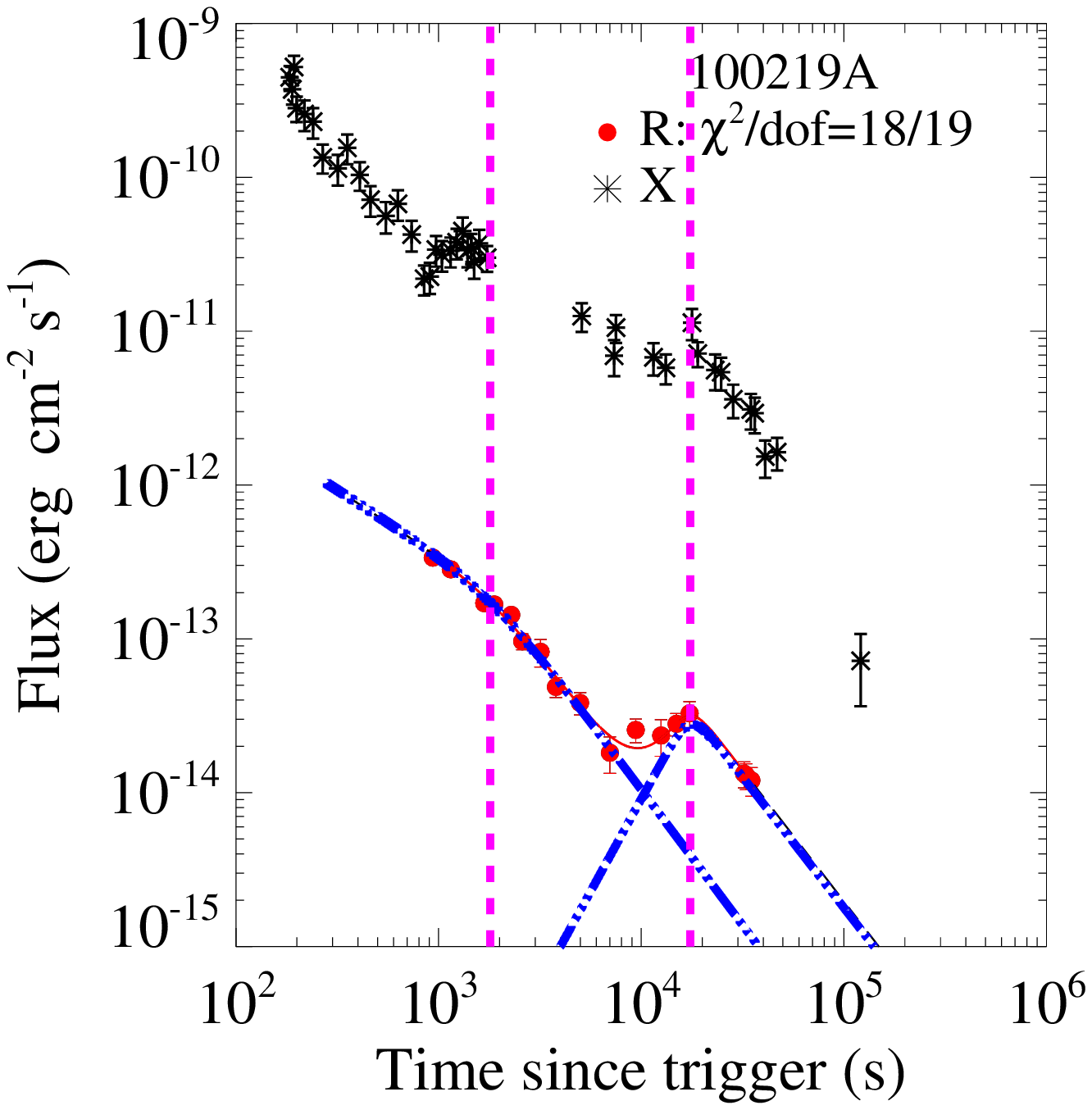}
\includegraphics[angle=0,scale=0.2,width=0.19\textwidth,height=0.18\textheight]{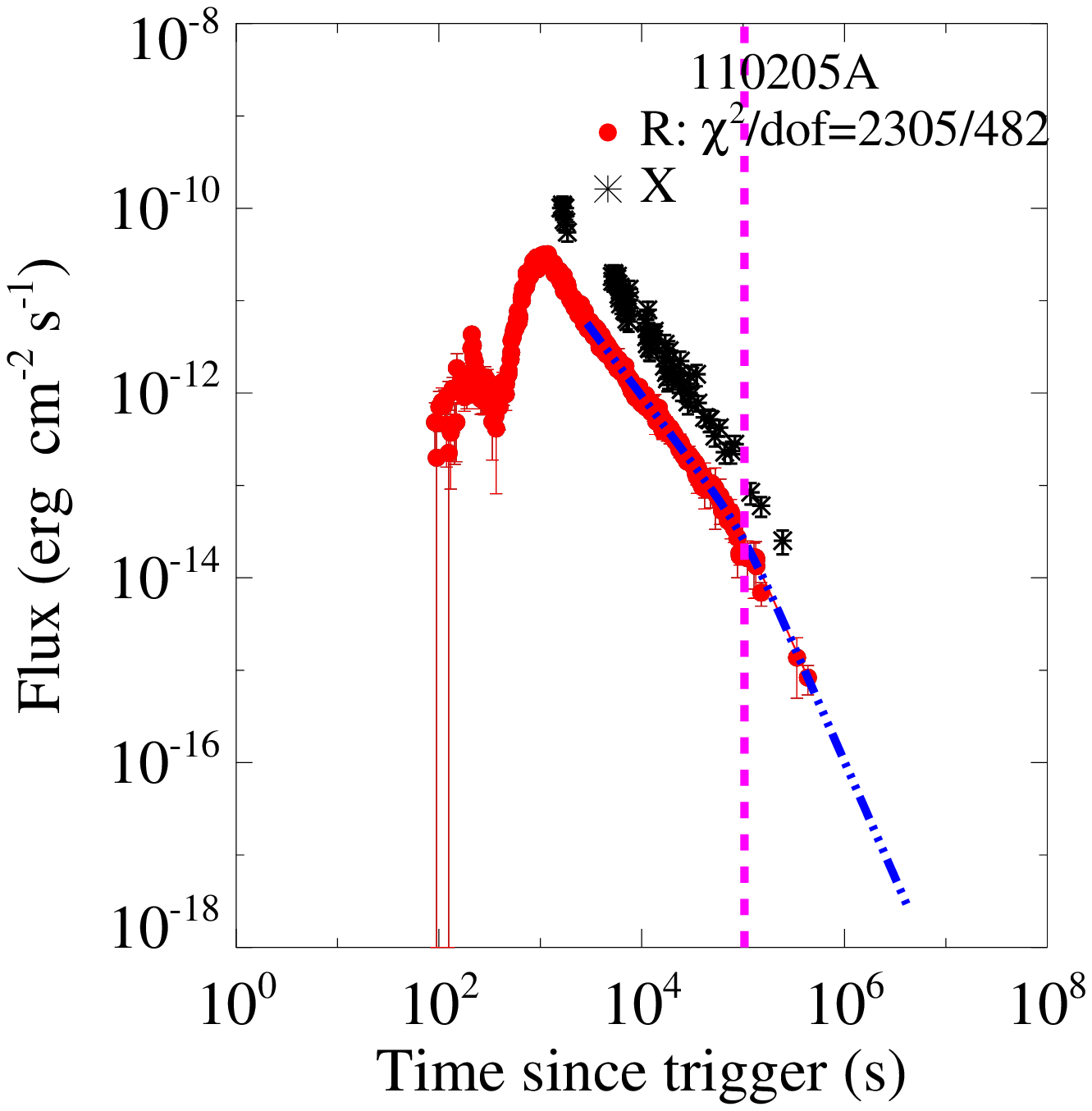}
\includegraphics[angle=0,scale=0.2,width=0.19\textwidth,height=0.18\textheight]{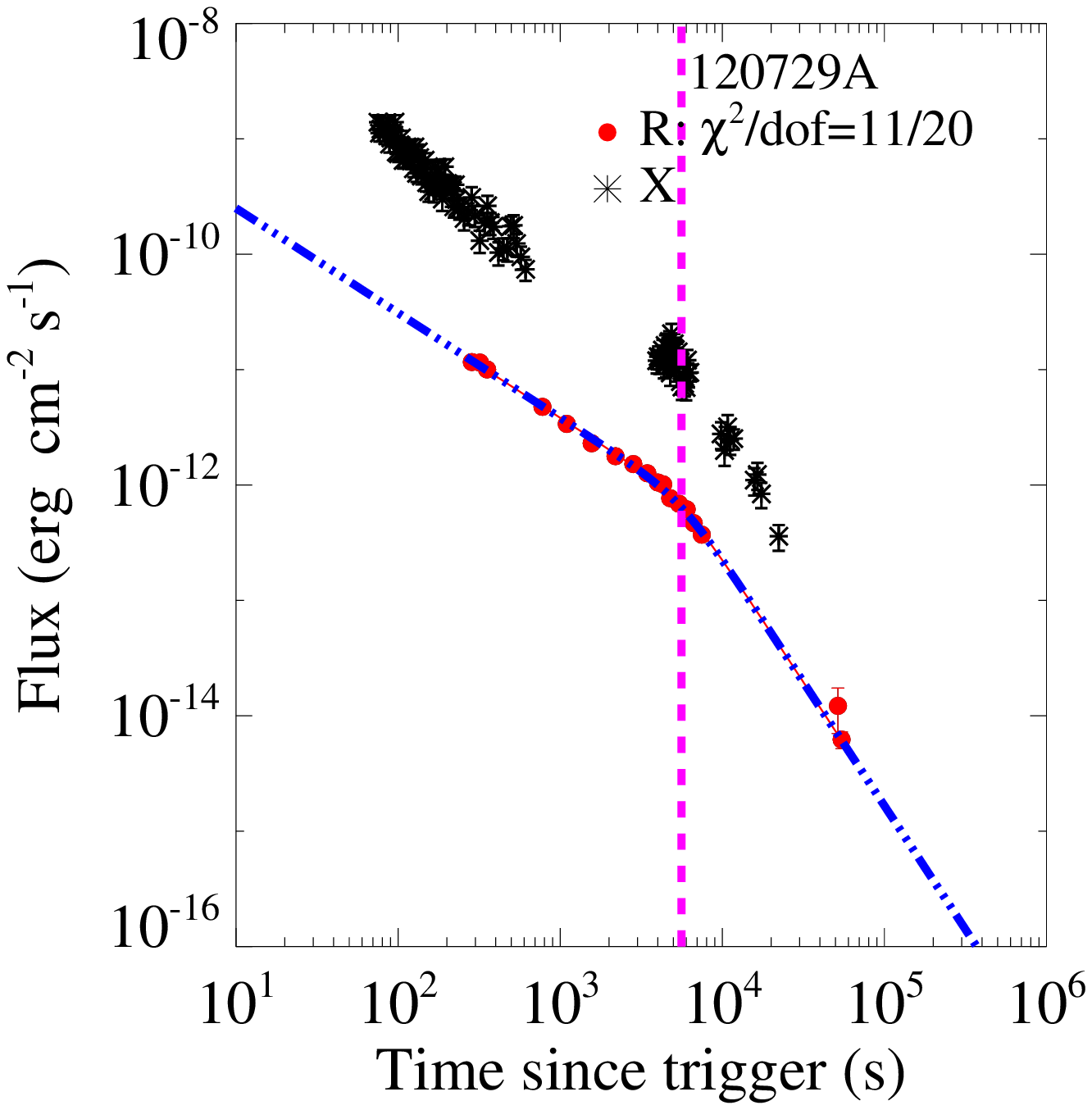}
\includegraphics[angle=0,scale=0.2,width=0.19\textwidth,height=0.18\textheight]{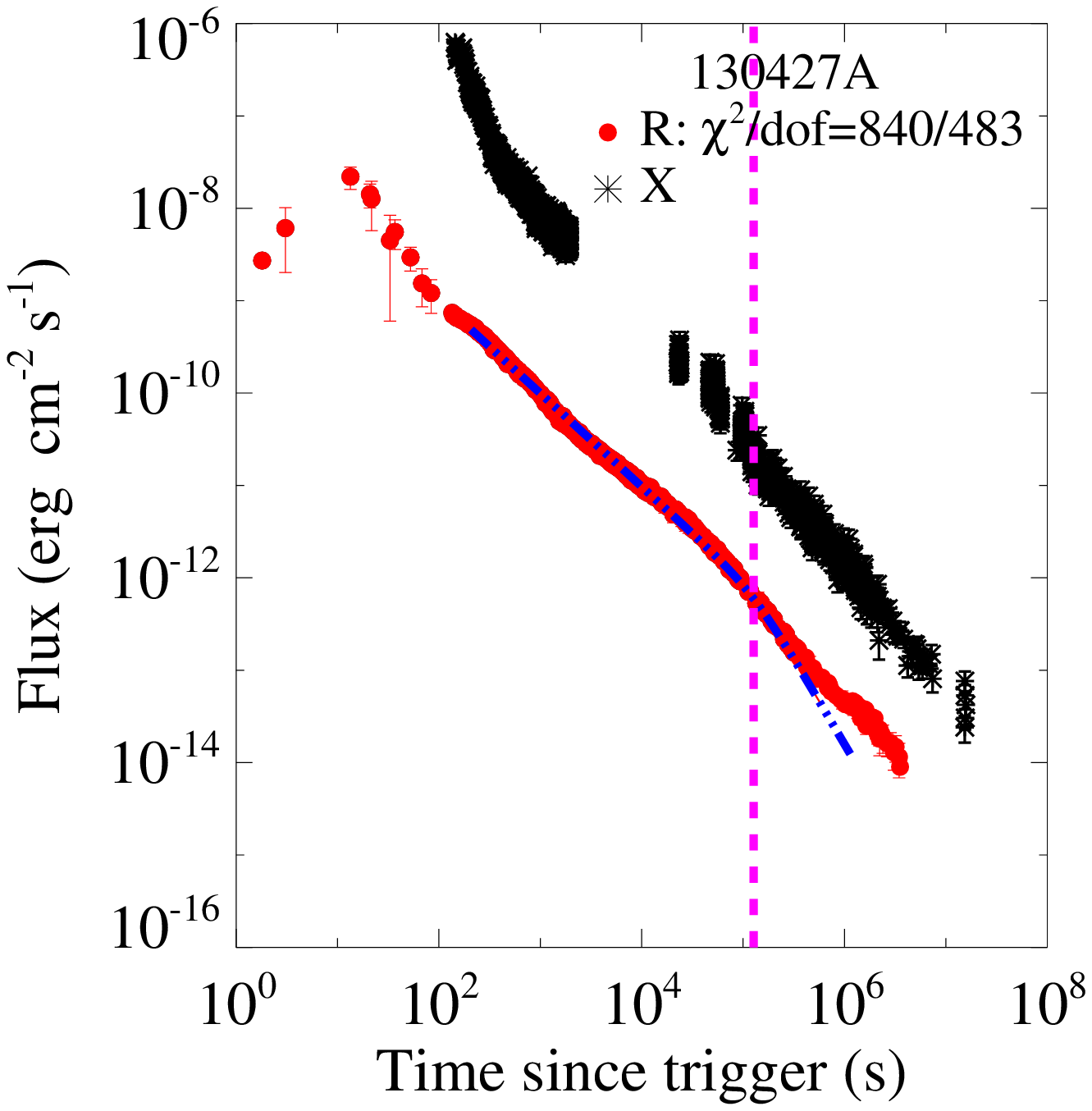}
\includegraphics[angle=0,scale=0.2,width=0.19\textwidth,height=0.18\textheight]{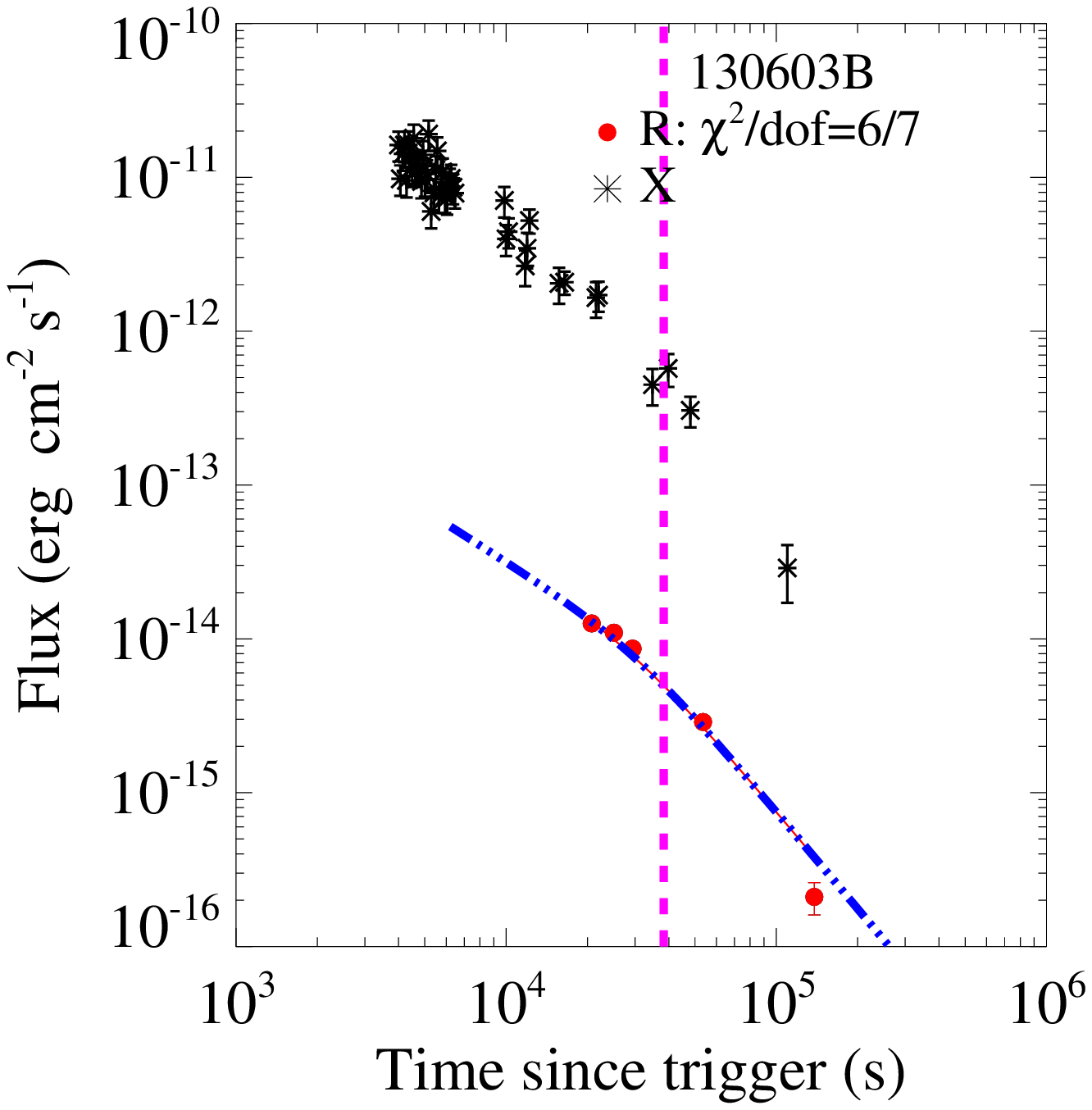}
\center{Figure. \ref{jetgrade}(Continued)}
%\caption{(Continued)} \label{jetgrade}
\end{figure*}

\clearpage
\setlength{\voffset}{-18mm}
\begin{figure*}
\includegraphics[angle=0,scale=0.2,width=0.19\textwidth,height=0.18\textheight]{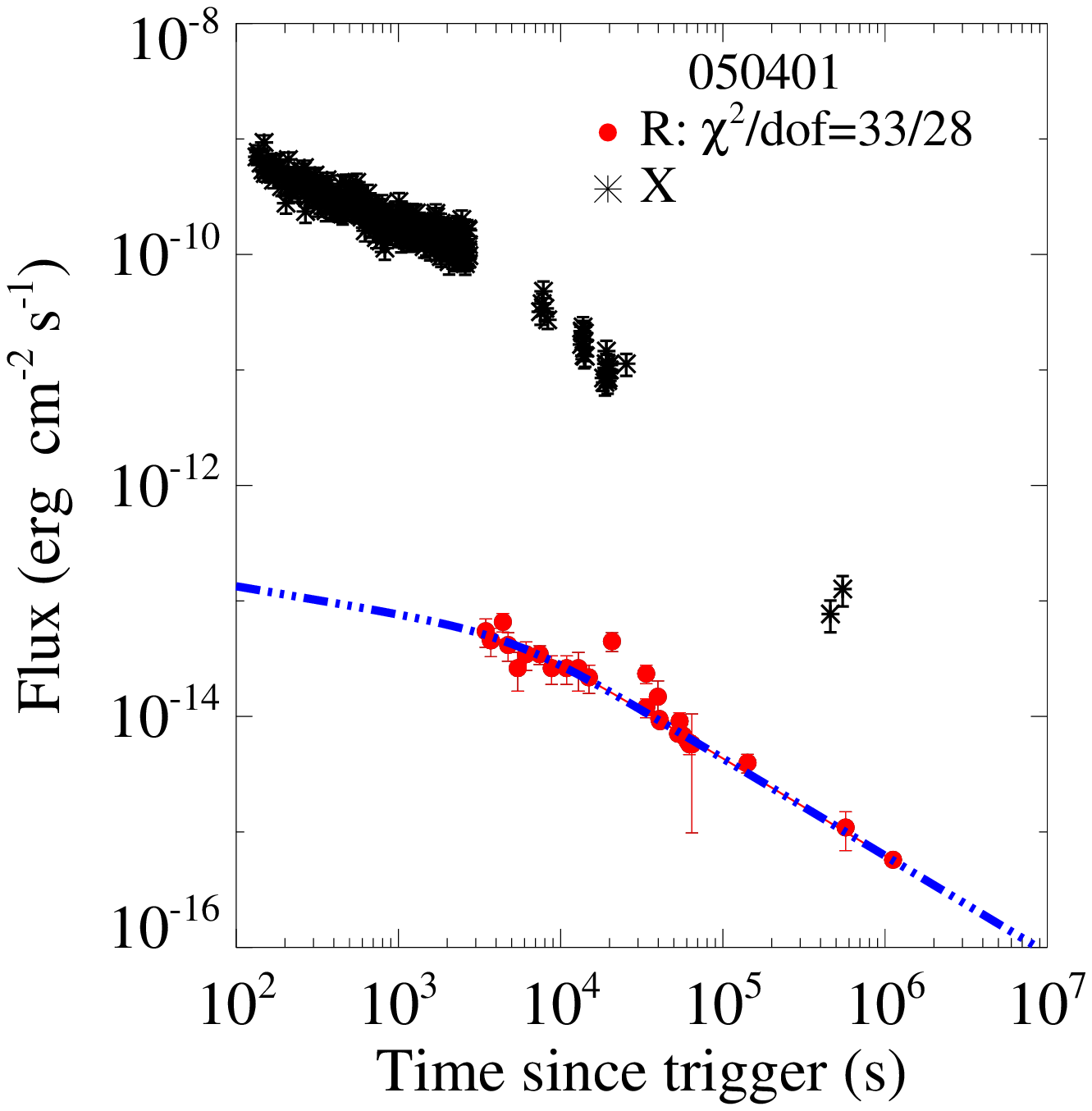}
\includegraphics[angle=0,scale=0.2,width=0.19\textwidth,height=0.18\textheight]{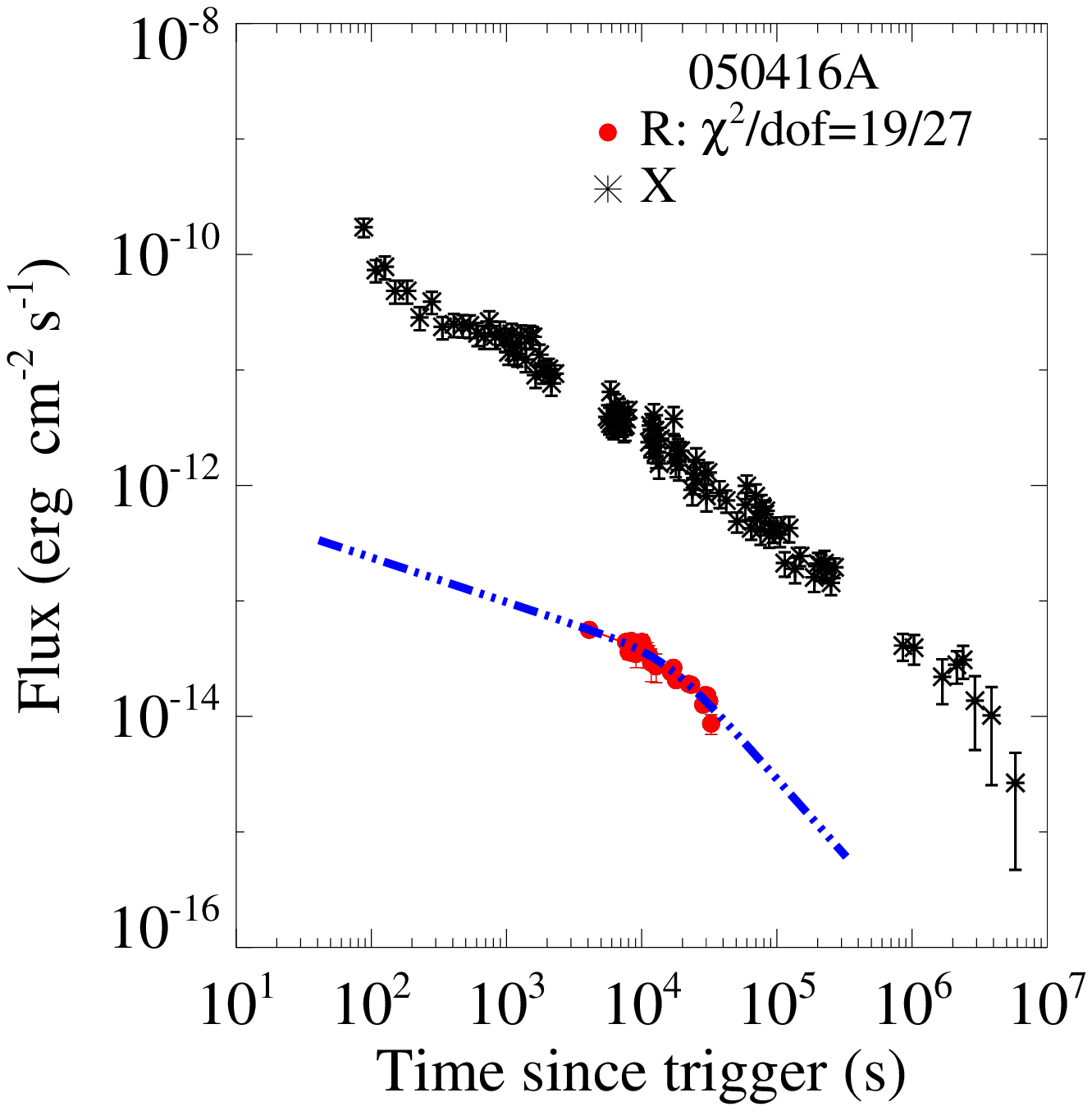}
\includegraphics[angle=0,scale=0.2,width=0.19\textwidth,height=0.18\textheight]{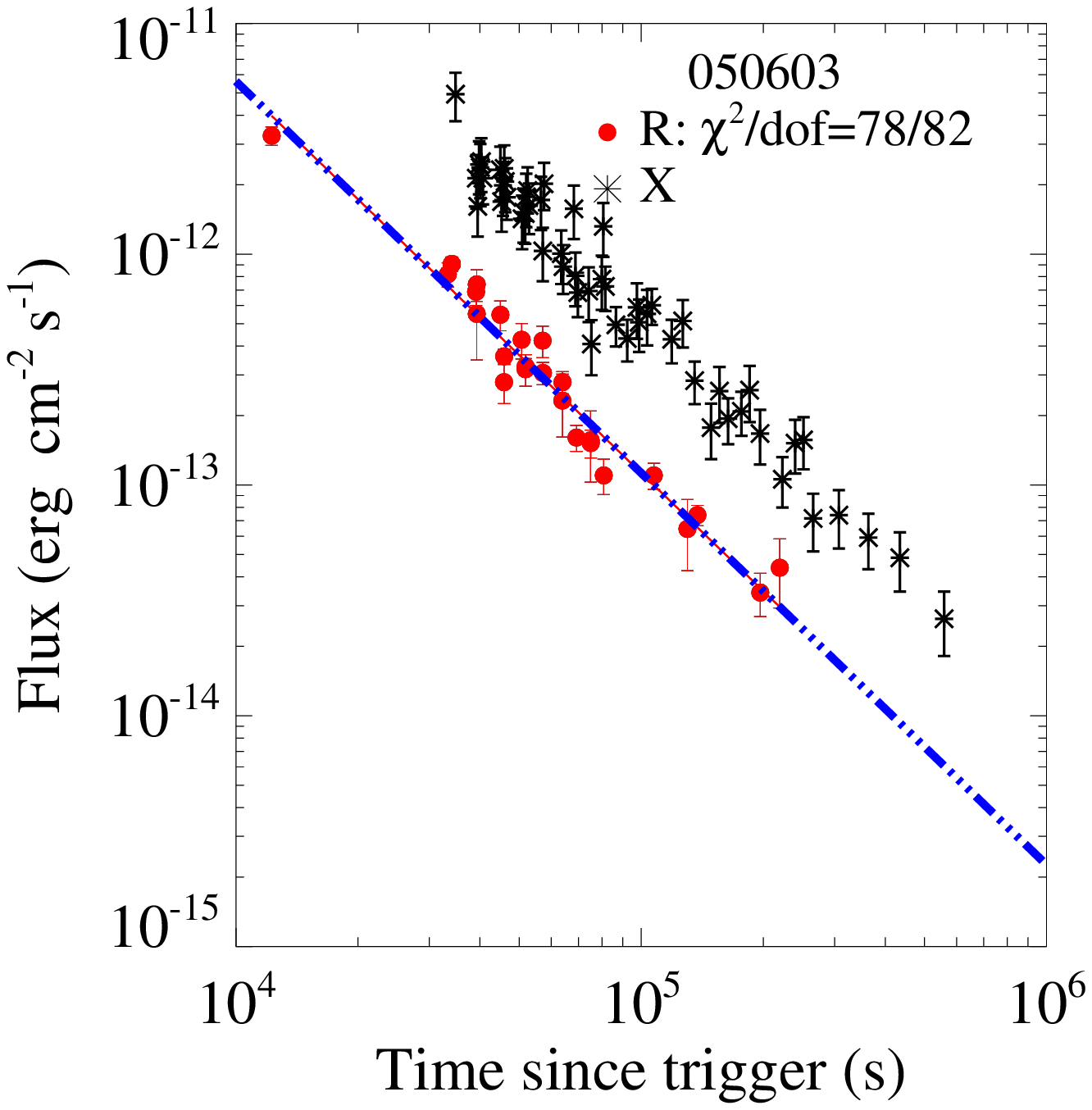}
\includegraphics[angle=0,scale=0.2,width=0.19\textwidth,height=0.18\textheight]{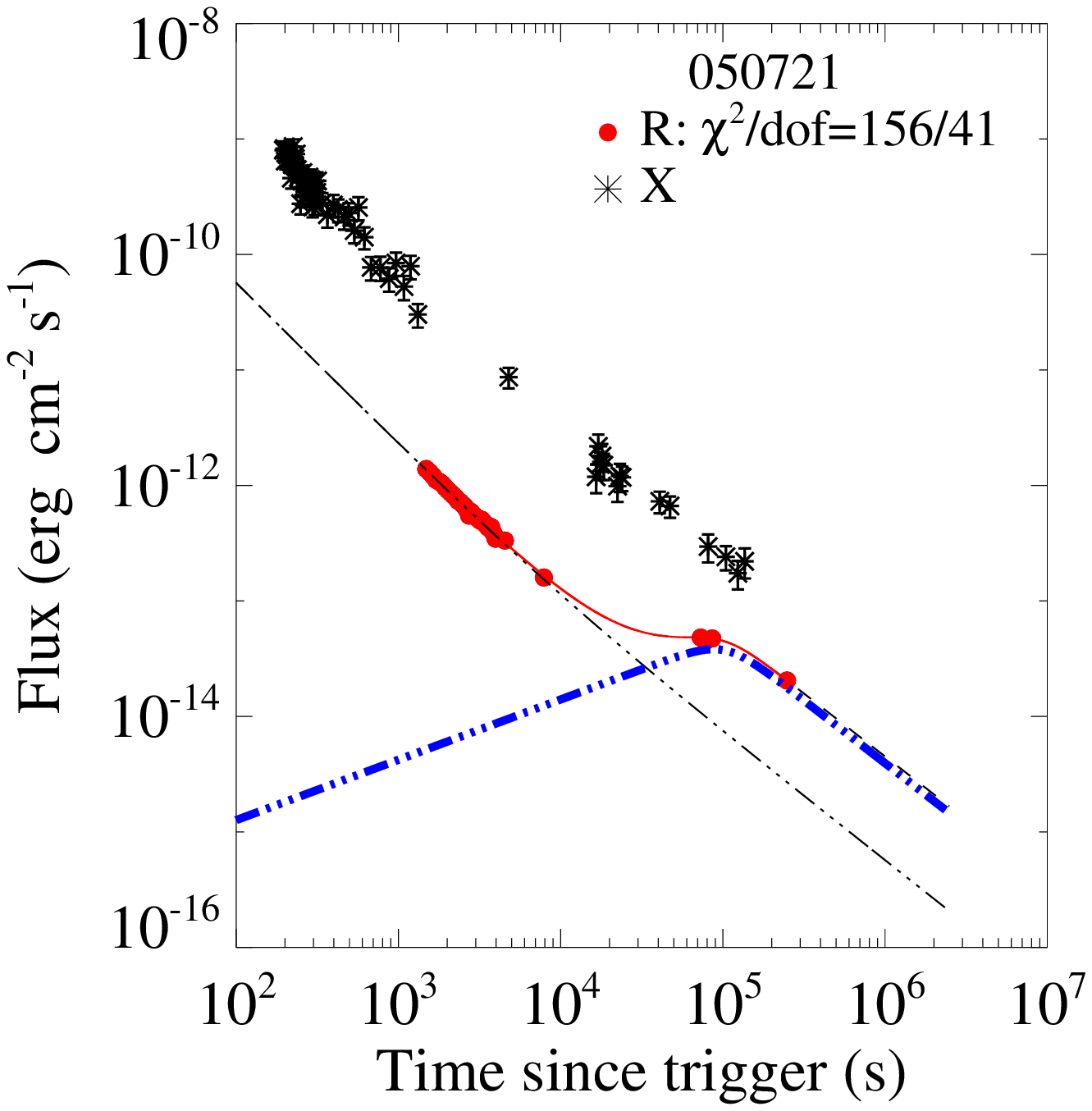}
\includegraphics[angle=0,scale=0.2,width=0.19\textwidth,height=0.18\textheight]{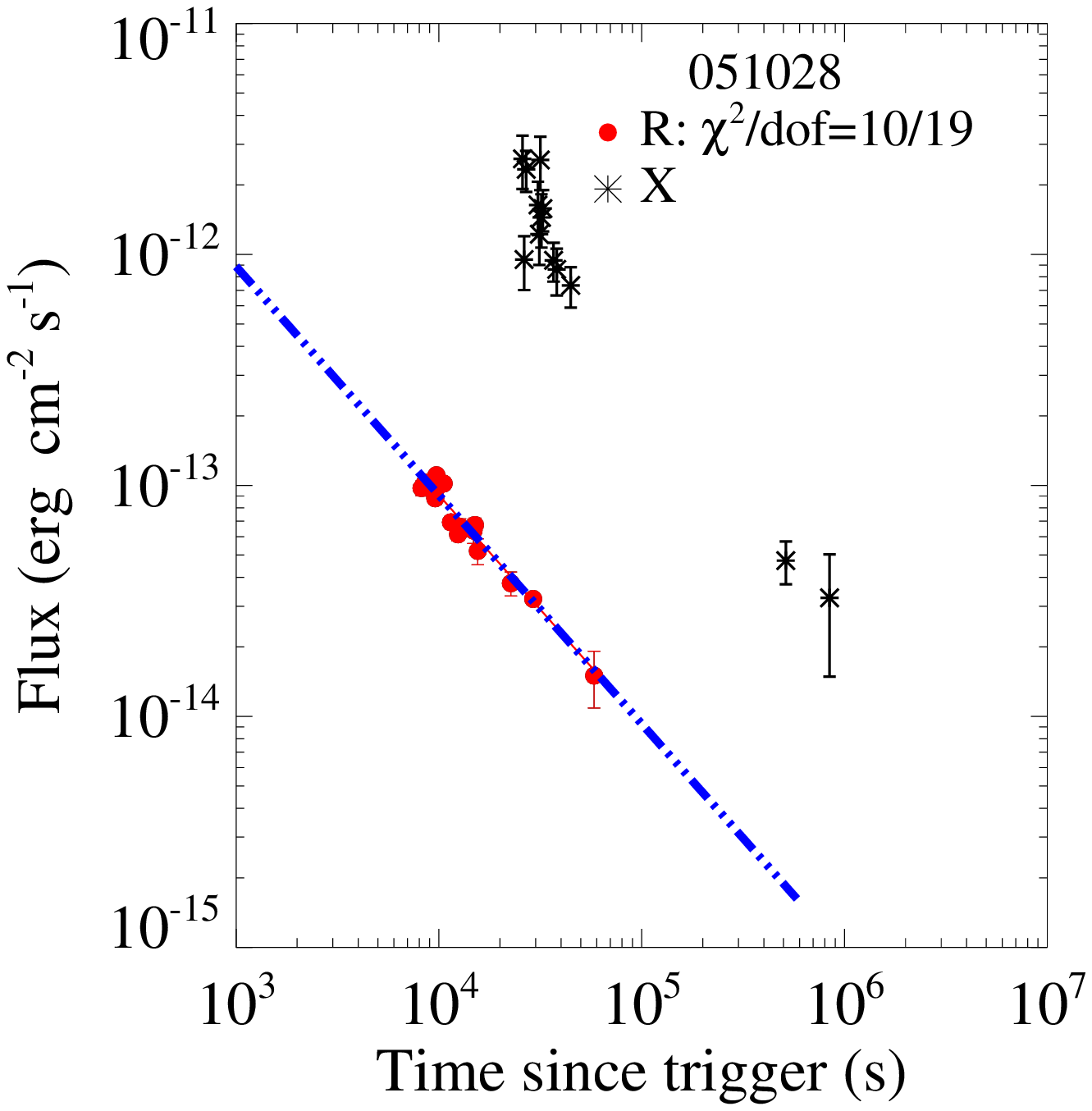}
\includegraphics[angle=0,scale=0.2,width=0.19\textwidth,height=0.18\textheight]{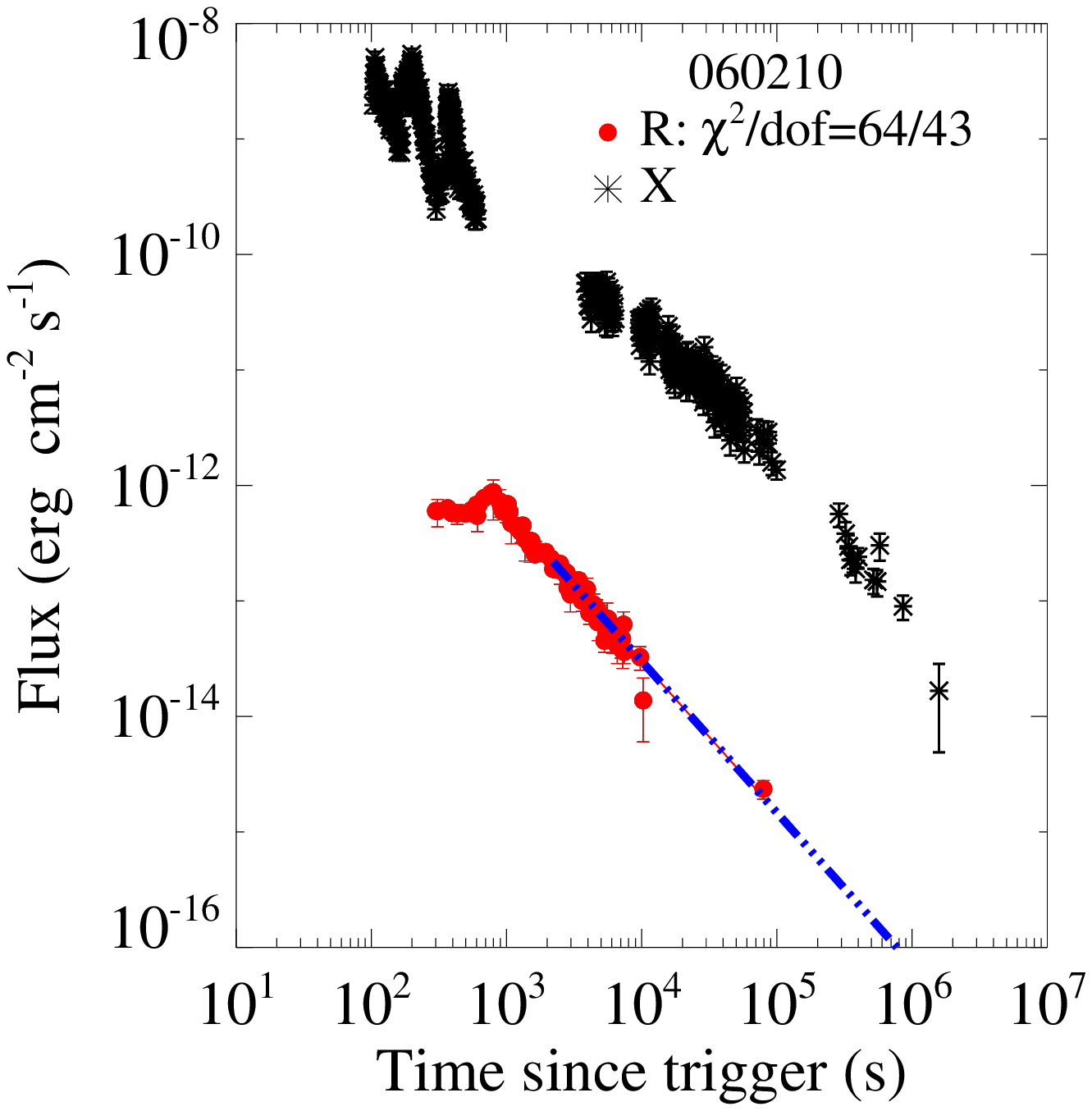}
\includegraphics[angle=0,scale=0.2,width=0.19\textwidth,height=0.18\textheight]{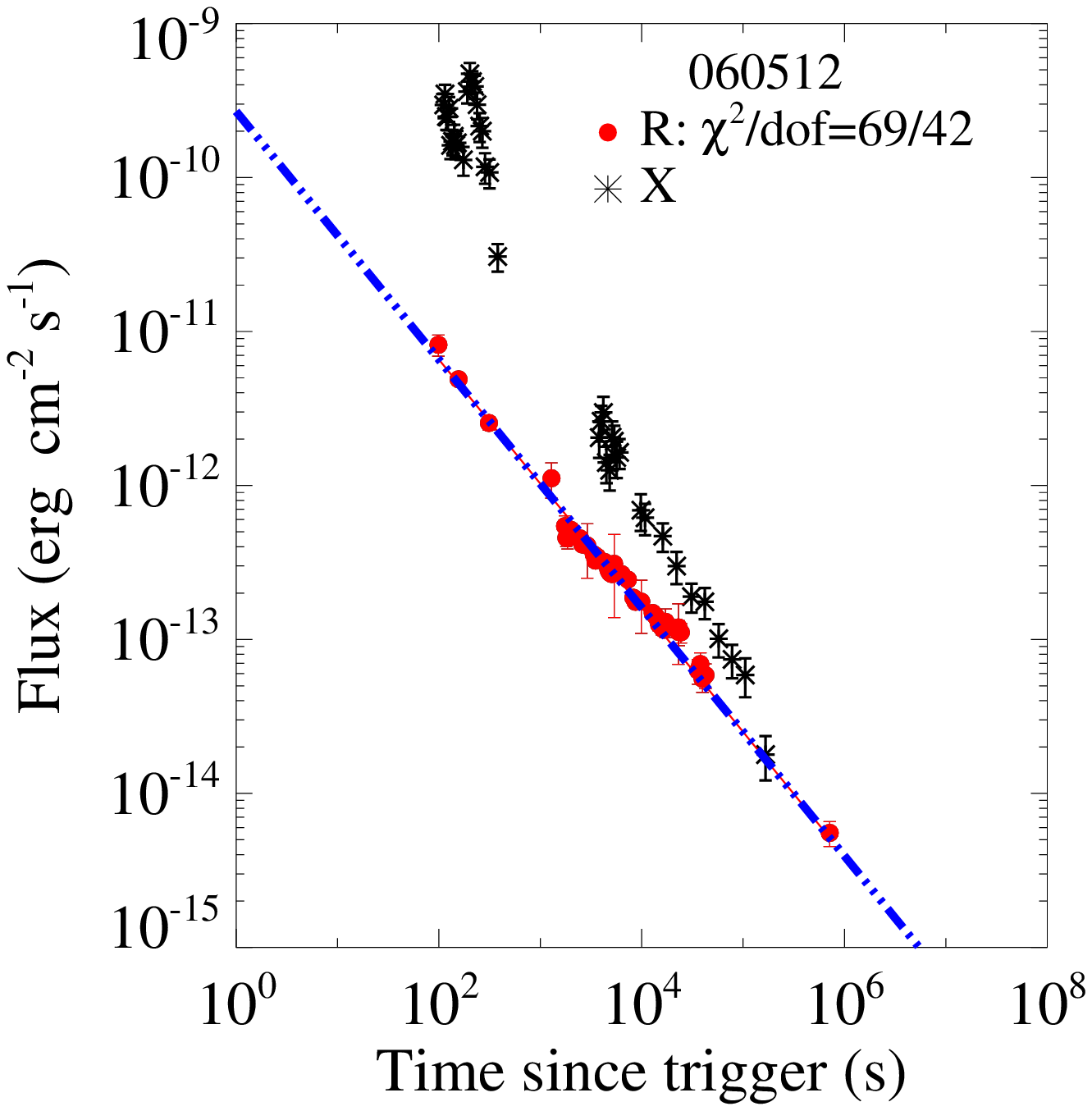}
\includegraphics[angle=0,scale=0.2,width=0.19\textwidth,height=0.18\textheight]{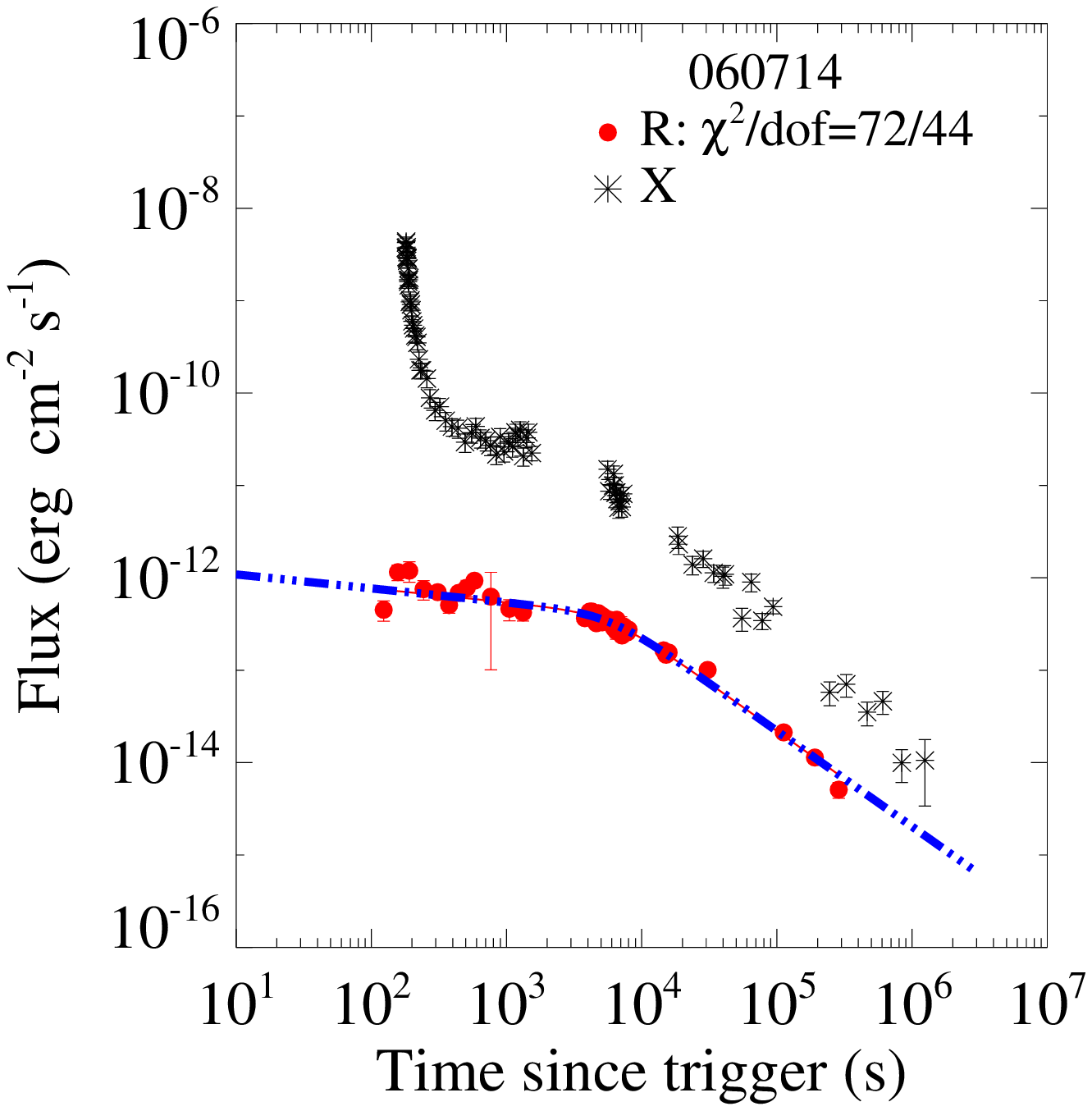}
\includegraphics[angle=0,scale=0.2,width=0.19\textwidth,height=0.18\textheight]{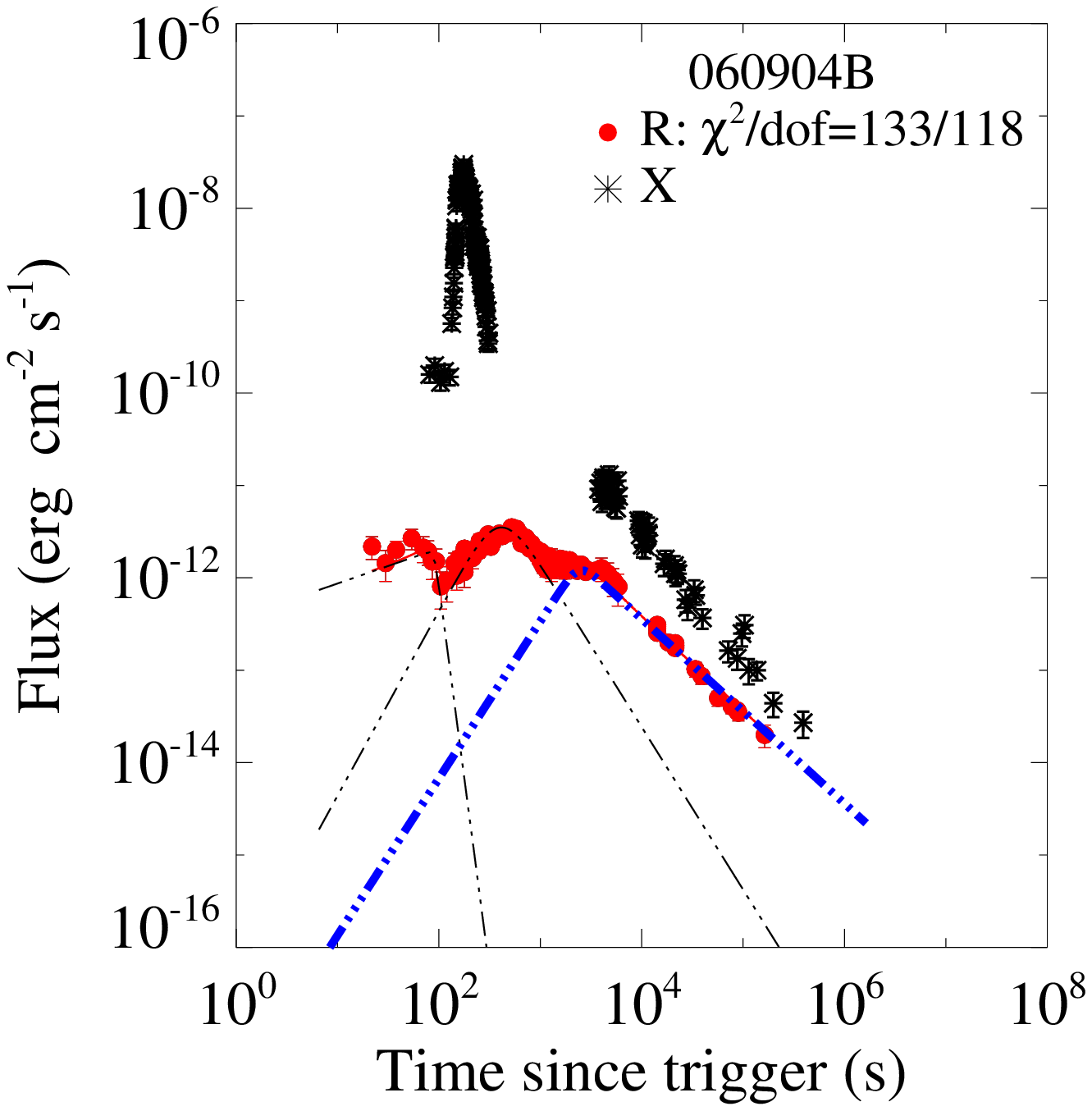}
\includegraphics[angle=0,scale=0.2,width=0.19\textwidth,height=0.18\textheight]{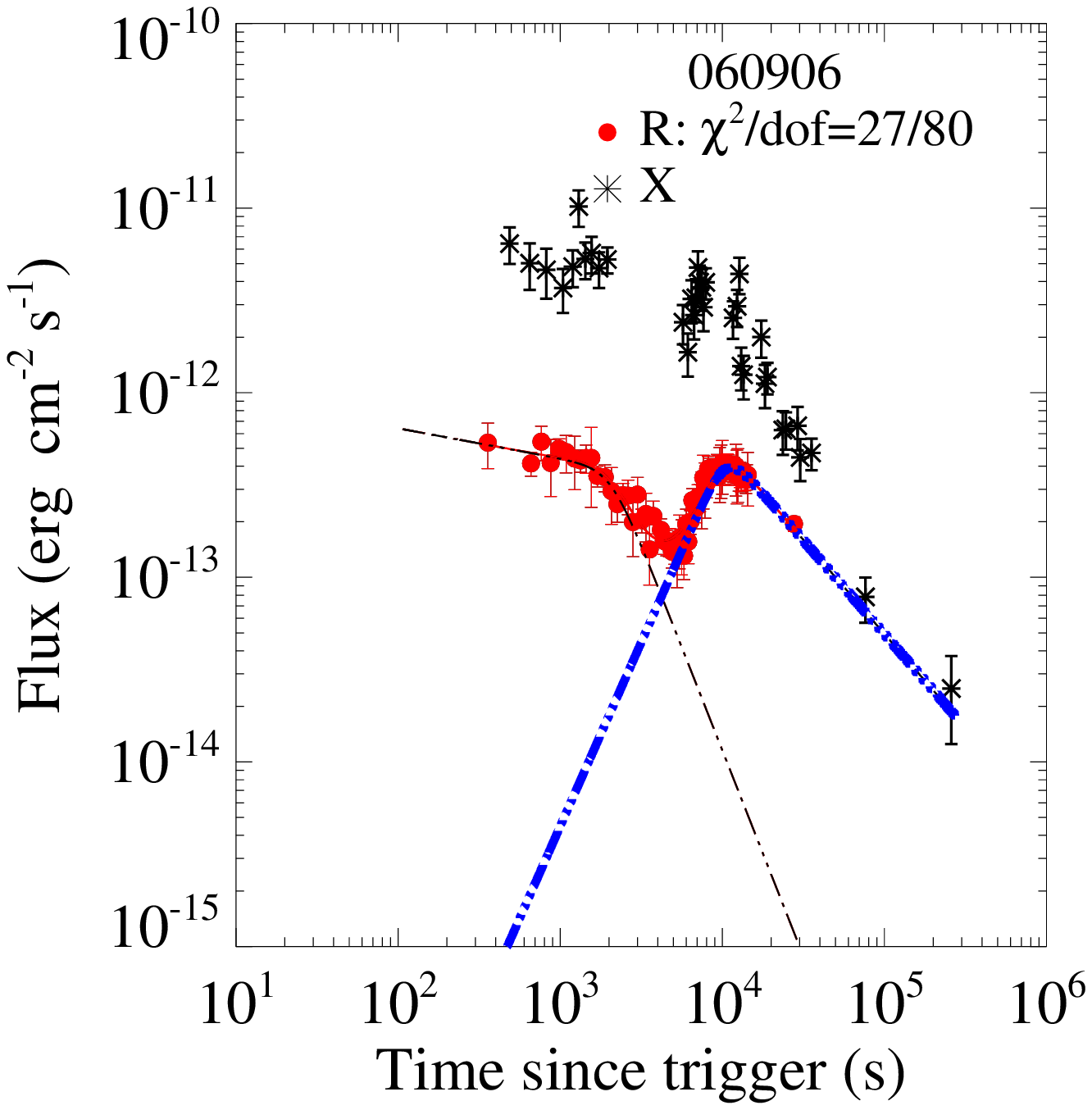}
\includegraphics[angle=0,scale=0.2,width=0.19\textwidth,height=0.18\textheight]{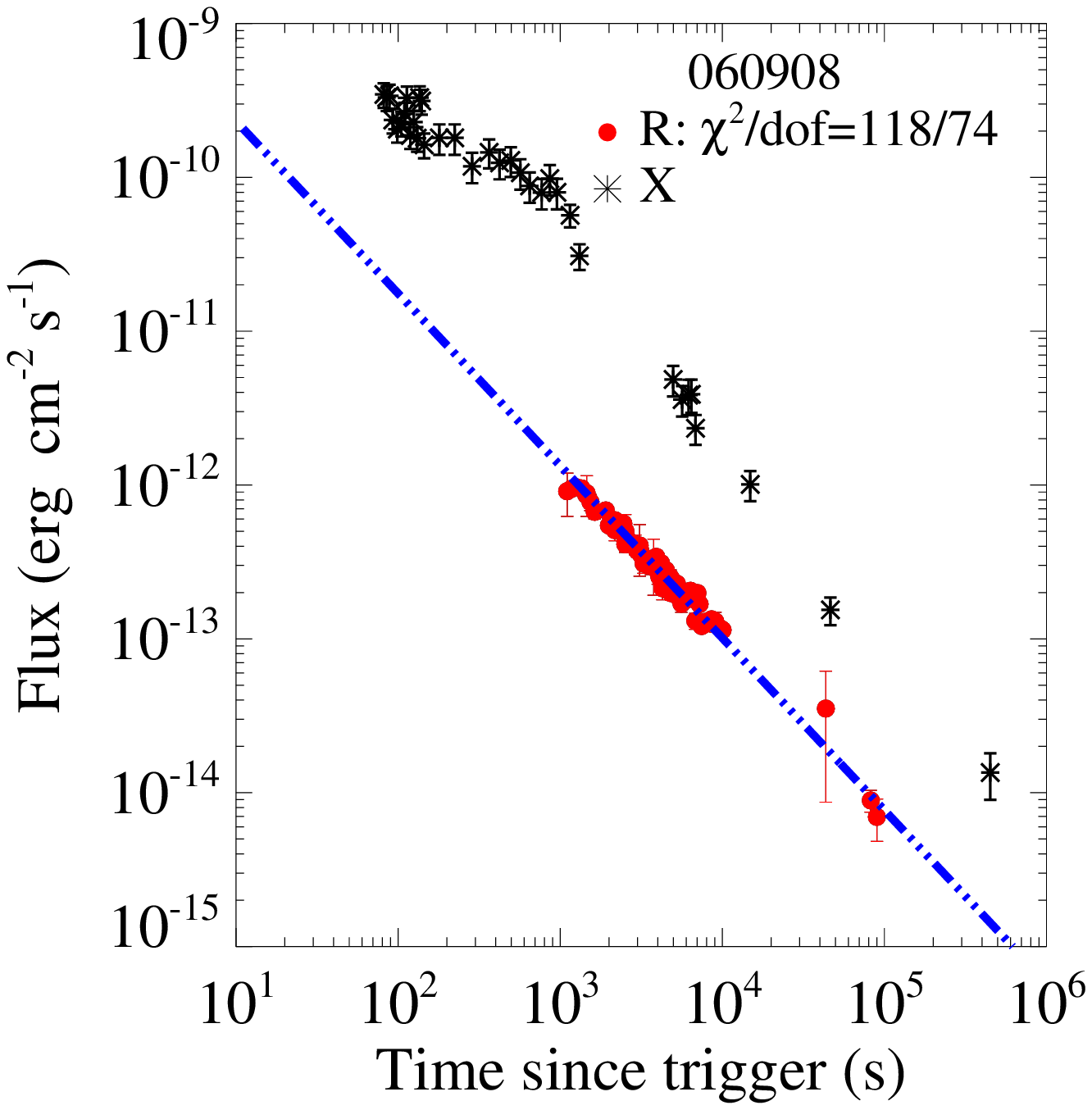}
\includegraphics[angle=0,scale=0.2,width=0.19\textwidth,height=0.18\textheight]{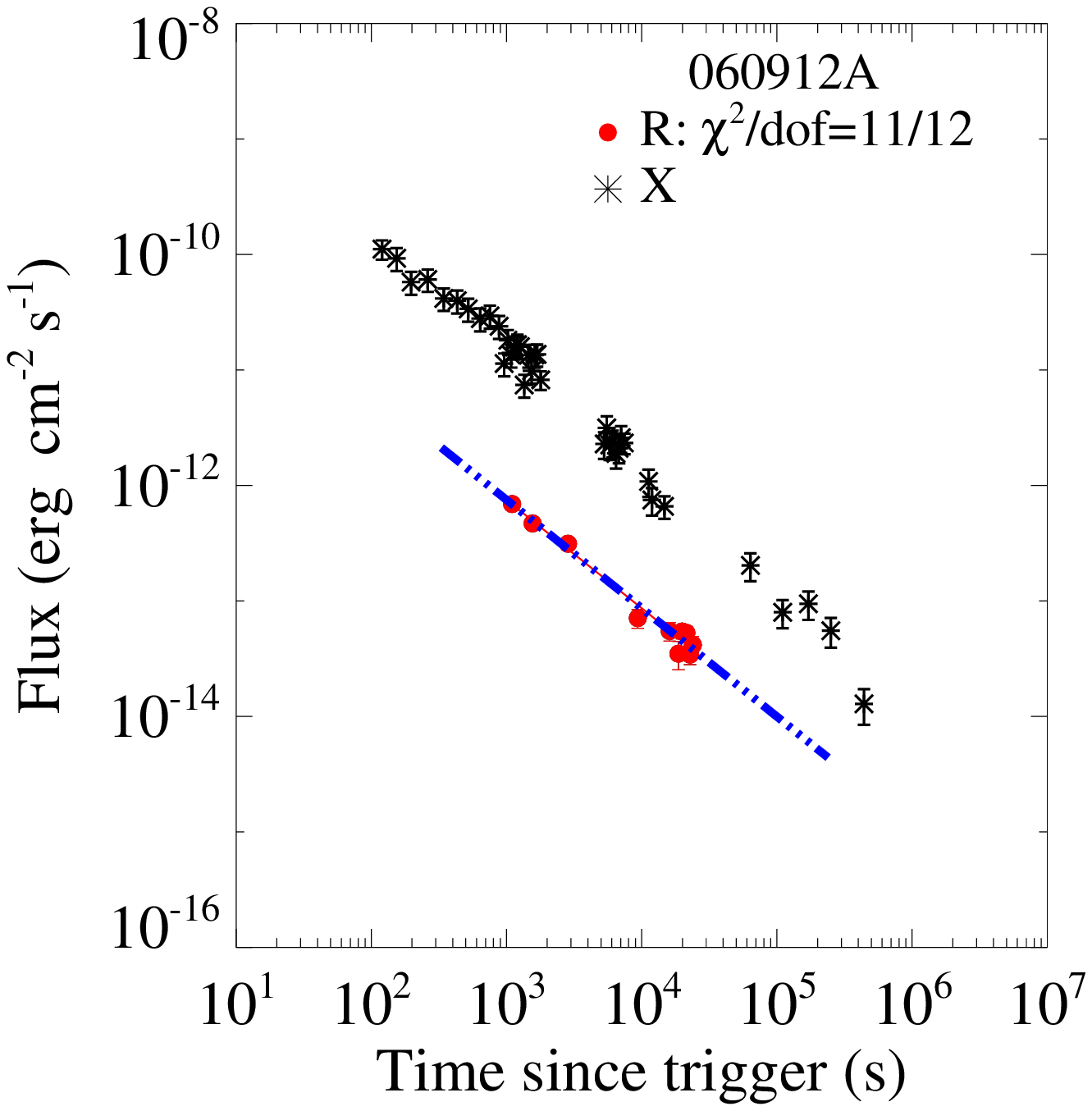}
\includegraphics[angle=0,scale=0.2,width=0.19\textwidth,height=0.18\textheight]{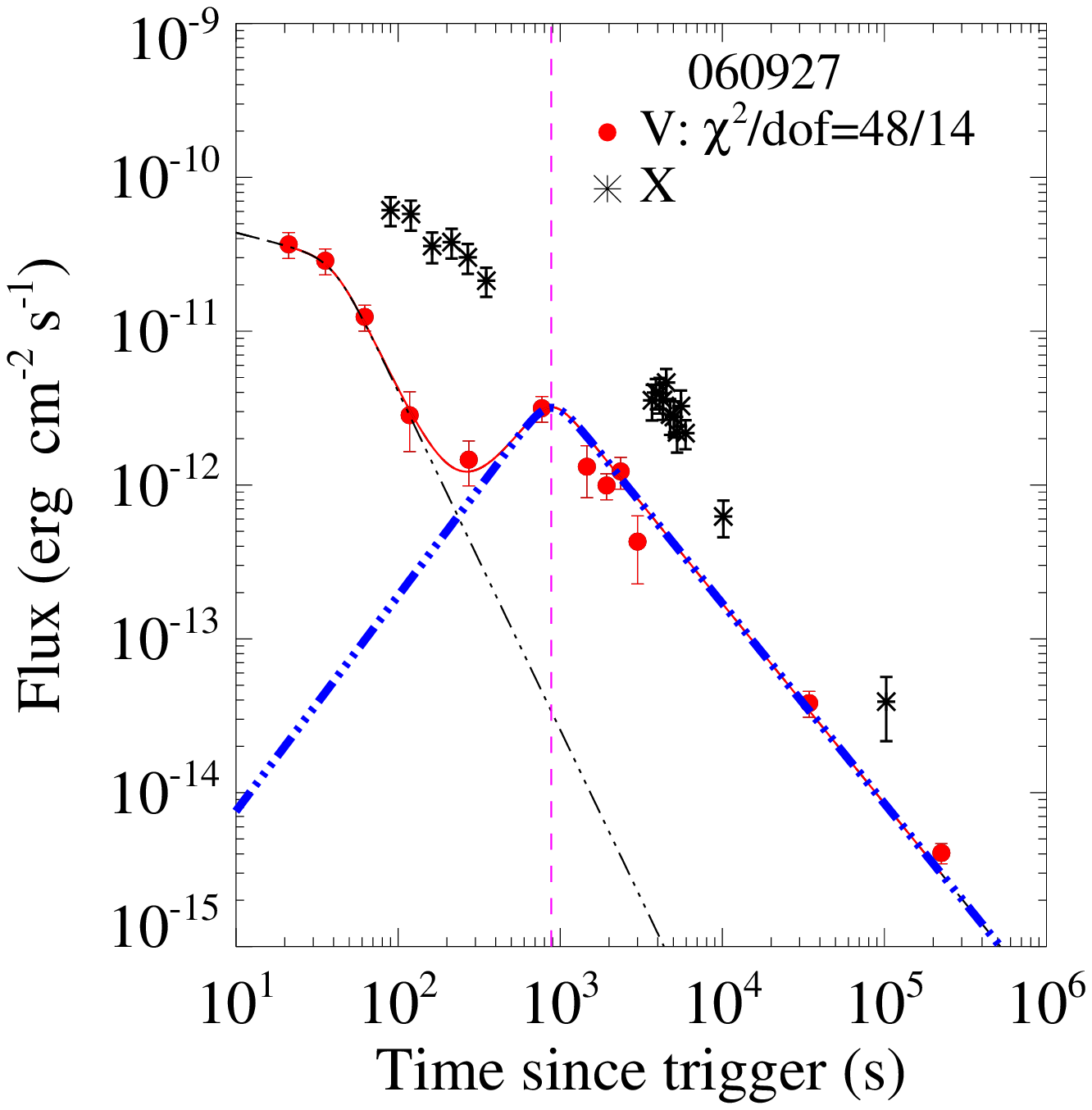}
\includegraphics[angle=0,scale=0.2,width=0.19\textwidth,height=0.18\textheight]{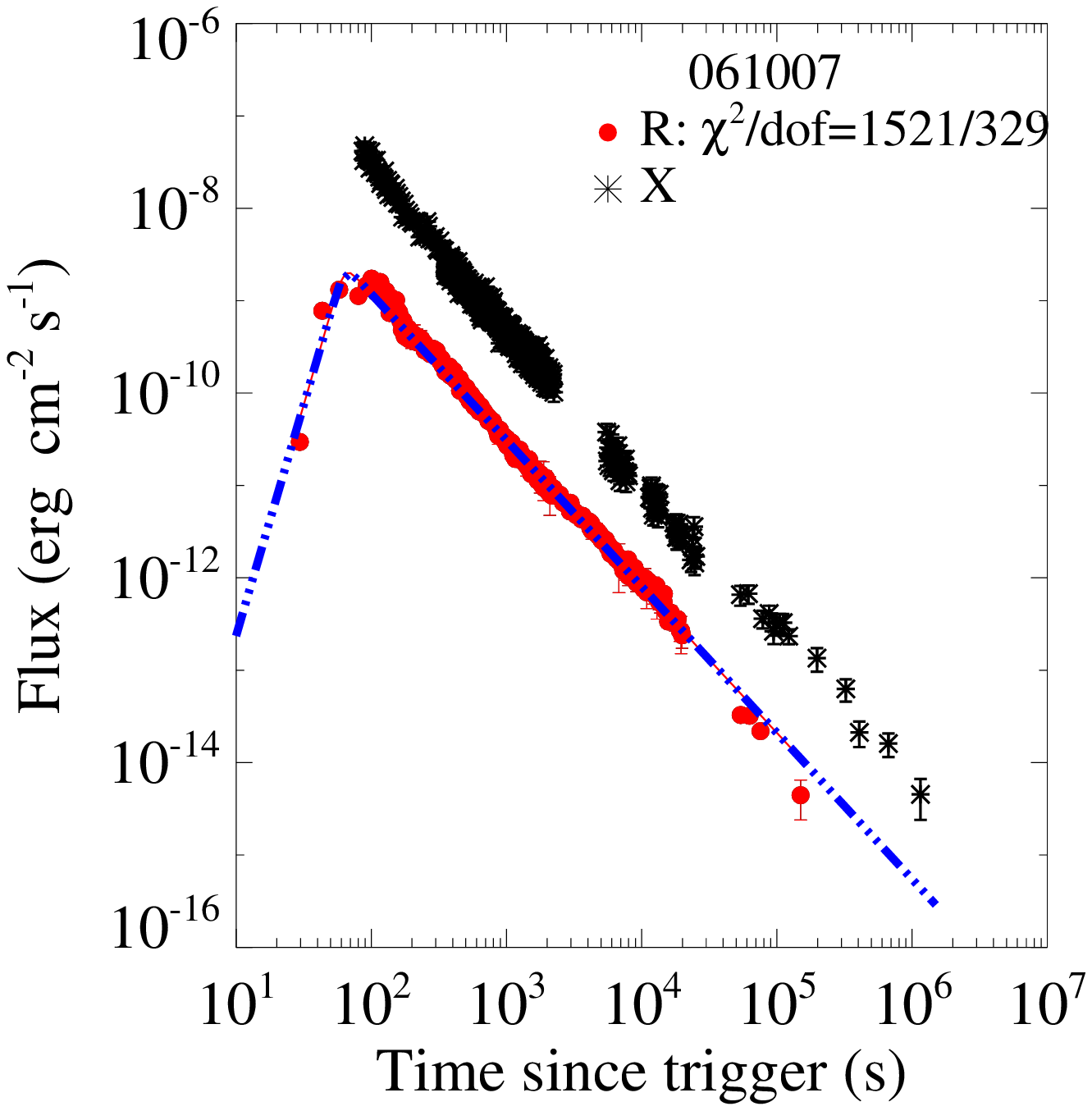}
\includegraphics[angle=0,scale=0.2,width=0.19\textwidth,height=0.18\textheight]{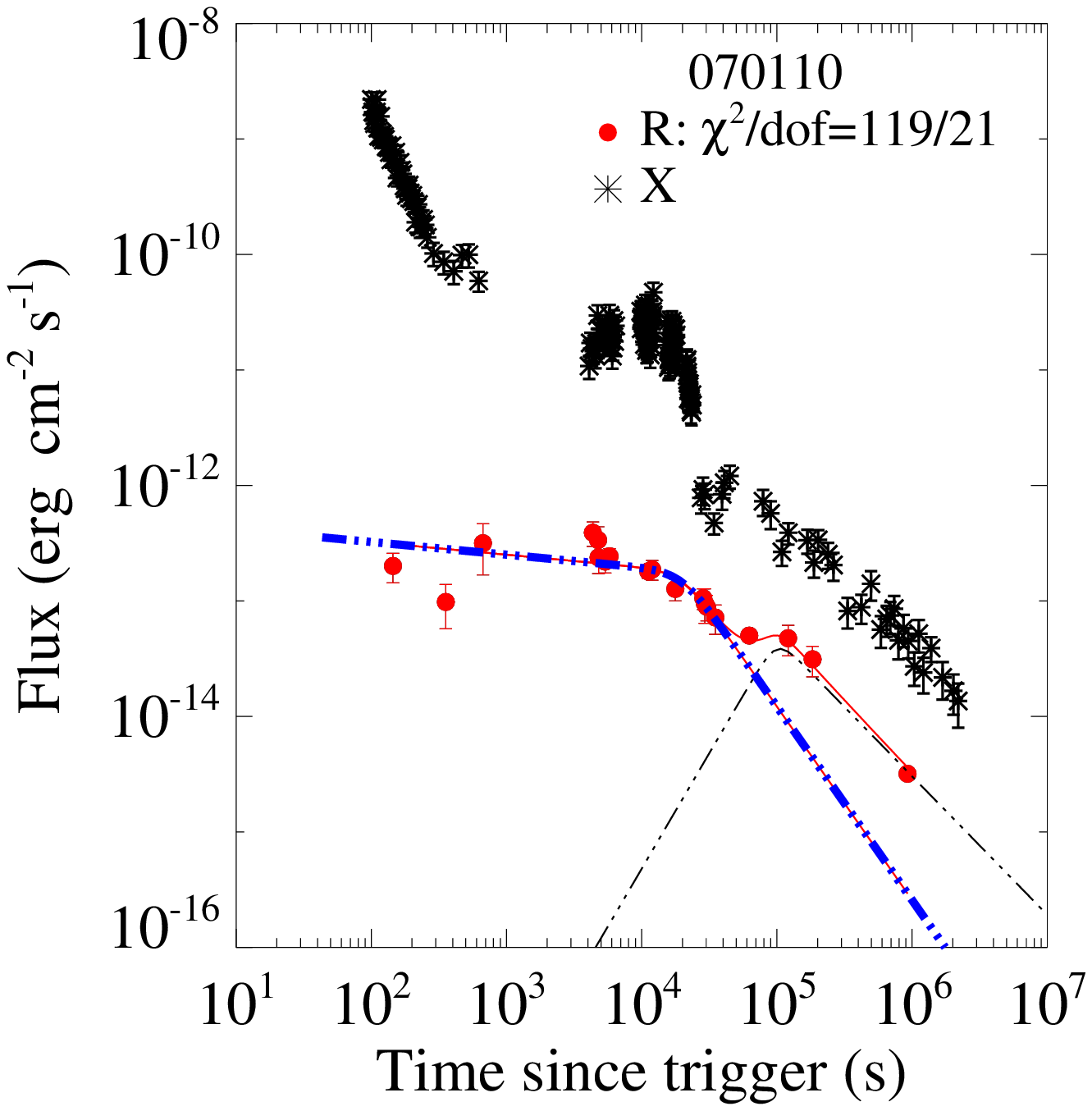}
\includegraphics[angle=0,scale=0.2,width=0.19\textwidth,height=0.18\textheight]{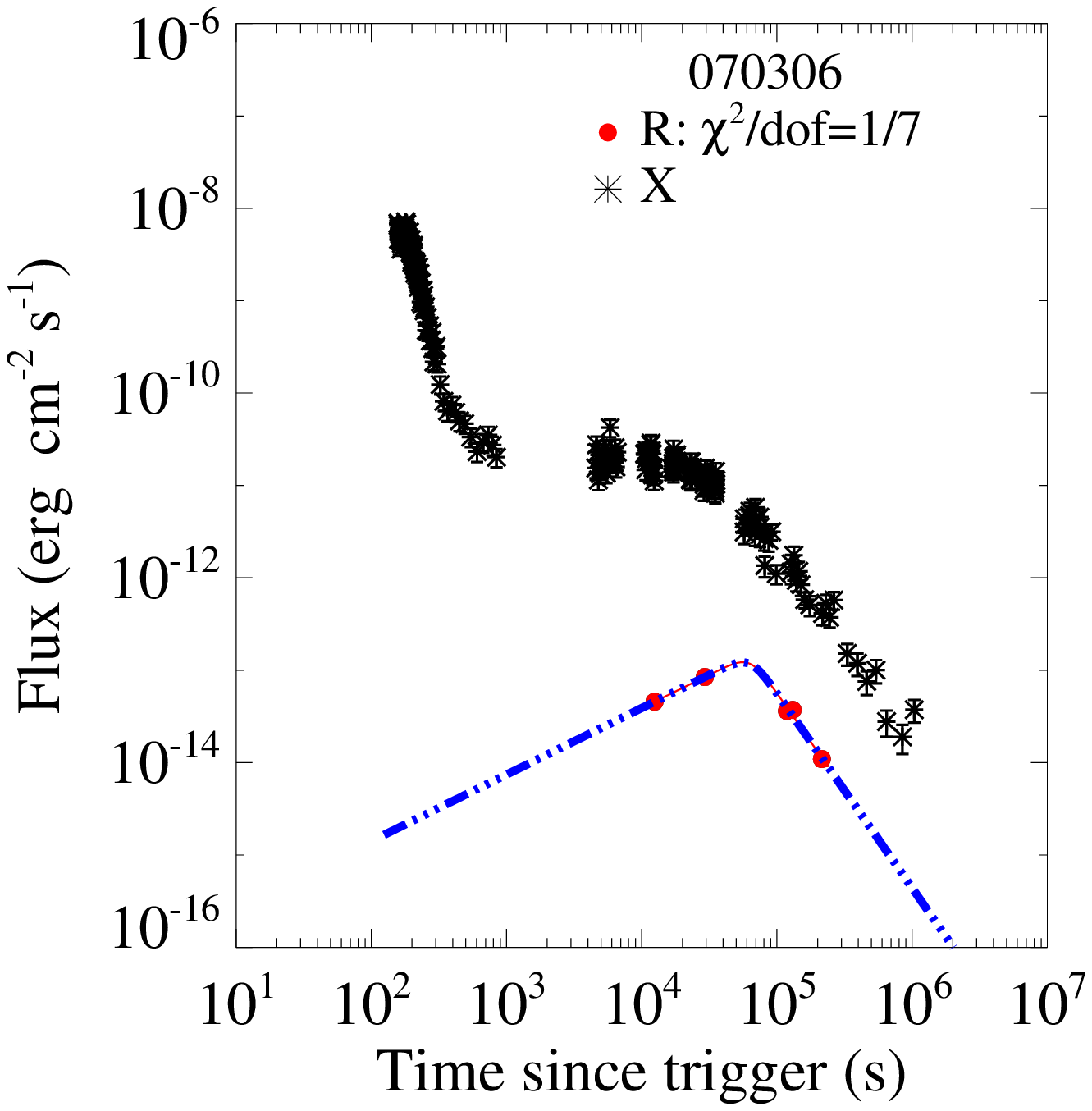}
\includegraphics[angle=0,scale=0.2,width=0.19\textwidth,height=0.18\textheight]{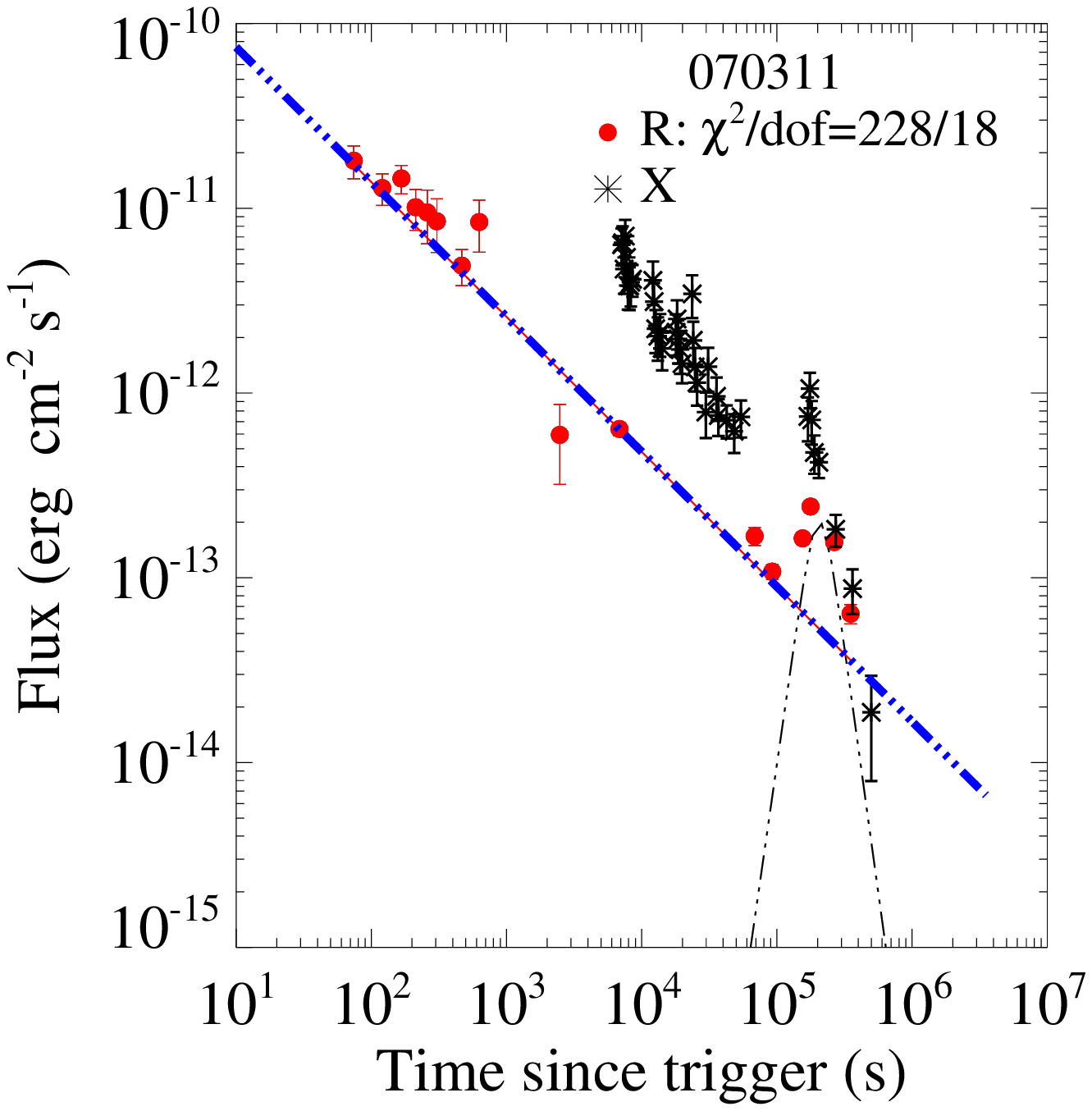}
\includegraphics[angle=0,scale=0.2,width=0.19\textwidth,height=0.18\textheight]{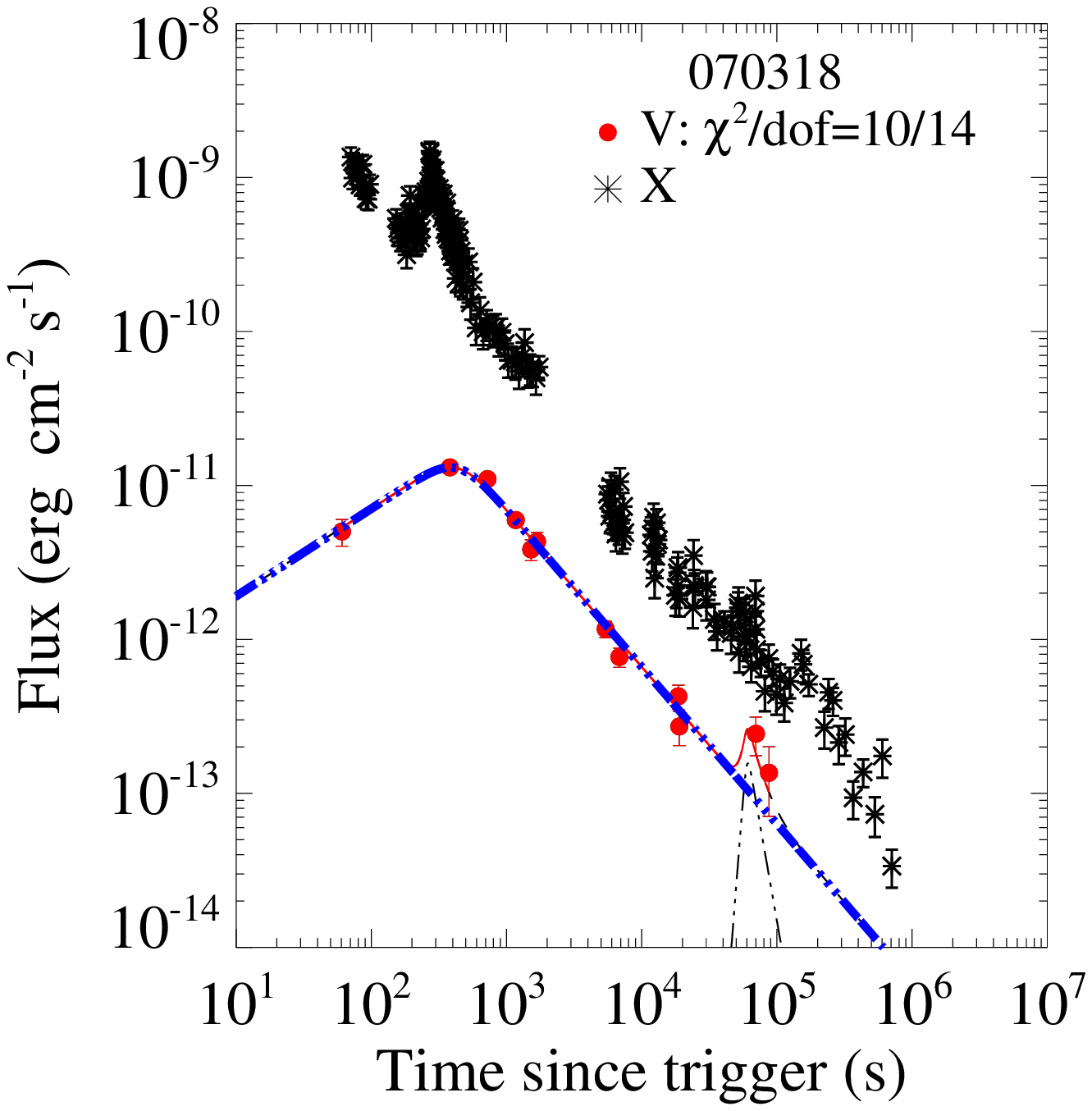}
\includegraphics[angle=0,scale=0.2,width=0.19\textwidth,height=0.18\textheight]{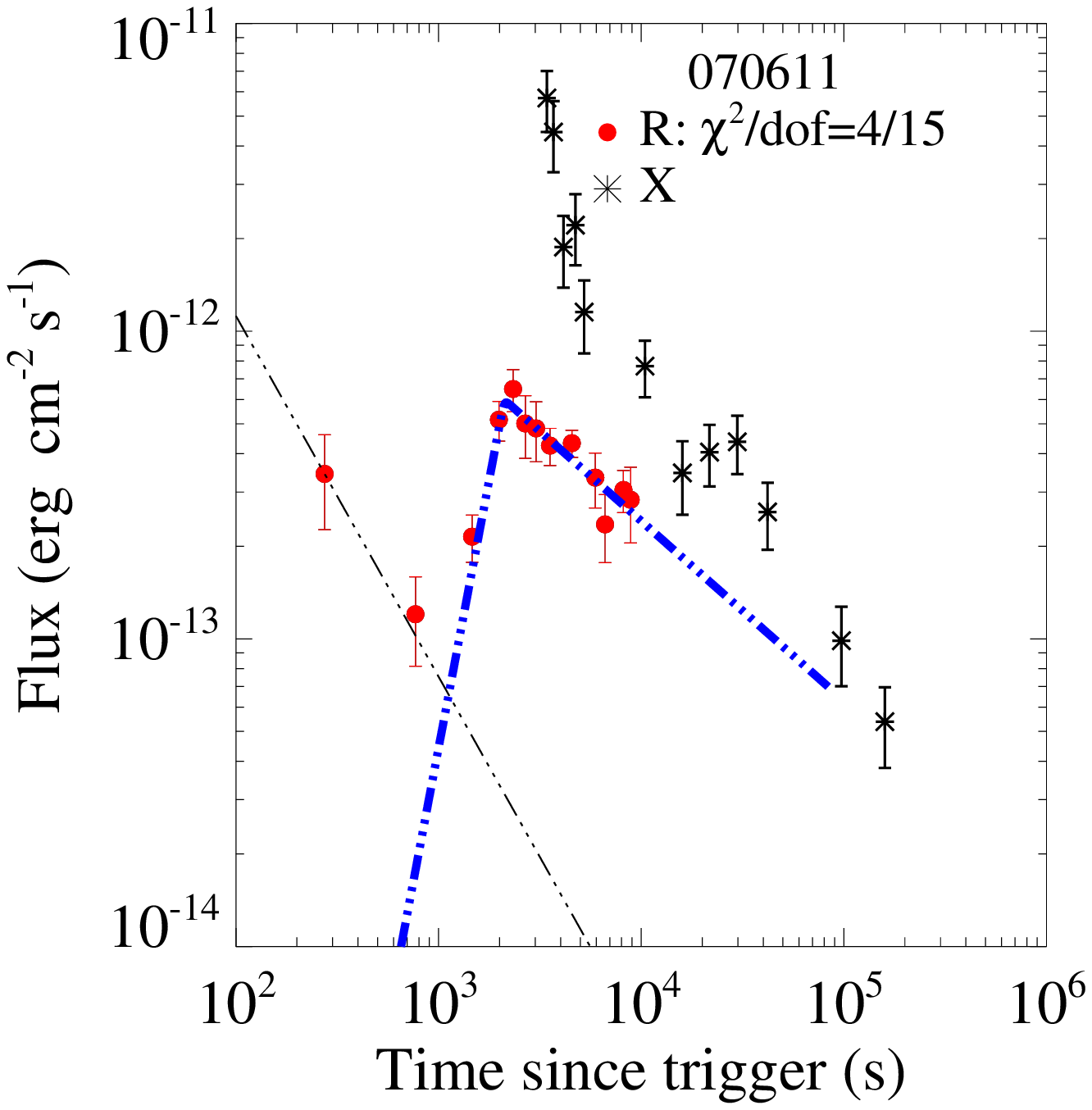}
\includegraphics[angle=0,scale=0.2,width=0.19\textwidth,height=0.18\textheight]{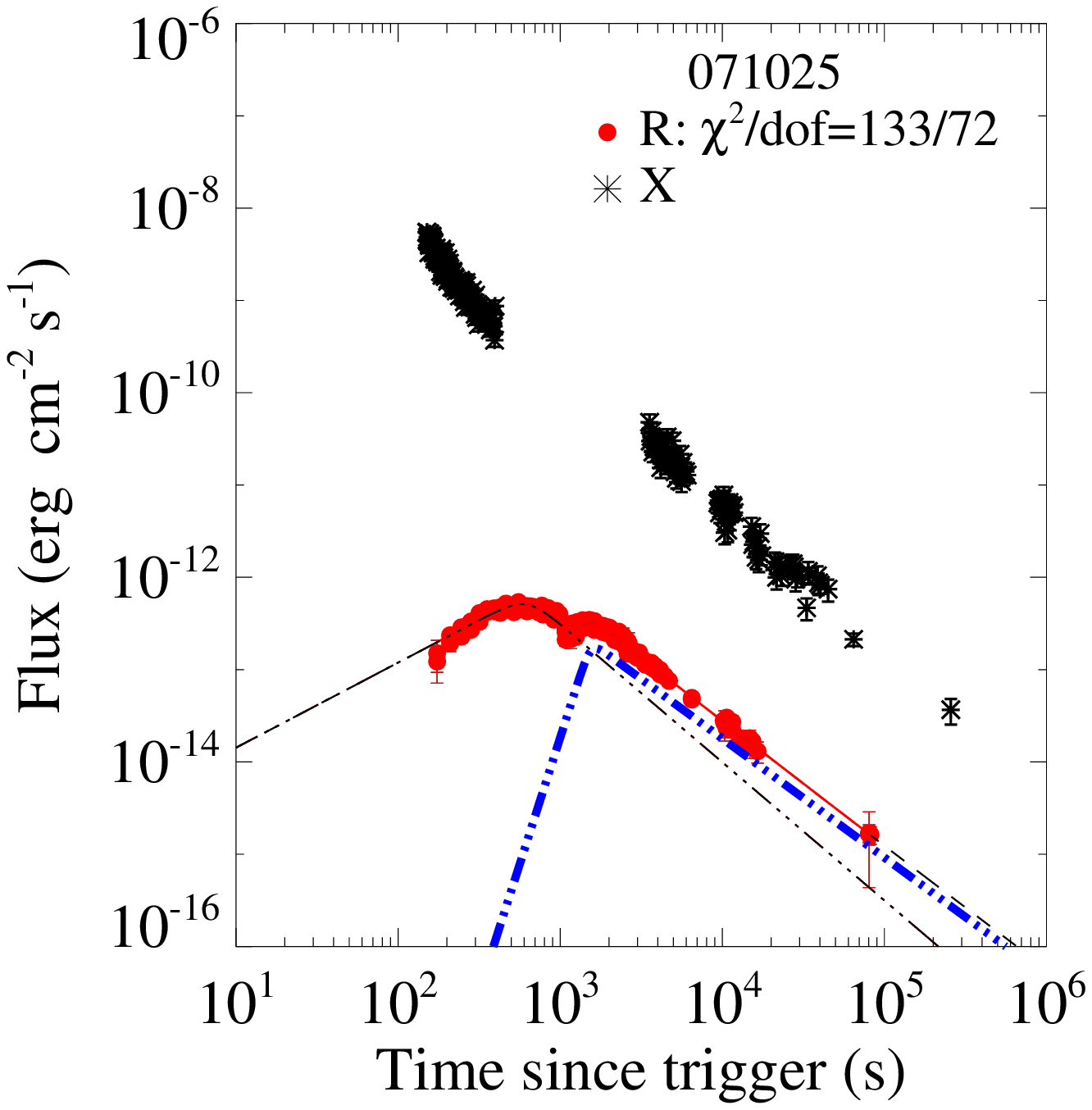}
\includegraphics[angle=0,scale=0.2,width=0.19\textwidth,height=0.18\textheight]{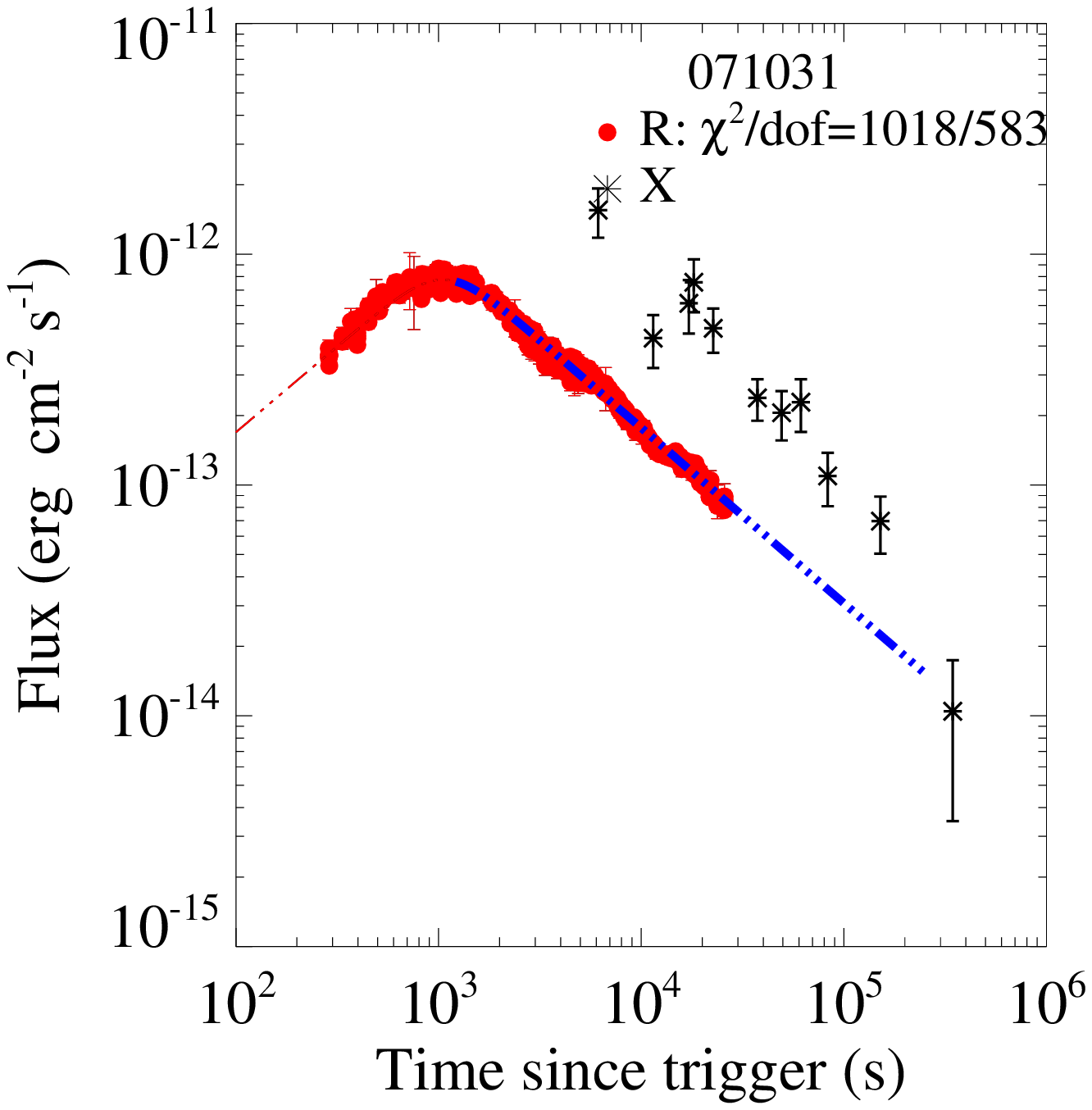}
\includegraphics[angle=0,scale=0.2,width=0.19\textwidth,height=0.18\textheight]{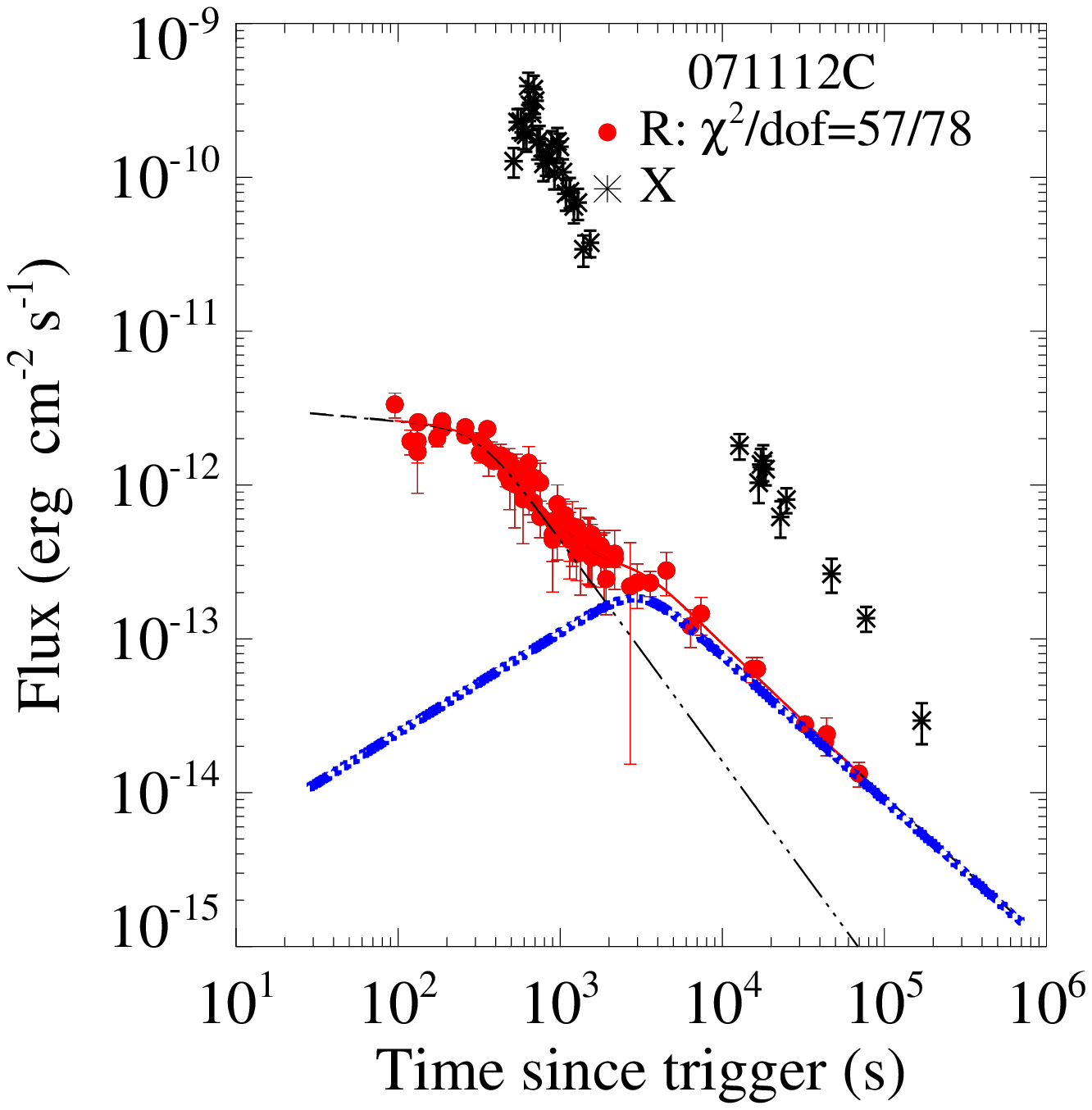}
\includegraphics[angle=0,scale=0.2,width=0.19\textwidth,height=0.18\textheight]{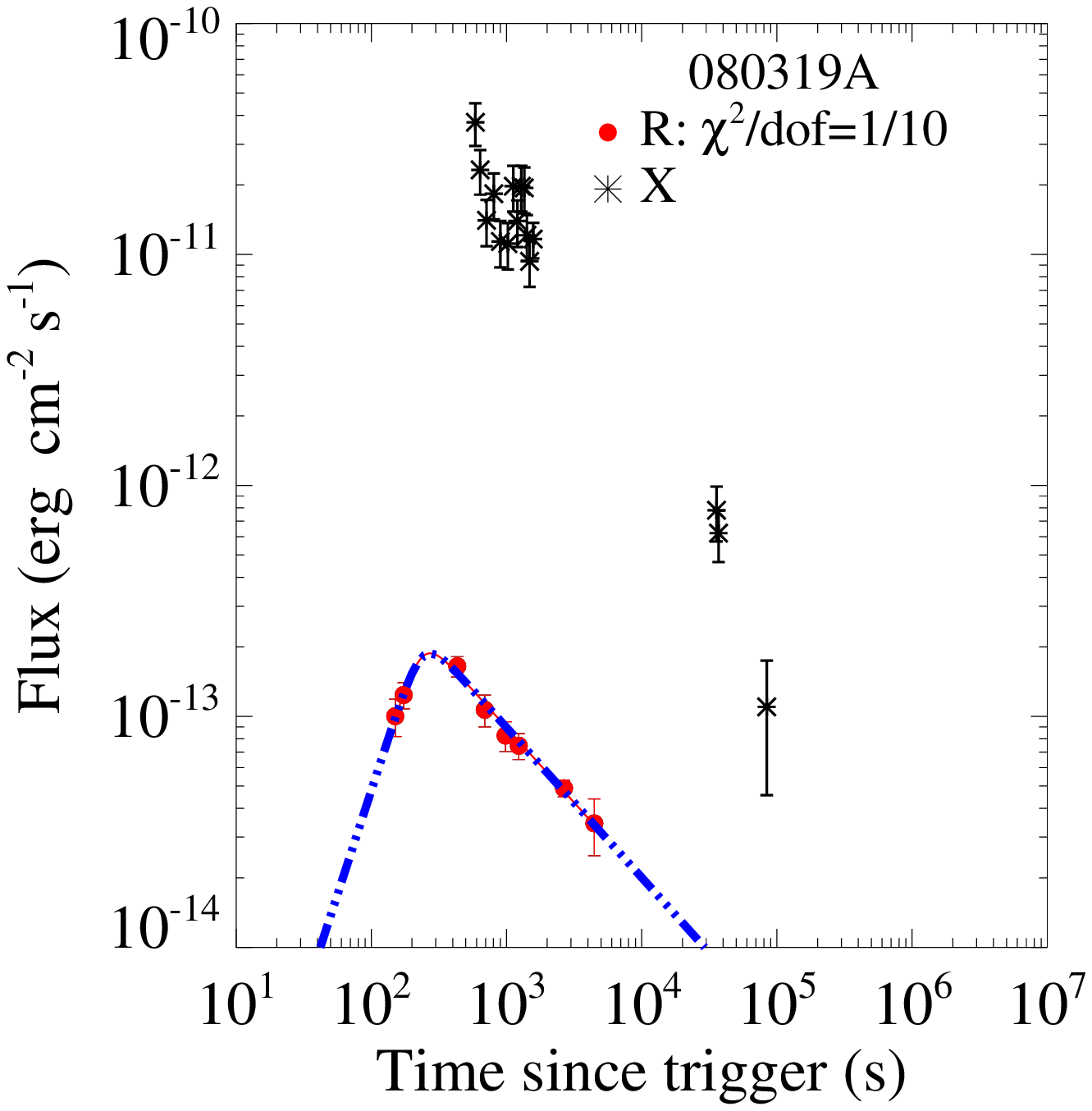}
\includegraphics[angle=0,scale=0.2,width=0.19\textwidth,height=0.18\textheight]{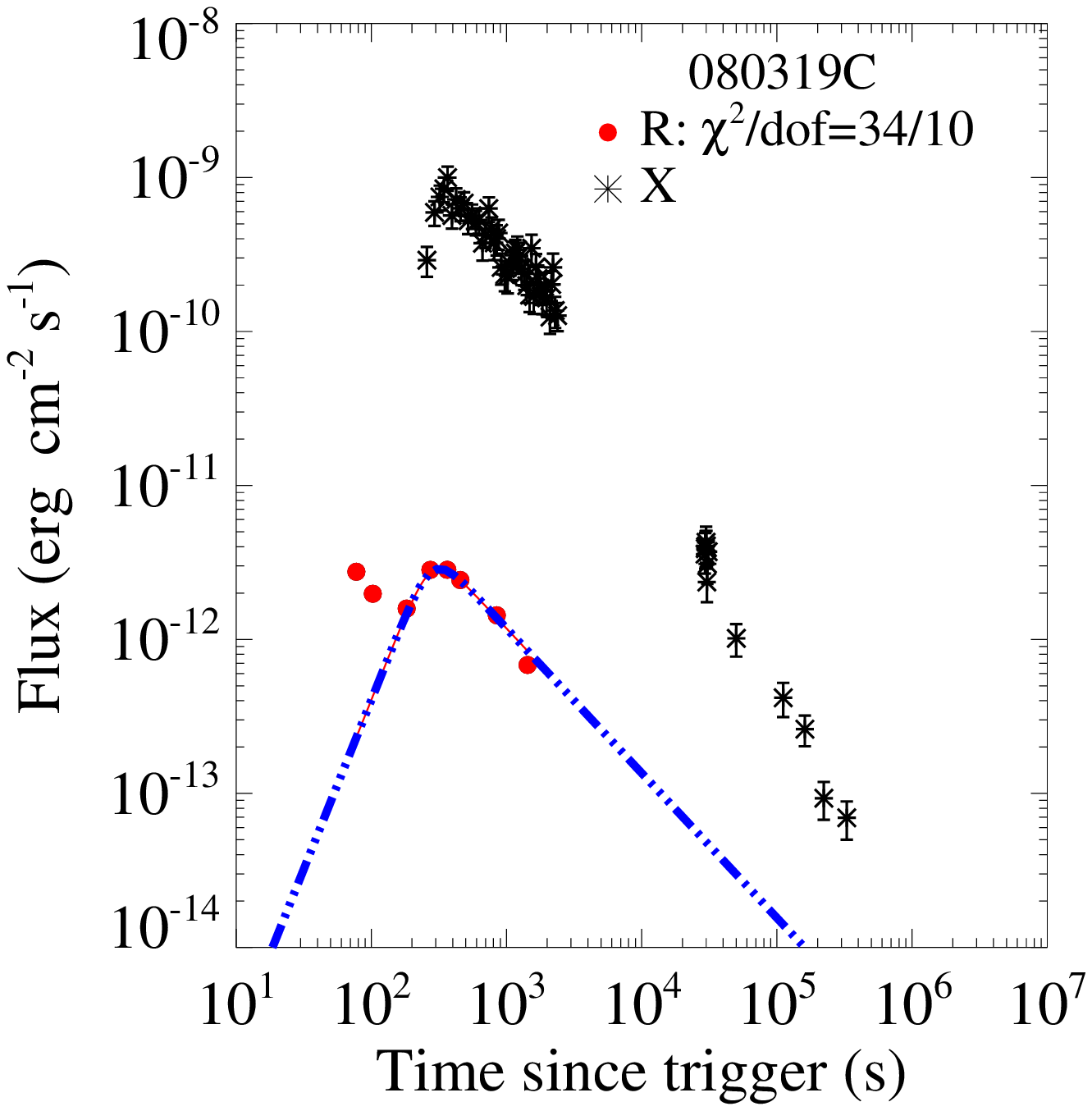}
\includegraphics[angle=0,scale=0.2,width=0.19\textwidth,height=0.18\textheight]{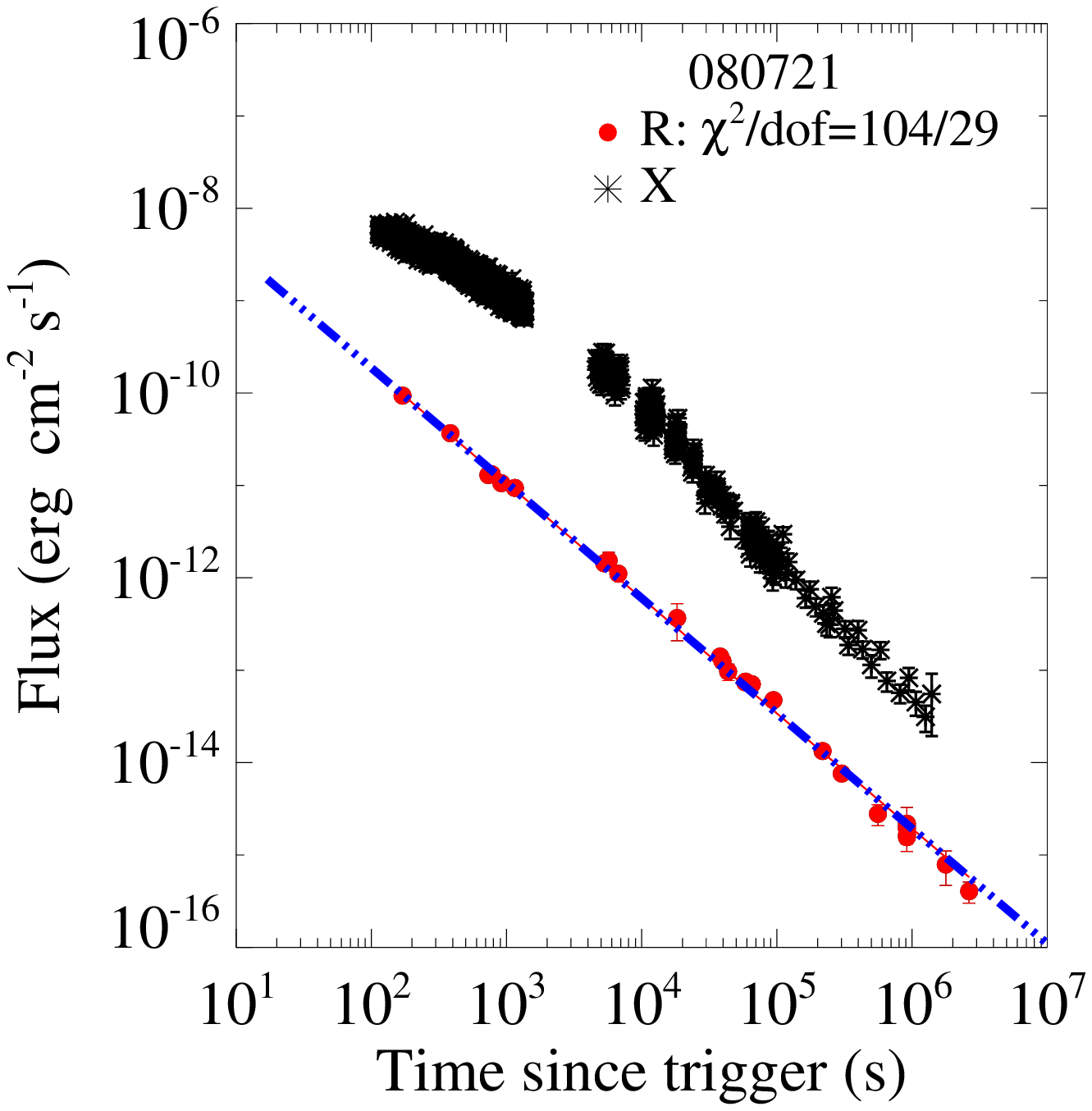}
\caption{Same as Figure \ref{jetgrade}, but for the lower limit sample.} \label{jetlower}
\end{figure*}

\clearpage
\setlength{\voffset}{-18mm}
\begin{figure*}
\includegraphics[angle=0,scale=0.2,width=0.19\textwidth,height=0.18\textheight]{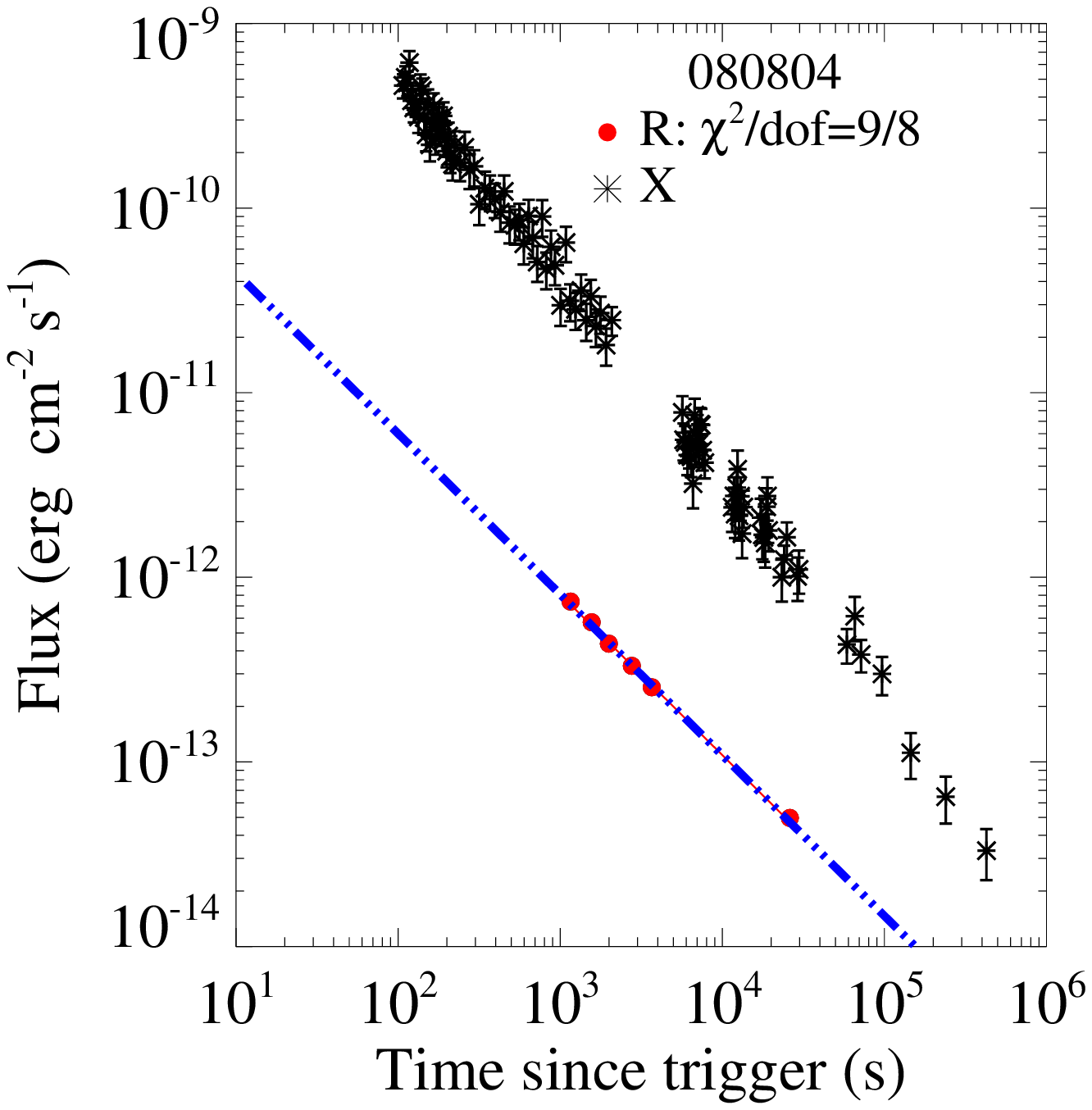}
\includegraphics[angle=0,scale=0.2,width=0.19\textwidth,height=0.18\textheight]{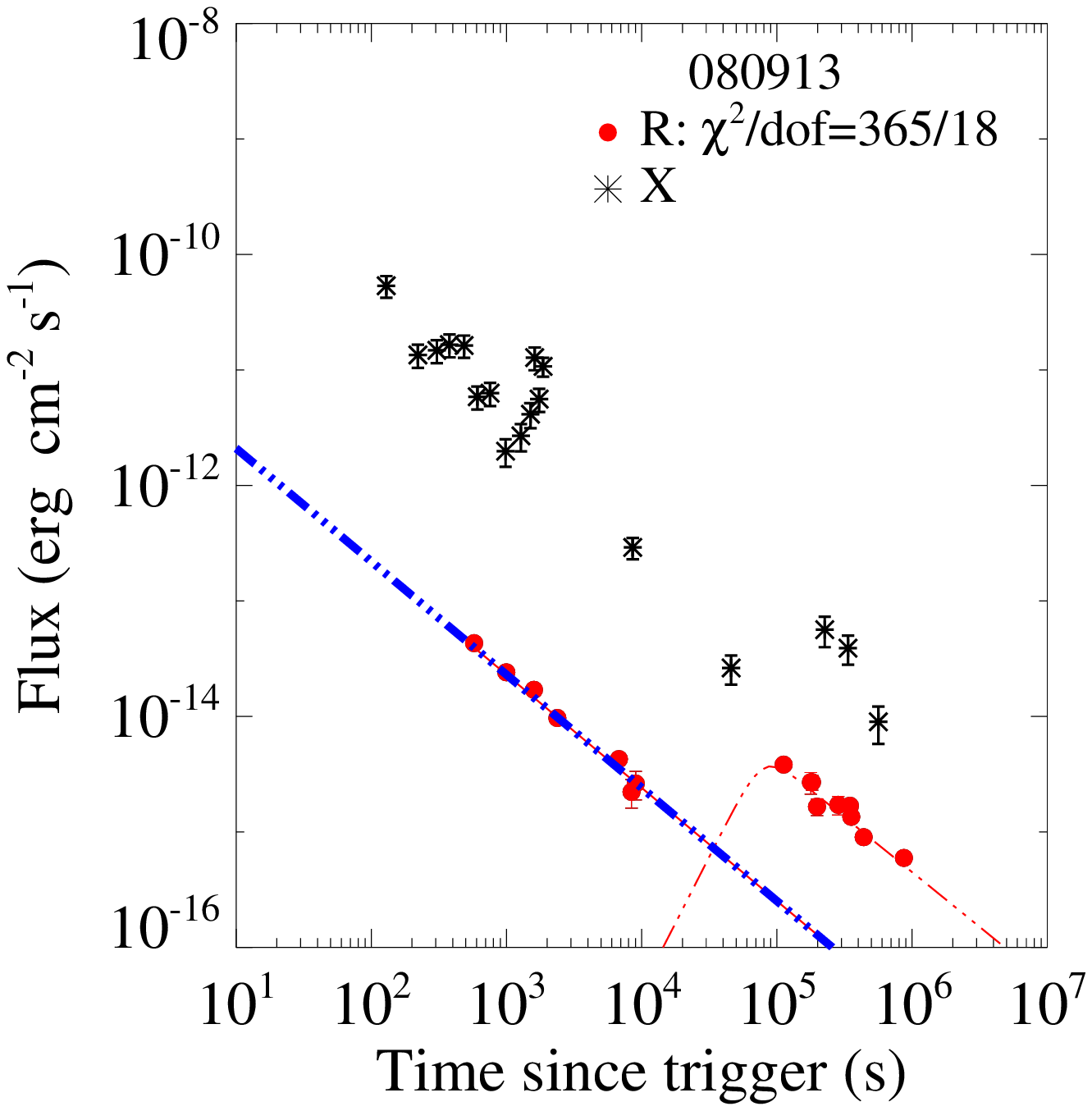}
\includegraphics[angle=0,scale=0.2,width=0.19\textwidth,height=0.18\textheight]{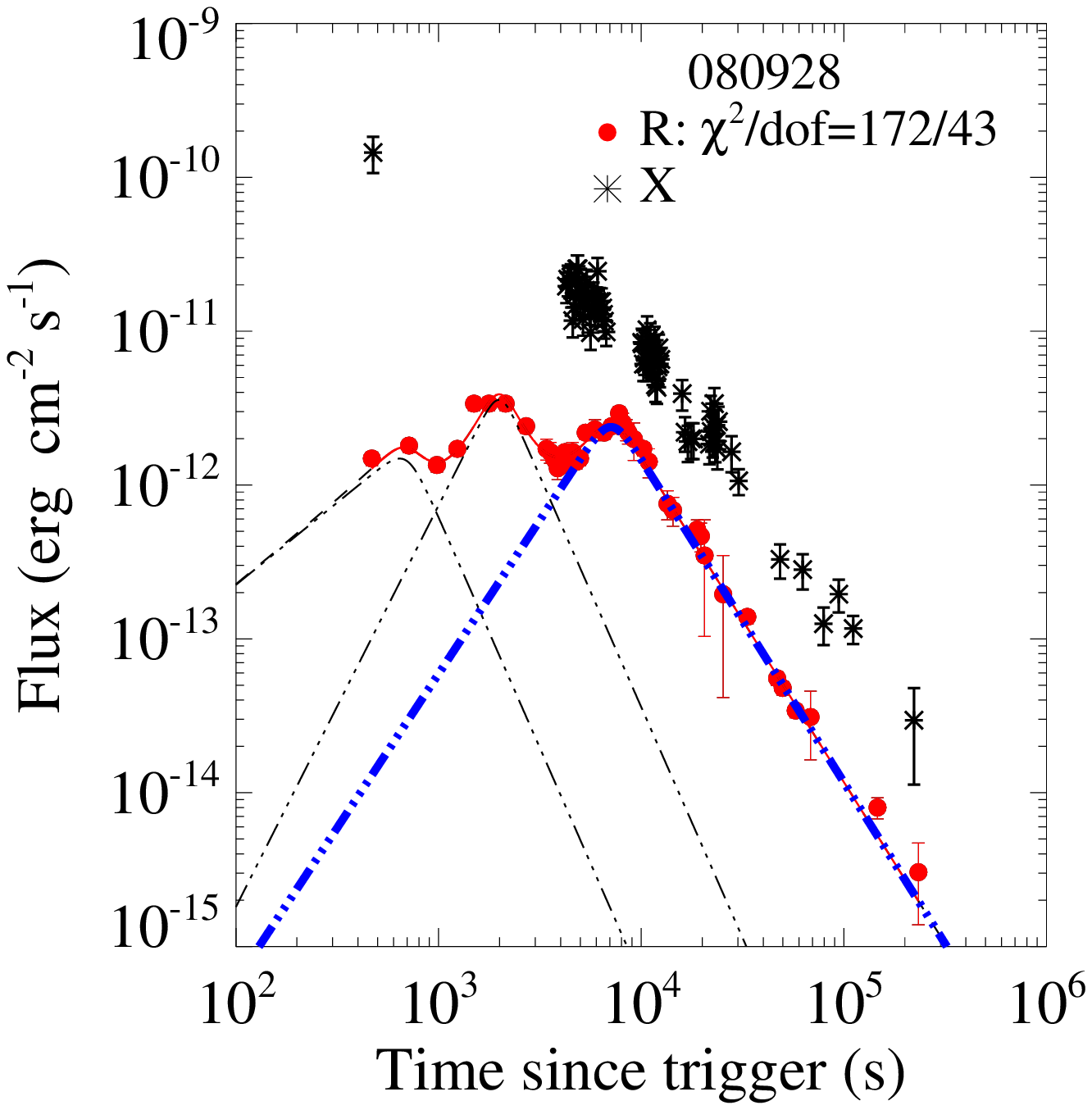}
\includegraphics[angle=0,scale=0.2,width=0.19\textwidth,height=0.18\textheight]{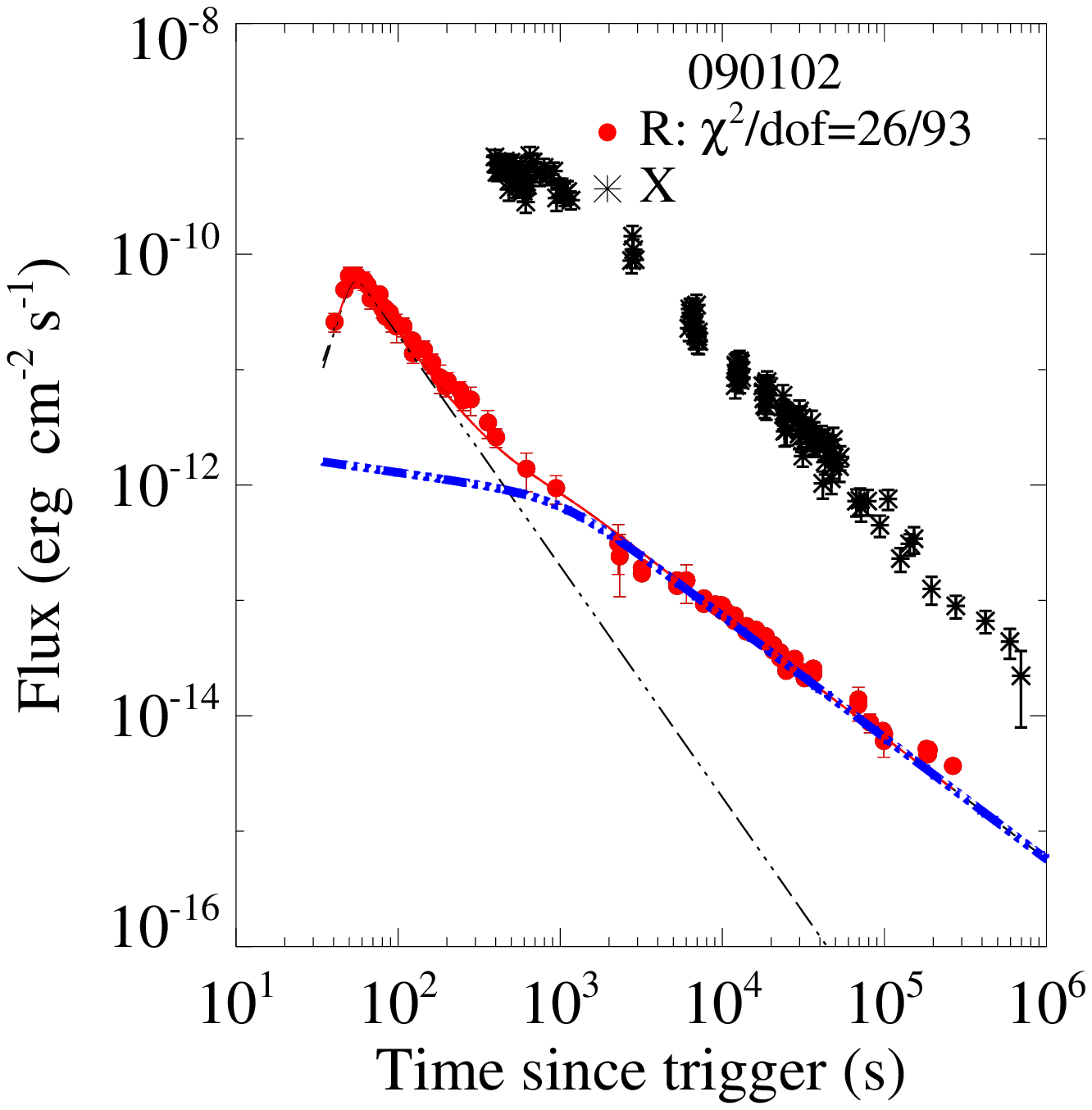}
\includegraphics[angle=0,scale=0.2,width=0.19\textwidth,height=0.18\textheight]{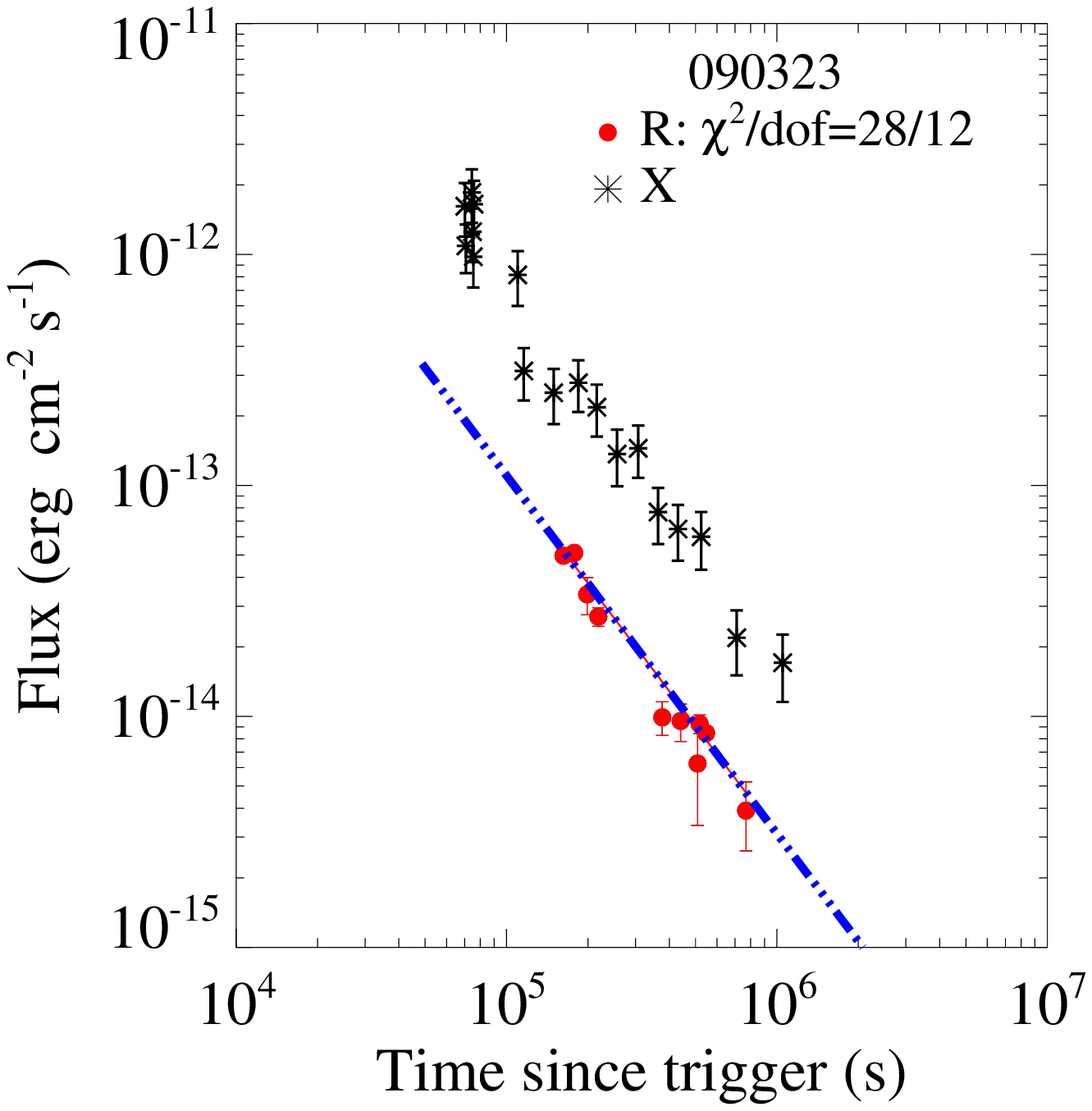}
\includegraphics[angle=0,scale=0.2,width=0.19\textwidth,height=0.18\textheight]{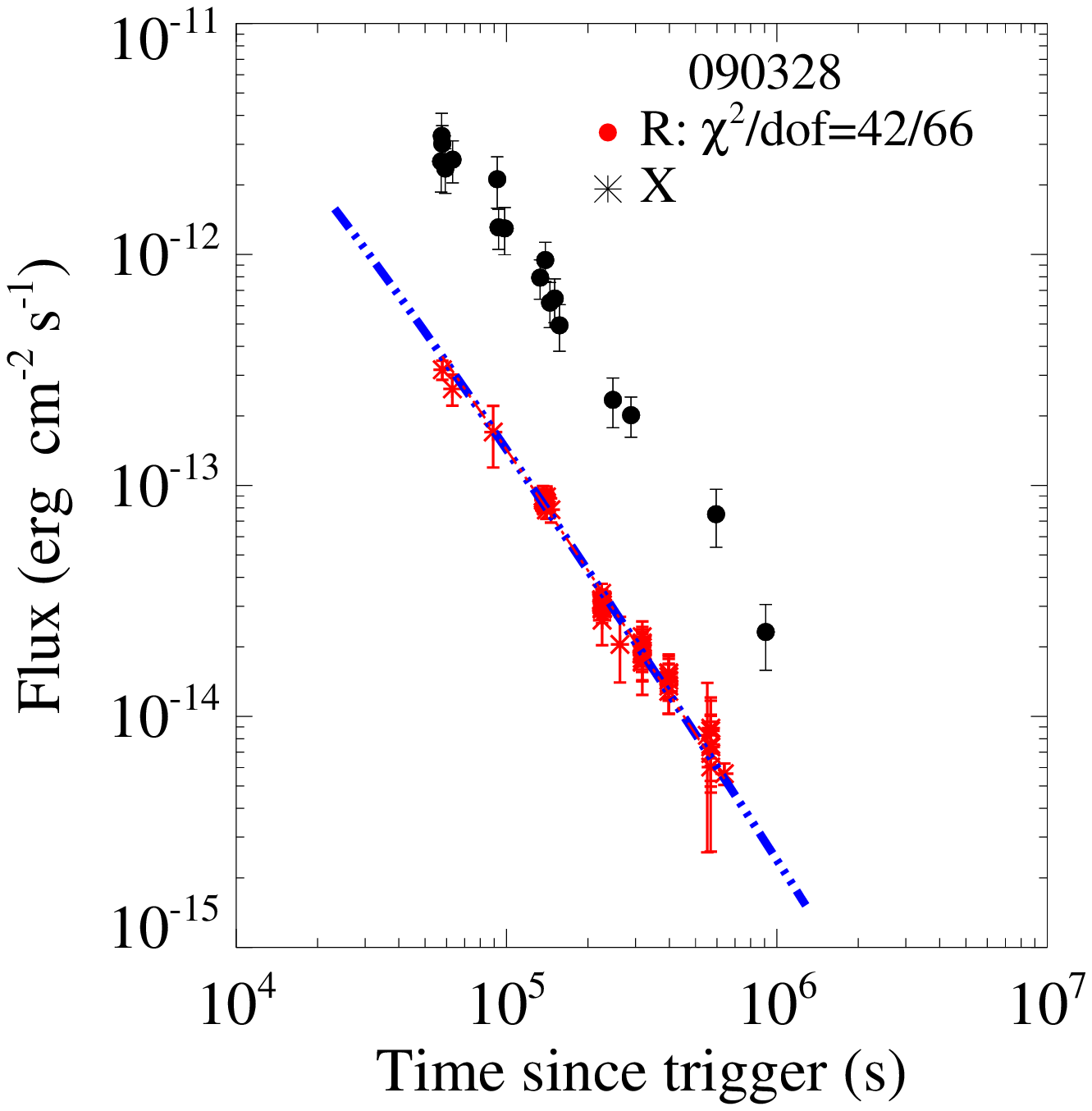}
\includegraphics[angle=0,scale=0.2,width=0.19\textwidth,height=0.18\textheight]{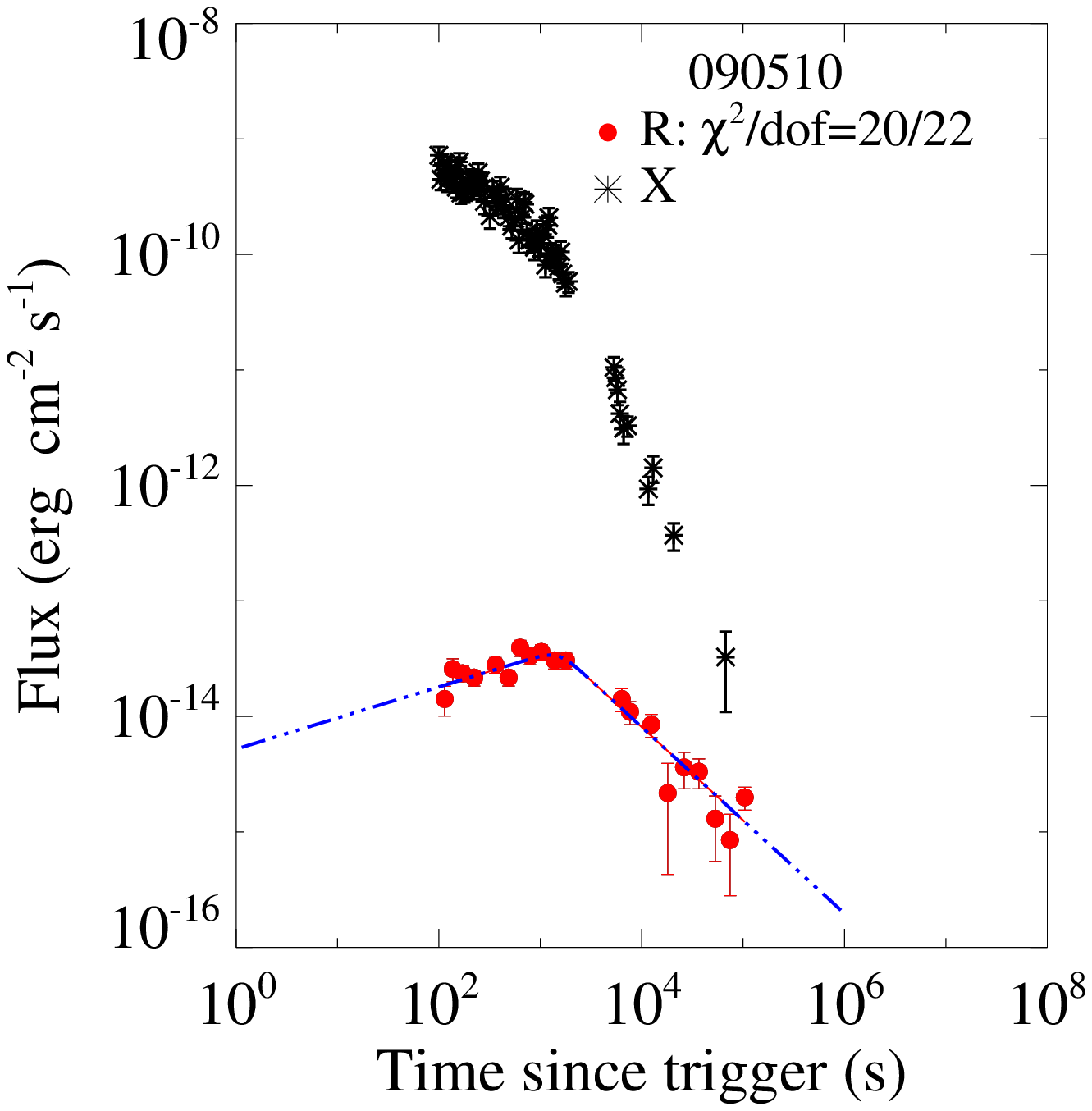}
\includegraphics[angle=0,scale=0.2,width=0.19\textwidth,height=0.18\textheight]{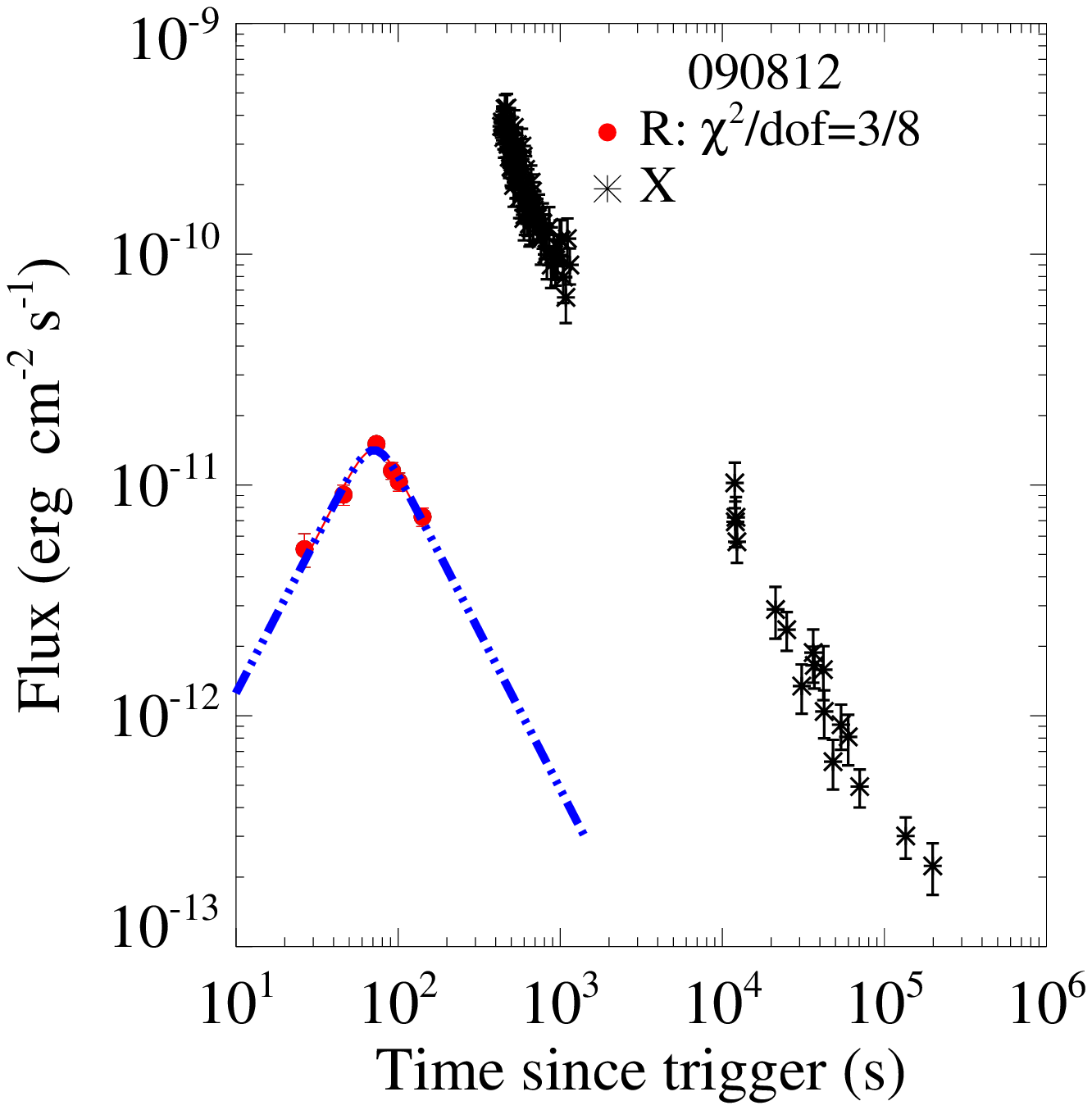}
\includegraphics[angle=0,scale=0.2,width=0.19\textwidth,height=0.18\textheight]{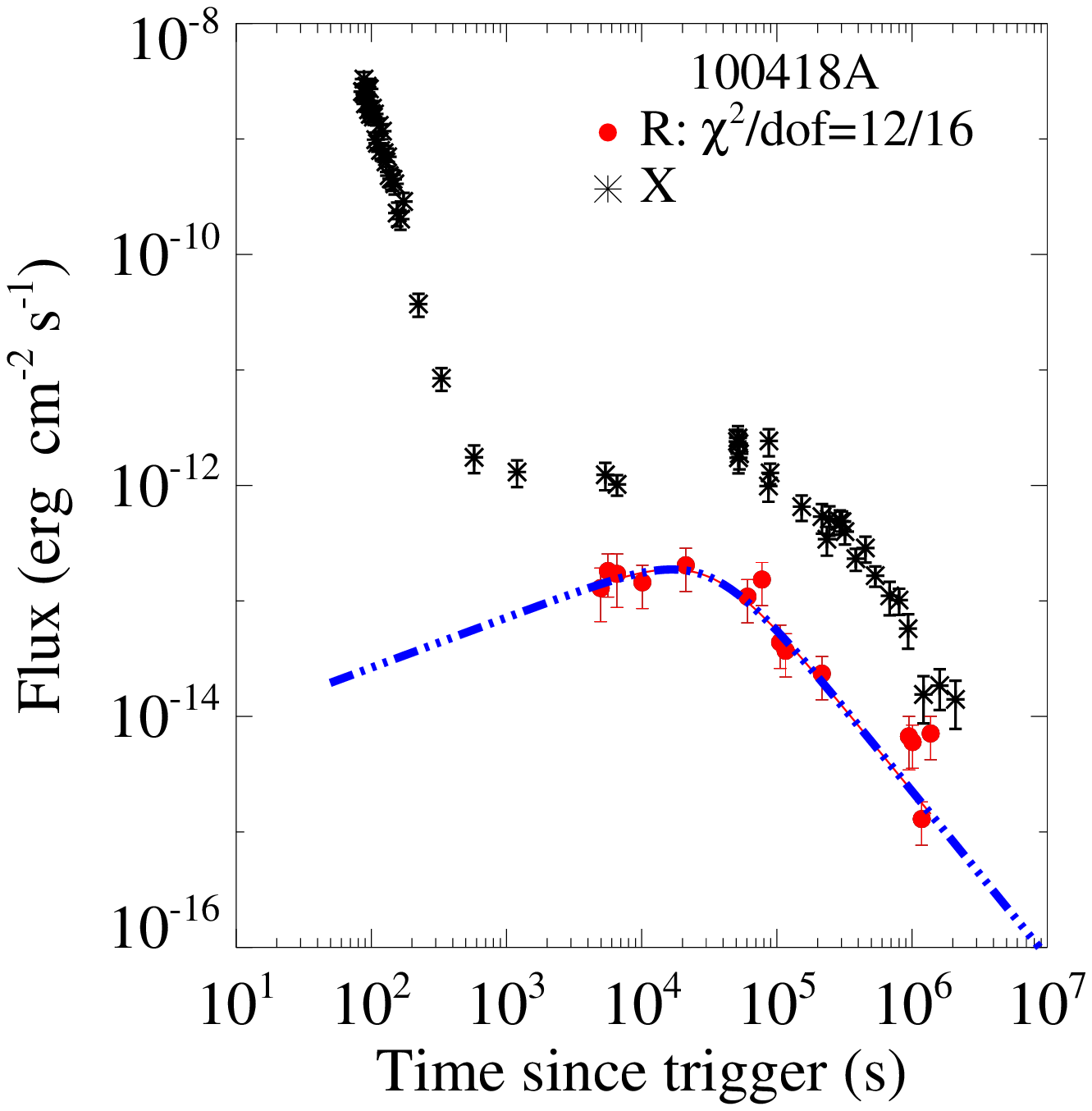}
\includegraphics[angle=0,scale=0.2,width=0.19\textwidth,height=0.18\textheight]{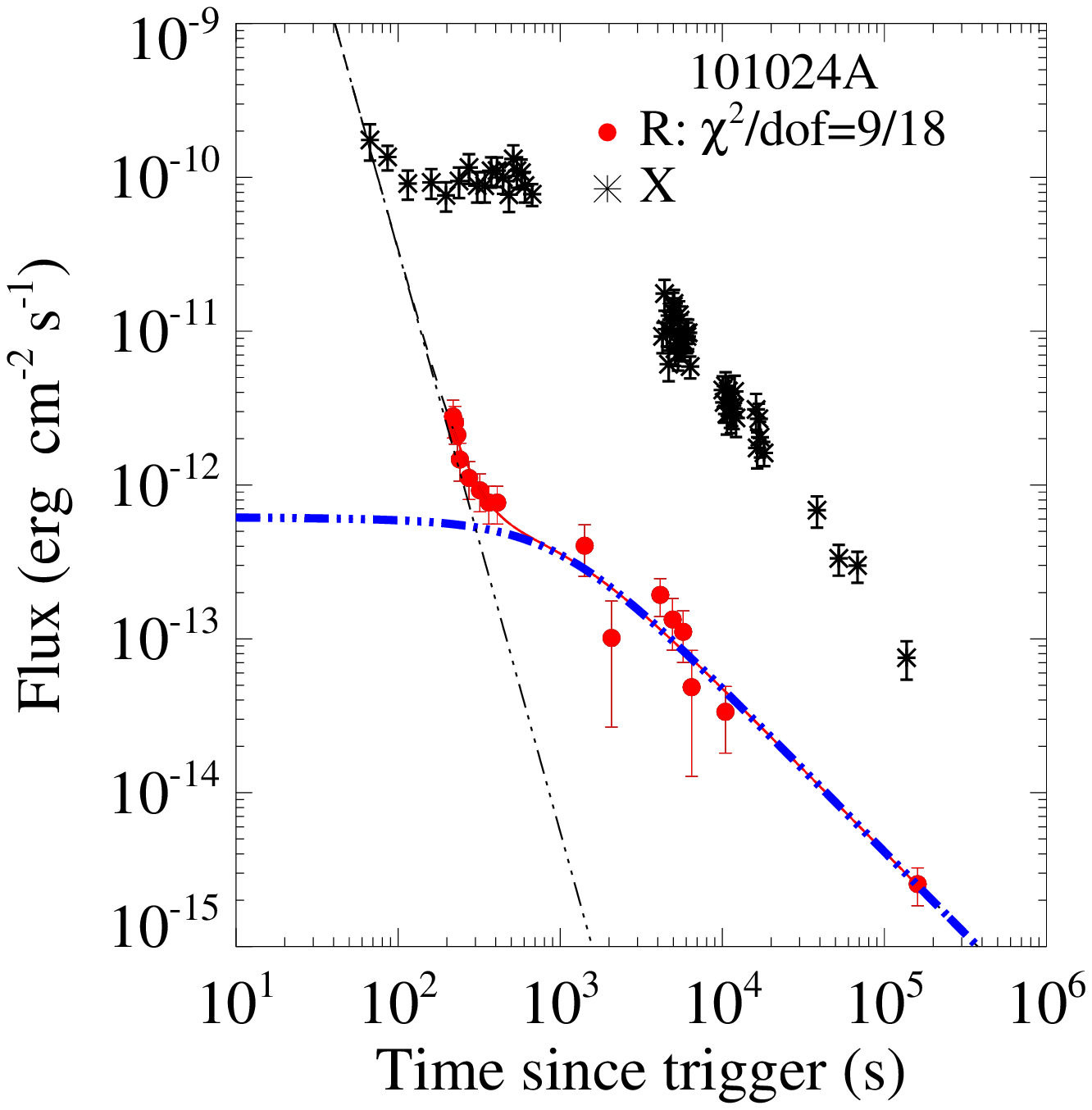}
\includegraphics[angle=0,scale=0.2,width=0.19\textwidth,height=0.18\textheight]{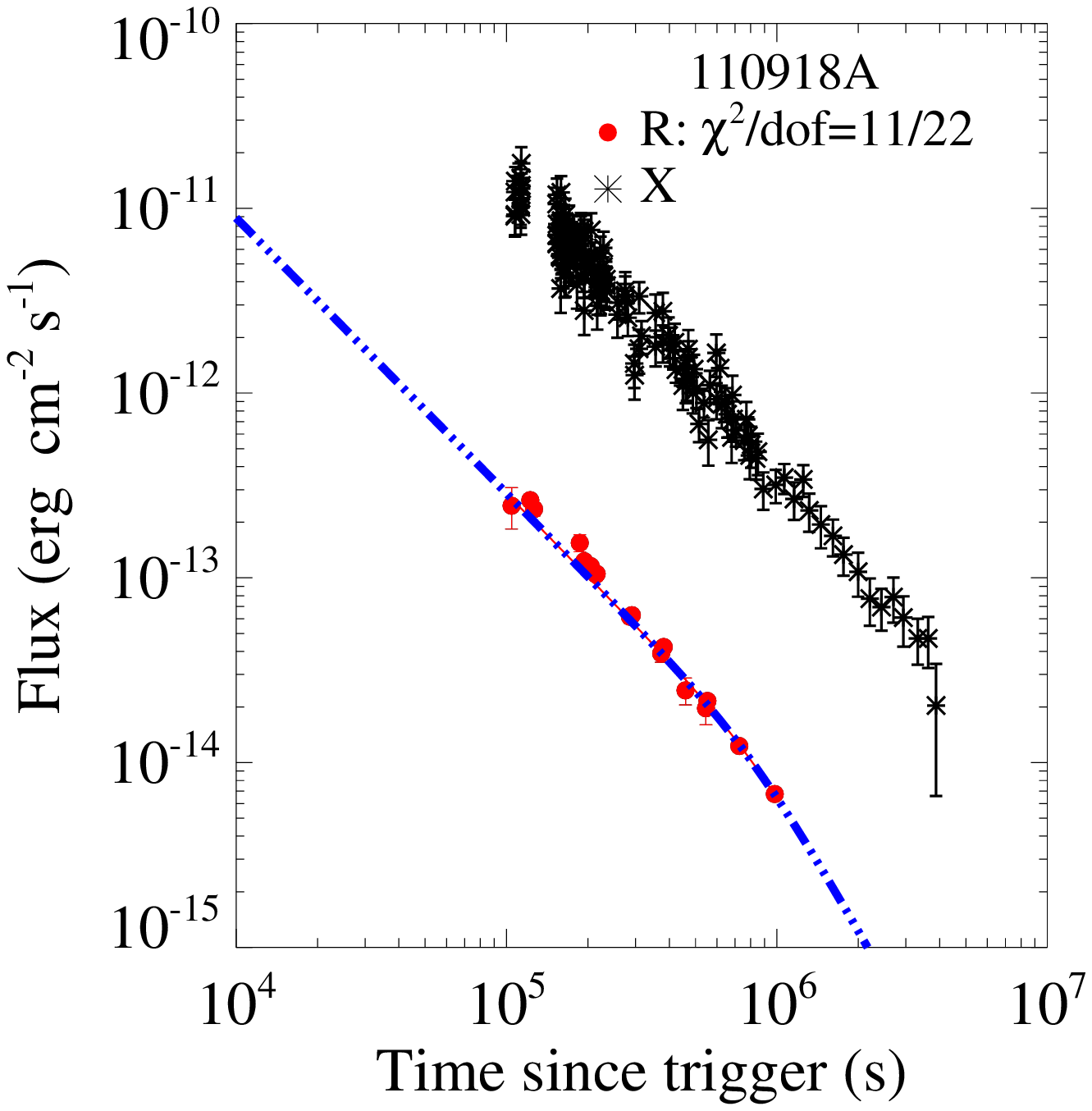}
\includegraphics[angle=0,scale=0.2,width=0.19\textwidth,height=0.18\textheight]{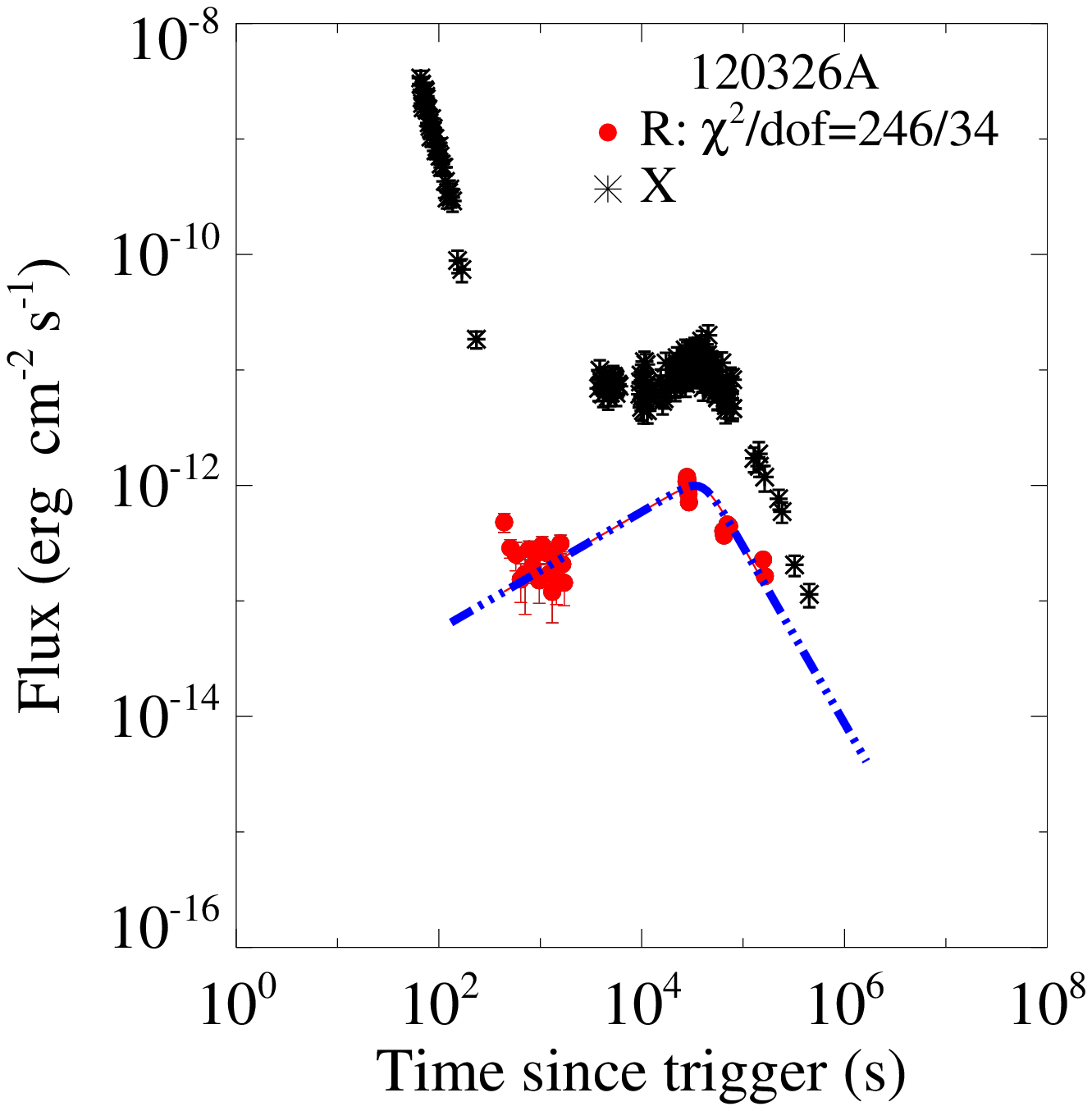}
\includegraphics[angle=0,scale=0.2,width=0.19\textwidth,height=0.18\textheight]{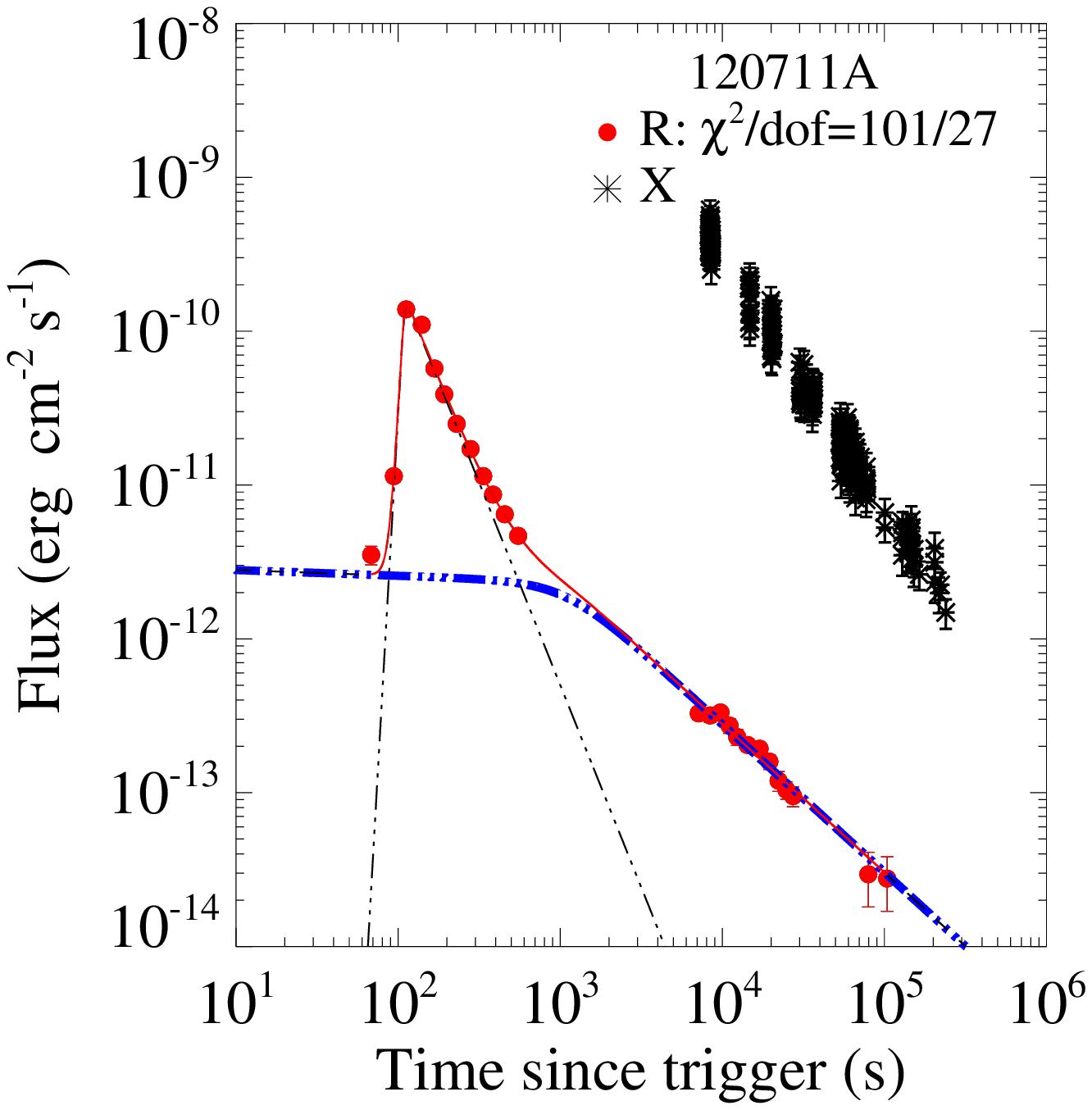}
\includegraphics[angle=0,scale=0.2,width=0.19\textwidth,height=0.18\textheight]{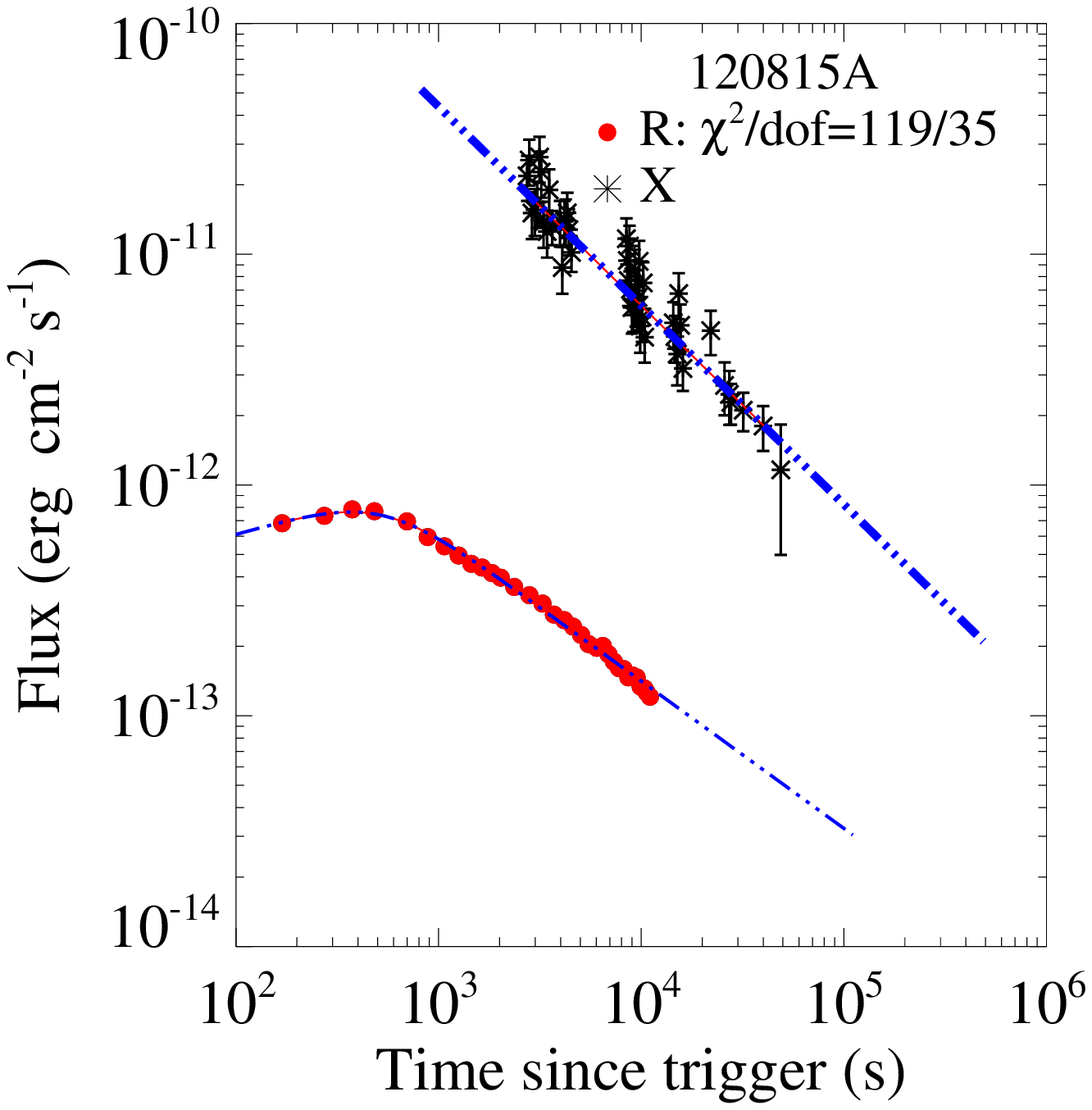}
\center{Figure. \ref{jetlower}(Continued)}
%\caption{Same as Figure \ref{jetgrade4}} \label{jetlower2}
\end{figure*}

%\clearpage
%\thispagestyle{empty}
\setlength{\voffset}{-18mm}
\begin{figure*}
\centering
\includegraphics[angle=0,scale=0.2,width=0.19\textwidth,height=0.18\textheight]{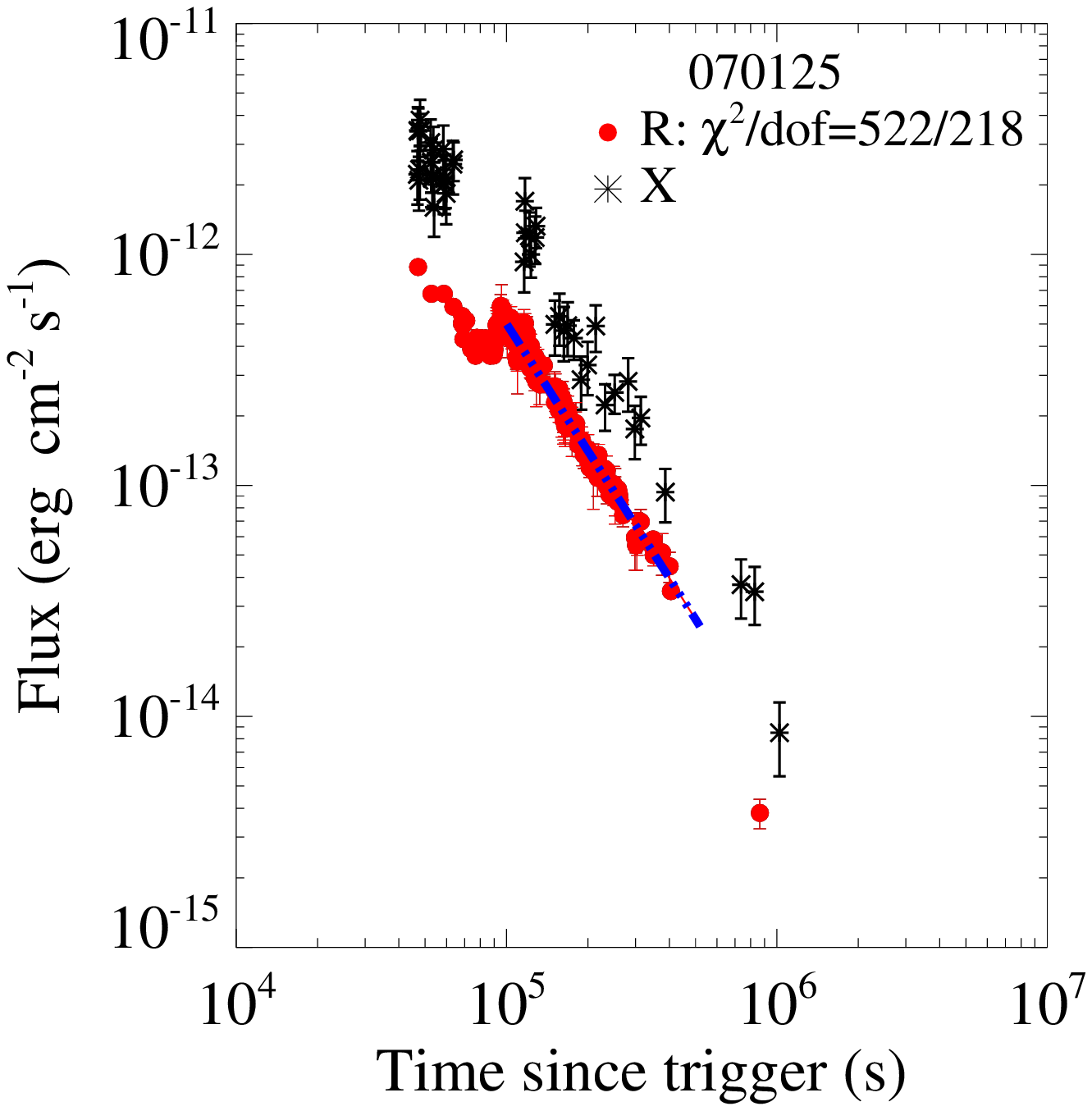}
\includegraphics[angle=0,scale=0.2,width=0.19\textwidth,height=0.18\textheight]{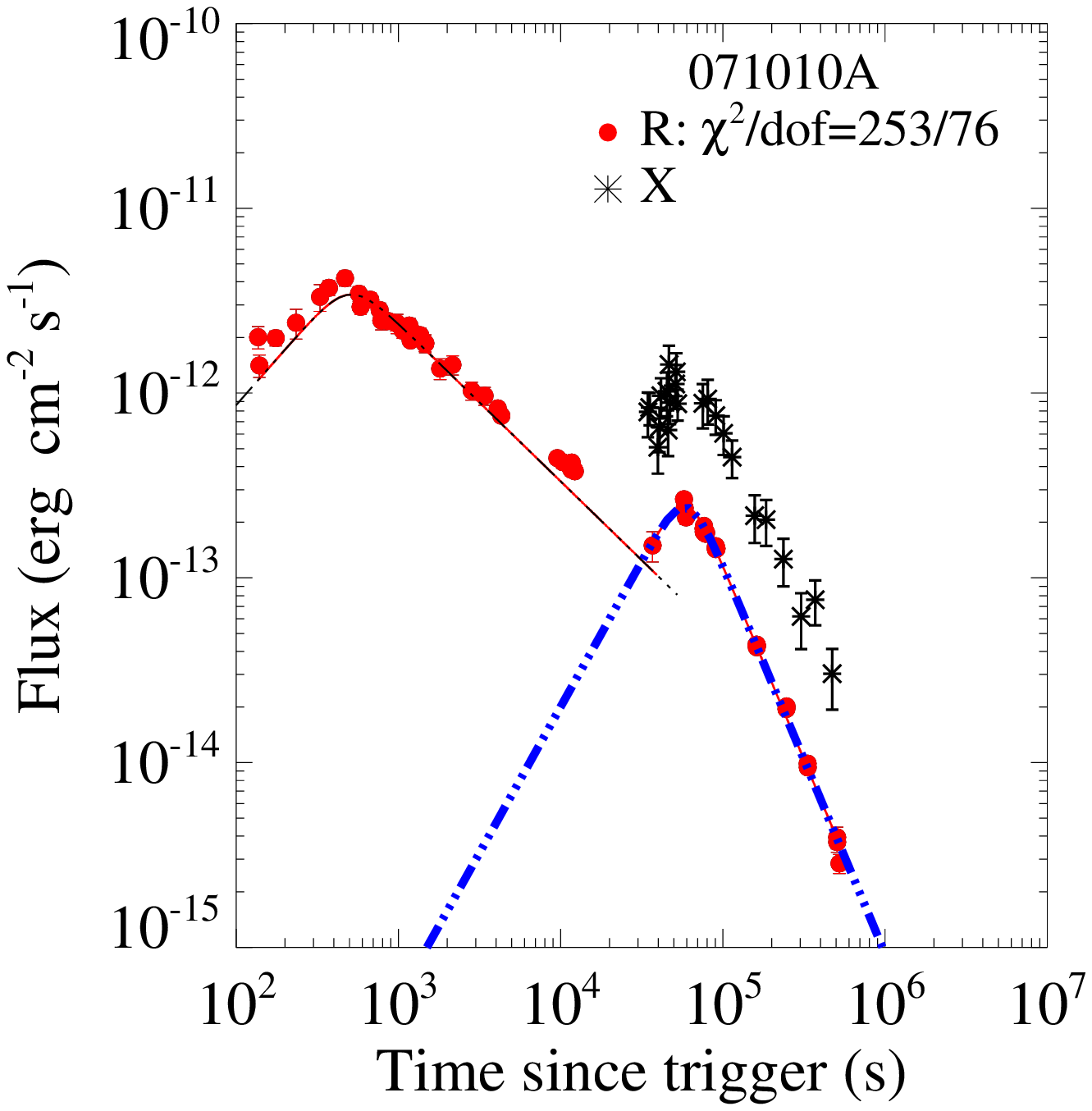}
\includegraphics[angle=0,scale=0.2,width=0.19\textwidth,height=0.18\textheight]{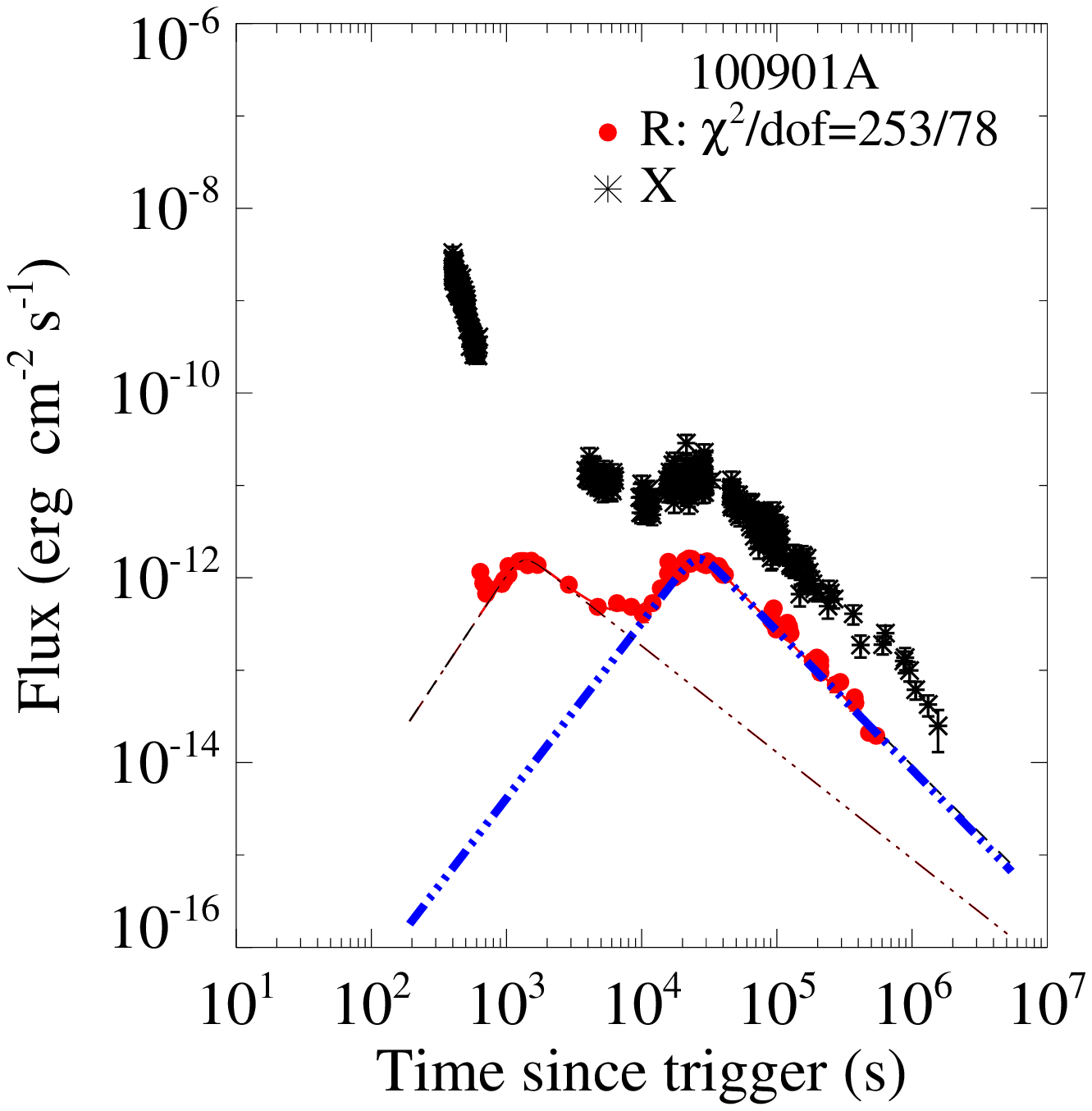}

\caption{Same as Figure \ref{jetgrade}, but for the upper limit sample.} \label{jetupper}
\end{figure*}

\clearpage
\setlength{\voffset}{-18mm}
\begin{figure*}
\centering
\includegraphics[angle=0,scale=0.8]{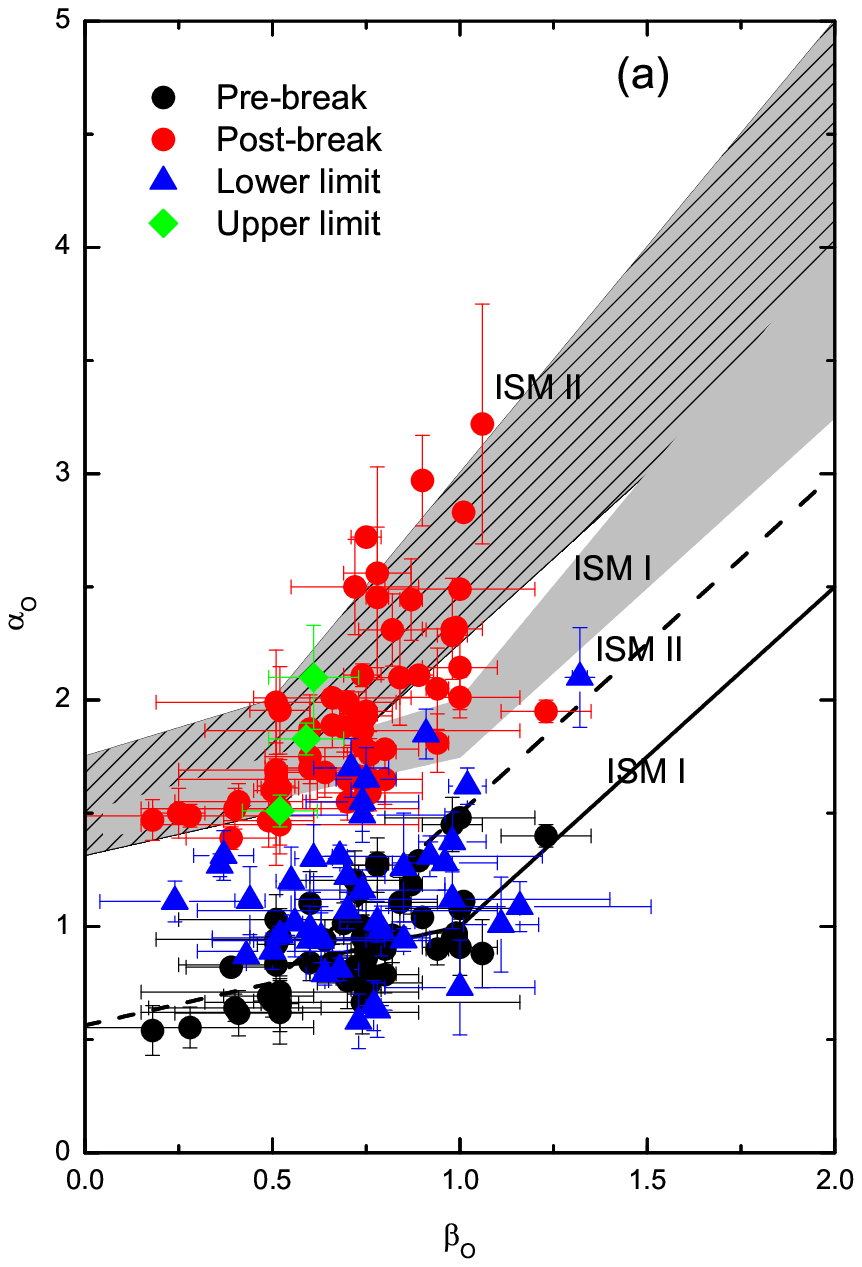}
\includegraphics[angle=0,scale=0.8]{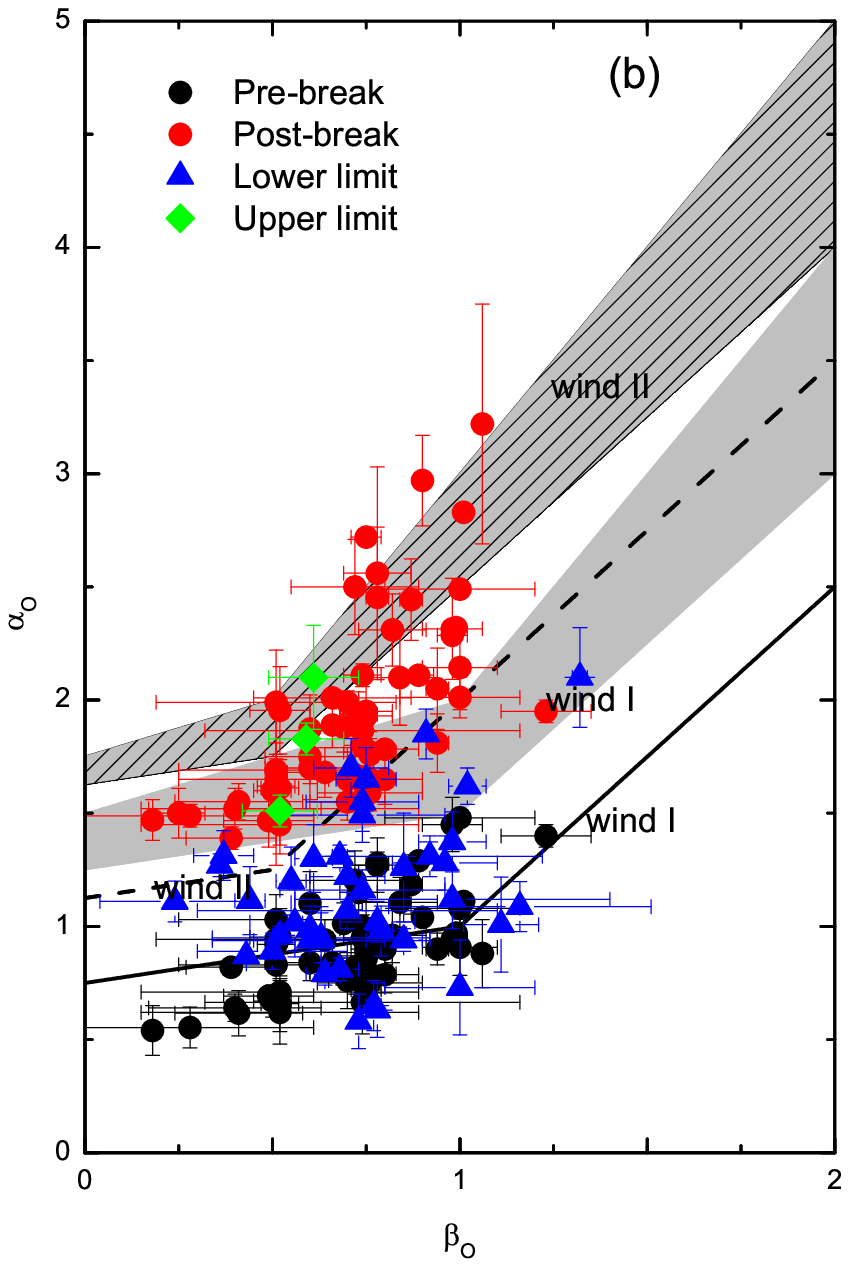}

\caption{The measured afterglow $\alpha$ and $\beta$ values compared against the closure
relations of jet break in the external forward shock model. The thick solid lines and solid shaded regions indicate the closure relations for the pre-and post-break segments in spectral regime I ($\nu>\nu_c$). The lower and upper boundaries of the regions are defined with closure relation, respectively, without and with sideways expansion taken into account. Similarly, the thick dashed lines and hatched regions are for the emission in the spectral regime II ($\nu_m<\nu<\nu_c$). The black and red filled circles symbols represent the segments of pre- and  post-break. The lower and upper limit sample marked as blue triangle and green diamond symbols, respectively. (a) ISM model; (b) wind model.}
\label{Jetclosure}
\end{figure*}

\clearpage %\thispagestyle{empty} \setlength{\voffset}{-18mm}
\begin{figure*}
\includegraphics[angle=0,scale=1.0]{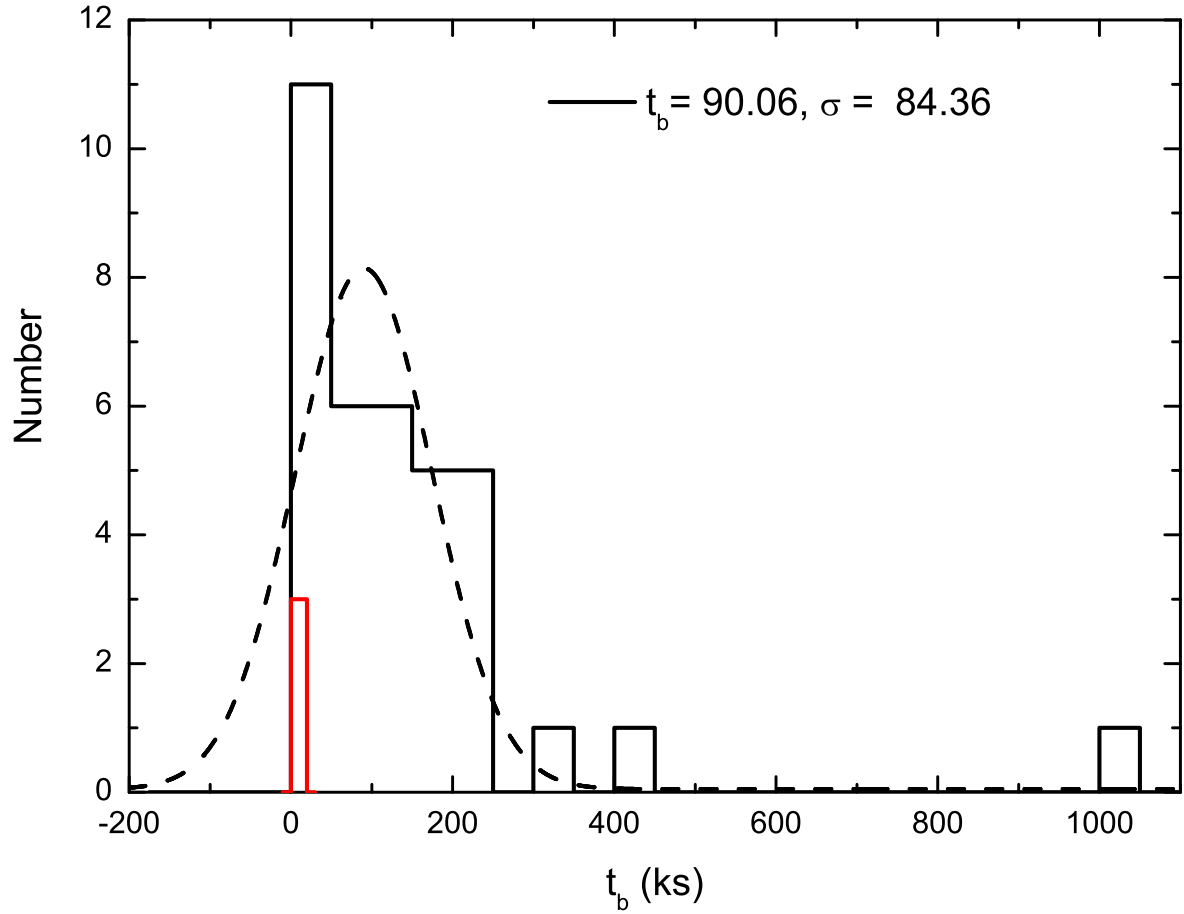}
\caption{The distributions of the jet break time $t_{b}$ (solid lines) and their best Gaussian fits (dash line). The typical values of the entire jet beak sample (black) and the first jet of the two-component jet GRBs (red) {\bf are $t_{b}=90.06\pm84.36$ ks and $t_{b}=0.2\sim 2$ ks,} respectively.}
\label{breaktime}
\end{figure*}

\clearpage %\thispagestyle{empty} \setlength{\voffset}{-18mm}
\begin{figure*}
\includegraphics[angle=0,scale=1.0]{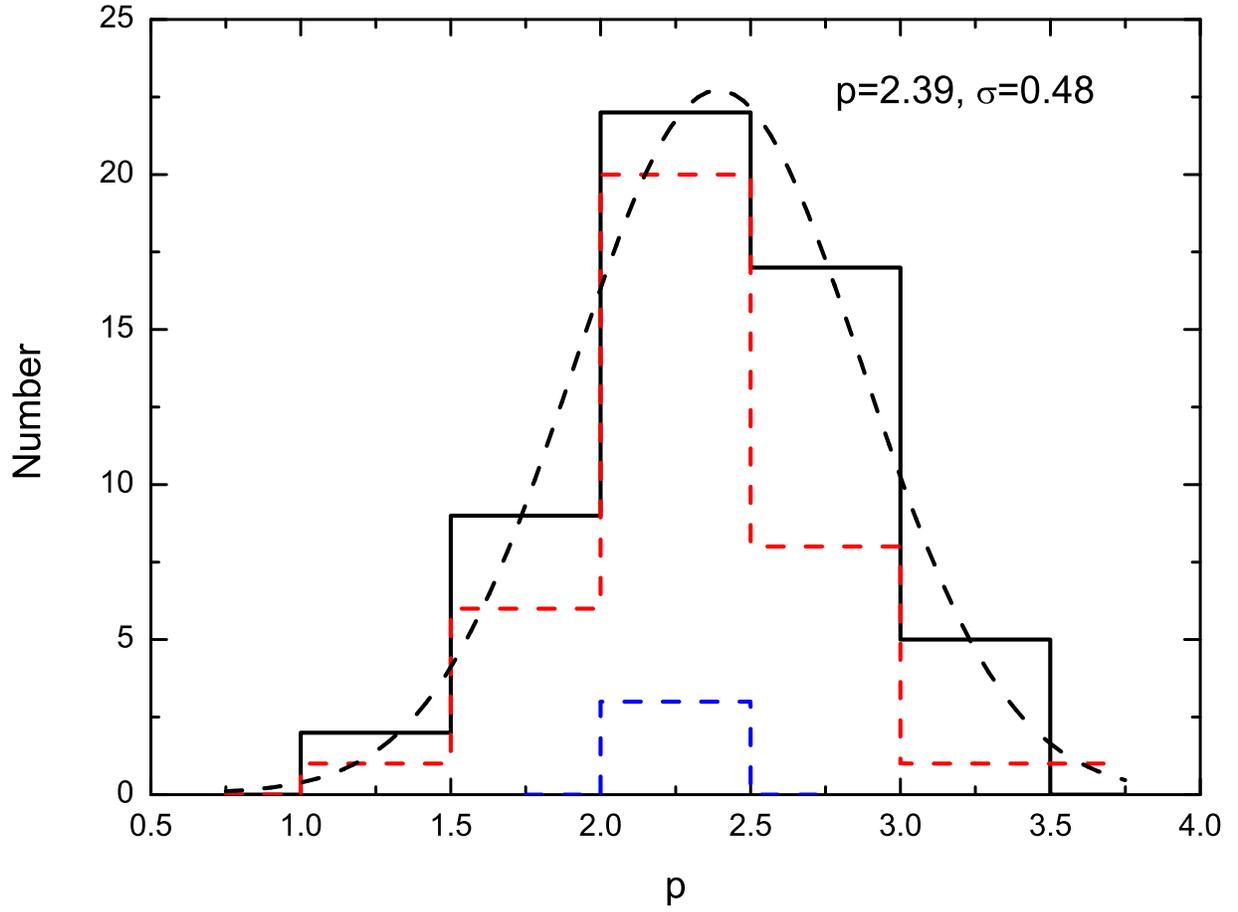}

\caption{The distribution of the inferred electron spectral index $p$ from jet break sample (black line), lower limit sample (red dashed line) and upper limit sample (blue dashed line). The black dashed line is the best Gaussian fit of the jet break sample, with $p=2.39\pm0.48$.}\label{pvalue}
\end{figure*}

\clearpage %\thispagestyle{empty} \setlength{\voffset}{-18mm}
\begin{figure*}
\includegraphics[angle=0,scale=1.0]{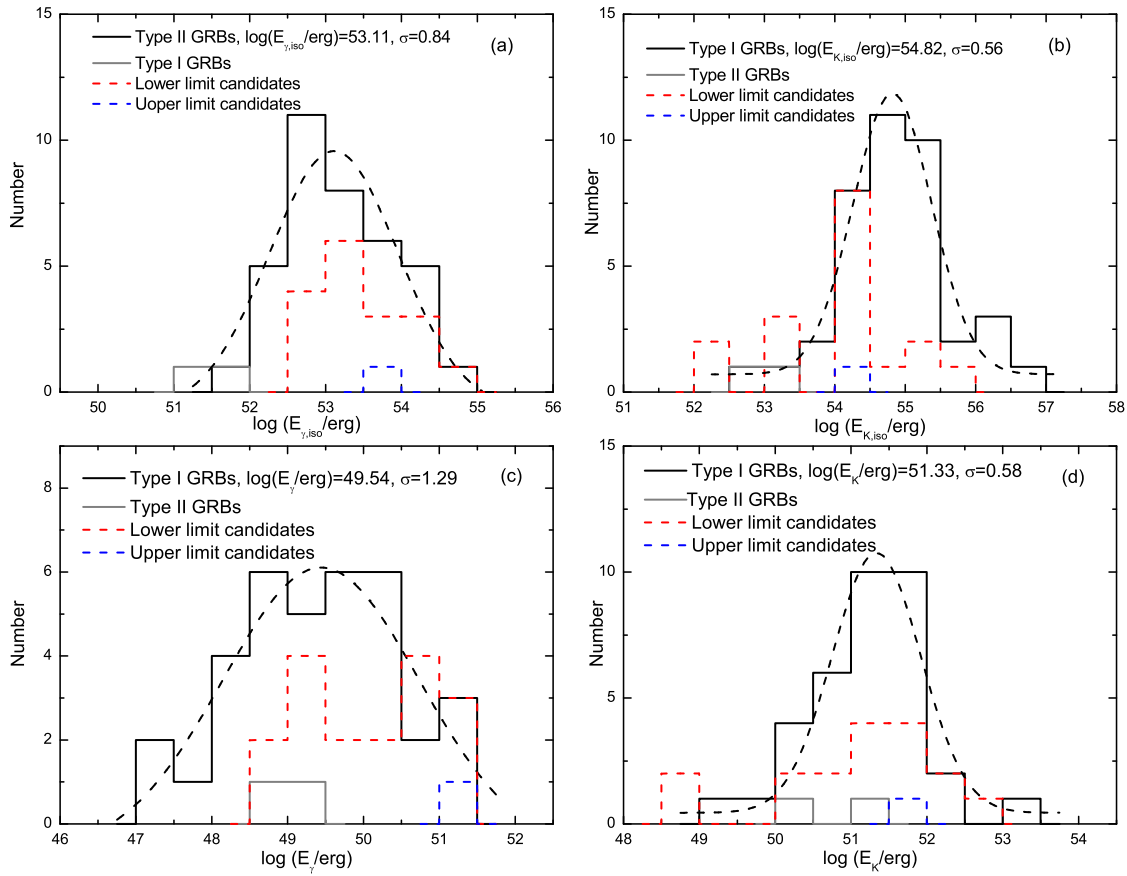}
\caption{The distributions of $\gamma$-ray and kinetic energies derived from the jet break sample and the lower limit sample. The Type I and II GRBs in the jet break sample are donated by gray and black line histograms, respectively. The lower and upper limit sample are represented as red and blue dashed line histogram. The black dashed lines are the best Gaussian fits for Type II GRBs: (a) isotropic $\gamma$-ray energy, $E_{\rm \gamma,iso}$, with a typical value $\log (E_{\rm \gamma,iso} \rm /erg)=(53.11\pm0.84)$; (b) isotropic kinetic energy, with a typical value $\log (E_{\rm K,end} \rm /erg )=(54.82\pm0.56)$; (c) geometrically corrected $\gamma$-ray energy ($E_{\rm \gamma}$), with $\log (E_{\rm \gamma} \rm /erg)=(49.54\pm1.29)$; (d) geometrically corrected kinetic energy ($E_{\rm K}$), with  $\log (E_{\rm K} \rm /erg)=(51.33\pm0.58)$.}
\label{Eiso}
\end{figure*}

\clearpage %\thispagestyle{empty} \setlength{\voffset}{-18mm}
\begin{figure*}
\includegraphics[angle=0,scale=1.0]{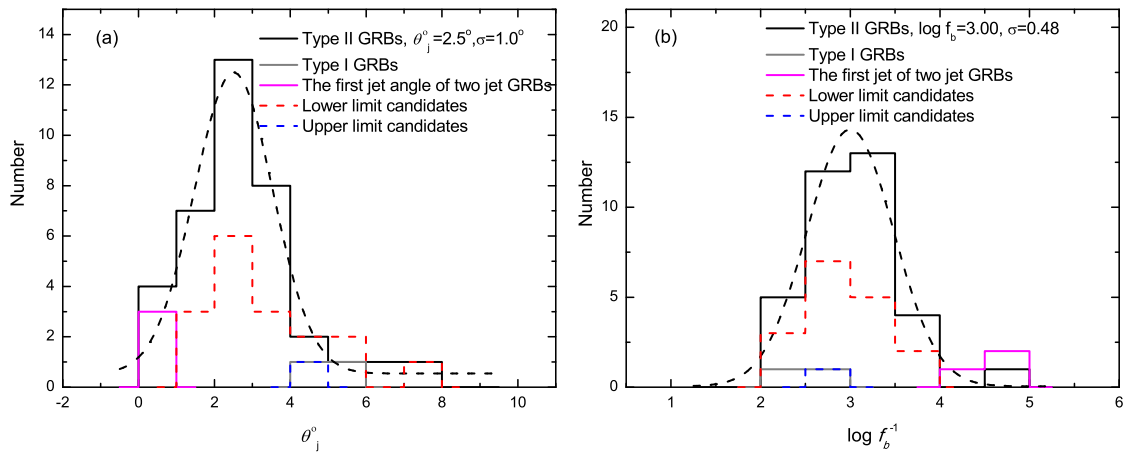}
\caption{The distributions of (a)jet opening angle $\theta_{j}$ and (b) beaming factor  $f_{b}^{-1}$.  The Type I and II GRBs in the jet break sample are marked by gray and black line histograms, respectively. The lower and upper limit sample are represented as red and blue dashed line histogram. The opening angle and beaming factor of the first jet in the two jet breaks also marked with the magenta line. The black dashed lines are the best Gaussian fits of Type II GRBs, with $\theta_{j}=(2.5\pm1.0)^{\rm o}$ and $log f_{b}^{-1} = 3.00\pm0.48$.}\label{jetangle}
\end{figure*}

\clearpage %\thispagestyle{empty} \setlength{\voffset}{-18mm}
\begin{figure*}
\includegraphics[angle=0,scale=0.8]{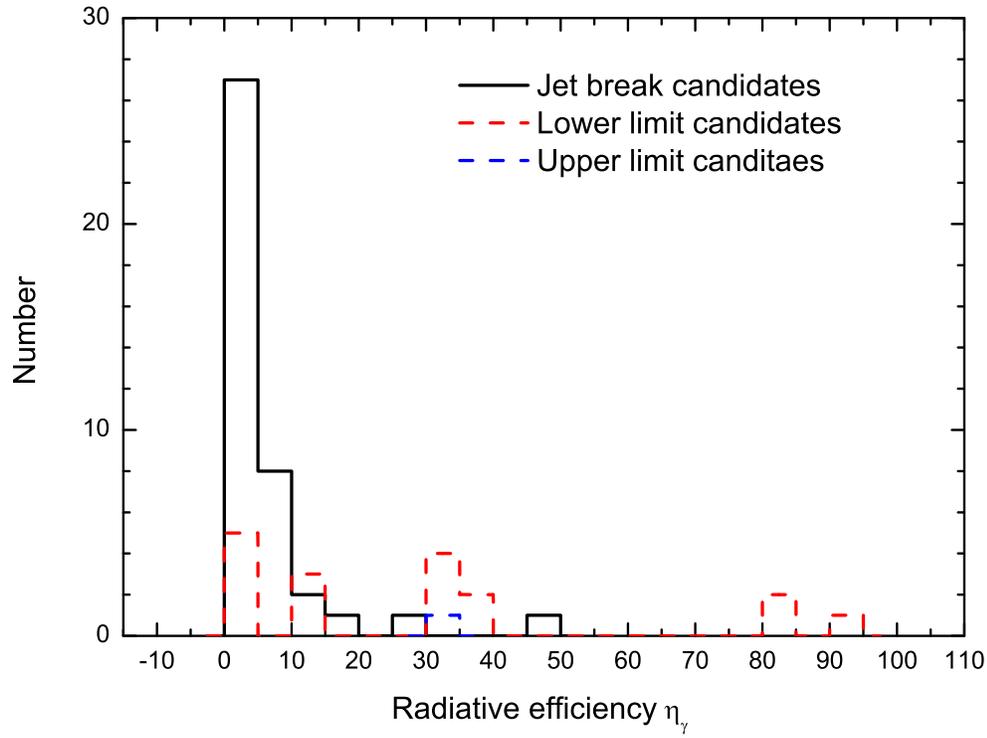}
\caption{The distributions of radiative efficiency $\eta_{\gamma}$.}\label{efficiency}
\end{figure*}

\clearpage \thispagestyle{empty} \setlength{\voffset}{-18mm}
\begin{figure*}
\includegraphics[angle=0,scale=0.4,width=1\textwidth,height=0.6\textheight]{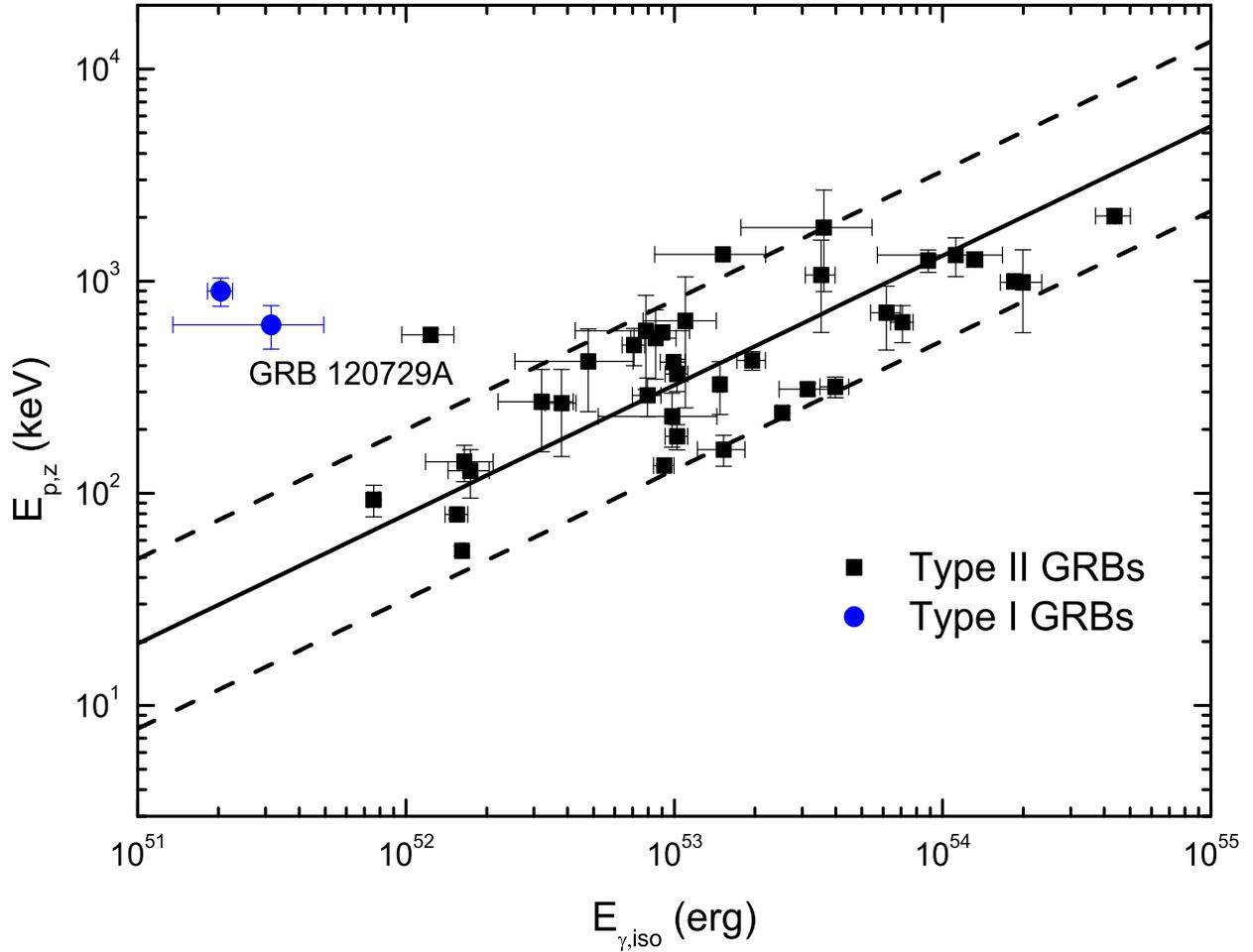}
\caption{GRBs in $E_{\rm \gamma,iso}-E_{\rm p,z}$ plane (\emph{Amati} relation). The Type II and I GRBs are marked with black squares and blue circles, respectively. The solid line is the best fit with $\frac{E_{\rm p,z}}{\rm 100 keV}\simeq(0.63\pm0.31)(\frac{E_{\rm \gamma,iso}}{10^{52} \rm erg})^{(0.69\pm0.07)}$, and their $2\sigma$ dispersion regions are shown with the dashed lines.}
\label{Amatirelation}
\end{figure*}

\begin{figure*}
\includegraphics[angle=0,scale=0.8,width=1\textwidth,height=0.6\textheight]{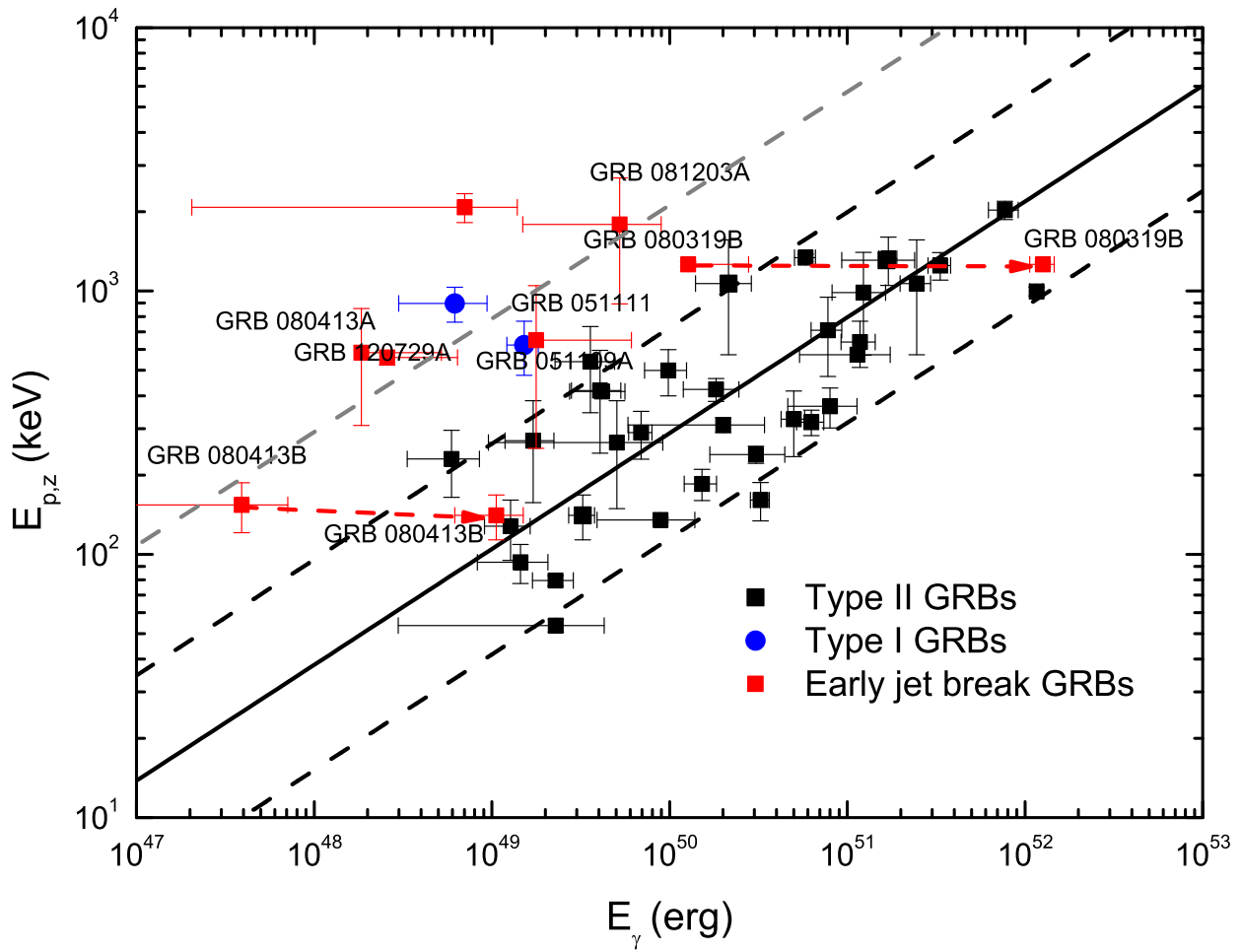}
\caption{GRBs in $E_{\rm \gamma}-E_{\rm p,z}$ plane (\emph{Ghirlanda} relation). The squares and circles symbols represent the Type II and I GRBs, respectively. The early jet breaks of Type II GRBs are marked with red squares. The red dashed lines with arrow are used to demonstrate the relation in two jet breaks from early one to the late one. The best fit line (solid) is $\frac{E_{\rm p,z}}{\rm 100 keV}\simeq(7.9\pm4.8)(\frac{E_{\rm \gamma}}{10^{51} \rm erg})^{(0.44\pm0.07)}$, and their $2\sigma$ dispersion regions are shown with the black dashed lines. The Type I GRBs and early jet break are located around a gray dashed line.}
\label{Ghirlandarelation}
\end{figure*}

\clearpage %\thispagestyle{empty} \setlength{\voffset}{-18mm}
\begin{figure*}
\includegraphics[angle=0,scale=1.0]{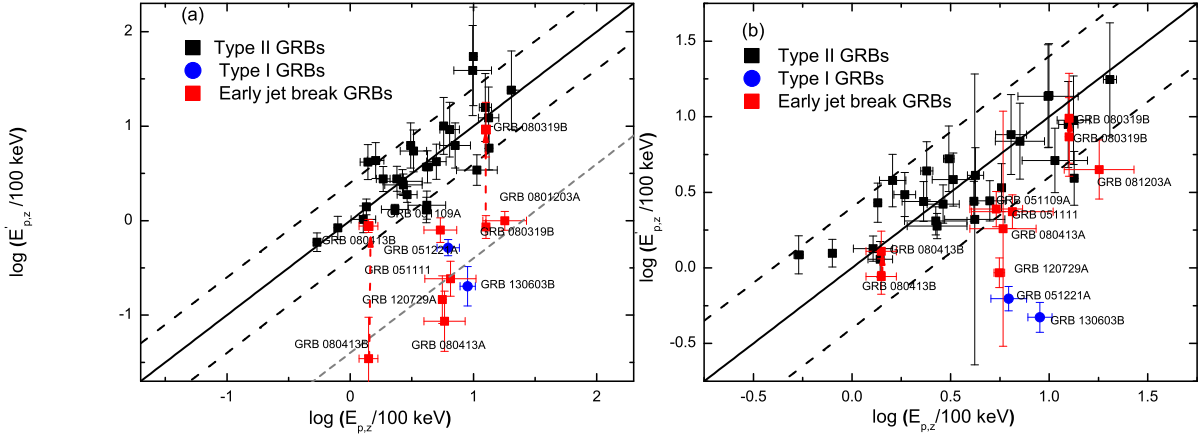}
\caption{The comparisons of the $E_{\rm p,z}$ and  $E_{\rm p,z}^{'}$ calculated from the  $E_{\rm p,z}-E_{\rm \gamma,iso}-t_{\rm b,z}$ (\emph{Liang-Zhang} relation) based on our sample. The squares and circles symbols represent the Type II and I GRBs, respectively. The early jet breaks of Type II GRBs are marked with red squares. The red dashed lines with arrow are used to demonstrate the relation in two jet breaks from early one to the late one. The best fit line (solid) and their $2\sigma$ dispersion regions (black dashed lines) are marked. (a) $\frac{E_{\rm p,z}}{\rm 100 keV}=(1.2\pm0.3)(\frac{E_{\rm \gamma,iso}}{10^{52} \rm erg})^{(0.56\pm0.07)}(\frac{t_{\rm b,z}}{\rm day})^{(0.67\pm0.08)}$ for the late jet break sample only. The Type I GRBs and early jet break are located around a gray dashed line. (b) $\frac{E_{\rm p,z}}{\rm 100 keV}=(1.3\pm0.4) (\frac{E_{\rm \gamma,iso}}{10^{52} \rm erg})^{(0.49\pm0.07)}(\frac{t_{\rm b,z}}{\rm day})^{(-0.08\pm0.05)}$ for all the Type II GRBs.}
\label{Liang-zhangrelation}
\end{figure*}

\clearpage %\thispagestyle{empty} \setlength{\voffset}{-18mm}

\begin{figure*}
\includegraphics[angle=0,scale=1.0]{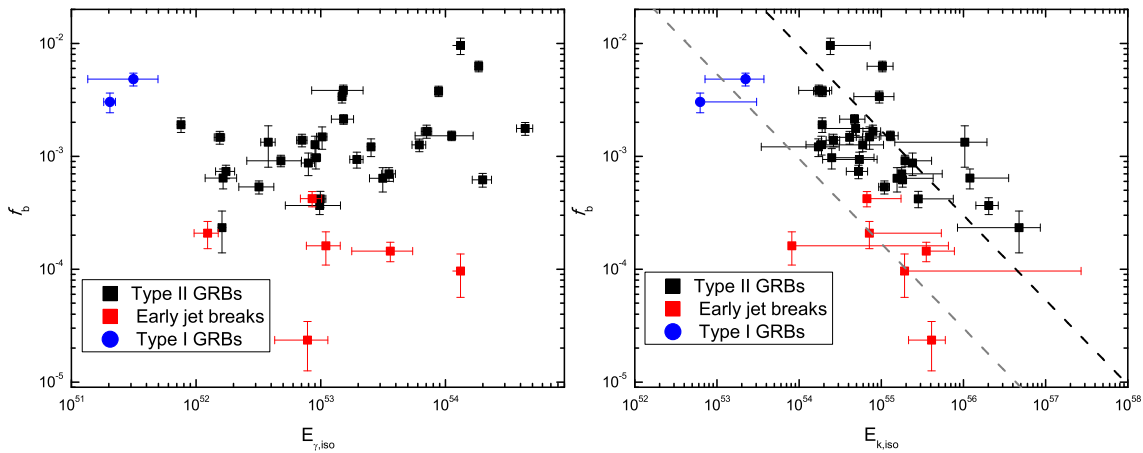}
\caption{GRBs in the $E_{\rm \gamma,iso}(E_{\rm K,iso})-f_{\rm b}$ plane (\emph{Frail} relation).The Type II and I GRBs are marked with black squares and blue circles, respectively.  The early jet breaks of Type II GRBs are marked with red squares. (a) the comparisons of the $E_{\rm \gamma,iso}$ and $f_{\rm b}$. (b) the comparisons of the $E_{\rm k,iso}$ and $f_{\rm b}$, The Type II GRBs can be fitted with $E_{\rm K,iso}\propto \sim f_{\rm b}^{-0.8}$ (black dashed line). The Type I GRBs and early jet break are located around a gray dashed line.}
\label{Frailrelation}
\end{figure*}

\end{document}